# Light's symmetry, asymmetry, and their role in nonlinear optics and ultrafast phenomena

OFER NEUFELD,[1,†] MATAN EVEN TZUR,[2] OFER KFIR,[3] AVNER FLEISCHER[4] AND OREN COHEN[2,‡]

*[1]Technion Israel Institute of Technology, Faculty of Chemistry, Haifa 3200003, Israel.*

*[2]Technion Israel Institute of Technology, Physics Department and Solid State Institute, Haifa 3200003, Israel.*

*[3]School of Electrical Engineering, Tel Aviv University, Tel Aviv 6997801, Israel*

*[4]Raymond and Beverly Sackler Faculty of Exact Sciences, School of Chemistry and Center for Light Matter Interaction, Tel Aviv University, Tel Aviv 6997801, Israel*

*[†]ofern@technion.ac.il*

*[‡]oren@technion.ac.il*

**Abstract:** The analysis of symmetries is extremely useful across all fields of science. In physics, symmetries are used to derive conservation laws, and to formulate selection rules for transitions in interacting systems. In the early days of nonlinear optics (NLO), symmetries were used to formulate a set of rules for nonlinear photonic processes according to the medium's symmetries that are reflected in the NLO coefficient tensor. While this approach was believed to be complete and closed, the field of symmetries and selection rules in NLO has recently reignited as multi-color ultrashort laser pulses with tailored polarization and spatiotemporal structures become standard in NLO processes. A more complete theory has been recently emerging, which aims to incorporate all possible dynamical degrees of freedom of light: spin and orbital angular momentum, spatial structure, time-dependent polarizations, temporal envelopes, etc., in addition to the symmetries of the medium. This theoretical development is also accompanied by experimental advances that rely on tailored intense light beams. Such beams can now be generated with ever-increasing complexity, including topologies in real and a variety of synthetic dimensions, carrying poly-chromatic carrier waves, time-dependent varying angular momenta, local-chirality, and more. The nonlinear interaction between light fields with unique symmetries (or asymmetries) and matter is especially appealing, since that holds the key for developing new ultrafast spectroscopies with sub-femtosecond resolution, for exerting exact control over matter, as well as improving our fundamental understanding of how light and matter interact. We here review these recent advances in this expanding field, focusing on the theory, its implications, and seminal experiments. As outlined in the outlook, we aim to establish a comprehensive database of symmetries and selection rules governing nonlinear light-matter interactions within the emerging new formalism and invite the scientific community to contribute to this effort.

## 1.   Introduction

Nonlinear optical processes often exhibit optical selection rules. These rules determine which wave-mixing channels are allowed/forbidden in given settings, and describe the properties of the emitted signals (e.g. polarization properties of generated high harmonics, phase relations between waves, etc.). Selection rules and light and matter symmetries therefore play a pivotal role in various scientific and technological applications of nonlinear optics (NLO). For example, some well-known selection rules in nonlinear optics include forbidden second harmonic generation (SHG) [1] in centrosymmetric media illuminated by monochromatic light [2], or circular polarization constraints for harmonics generated in rotationally-symmetric media driven by circular light fields [3] [4]. As a result of these symmetry-induced constraints, NLO exhibits a particularly high sensitivity to symmetry breaking phenomena, a property which has for instance been utilized for exploring structural and electronic properties of surfaces and interfaces [5] and nonlinear spectroscopy of dynamical processes that break symmetries, e.g. molecular vibrations [6], coherent phonons [7], chirality [8] [9] [6], etc. (see general illustration in Fig. 1). Essentially, by utilizing the symmetry-imposed selection rule as a blueprint for the expected NLO response, symmetry breaking spectroscopy can be employed to monitor deviations for the selection rule and track intrinsic properties of the medium [10] [11] [12] [13] [14]. This approach is especially appealing in emerging quantum [15] [16] [17] [18] [19] [20] [21] [22], chiral [6], and ultrafast evolving systems [23] [24] [25]. A complementary and well-established approach is instead to induce symmetry breaking directly within the driving field itself (i.e. engineer an asymmetric laser field), which can probe analogous broken symmetries in the medium. This is often employed e.g. for probing chirality or magnetism with circular-dichroic responses (which break time-reversal symmetry), or for investigating dynamical phenomena such as phase transitions and chemical reactions on ultrafast timescales.

Two traditional methods have been historically used to describe and derive selection rules in NLO ever since the very first observation of SHG [1]. In the standard approach, which appears in NLO text books [2] [26] [27] [28], tables for nonlinear optical coefficient tensors are classified according to the symmetry properties of the nonlinear medium (which can be readily derived from point groups and space groups [29] [30] [31], Fig. 1(b) left panel). Any constraints amongst the various tensorial coefficients (e.g. some terms being degenerate, or vanishing), directly affect the emerging nonlinear response, which can be written in the typical perturbative expansion for the nonlinear induced polarization [2]. This approach is very effective when dealing with perturbative processes driven by monochromatic fields, but becomes very cumbersome and even ineffective either if the response is highly nonlinear and inherently non-perturbative (e.g. high harmonic generation (HHG) [24] [32]), or if the processes are driven by complex forms of tailored light possibly containing many carrier waves [6] [33].

The second historical approach for formulating optical selection rules relies on photonic conservation laws (Fig. 1(b) bottom panel). In *parametric* processes, the medium returns to its initial state at the end of the light-matter interaction, hence the total radiation (i.e. photonic) energy, momentum, and parity, are conserved [2] [34] [28]. In harmonic generation and wave mixing phenomena for instance, energy conservation coerces the sum of energies of any generated photons to equate exactly to the total energy of annihilated photons. Thus, if the pump fields consist of only photons at carrier frequency $\omega$, then emission of non-integer harmonics is forbidden. In another example, within electric dipole interaction, only odd harmonics are emitted from HHG driven by monochromatic fields in inversion symmetric media, which can be derived from photonic parity conservation.

More recently, the field has been reignited, and several works established more general symmetry theories that take into account not only the nonlinear medium's symmetries, but also that of the driving laser field [35] [36] [37] [38] [39]. The novelty is that this approach is generally non-perturbative (applying to situations like HHG), and can also be readily applied for cases with tailored and structured laser beams that have complex time-dependent polarization and other symmetry relations. These theories have been used also to derive new selection rules that were not previously known in NLO, as well as new photonic conservation laws [40]. They can even be extended to systems that have broken symmetry [41], or symmetry in synthetic dimensions of the Hamiltonian [42]. Amongst the various applications of this new paradigm, it has given rise to a wave of new ultrafast and nonlinear optical spectroscopies, mostly relying on symmetry breaking and engineering symmetries and asymmetries in the tailored driving beam. It also suggests novel tools to tailor material responses such as photocurrents [43] [44] [45] [46] [47], magnetism [48] [49] [50], and dressed phases [51] [52] [53] [54] [55] [56] [57] [58], and can be extended into neighboring fields of physics and science.

In this review, our goal is to introduce the new symmetry theories from the point of view of electromagnetic (EM) theory, initially (section 2). This is accomplished by a comprehensive analysis of the history of the field, logic of the concepts, and the directions the field is expected to take, while employing multiple practical examples from experiments performed in recent years. Following (section 3), we introduce the notion of NLO selection rules as a result of symmetries, conservation laws, generalized symmetries, and related concepts. Connections to neighboring fields are drawn out, and several test examples and seminal works are highlighted. Throughout the review we highlight not only the role of light's symmetric states, but also symmetry-broken states, including the concept of chiral light and chiral light-matter interactions. Lastly, we discuss applications of these symmetry theorems, especially in the fields of ultrafast spectroscopy, light-matter interaction control, and novel light sources (section 4). Section 5 summarizes our work and proposes an outlook of what's to come, as well as calls the community to participate in establishing a comprehensive database for dynamical symmetries and selection rules in nonlinear light matter interaction.

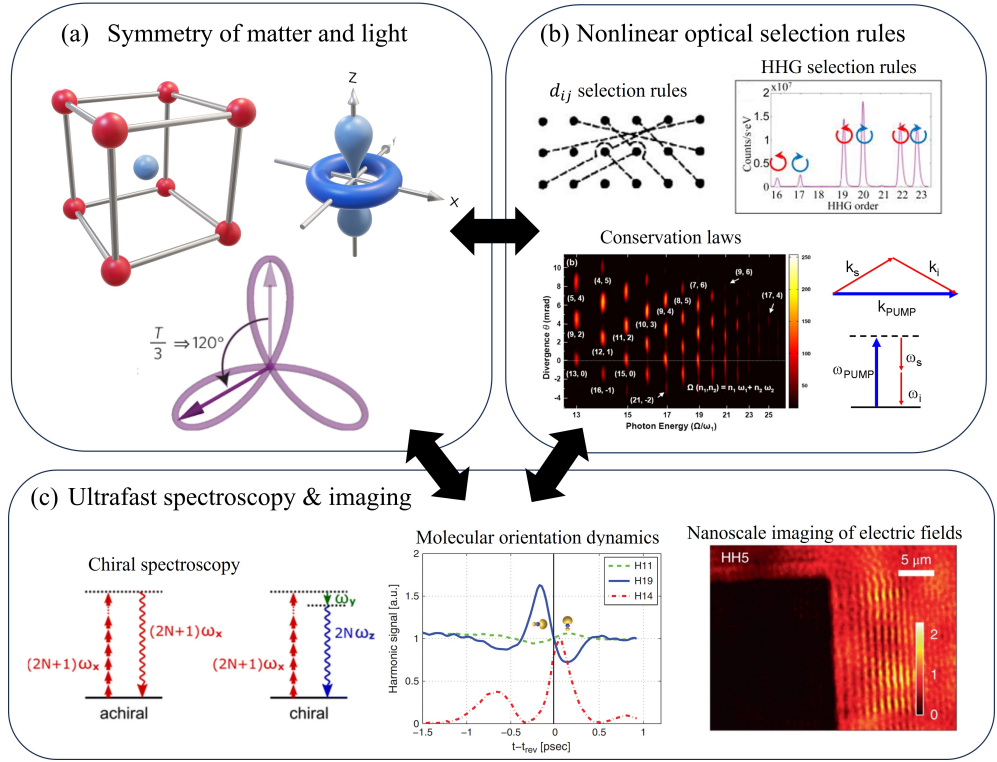

**Fig. 1**. Introductory figure illustrating the interplay between symmetry, selection rules, spectroscopy, and imaging, in nonlinear optics. (a) In nonlinear optics, symmetry may arise from the interacting medium (e.g., molecular or lattice symmetry) or from light's degrees of freedom. For example, the optical field depicted in purple exhibits threefold symmetry. (b) Symmetry dictates selection rules and conservation laws that govern nonlinear optical processes. This includes selection rules for the nonlinear susceptibility tensor $d_{ij}$, forbidden harmonics in HHG (repurposed from ref. [59]), and the conservation of photonic energy and momentum in high-harmonic generation (repurposed from ref. [60]). (c) Spectroscopy and imaging applications in nonlinear optics leverage symmetry, selection rules, and conservation laws. Based on these principles, tailored light fields enable ultrasensitive chiral spectroscopy of molecules (repurposed from ref. [23]), symmetry breaking allows time-resolved studies of molecular orientation dynamics [61], and nanoscale vectorial electric field imaging is achieved by exploiting polarization and frequency selection rules (repurposed from ref. [62]).

## 2. Symmetries of light

### 2.1 Dynamical symmetries of light

#### 2.1.1 Symmetries in electromagnetic theory

This section will introduce the basic concepts for describing symmetries of EM waves. In particular, we will review the theory of dynamical symmetries (DS), which describe the complex spatio-temporal relationships between different degrees of freedom (DOF) of multi-dimensional functions. It is worthwhile to first consider some basic concepts, as well as to exactly define the scope of this endeavor. In this respect, our objective is solely to describe the symmetries of light, and not to cover the more general mathematical formulation of symmetry groups for any type of function or equation (even though a great deal of what will be presented is transferable). Light, in the context discussed in this review, is any function that solves Maxwell's equations in vacuum [63] [2]:

$$\nabla \cdot \mathbf{E} = 0$$
$$\nabla \cdot \mathbf{B} = 0 \tag{1}$$
$$\nabla \times \mathbf{E} = - \partial_t \mathbf{B}$$

$$\nabla \times \mathbf{B} = \frac{\partial_t \mathbf{E}}{c^2}$$

where $\mathbf{E}$ and $\mathbf{B}$ are the electric and magnetic fields, respectively, and $c$ is the speed of light. We do not consider at this stage the presence of matter, which will be introduced later on when addressing how such fields interact with matter and propagate through it. From Eq. (1) one may also write the EM wave equations in vacuum:

$$\partial_t^2 \mathbf{E} - c^2 \nabla^2 \mathbf{E} = 0$$
$$\partial_t^2 \mathbf{B} - c^2 \nabla^2 \mathbf{B} = 0 \tag{2}$$

Since the objective is to discuss the symmetries of solutions to Maxwell's equations, it is worth to first point out that the solutions (including their symmetry properties) are affected by the underlying symmetries of the equations themselves. Such relations have been studied for over a century, and have a variety of interesting consequences [64]. For instance, Maxwell's equations are notoriously Lorenz invariant, leading to a multitude of physical consequences, including energy, momentum, and angular momentum conservation, as well as suggesting a constant speed of light which triggered the development of special relativity. A slightly more elaborate example is EM duality, where Eq. (1) is invariant under the transformation $\mathbf{E} \rightarrow \cos(\alpha)\,\mathbf{E} - \sin(\alpha)\,\mathbf{B}$, $\mathbf{B} \rightarrow \cos(\alpha)\,\mathbf{B} + \sin(\alpha)\,\mathbf{E}$ (subject to a symmetric representation of Eq. (1)) [65]. Being a continuous symmetry with respect to $\alpha$, the transformation leads to a conserved charge *via* Noether's theorem [66], which in this case is the EM helicity [67] [68] [69]. In fact, this duality transformation is just one private example from a larger family of duality symmetries [70] [71]. These examples hint towards the already very rich literature and knowledge base in the field.

Naturally, all EM waves obey these internal symmetry relations of Maxwell's equations. In other words, $\mathbf{E}$ and $\mathbf{B}$ functions that do not solve Eq. (1) cannot exist in vacuum. Here we review a class of light fields that exhibit additional symmetries that are decoupled from the equations of motion, i.e., symmetries on top of the symmetries that result from Maxwell equations. Let us consider a single plane wave as an example solution of the form:

$$\mathbf{E}(\mathbf{r}, t) = \mathbf{E_0}\cos(\omega t - \mathbf{k} \cdot \mathbf{r})$$
$$\mathbf{B}(\mathbf{r}, t) = \mathbf{B_0}\cos(\omega t - \mathbf{k} \cdot \mathbf{r}) \tag{3}$$

where $\mathbf{k}$ is the wave vector, $\omega$ is the carrier frequency (connected to the temporal period $T = 2\pi/\omega$), $\mathbf{r}$ represents the spatial DOF, and $\mathbf{E_0}$ and $\mathbf{B_0}$ are respectively the electric and magnetic field amplitudes and polarization vectors. This wave exhibits several internal obvious symmetry relations, such as a discrete spatial translational symmetry $\mathbf{r} \rightarrow \mathbf{r} + 2\pi\hat{\mathbf{k}}/|\mathbf{k}|$, a discrete temporal translation $t \rightarrow t + T$, and combinations thereof. It also exhibits continuous rotational symmetries about the $\hat{\mathbf{k}}$ axis in the spatial domain, as well as time-reversal symmetry, and several other symmetries.

On the other hand, some other symmetries may not be as trivial to uncover. For instance, further inspection shows that Eq. (3) exhibits the symmetry relation $\mathbf{E}(\mathbf{r}, t) = -\mathbf{E}(\mathbf{r}, t + T/2)$ (with a similar equation for the magnetic field). It also upholds a symmetry that couples time reversal with spatial inversion (parity-time symmetry, so-called PT symmetry [72]), i.e., invariance under $t \rightarrow -t$, $\mathbf{r} \rightarrow -\mathbf{r}$. We can already note some internal logic in this structure, where the presence of one symmetry is sufficient to determine some of the other symmetries. For instance, if $\mathbf{E}(\mathbf{r}, t) = -\mathbf{E}(\mathbf{r}, t + T/2)$, then that means $\mathbf{E}(\mathbf{r}, t + T) = -\mathbf{E}(\mathbf{r}, t + 3T/2) = \mathbf{E}(\mathbf{r}, t + T) = -\mathbf{E}(\mathbf{r}, t + T/2) = \mathbf{E}(\mathbf{r}, t)$. That is, because of the presence of one type of symmetry, the field also exhibits another - time-periodicity. This hints towards the inherent mathematical group structure of these relations – some symmetries can act as generators for a symmetry group, such that squaring them or multiplying them by other generators produces the remaining symmetry relations characterizing the specific EM field.

The abundance of symmetries within a plane wave is not very surprising, considering it has a highly simple functional form. However, due to the linearity of the differential equations, superpositions of different plane waves are also a solution, which allows a much richer space for observing different symmetries. This is apparent even if one considers only two plane waves, but with different carrier frequencies:

$$\mathbf{E}(\mathbf{r}, t) = \mathbf{E_1} \cos(\omega_1 t - \mathbf{k_1} \cdot \mathbf{r}) + \mathbf{E_2} \cos(\omega_2 t - \mathbf{k_2} \cdot \mathbf{r})$$
$$\mathbf{B}(\mathbf{r}, t) = \mathbf{B_1} \cos(\omega_1 t - \mathbf{k_1} \cdot \mathbf{r}) + \mathbf{B_2} \cos(\omega_2 t - \mathbf{k_2} \cdot \mathbf{r})$$

(4)

Generally, for $\omega_1 \neq \omega_2$ most of the symmetry relations presented above no longer hold, i.e., the field has now a somewhat reduced symmetry group. Practically though, this presents the opportunity for controlling the breaking and formation of symmetries in light simply by tuning the EM field's functional form. It is important to note that the addition of the second frequency component also opens the path to additional symmetries that were not possible before, e.g., by allowing for a three-dimensional polarization space, and a three-dimensional spatial dependence. Expectedly, by involving also the polarization DOF and possible spatial and temporal envelopes into this discussion, one generates a very large phase space for symmetries.

Altogether, we can enumerate the different DOF that one expects to play a role: (i) the temporal dependence of the field (one dynamical dimension, $t$), (ii) the macroscopic spatial behavior of the field (up to three dimensions, $\mathbf{r}$), (iii) the field's microscopic polarization dependence embedded in its directionality (up to three dimensions). This sets the ground for a symmetry group theory of physical dimensions up to 3+3+1, which contains a huge potential for exploring the properties of light. The basic symmetry relationships illustrated above form the basis for so-called DS, which describe dynamical or 'Floquet groups', as was derived in [35] and will be described below. Before moving on to address these DS, we acknowledge that our main interest is in freely propagating solutions to Maxwell's equations, i.e., coherently propagating light beams. The main reason is that these beams practically pose the main source for intense coherent light sources for nonlinear light-matter interactions. Notably though, for freely propagating waves, the electric and magnetic fields typically have an identical functional structure, due to which we will only analyze the symmetries of the electric field component.

### 2.1.2 Dynamical symmetries in the dipole approximation

We shall start by describing the group theory for DS of light, closely following refs. [73] [74] [75] [76] [41] [77] [35] [36] [42] [78] [79] [80] [81] [82]. To systematically introduce the theory, we begin by only considering the microscopic DOF of the EM field. In other words, we assume for now the dipole approximation and neglect the spatial dependence in $\mathbf{E}(\mathbf{r}, t)$, $\mathbf{E}(\mathbf{r}, t) \approx \mathbf{E}(0, t)$. What is left are possible symmetry relations between the temporal dependence of the EM field and its polarization components, which can be spanned out in a three-dimensional Cartesian space:

$$\mathbf{E}(t) = E_x(t)\hat{\mathbf{x}} + E_y(t)\hat{\mathbf{y}} + E_z(t)\hat{\mathbf{z}}$$

(5)

DS then arise as intricate connections amongst the temporal evolution of the different polarization components. Let's take an illustrative example, and imagine a field that is invariant under a discrete rotational symmetry of $120^0$, but which is coupled to a temporal translation: $\mathbf{E}(t) = \hat{r}_3 \mathbf{E}(t + T/3)$. Here $\hat{r}_3$ denotes a rotation operator by $2\pi/3$ in the polarization space, acting on the electric field's polarization components and coupling them:

$$\mathbf{E}\left(t - \frac{T}{3}\right) = \hat{r}_3 \mathbf{E}(t) = \begin{pmatrix} \cos\left(\dfrac{2\pi}{3}\right) & -\sin\left(\dfrac{2\pi}{3}\right) & 0 \\ \sin\left(\dfrac{2\pi}{3}\right) & \cos\left(\dfrac{2\pi}{3}\right) & 0 \\ 0 & 0 & 1 \end{pmatrix} \mathbf{E}(t)$$

(6)

$$= \left[ -\frac{1}{2} E_x(t) - \frac{\sqrt{3}}{2} E_y(t) \right] \hat{\mathbf{x}} + \left[ \frac{\sqrt{3}}{2} E_x(t) - \frac{1}{2} E_y(t) \right] \hat{\mathbf{y}} + E_z(t) \hat{\mathbf{z}}$$

From Eq. (6), this EM field exhibits a symmetry relation where the polarization components in the $xy$ plane are interconnected *via* the time axis (see illustration in Fig. 2(a)). Another way to think of it is that this DS characterizes a particular structure in the field's evolving time-dependent polarization, or its Lissajous plot. As it turns out, such a field can be generated by superimposing just two counter-rotating circularly-polarized (CP) plane waves with different carrier frequencies, as was explored in refs. [83] [84] [85] [86] [59] [87] [88] and is quite useful for a variety of applications that will be later addressed. In fact, even higher order rotational symmetries in EM fields have been recently generated by combining well-selected bi-chromatic frequencies which are polarization controlled [89] [90] [91] [92], which have been used for shaping photoelectron vortices [93] (see Fig. 2(f)).

To move forward, we need to precisely define the space in which these symmetries act. In the polarization space, since we have three-dimensional vectors, it is natural to consider point group operations [94] [95] [29], including rotations $\hat{r}_n$, improper rotations $\hat{s}_n$, reflections $\hat{\sigma}$, and inversion $\hat{\imath}$. Since the time domain is a one-dimensional object, we need only consider temporal translations that we denote as $\hat{\tau}_n$, corresponding to $t \to t + T/n$, and time reversal that we denote as $\hat{T}$, corresponding to $t \to -t$ [96] [97] [98]. A DS of the EM field is then any combination of such symmetry operators that the field is invariant under. For instance, $\hat{X}$ is a symmetry of $\mathbf{E}(t)$ if $\mathbf{E}(t) = \hat{X}\mathbf{E}(t)$, at which point we can drop the equation notations for DS relations and only consider the operators themselves. $\hat{X}$ can be comprised of any multiplication of either temporal or point group operations [29] [95], where due to a fundamental separation of the time and polarization dimensions of $\mathbf{E}(t)$, temporal and polarization-based operators always commute. In order for various operators $\hat{X}$ to form mathematical groups, four underlying properties must be satisfied: (i) sequential operations must be associative, (ii) there must exist an identity operator, (iii) each operator must have an inverse, and (iv) the group must be closed such that multiplication of any two elements in the group also leads to an element in the group.

We will now show directly that all of these are upheld for DS that describe symmetries of EM fields in the dipole approximation. First, associativity is guaranteed by construction – the point group operators act as matrices on polarization vectors, and are hence associative, and the temporal operators act similarly but in a more compact one-dimensional space. Second, one can define an identity just as for point groups, with the addition of a temporal identity operator. Here we note a potential complication that arises for time-periodic EM fields that are our main interest (i.e., neglecting any temporal envelopes) – just as a rotation by $360^0$ $\hat{r}_1$, acts as an identity operator for point groups, in the temporal domain if the field is time-periodic with a fundamental period $T$, the temporal identity can be defined as $\hat{\tau}_1$. While this is not strictly necessary, most interesting cases that involve coupling of the temporal and polarization domains involve temporal periodicity. Because of this, from this point on we only discuss time-periodic fields and define the identity as $\hat{1} \equiv \hat{\tau}_1\hat{r}_1$, which assumes that the EM field exhibits an underlying periodicity of $T$. Third, each point group operator by definition has an inverse being part of a group, and for the time domain one can easily see that $\hat{T}$ is its own inverse ($\hat{T}^2 = \hat{1}$), while $\hat{\tau}_{-n} = (\hat{\tau}_n)^{n-1}$ is the inverse of $\hat{\tau}_n$. Thus, due to the commutativity of the temporal and polarization operations, any DS also has an inverse operator. Four, and last, closure is maintained with this definition because it is separately maintained for point group operators, and for temporal operators, and those commute. Here some caution is needed – even though closure is upheld, this does not mean that all possible DS groups describe a physically realizable EM field – it is possible also that some groups cannot be realized by non-trivial solutions of Maxwell's equations. To illustrate this, we can consider the exemplary DS $\hat{X} = \hat{r}_5\hat{\tau}_2$. This DS is clearly contained in our space of exploration and adheres to the four principles that define mathematical groups. If $\hat{X}$ is an element in some yet unknown group $G$, then due to closure we must have that also $\hat{X}^2 = (\hat{r}_5 \hat{t}_2)^2 = \hat{t}_5^2$ is an element (note throughout we employ the notation $\hat{r}_{n,m} \equiv (\hat{r}_n)^m$). Consequently, we must also have that $\left(\hat{X}^2\right)^2 = (\hat{r}_5^2)^2 = \hat{r}_{-5}$ is an element. Thus, the EM field that upholds $\hat{X}$ must also be 5-fold symmetric in the polarization domain without any additional temporal operator, simply because it is described by the group $G$. Physically, and assuming without loss of generality that the rotational symmetry is along the $z$-axis, this condition requires that $\hat{r}_5 \mathbf{E}(t) = \mathbf{E}(t)$, which can only be solved for $E_x(t) = E_y(t) = 0$. In other words, only a trivial solution of a zero electric field vector can be described by $G$. The situation is worsened by noting that $G$ also contains the operator $\hat{Y} = \hat{r}_{-5} \hat{X} = \hat{t}_2$, i.e., just translating by $T/2$ in time keeps the field invariant, which contradicts the definition of $T$ as the minimal period. The physical conclusion from this discussion is that whenever DS have incommensurate operation orders for their temporal and point group parts, one directly also has DS with temporal-only and polarization-only terms within the group, because one can always raise the operator to the order of its individual components to obtain other elements in the group. For instance, in the DS $\hat{r}_4 \hat{t}_4$ both $\hat{r}_4$ and $\hat{t}_4$ are 4th order operators that give the identity when raised to the power of 4, which does not lead to such effects. $\hat{r}_2 \hat{t}_4$ on the other hand, inherently means that $\hat{t}_2$ and $\hat{r}_2$ are also part of the symmetry group. For intents of physical purposes, we should thus exclude DS that leads to a temporal-only DS of the type $\hat{t}_n$ for $n \neq 1$, because that contradicts our fundamental assumptions. In practical terms, this forbids the appearance of DS that have incommensurate operation orders where $\hat{t}$ is the temporal part of the DS.

With these finer technical points out of the way, we are left with systematically enumerating all possible DS operations (from all possible blocks enumerated in Table 1), which simply requires coupling all different options of temporals and polarization-based operators, as has been presented in ref. [35]. Some exemplary Lissajous for EM fields that exhibit interesting DS can be seen in Fig. 2. From the set of these fundamental DS, one can construct the dynamical groups by assigning generators to each group. This includes both abelian and non-abelian groups, depending on if the generators commute or not. All in all, every single EM field that solves Maxwell's equations (within the dipole approximation) can be cataloged into one such dynamical group, just as crystals are cataloged into space groups, and molecules into point groups.

**Table 1.** Summary of all relevant symmetry operators (individual blocks) for multi-scale dynamical symmetries, cataloged based on the symbol we employ in this review, and the operator type (i.e. the space it acts in).

| Symmetry operator | Active space | Operation order | Practical action for the EM field |
|---|---|---|---|
| $\hat{r}_n$ | EM polarization | $n$ | Rotation of EM polarization by $2\pi/n$ |
| $\hat{\sigma}$ | EM polarization | 2 | Reflection of EM polarization along a specific plane. '$h$' indicates a mirror plane transverse to a connected rotational axis, and '$v$' a mirror plane containing the rotational axis |
| $\hat{s}_{2n} = \hat{\sigma}_h \hat{r}_{2n}$ | EM polarization | $2n$ | Improper rotation of EM polarization by $2\pi/(2n)$ |
| $\hat{s}_{2n+1} = \hat{\sigma}_h \hat{r}_{2n+1}$ | EM polarization | $2(2n+1)$ | Improper rotation of EM polarization by $2\pi/(2n+1)$ |
| $\hat{i}$ | EM polarization | 2 | Inversion of EM polarization |
| $\hat{b}_\varepsilon$ | EM polarization | — | Scaling of EM polarization by factor $\varepsilon$ along a particular axis in space. |
| $\hat{\tau}_n$ | Time | $n$ | Translation of time along EM field time axis by $T/n$, with T the fundamental period. |
| $\hat{T}$ | Time | 2 | Time-reversal along EM field time-axis. |
| $\hat{R}_n$ | Spatial domain | $n$ | Rotation of EM field in space (i.e. of its spatial dependence alone) by $2\pi/n$ |
| $\hat{\Sigma}$ | Spatial domain | 2 | Reflection of EM field in space (i.e. of its spatial dependence alone) along a specific plane. |

| $\hat{S}_{2n} = \hat{\Sigma}_h \hat{R}_{2n}$ | Spatial domain | $2n$ | Improper rotation of EM field in space (i.e. of its spatial dependence alone) by $2\pi/(2n)$ |
|---|---|---|---|
| $\hat{S}_{2n+1} = \hat{\Sigma}_h \hat{R}_{2n+1}$ | Spatial domain | $2(2n+1)$ | Improper rotation of EM field in space (i.e. of its spatial dependence alone) by $2\pi/(2n+1)$ |
| $\hat{I}$ | Spatial domain | $2$ | Inversion of EM field in space (i.e. of its spatial dependence alone) |
| $\hat{L}_n$ | Spatial domain | $n$ | Spatial translation of the EM field along a particular axis in space by $\lambda/n$, where $\lambda$ is the fundamental wavelength |

One additional noteworthy point is that the symmetries of EM fields can also include elements that are not found in molecular groups. For instance, superpositions of elliptically-polarized light (where the ellipticity is neither zero nor unity) can exhibit elliptical DS [35], which are analogous to the rotational $\hat{r}_n \hat{\tau}_n$ case, but where the standard rotational operator $\hat{r}_n$ is replaced by an elliptical rotation along an ellipse of ellipticity $\varepsilon$. The mathematical form of this operator can be expressed with additional scaling operators along the elliptical main axis (e.g. assumed to be $x$-axis here):

$$\hat{e}_n = \hat{b}_\varepsilon \, \hat{r}_n \hat{b}_{1/\varepsilon} \tag{7}$$

where $\hat{b}_\varepsilon$ is a scaling operation along the elliptical major axis in the plane transverse to the rotational axis, acting in 3D polarization space as:

$$\hat{b}_\varepsilon = \begin{pmatrix} 1 & & \\ & \varepsilon & \\ & & 1 \end{pmatrix} \tag{8}$$

The DS $\hat{e}_n \hat{\tau}_n$ can then be thought as a generalization of $\hat{r}_n \hat{\tau}_n$, and one obtains $\hat{e}_n \hat{\tau}_n = \hat{r}_n \hat{\tau}_n$ for $\varepsilon = 1$.

At this stage, we discuss some consequences and implications of this derivation. First, we note that while every EM field belongs to some dynamical group, it is not always straightforward to identify the group. For molecules, this is usually done with a flow-chart type approach, while for solids numerical codes are often required. Here, and in the case of time-periodic EM fields within the dipole approximation, we will show later that a similar flow-chart type approach is applicable that relies on the selection-rules that such EM fields entail. Second, while this exercise might seem purely academic, we acknowledge that the particular dynamical group that characterizes the EM field carries physical significance in how that field interacts with matter, just as the molecular or crystal group carries similar meaning for other physical observables. It also establishes which types of symmetries of matter can be broken by light, which is crucial for spectroscopic purposes. These applications and more will be discussed in following sections.

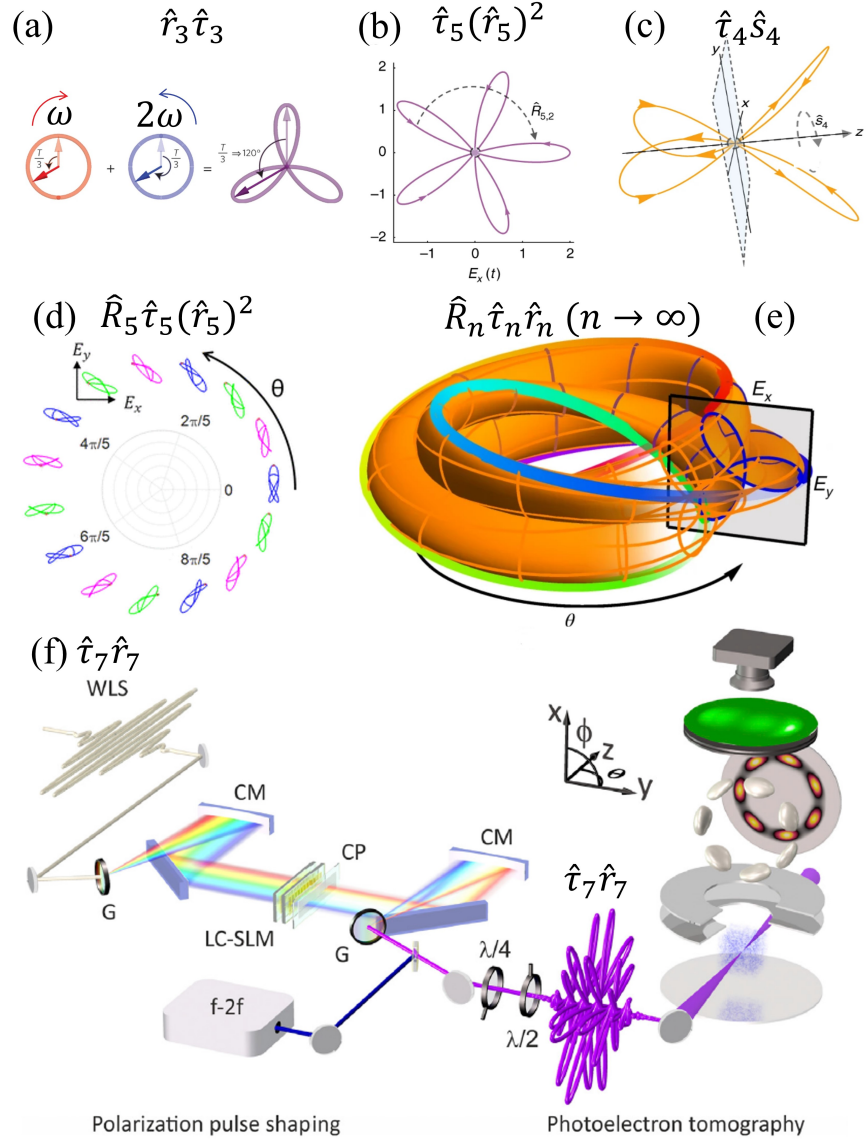

**Fig. 2**. Optical fields with tailored spatial and polarization profiles that exhibit dynamical symmetries. (a) The combination of counter-rotating circularly polarized fields with frequencies ω and 2ω results in an optical field with a three-fold Lissajous curve, i.e. third-order rotational DS. This bi-circular field is useful for generating circularly polarized high-order harmonics (repurposed from ref. [88]). (b) By combining co-/counter-rotating fields of different frequencies, optical fields with higher-order rotational symmetries can be realized. For example, a field with fifth-order rotational dynamical symmetry is generated by combining counter-rotating fields with frequencies 2ω and 3ω. (c) The combination of non-collinear beams propagating along different axes produces a 3D polarization, which can exhibit dynamical symmetries unique to 3D such as dynamical improper rotations (figures in (b,c) repurposed from ref. [35]). (d) Multiscale dynamical symmetries arise from the superposition of optical beams carrying both orbital and spin angular momentum while differing in frequency. The depicted example shows Lissajous curves plotted locally across the transverse wavefront, revealing a repeating polarization pattern every 2π/5 rotation with a polarization rotation of -4π/5 and a time shift of -T/5 where T is the period of the combined field. However, at a fixed angle, the field lacks symmetry (repurposed from ref. [36]). (e) The superposition of a right-circularly polarized beam with no OAM with its left-circularly polarized second harmonic with OAM creates a trefoil-shaped polarization pattern. When mapped onto a twisted cylindrical topology, it reveals a 3rd-order torus knot structure in its polarization and spatial dependence (repurposed from ref. [37]). (f) Experimental setup for multichromatic polarization shaping to control the dynamical symmetry of an optical field. Velocity map imaging photoelectron tomography is used to reconstruct electron dynamics imparted by the optical field, which in this case exhibits 7-fold rotational DS (repurposed from ref. [92]).

### 2.1.3 Multi-Scale Dynamical symmetries (3+3+1)D

Next, we allow the EM field to exhibit spatial dependence as well, i.e., lifting the dipole-approximation. In other words, we allow the EM field to have both temporal, polarization, and spatial DOF: $\mathbf{E} = \mathbf{E}(\mathbf{r}, t)$.

This gives rise to a significantly larger space of DS that couple space, time, and polarization. Within this discussion, we should separate the main physical effects of two limiting cases: (i) purely spatial symmetries of the EM field (involving the spatial dependence $\mathbf{r}$, e.g. an inversion symmetry such as $\mathbf{E}(\mathbf{r}, t) = \mathbf{E}(-\mathbf{r}, t)$), (ii) DS that couple spatial symmetries with temporal ones, and (iii) what we denote as multi-scale symmetries, which non-trivially couple the EM field's polarization and its spatial behavior. Purely spatial symmetries are quite well understood as those captured by point groups and/or space groups. Since they do not involve a temporal or polarization part, they can be uncovered by setting $t = 0$ in the EM wave. Such symmetries have already been covered extensively, e.g. in refs. [95] [94]. For DS that couple the spatial and temporal dimensions, but leave out the polarization dimensions, we emphasize an important analogy: such DS are essentially copies of the DS group theory derived above within the dipole approximation, but where all of the polarization-based operators are replaced by equivalent operators that act in real space (see all options laid out in Table 1). In this respect, the DS themselves need not be rederived, but we should simply replace lower-case notation for point group operators in polarization space, with capital letter notation that indicates real-space operations. Thus, in real-space we employ $\hat{R}_n$ for rotations, $\hat{S}_n$ for improper-rotations, $\hat{\Sigma}$ for reflections, and $\hat{I}$ for inversion. The only practical difference is that we must also include spatial translation operators that are denoted by $\hat{L}_{i,n}$ which takes $r_i \to r_i + L_i/n$, with $L_i$ being the fundamental period along the $i$'th spatial axis (e.g., $L_i = \lambda$ along a plane wave propagation direction, where $\lambda = 2\pi/k$ is the wavelength that is connected to carrier frequency in vacuum by $k = \omega/c$). This addition upgrades the point-group operators to space group operators [95] [99] [100] [101] [102], and allows a larger variety of DS (i.e. where spatial translation can be coupled to various other operators making glide symmetries). Lastly, for the multi-scale operations, we have the option of EM fields with intricate coupled relationships between the macroscopic scale (real-space, $\mathbf{r}$), the polarization space, and the temporal domain.

We will not formulate this full multi-scale dynamical symmetry theory here. Instead, we refer interested readers to ref. [36], where it is rigorously derived. Let us however contemplate the configuration space of such a symmetry theory, and also consider some examples which illustrates the strength and generality of this concept. Given an EM field, $\mathbf{E}(\mathbf{r}, t)$, the most general multi-scale DS the field could exhibit is of the form $\mathbf{E}(\mathbf{r}, t) = \hat{X}\hat{x}\hat{t}_n\hat{\mathbf{T}}^m\mathbf{E}(\mathbf{r}, t)$. In this notation $\hat{X}$ is any spatial operator that acts on the $\mathbf{r}$-dependence of $\mathbf{E}(\mathbf{r}, t)$ such as rotations translations and mirror operators, $\hat{x}$ is a microscopic operator that acts on the field's polarization components just as introduced in the section above, $\hat{t}_n$ is a time-translation operator (for $n = 1$ there is no time translation), and $\hat{\mathbf{T}}^m$ denotes time reversal which could either apply if $m$ is odd, or not if $m$ is even. In terms of dimensionality, this notation permits up to (3+3+1)D for complex spatio-polarization-temporal symmetries. For instance, we might envision an EM field that is symmetric under a complex multi-scale operation of 2-fold rotations in space, coupled to an inversion operation of its polarization components, and translation by half an optical cycle along with time-reversal: $\mathbf{E}(\mathbf{r}, t) = \hat{R}_2\hat{\imath}\hat{t}_2\hat{\mathbf{T}}\mathbf{E}(\mathbf{r}, t) = -\mathbf{E}(\hat{R}_2 \cdot \mathbf{r}, -t + T/2)$. How to obtain such a field is currently irrelevant, but the most important thing is that in principle it can be generated, and its engineered symmetry might be employed for various applications from enhanced spectroscopy and imaging, to tailoring properties of matter on nanometric length scales (which will be discussed in section 4.2). Note that just as for the microscopic case, the operation order of the various components must be commensurate or no such EM fields could exist. Furthermore, the operators acting in different sub-spaces of the EM field inherently commute (meaning it does not matter if one first translates in time, and then rotates in space, or vise versa).

Let us consider some concrete examples. For instance, beams that intrinsically carry orbital angular momentum (OAM) [103] are symmetric under concerted spatial rotations coupled to time translation, and/or spatial translations (which will be discussed in more detail below). This is because they have a rotating phase front that evolves in space and time. Similarly, vector beams can exhibit complex relationships between their polarization and temporal DOF as illustrated by beams carrying torus-knot angular momentum as discussed in refs. [37] [39], which might be symmetric under concerted rotations in polarization space, rotations in real space, and time translations (see Fig. 2(e)). It has also been shown that such multi-scale dynamical symmetries yield light-matter interactions selection rules that will be discussed in upcoming sections [36], which might become apparent in nonlinear optical responses such as HHG [36], or even in nonlinear diffraction patterns [104]. Besides these cases, in ref. [36] it was shown that a coherent superposition of a Bessel beam and standard Gaussian $\omega$-$2\omega$ counter-rotating bi-circular beam (similar to that described in Eq. (6), see Fig. 2(d)) also generates unique multi-scale symmetries. Due to the Bessel beam having a different group velocity, the beam's polarization behavior evolves along its propagation axis in a consorted manner that is invariant under a combined $\hat{\tau}_3\hat{r}_3\hat{L}_{z,3}$ operator. That is, the evolving laser field has a typical behavior over a length scale $\lambda/3$ due to the phase slip between the Bessel and Gaussian $\omega$-$2\omega$ beams, and this behavior is mimicked in the other parts of the propagation (from $\lambda/3$ to $2\lambda/3$, and then again from $2\lambda/3$ to $\lambda$), but in a manner where the field polarization is rotated and it's temporal axis is shifted. This field was experimentally shown to lead to unique symmetry-induced selection rules in HHG, forbidding the generation of certain harmonic orders (see Fig. 6(g)).

## 2.2 Topological light
### 2.2.1 OAM beams

Having introduced DS and dynamical groups for characterizing light in the previous section, we now focus on some examples that are particularly interesting from an experimental standpoint. These are light fields that carry a so-called topological charge, which is connected to the wave's spatially-varying structure. Surprisingly, such spatial structure and its connection to angular momentum of light (including its separation to spin and orbital terms in the paraxial approximation) was discovered only in the 90's [103] [105]. In this section we will introduce the background to some of these concepts, and focus on novel experimental and theoretical advances where evermore complex waveforms are generated by combining the various DOF in EM waves. We will also re-formulate some of the main physical properties of such waves in the language of DS to connect with the group theory discussion above.

We begin with perhaps the simplest example of a topological light beam, which is a Gauss-Laguerre (GL) beam that carries non-zero orbital angular momentum (OAM) of +1 (in units of $\hbar$), whose electric field can be written as:

$$E(\mathbf{r},t) = \text{Re}\{\mathbf{E_0}\psi(\mathbf{r})\exp[i\omega t - ikz]\} \tag{9}$$

where the main fast oscillatory behavior of the EM field was separated out for convenience within the paraxial approximation, leaving just the spatial dependence of its envelope, $\psi(\mathbf{r})$ (note that from this point on we drop the "Re{}" operator for brevity, but for practical purposes it is always applied to the final form of EM fields). The spatial envelope encodes the main structure of interest of the form:

$$\psi(\mathbf{r}) = \frac{2\rho}{\sqrt{\pi}w^2(z)}\exp\left[-\frac{\rho^2}{w^2(z)} - \frac{ik\rho^2 z}{2(z^2 + z_R^2)} + 2i\zeta(z) + i\theta\right] \tag{10}$$

where $\rho = \sqrt{x^2 + y^2}$ denotes a radial coordinate in the $xy$ plane, the beam's propagation axis is taken along the $z$-axis, $\theta$ is a radial coordinate within the $xy$ plane, $w(z) = w_0\sqrt{1 + z/z_R}$ is the beams width along the $z$-axis with $w_0$ the beam's waist, $z_R = \pi w_0^2/\lambda$ is the Rayleigh range, and $\zeta(z)$ is the typical Gouy phase:

$$\zeta(z) = \arctan\left(\frac{z}{z_R}\right) \tag{11}$$

Most importantly, this solution exhibits a phase front that evolves continuously in $\theta$ for any given $z$-coordinate. One is tempted to argue that the solution is cylindrically symmetric, but the $\theta$-dependent phase spoils this symmetry. In other words, $E(\mathbf{r}, t)$ is not invariant under the symmetry operation $\hat{R}_n$, but instead is invariant under a more complex relation where this rotation is compensated by an additional phase factor. In the language of DS, the phase can be absorbed in the oscillatory part of the wave by a temporal translation, or by a spatial translation along the $z$-axis, i.e., we have that $\hat{R}_n\hat{\tau}_n$ and $\hat{R}_n\hat{L}_{z,n}$ are DS of $E(\mathbf{r}, t)$ in Eqs. (9)-(11). Fig. 3(a) illustrates this concept showing that the phase relation exists within each $z$-pane, and results in a screw-like evolution of the phase front as the beam propagates. Note that $\hat{R}_n\hat{L}_{z,n}$ here is only an approximate symmetry of the beam due to its width evolving along the z-axis.

Notably, $\hat{R}_n\hat{\tau}_n$ is a symmetry of $E(\mathbf{r}, t)$ for any value of $n$, such that one may take $n \to \infty$ corresponding to an infinitesimal rotation by $\delta\theta$, accompanied by an infinitesimal temporal (or spatial) translation. The symmetry is continuous, connecting it through Noether's theorem with a conservation law, which in this case represents the conserved OAM of the EM field.

The connection of continuous DSs to conservation laws will be discussed more extensively in later sections of this review. In the meantime, we can contemplate on the physical significance of this EM field: It established a new path for transferring angular momentum between light and matter that is separate from the spin angular momentum (SAM) carried by the polarization state of the EM field (explicitly, we could have that $\mathbf{E_0}$ in Eq. (9) is linearly polarized everywhere such that the beam carries zero SAM). More general forms for the GL solutions can carry arbitrary integer OAM of integer value $m$:

$$\psi_{GL}^{(p,m)}(\mathbf{r}) = \frac{C_{GL}^{(p,m)}}{w(z)}\left(\frac{\sqrt{2}\rho}{w(z)}\right)^{|m|} L_p^{(|m|)}\left(\frac{2\rho^2}{w^2(z)}\right) \exp\begin{bmatrix}-\dfrac{\rho^2}{w^2(z)} - \dfrac{ik\rho^2 z}{2(z^2+z_R^2)} \\ +i(2p+|m|+1)\zeta(z) \\ +im\theta\end{bmatrix} \tag{12}$$

where $p$ is an integer characterizing the beam's radial index, $C_{GL}^{(p,m)} = \sqrt{\frac{2p!}{\pi(p+|m|)!}}$ is a normalization constant, and $L_p^{(m)}(x)$ is a generalized Laguerre polynomial with an angular mode of $m$ and a radial index $p$. The presence of nonzero angular momenta is also associated with a vortex in the light intensity at $\rho = 0$ such that $\psi_{GL}^{(p,m)}(\rho = 0) = 0$ if $m \neq 0$, creating a type of artificial phase singularity around the beam's propagation axis (see illustration in Fig. 3(a,c)). In this respect, beams carrying OAM are often regarded as 'topological' – the vortex cannot be removed from the system unless it is 'annihilated' with angular momentum of the opposite sign [106] [107], providing some analogous (though not equivalent [108]) form of topological protection to that discovered in condensed matter systems [109] [110] [111] [112]. The OAM index, $m$, acts as a topological invariant that counts the number of times the phase winds around the center with a wavelength. Importantly though, there is no true singularity within topological beams of light, as the intensity of the beam always vanishes at the vortex. Mathematically, this is important, because it means no special treatment is required around the beam center. We should also point out that the EM fields of GL beams exhibits many other DS, both in polarization, in space, and multi-scale ones, but we focus on the phase-front continuous symmetry for now which is responsible for the OAM and main topological character of the beam.

The particular GL form is not a unique case for topological light carrying OAM [113]. It has been shown also that other generalizations exist, including Bessel beams [114] [115], Airy beams [116], Hermite-Gauss beams [117], and Ince-Gauss beams [118]. These can offer some advantages over the GL counterparts, e.g. self-healing properties of Bessel and Airy beams [119] [120] [121]. At the same time, the experimental realization of GL beams with OAM is nowadays quite straightforward, and can be implement with relatively simple optical elements even for states carrying a high OAM index [122] [123] [124] [125] [126]. Such implementations exist even in the femtosecond and intense light regime, which opens the door to the utilization of such light forms in nonlinear optics.

We should also note that there are other forms of light carrying topological properties, some of which will be discussed in later sections. However, the concept of having a vortex in a beams center, around which some property unwinds continuously (which can therefore carry a topological invariant in some phase space), is very general. For instance, one can generate vector beams in which the polarization state of the beams rotates around the vortex, but the beam does not necessarily carry OAM (e.g. see refs. [127] [128] [129] and therein).

We will not further review here other theoretical aspects of topological beams, but rather focus on their implementation and utilization. While the linear-optical domain is generally not within our scope, we still wish to point out several interesting applications of topological light in it, as those form the technical basis also for nonlinear phenomena, and should cross-fertilize ideas between the fields. These include photonically-controlled interactions between the spin and orbital angular momenta of the wave, and transferring momenta back and forth [130] [131] [132] [133] [134] [135], creating entanglement or encoding information within the beams [136] [137] [138] [139] [140] [141], creating and imaging plasmons [142] [143], as well as applications for enhanced imaging techniques [140] [144] [145] [146] [147]. Applications connected to chirality sensing will be discussed separately in section 4.2.3.1.3.

The importance of the topologically-protected node surrounded by a winding phase can become even more significant in NLO applications. For instance, in nanoscopic bio-imaging, the Nobel-winning stimulated emission depletion (STED) microscopy uses a GL beam to nonlinearly suppress photo-excited bio-marker dyes, allowing for the positioning of any residual fluorescence to the beam center with deep-subwavelength precision [148] [149]. In atomic physics, topological beams can also be used to trap single atoms and molecules within their center, subjected to topologically protected regions with vanishingly small optical fields [150] [151].

In the regime of extreme nonlinear optics (which we will discuss in greater details in upcoming sections), seminal works from several groups working in recent years showed that topological light can be employed to drive HHG [152] [153] [154] [155] [156] [157] [158] [37] [39] [159] [38]. The OAM within the driving beam then gets up-converted following conservation laws (which we will show below can also be understood from a more general selection rules perspective), generating XUV beams with extremely high OAM numbers. Such beams may facilitate optical attosecond pulses with high OAM, as well as additional generation of X-ray light carrying OAM for probing chiral and magnetic phenomena (see Fig. 3 for illustrations).

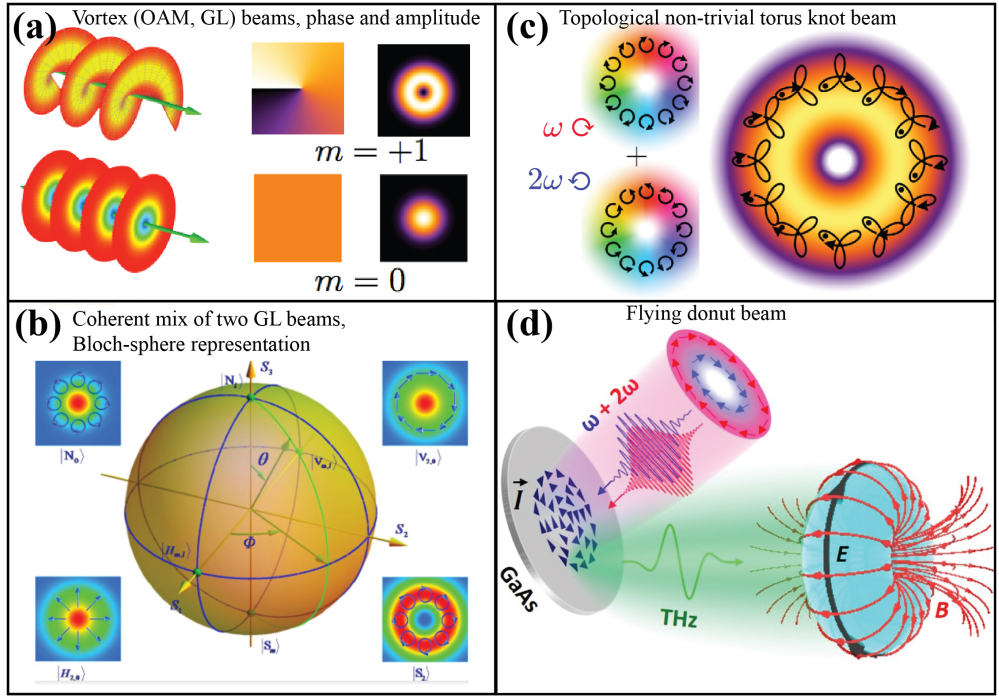

**Fig. 3. Topological light beams and their connection to multi-scale dynamical symmetries.** (a) (left) 3D equi-phase surfaces, phase- and intensity map of GL beams with/without OAM, $m = 1,0$ (Figure reproduced from Wikimedia). (b) Bloch sphere representation (amplitude and phase) of a hybrid vector beam from a superposition of two GL beams, generating various kinds of vector beams with polarization singularities and multi-scale DS. Figure repursed from ref. [160]. (c) Torus-knot beam comprising bi-chromatic GL beams with opposing circular polarization. The fractional knot forms due to the 3-fold rotational microscopic DS of the local field, combined with its continuous symmetry under rotation and time translation. Figure repursed from ref. [39]. (d) Flying doughnut in the THz regime formed by a bi-chromatic GL beam (vector beam) incident on a semiconductor. Figure repursed from ref. [161].

### 2.2.2 Hybrid beams with topological charge and DS

Following the initial application of light with OAM to highly nonlinear optics (which typically involved using the standard structure of GL beams, but in the femtosecond intense light regime), more complex field settings have been developed. Let us consider now some exotic examples of fields carrying a combination of topological and other symmetry properties. In ref. [157] Rego *et al.* introduced the concept of time-varying OAM beams, which are driven by temporally overlapping sets of pulses with varying topological charges. In such fields one can imagine that the topological charge additionally couples to the time domain through temporal translations, and to the spatial domain through translations along the beam's propagation axis. In that respect, the OAM carried by the light field can be thought of as evolving in time. Such effects can be nicely described within the group theory description presented above in the form of a multi-scale DS, as they reflect coordinated rotations of the beam's phase front and how those evolve in time and space.

One peculiarity about such beams of light is that since the OAM changes in time, the beam can be thought of as exhibiting a sort-of torque upon itself, i.e. a self-torque, generating the illusion that the OAM is not conserved. The origin of this behavior lies in interference between coherent superposition states of light carrying different topological charges and generated by separate photonic channels (which also introduce a time delay between those OAM components). Importantly though, when the entire light field is considered as whole, OAM conservation is recovered. The self-torque itself can be controlled by tuning the properties of the various component beams, and it was also shown to imprint a frequency chirp along the beam spatial propagation axis, which serves both for applications and as telltale signs for measuring time-varying OAM. Such chirped components could generally also be described within a group theoretical approach for light; however, that would practically require removing the constraint of time-periodic light fields (see details in ref. [36]).

Another interesting application is potential use of OAM-carrying beams for controlling electronic currents in space and time for generating magnetic light sources. In ref. [43], Sederberg *et al.* employed a bi-chromatic, co-linearly-polarized, vector-beam carrying a topological charge. These beams have a singularity that is very similar to that in the GL beams, but where the topological charge is not caried by a continuously rotating phase front, but rather, by a continuously rotating polarization direction of the bi-chromatic field around the vortex (see illustrations in Fig. 3 for various cases). Locally, in every point in space the co-linearly polarized $\omega$ -2 $\omega$ beam breaks spatial inversion symmetry and time-reversal symmetry (depending on their relative phase), allowing the generation of injection and/or shift currents through second-order (or higher even order) optical nonlinearity [162] [163] [164] [44] [165]. Spatially, because the vortex beam carries a circular continuously rotating polarization state, the locally-generated currents are driven helically in a circle around the beam's vortex (at the beam waist scale, which is macroscopic to the lattice). This allows not only directly controlling the spatial and temporal dynamics of the currents by tuning the beam parameters, but using the unique configuration to generate magnetic impulses as a result of Bio-Savart law (more generally given by Jefimenko's equations [166]). Such currents have been previously analyzed in light-matter interactions between atoms and molecules and intense laser pulses with helical polarization states [167] [168], but the vector beam polarization-associated topological charge allows larger degrees of control, flux, and applicability. The concept has indeed been further applied to atomic systems [168], as well as for optoelectronics and spatially-tailored magnetic fields [169]. The use of combinations of beams with complex DS and polarization states with vector beams carrying different types of topological charges thus could have immense applications in nonlinear optics, both for spectroscopy and for generation and control of novel light sources.

In ref. [37], Pisanty *et al.* introduced the concept of torus knot angular momentum [170] to bichromatic shaped pulses. This light exhibits a form of a continuous multi-scale DS, which carries an additional topological charge in light's phase space (see Fig. 2(e)). By combining co-propagating bi-chromatic counter-rotating circularly-polarized light fields that also carry OAM, the total EM field can be made to exhibit the following continuous symmetry relation:

$$R(\gamma\alpha)\mathbf{E}(R^{-1}(\alpha)\mathbf{r}, t) = \mathbf{E}(\mathbf{r}, t + \tau\alpha) \tag{13}$$

where $R(\alpha)$ is a rotation by angle $\alpha$ about the beam propagation axis (assumed z-axis here), $\tau$ is a constant that is connected with $\gamma$, and $\gamma$ is a fractional charge associated with the order of coordinated rotational symmetry. For instance, in ref. [37] a beam with $\gamma = 1/3$ was generated, corresponding to a three-fold spatial structure. In other words, the beam in this case exhibited a discrete 3-fold DS in every point in space due to the bi-circular components; but moreover, as a whole it exhibited the symmetry relation in Eq. (21) which corresponds to a continuous DS by any $\alpha$ – rotation by $\delta\alpha$ in real-space is compensated for by a rotation in polarization space, as well as a time translation which makes sure to shift to relative phase between the beam components. This is a clear form of a multi-scale DS as derived in the general group theory in ref. [36]. The unique structure of this beam and its continuous symmetry nature means that it carries a topological charge in the dimension associated with the relative phase of the field. In fact, the beam is a direct eigenstate of the torus-knot angular momentum operator derived in ref. [37], which is a sum of the OAM and SAM generators but with a relative amplitude of $\gamma$, inducing a fractional topological charge. This charge was predicted to be conserved in HHG [39] (which can also be understood since it is a sum of OAM and SAM generators that were shown to separately be conserved in HHG [153] [158] [86] [171]), and was recently measured experimentally [38].

The above-mentioned topological forms of light generally employ the paraxial approximation for the involved light beams. Indeed, that is the standard conditions in which the separation of light's total angular momentum into SAM and OAM is allowed (with some exceptions around the vortex where azimuthal components sometimes cannot be neglected). However, adding a third spatial dimension gives rise to optical beams with toroidal (doughnut-like) topologies, which can carry angular momentum also along other spatial axes, and can in general have very complex spatiotemporal mathematical forms. A pure spatial example is radially polarized continuous-beams that expresses an electric- or magnetic-field phase winding at their focal spot [172]. Spatiotemporally, the wide-band frequency range of pulsed lasers enables the closing of a vortex-loop along the extra dimension, creating a propagating doughnut-pulse [161] [173] [174] [175] [176] (e.g. in Fig. 3(d)), or pancakes-like pulses [177] [178]. It is indeed not our intention to fully characterize all such unique light forms, but we nonetheless wish to point out that the complex spatiotemporal symmetry relations within the EM field that give rise to the unique topological nature fall within the scope of group theory, which can be employed for their analysis. We foresee such waveforms gaining use in nonlinear optics in coming years, with some suggestions already published [179] [180]. We refer interested readers to recent reviews on structured light and its applications [181] [182] [183] [184] [185].

### 2.2.3   Other related concepts of light's topology

Lastly, we also mention some other exotic fields of light that carry other types of topological properties. To our knowledge, all of the following examples have only been implemented within the linear optics domain, but their utilization in nonlinear optics could give rise to interesting developments.

First, we note the work by Tsesses *et al.*, which have shown that evanescent waves of EM fields can be generated in a manner that mimics Skyrmions [186], which are general topological entities that arise in solids and magnetic systems [187] [188] [189]. Besides the fundamental importance of generating ever more complex states of light, optical Skyrmions could pave the way to novel forms of information processing. They also allow mimicking certain properties of solid systems with light (such as magnetism), which can be used for exploring fundamental interactions [190]. This work follows a more general trend of topological photonics, where the optical and photonics community is attempting to transfer concepts developed from condensed matter theory into photonics, and vice versa, allowing both fields to gain from knowledge transfer and development of new ideas [191] [192] [193]. It in fact is in accordance with similar efforts in the extremely nonlinear optics community that is attempting a similar two-way transfer of knowledge between the fields of atomic molecular and optical physics, and condensed matter [18] [194] [195]. Therefore, we believe complex topological light forms will soon play a role in highly nonlinear optics.

Second, we briefly discuss a unique family of EM fields known as 'knotted fields', which has been gaining a lot of attention in recent years [196] [197] [198] [199] [200] [201] [202] [203] [204] [205] [206] [203] [207] [208]. Knotted fields comprise complex spatial and polarization arrangements of EM waves that exhibit a set of topological invariants associated with knot theory. In essence, one can imagine EM field lines that form closed loops, but where the loops have a non-trivial knotted character that cannot simply be gauged away (see illustrations in Fig. 4, which one might oddly see as analogous to the bi-circular Lissajous plotted in Fig. 2(a), or the trefoil torus-knot in Fig. 3(c)). From a topological standpoint, two knots are distinct if they cannot be continuously deformed into one another without breaking the knot, resembling similar notions of topology as in vortex beams. Each particular knotted field configuration is uniquely associated with a conserved quantity in the form of a linking number that characterizes its state (along with other potential conserved quantities). Such fields can also be associated with measures of light's chirality, which we will discuss in detail in the next section. From a practical point of view, they are proposed for information processing and storage due to the protected nature of their topology, and we refer readers to the recent review ref. [209].

The most remarkable thing perhaps, at least from a historical perspective, is that the seminal works in refs. [196] [197] predate the discovery of light's OAM by two to three years. Although refs. [196] [197] did not discuss solutions to Maxwell's equations with quantized OAM, they formulated a theory of light's topological structure and suggested solutions to Maxwell's equations with quantized optical helicity, ttwhich is somewhat analogous to the seminal work of Allen et. al. on OAM [103]. However, potentially because of the more complex mathematical requirements for realizing such fields in experiments (as well as for applications), the work did not gain much attention until recently. Moreover, it seems that the field of knots in EM theory has had a relatively minimal penetration and influence into other fields of nonlinear optics, including HHG and chiral sensing. In our opinion this example is worth noting simply because we envision such mutual fertilization of ideas in the coming years.

We should also mention, though it is out of our scope, that similar topological beams can also be generated for particles other than photons, e.g. electrons [210] [211] [212] and even atoms and molecules [213].

(a) EM field lines of 3-fold symmetric knotted light

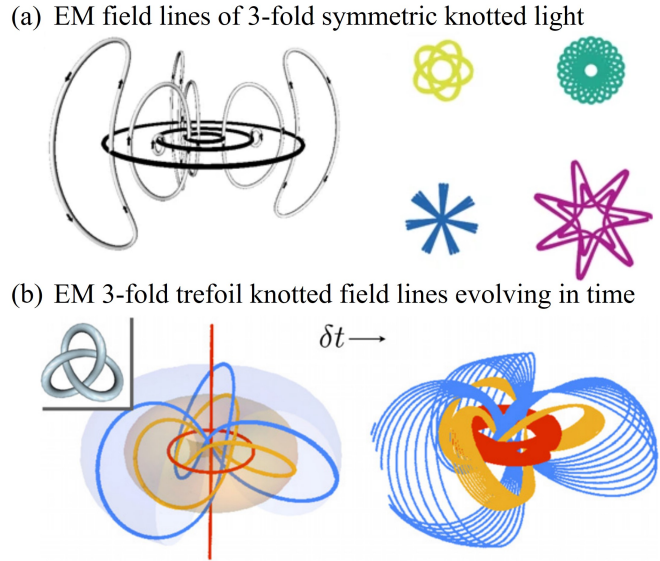

(b) EM 3-fold trefoil knotted field lines evolving in time

**Fig. 4. Knotted topological light.** (a) (left) EM knotted field lines in a rotationally 3-fold symmetric case. (right) Resulting field lines along different tori with different winding ratios. The EM field lines wrap around each other with a nonzero winding, yielding an inherently topological structure in light, which can also exhibit additional DS. Figure repurposed from ref. [198]. (b) Magnetic field lines for a knotted field whose electric and magnetic parts are not transverse everywhere. The field is 3-fold symmetric with a trefoil knot (nonzero winding number). EM helicity (and knottedness) is conserved over temporal evolution, even with the field lines evolving over time and deforming (the topological vortex is conserved). Figure repurposed from ref. [203].

## 2.3 Chiral light

Chirality is one of the most fundamental properties observed in nature. It manifests across scientific fields, including chemistry, biology, optics, particle physics, astrophysics, and even atomic physics. At its heart, the study of chirality is inherently a study of symmetries and asymmetries; thus, it is intimately connected with the concepts discussed in this review, both on the fundamental side, as well as the applications side. The importance of this field in the greater context of probing and understanding chiral phenomena will be discussed in detail in section 4.2.3. In this section, we focus on chiral light, and discuss the concept of chirality in EM theory and its relationship with the closely related field of molecular chirality. We explore different types of chiral charges carried by light, and measures for describing such phenomena.

The section is organized into sub-sections according to different hierarchies of chirality in EM theory.

### 2.3.1 Monochromatic chiral light

As a first and simple case for light carrying an intrinsic chiral charge, let us discuss monochromatic circularly polarized plane waves. This paradigmatic example already contains rich physics, and forms the basis for the field of chiroptical effects for measuring chirality [27]. For mathematical concreteness, we consider the field in Eq. (3), with the polarization components of the EM field comprising circularly-polarized (CP) unit vectors (i.e., $\hat{e}_{\pm} = \frac{1}{\sqrt{2}}(\hat{x} \pm i\hat{y})$ where $\pm$ refer the either left or right CP components). Such a field plots a helical screw-like shape in space and time that naturally exhibits a chiral structure (see Fig. 5(a)).

At this stage, it's worth introducing the mathematical symmetry-based definition of chirality – it is the absence of mirror-type symmetries in a physical system (note that this includes also inversion and improper-rotations). This definition is widely employed in molecular chemistry, as well as in solid state physics [95] [29]. It in essence translates to all scientific fields studying chiral phenomena, because it is a rigorous formulation of the basic property that any chiral object must uphold – be distinct from its mirror image, which simply cannot occur in the presence of mirror symmetries. This is true whether the fundamental object is a particle, a molecule, or an EM field. With this in mind, the helical structure of CPL clearly respects this asymmetry, because any mirror operation (including those involving DS) would result in flipping the helicity of the light field, such that it is purely odd with respect to parity. Note however that this is only correct when considering the full wave form including its spatial structure. Indeed, within the dipole approximation the time-dependent polarization of CPL traces a circle, which exhibits a mirror symmetry in the plane of polarization, and is therefore achiral. From this observation, we expect that most chirality-related phenomena arising from CPL require beyond-electric-dipole light-matter interactions; and indeed, a combination of electric-quadrupole and magnetic-dipole phenomena form the basis for standard chiroptical techniques such as circular dichroism and optical rotation [27]. There are some unique cases that defy this paradigm (where vectorial and/or tensorial observables are measured) which will be discussed in section 4.2.3.

As we established that CPL is chiral, the next important order of business is to characterize its chirality. This can be accomplished using a variety of measures that have been developed over the years, some empirical, and some deeply rooted in EM theory. As a preliminary attempt, one could simply consider light's ellipticity-helicity, $\varepsilon$, or spin angular momentum (SAM) [103]. The ellipticity-helicity of light is directly computed from the stokes parameters [214], and is in one-to-one correspondence with the degree of circular polarization (DCP). It is strictly odd under parity and even under time-reversal, as required from chiral pseudoscalars [27]. It maximizes at unity for right-CPL, minimizes at minus one for left-CPL, and vanishes for linearly-polarized light, as expected. It thus smoothly maps the degree to which the light breaks mirror symmetry, and seems a quite natural way to estimate the chirality of polarized light. This approach works very well for monochromatic EM fields, but the trouble starts with poly-chromatic light. For instance, consider a counter-rotating bi-circular field with two main frequency components at $\omega$ and $2\omega$, which upholds the 3-fold DS in Eq. (6), and has the following electric field form:

$$\mathbf{E}(\mathbf{r},t) = E_1\hat{e}_+ \cos(\omega t - \mathbf{k} \cdot \mathbf{r}) + E_2\hat{e}_- \cos(2\omega t - 2\mathbf{k} \cdot \mathbf{r}) \tag{14}$$

Such a field exhibits a clover-like time-dependent polarization that does not resemble an ellipse (see Fig. 5(b)), intrinsically defying the concept of ellipticity. If one tenaciously persists, the definition of the cycle-averaged ellipticity of the stokes parameters can still be employed, but it would lead to different results depending on the intensity ratios of the two fields. In particular, if the field power ratios are 1:2, one obtains a time-averaged ellipticity of zero, which does not do any justice to the field's intrinsic chirality, as it still does not exhibit mirror symmetries for this power ratio. Thus, ellipticity-helicity is not an effective measure of chirality beyond monochromatic fields.

Considering the origin of the issue, one could instead define a time-dependent ellipticity, i.e., allow for $\varepsilon = \varepsilon(t)$, adding more freedom for describing the light field [215]. This approach can be effective in certain cases [216], but for practical cases still requiring temporal averaging. Moreover, one could have a field with $\varepsilon(t) \neq 0$ that exhibits mirror symmetries. An obvious example would be an EM field which is right-CP for one half-cycle, and left-CP for the other half-cycle, such that it overall exhibits a mirror DS.

Alternatively, one could define a so-called spectral-chirality. Given a time-dependent EM field, e.g., with some temporally evolving polarization $\mathbf{E}(t)$, the field's spectrum can be obtained by Fourier transforming: $\tilde{\mathbf{E}}(\Omega) = \mathcal{F}\{\mathbf{E}(t)\}$. One then decomposes the spectral power into right- and left-CP components: $\tilde{I}_{\pm}(\Omega) = \left| \tilde{E}_x(\Omega) \pm i\tilde{E}_y(\Omega) \right|^2$. The spectral chirality is constructed as the normalized difference in spectral power between left and right components, $\varepsilon_\Omega = \frac{\int (\tilde{I}_+(\Omega) - \tilde{I}_-(\Omega))d\Omega}{\int (\tilde{I}_+(\Omega) + \tilde{I}_-(\Omega))d\Omega}$. It represents a frequency-weighted average for the ellipticity of individual frequency components, and can be quite useful in characterizing certain type of polychromatic chiral light fields. For instance, it can be used as an indicator for whether a broad-spectrum field is overall helical [217]. Unfortunately, it still suffers from similar deficiencies to the time-averaged ellipticity, e.g. it might vanish for EM fields that are asymmetric, and is therefore not in one-to-one correspondence with chirality. Needless to say, such measures are inherently based on light's SAM, and therefore do not capture chirality carries by its spatial arrangement (e.g. nonzero OAM).

The next set of chirality indicators are based on rigorous EM theory (whereas those presented above are more empirical in nature). Firstly, let us introduce the concept of optical chirality (OC), which was re-formulated in the seminal work of Tang and Cohen [218]. OC in vacuum takes the form:

$$\text{OC} = \frac{\varepsilon_0}{2} \mathbf{E} \cdot \nabla \times \mathbf{E} + \frac{1}{2\mu_0} \mathbf{B} \cdot \nabla \times \mathbf{B} \tag{15}$$

where $\varepsilon_0$ and $\mu_0$ are the vacuum permittivity and permeability, respectively. This form is symmetric between the electric and magnetic parts, and therefore in most standard cases the electric and magnetic parts both equally contribute to the OC of the EM field, in which case we analyze here only the electric field part as done in previous sections. The origin for the formula in Eq. (15) arises from a derivation by Lipkin [219], which identified a set of ten conserved quantities of EM fields, the so-called Lipkin's 'zilches'. It has been shown that one can derive an infinite number of such conserved quantities that don't necessarily carry physical meaning [220] [221] [222]. However, the zero-order zilch, or OC, can be associated with a chirality density carried by the EM field, and is physically connected to the angular momentum of the field's curl [69]. Indeed, for monochromatic fields in vacuum the OC is proportional to light's SAM, as well as to the optical helicity.

Let us briefly analyze the functional form of the OC in Eq. (15) and make some noteworthy points. First, within the dipole approximation (neglecting any spatial DOF) the OC vanishes, because the curls of the EM field vanish. Second, we note that in vacuum one can replace the field curls in Eq. (15) with a temporal derivative through the Faraday and Amper's laws and obtain:

$$\text{OC} = -\frac{\varepsilon_0}{2} \mathbf{E} \cdot \partial_t \mathbf{B} + \frac{\varepsilon_0}{2} \mathbf{B} \cdot \partial_t \mathbf{E} \tag{16}$$

It is then clear that the OC vanishes for linearly-polarized plane waves where the electric and magnetic fields are transverse. For circularly-polarized plane waves however, the result is nonzero because taking a temporal derivative of the electric (and magnetic) field flips the polarization axis by 90 degrees. Third, we note that there are generally two components contributing to the OC in Eq. (15) for freely propagating EM fields: (i) light's time-dependent polarization structure (i.e., SAM), and (ii) light's spatial dependence transverse to the propagation axis (i.e., OAM). Generally, the two contributions are mixed, but within the paraxial approximation (assuming propagation along the z-axis) it has been shown that they can be separated to the forms [223]:

$$
\begin{aligned}
\mathrm{OC_p} &= \frac{1}{c_0^2} \phi'(t) I(t) \\
\mathrm{OC_l} &= \frac{1}{c_0} \left[ I_x(t) \partial_y \left( \frac{E_z(t)}{E_x(t)} \right) - I_y(t) \partial_x \left( \frac{E_z(t)}{E_y(t)} \right) \right]
\end{aligned}
\tag{17}
$$

where $\mathrm{OC_p}$ represents the polarization-associated contribution to the OC, and $\mathrm{OC_l}$ the contribution coming from OAM-related dependencies in the field, and where we only analyze the electric field part. Here $I(t)$ is the field's instantaneous power, defined as $I(t) = \frac{c_0 \varepsilon_0}{2} \mathbf{E}(t)^2 + \frac{c_0}{2\mu_0} \mathbf{B}(t)^2$, and the corresponding cartesian component $I_j(t)$ is the projection of the power along the $j$'th axis. $\phi(t) = \arctan\left( \frac{E_y(t)}{E_x(t)} \right)$ is the time-dependent angle that the electric field vector points to, such that its temporal derivative $\phi'(t)$ essentially indicates the angular velocity of the EM field's polarization vector. Eq. (17) presents an intuitive interpretation for the origin of OC in EM fields – rotating field polarization components give rise to nonzero $\mathrm{OC_p}$, while spatial rotational behavior leads to nonzero $\mathrm{OC_l}$. Fourth, OC is generally a time-dependent function (except for simple monochromatic fields), as well as a spatially-dependent function, and therefore in the more complex tailored-light cases, one needs to resort to averaging procedures which are less physically grounded (see Fig. 5(b)). Lastly, we note that the OC has been shown proportional to optical activity in a chiral media, making it physically relevant [218] [224] [225]. It has been argued that one can engineer the optical beam parameters in order to try and maximize the OC in a given region in space, potentially enhancing chiral-responses and generating 'superchiral' fields [226] [227] [228] [229]. This can arise either from the polarization response (e.g. within nanophotonic or mirror geometries [228] [230] [231]), or the spatial distribution of the field. It is a widely debated and somewhat controversial topic from the fundamental standpoint of the origin of the enhancements [232] [233] [234]. The field has also in recent years partially combined with the field of metasurfaces and chiral photonic structures for potentially enhancing the OC density of light, which we will discuss in more detail in section 4.2.3.3. Regardless, what is undebatable is that OC presents an effective and physically-grounded measure for light's chirality in the form of a formally conserved charge of EM fields.

It is worth mentioning in this context that there exists a complex hierarchy of other measures for the intrinsic chirality of light, including the so-called optical helicity, optical spin, and helicity array [69] [235] [236] [237] [238], as well as some other suggestions [239] [68]. In general, while different measures are distinct, they are often similar in nature, and usually coincide for monochromatic light beams. Such methodologies have also been applied to quantized EM fields [232], which are beyond our scope. A common theme to all of these properties is that they vanish within the dipole approximation, suggesting that light does not carry chirality purely in its polarization (i.e., in a given finite point in space), which will be addressed in the next sections.

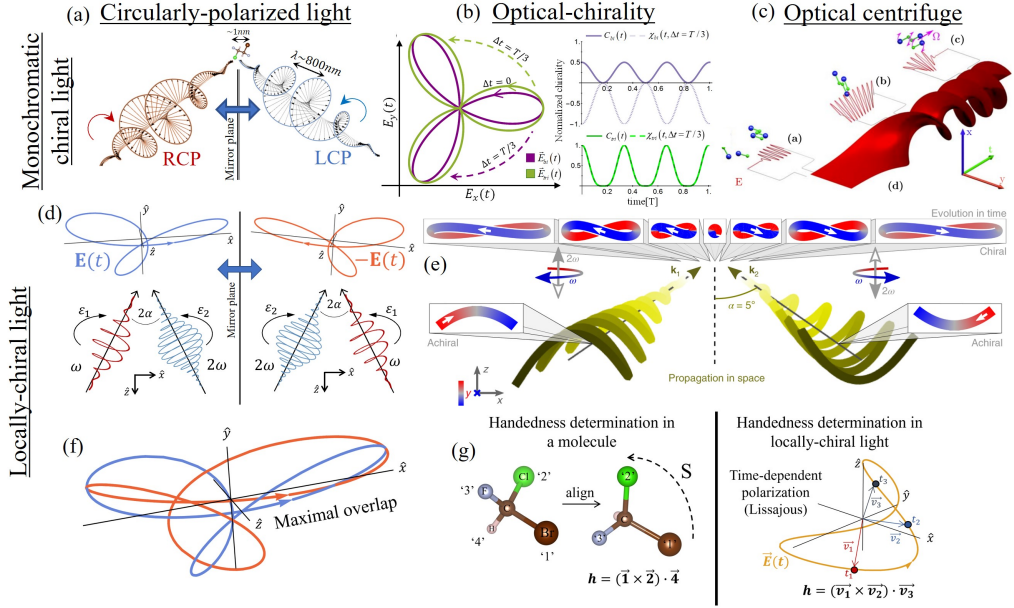

**Fig. 5. Light asymmetries embedded in chiral light wave forms.** (a) Chiral structure of monochromatic circular laser beams. The time-dependent polarization of the light evolves in a spatial helix on the scale of the wavelength (few hundred nanometers). This is opposed to the length scale of structural chirality, which is usually at few nanometers for molecules. A result of this mismatch is that circularly-polarized light can only probe chirality by including beyond-dipole terms, such as magnetic-dipole interactions. (b) Illustration of optical chirality in polarization-tailored bi-chromatic ω-2ω beams and ω-2ω-4ω tri-chromatic beams with rotational dynamical symmetry. The optical chirality can be separated to different timescales, e.g. instantaneous or longer, some of which may have different handedness, affecting chiral light-matter interactions. Figure re-purposed from ref. [223]. (c) Illustration of an optical centrifuge pulse – the pulse is linearly-polarize in short time intervals, but rotates over longer timescales, inducing a chiral asymmetric structure. This structure would instantaneously have vanishing optical chirality, but exhibits nonzero optical chirality on longer timescales. Figure repurposed from ref. [240]. (d) Illustration of locally-chiral light obtained from bi-elliptical non-collinear ω-2ω beams. The total time-dependent polarization of the electric field traces a shape that lacks mirror, inversion, and improper-rotational, dynamical symmetries, making it distinct from its mirror image. Such sources allow electric-dipole induced chiral signals. Figure repurposed from ref. [241]. (e) Locally-chiral light generated by a combination of linearly-polarized non-collinear ω-2ω beams, allowing also global chirality for proper relative phase choices. Figure repurposed from ref. [242]. (f) Illustration of degree of chirality calculation – the two mirror image fields (from (d)) are oriented to maximally overlap, from which their degree of overlap is evaluated, measuring the local-chirality of the waveform. Figure repurposed from ref. [241]. (g) Handedness definition in chiral molecules based on vector triple products (left), compared with the equivalent optical approach for locally-chiral light (see text). Figure repurposed form ref. [243].

### 2.3.2 Instantaneous and non-instantaneous optical chirality

One potential shortcoming of the optical chirality in Eq. (15) is concerned with the different timescales upon which mirror symmetries are broken, or in other words, the timescales under which light carries a chiral charge. This is evident from the form of the polarization-associated OC in Eq. (17), which only depends on time-local functions. That is, the OC would be nonzero at a given instance in time only if the field has a nonzero power in that moment, and its polarization vector is simultaneously rotating. However, consider tailored light pulses such as an optical centrifuge [244] [245] [246] [240] or sequences of linearly-polarized pumps [247] [248] [249] – these are sets of individually achiral light pulses with a polarization axis that rotates from pulse to pulse, generating a type of helical structure (see Fig. 5(c)). Within such a pulse the polarization rotations occur when the electric field is instantaneously zero, indicating the OC vanishes identically. Nonetheless, considered as a whole also on longer timescale, these pulses clearly break a mirror symmetry and should carry a chiral charge.

One potential remedy is to extend the definition of OC to multiple timescales, as formulated in ref. [223]. This approach is somewhat empirical, but provides some description for chiral phenomena occurring also on non-instantaneous longer timescales. Starting from Eq. (17) for the polarization-associated OC, one can extend the temporal derivative of $\phi(t)$ to include coherences in the EM field between different temporal intervals, i.e., between the field at $t$, to the field at $t + \Delta t$:

$$\chi(t, \Delta t) = \frac{1}{c_0^2} \frac{\phi(t + \Delta t) - \phi(t)}{\Delta t} \bar{I}(t, \Delta t) \tag{18}$$

where $\chi(t, \Delta t)$ is the polarization-associated OC on a timescale of $\Delta t$, with $\bar{I}(t, \Delta t) = 0.5(I(t) + I(t + \Delta t))$ is the average intensity between those moments in time. One could then weigh different timescales of chirality, $\Delta t$, depending on the particular field involved, as well as average over the time coordinate, $t$, in order to derive a single number indicating the overall chirality of a particular pulse (see Fig. 5.(b)). Ref. [223] showed that this approach can be useful for understanding the helicity of attosecond pulses emitted from atomic media driven by helical poly-chromatic laser pulses.

Notably, it remains a currently open question whether non-instantaneous chirality can also be derived from rigorous conserved quantities of maxwell's equations, rather than empirically. We should also clarify that the addition of timescales for chirality still requires a spatially-dependent EM field, such that just like OC, $\chi(t, \Delta t)$ also vanishes within the dipole approximation.

### 2.3.3 Locally-chiral light

#### 2.3.3.1 Conceptual idea

At this point, let us depart from the discussion above that considered EM fields with spatial structures, i.e., $\nabla \times \mathbf{E} \neq 0$. All of the chirality measures discussed up to now inevitably describe helical structures in the EM field's spatio-temporal nature. The practical implications for chiral sensing are that chiral signals from interactions of chiral light with chiral matter could only arise due to beyond-electric-dipole interactions, i.e., electric-quadrupole or magnetic-dipole contributions. While this is not always the case (sometimes unique measures can be taken in order to enhance the chiral signals in the electric-dipolar contributions, as will be discussed in section 4.2.3), it means that typical chiral signals are very weak, and much weaker than the dominant electric-dipole response of matter [27]. Thus, a natural point of progression would be to ask the following question – do chiral EM fields exist even in a single point in space, i.e., within the dipole approximation? In this section we will show that the answer is yes, and introduce the required math for describing such entities.

To start, it is instructive to revisit the definition of chirality in molecular chemistry – here a molecule is chiral if and only if its point group does not include improper-rotational symmetries (note that mirror and inversion symmetries are improper rotations of order, 1, and 2, respectively). As discussed above, this mathematical definition is equivalent to demanding a molecule is distinct from its mirror image. One is then tempted to apply a similar formalism to EM fields. That is, instead of following the traditional path of describing EM chirality using SAM, OAM, OC, or other properties arising from analysis of conserved quantities of Maxwell's equations, to try and apply chemical-intuition to determine the chirality of EM fields. From a mathematical perspective, one can then look at the dynamical group of a field in a given point in space (discussed in section 2.1.2), and check whether that group includes improper-rotations. If it does not, the field should be 'chiral' in that point in space. Alternatively, this approach can be taken as a guideline for engineering light that is chiral in a given point in space – choose all of the properties of the beam in a manner that breaks all of the desired DS and the resulting EM field is asymmetric. This technique leads us to define a new physical property of EM fields, so-called 'local-chirality', i.e., a chiral charge carried by EM fields even in a single local point [27]. What is not yet clear at this stage of the discussion is if non-trivial solutions to Maxwell's equations that carry local-chirality physically exist, and if so, how their chirality differs from that carried by e.g., OC.

In order to better understand these issues, it is helpful to analyze the particular symmetries that need to be broken. Let us consider freely-propagating time-periodic plane waves, which can coherently interfere in a given point in space to generate a time-dependent polarization, $\mathbf{E}(t)$. One can rigorously derive the physical constraints on $\mathbf{E}(t)$ for its dynamical group to be chiral [27]. For this, we enumerate the different types of symmetries and DS that include improper-rotational elements in a systematic manner, and formulate the constraints on EM field such that they would not exhibit them. First, since the group cannot include 'static' mirror planes, the field polarization must span the full three dimensions. Inherently, this means monochromatic fields are excluded (since their polarization is always contained in a plane). It also means that at least two non-colinear plane waves are required, because colinear waves can only span up to two polarization components (in the paraxial approximation, see discussion in section 2.2 on vector beams that permit this). Second, 'static' improper-rotations and inversions are inherently excluded in non-trivial time-periodic solutions to Maxwell's equations, as already shown in section 2.1.2. The next stage would be to exclude DS involving reflections, inversion, and improper-rotations. The technique to establish the constraints in those cases relies on the selection rule derivation that will be presented in section 3.2, but is otherwise similar in spirit to the constraints from the static symmetries (see ref. [27] for the derivation). Overall, the emerging physical requirements to generate locally-chiral light, assuming a time-periodic electric field, can be summarized as follows:

1. $\mathbf{E}(t)$ must be comprised of at least two optical beams that propagate non-collinearly.

2. $\mathbf{E}(t)$ cannot be monochromatic and must include both even and odd harmonics of its fundamental frequency.

3. If the even and odd harmonics in $\mathbf{E}(t)$ are not transversely-polarized, then $\mathbf{E}(t)$ is locally-chiral. If they are transversely-polarized, for $\mathbf{E}(t)$ to be locally-chiral at least one of the odd harmonics in it needs to have a non-linear polarization, and in addition, it is enough for one of the following constraints to apply for local-chirality:

   3.1. At least one of the odd harmonics is not circularly-polarized.

   3.2. The odd harmonics can all be circularly-polarized, but cannot have alternating helicity from order to order.

   3.3. The odd harmonics can all be circularly-polarized and have alternating helicities, but $\mathbf{E}(t)$ must contain odd harmonics of frequencies other than $(2nq \pm 1)\omega$, and/or even harmonics of frequencies other than $n(2q + 1)\omega$, where $q$ and $n$ are integers, and $n$ is even.

This approach provides a flow-chart type analysis for determining if a field is locally-chiral or not, just like one typically has in the molecular case [29]. It presents a conceptual leap from chirality measures such as OC, because it attempts to describe molecular chirality and chirality of EM fields in an equivalent mathematical platform. In the context of this review, it puts the symmetries (or asymmetries) of light on equal footing with those of matter. This analysis also clearly establishes that locally-chiral light is physically achievable. For instance, a relatively simple combination of two beams at frequencies $\omega$-$2\omega$ with generic elliptical polarizations that propagate non-collinearly is locally-chiral (see Fig. 5(d)), and has the form:

$$\mathbf{E}(\mathbf{r}, t) = \mathbf{E_1} \cos(\omega t - \mathbf{k_1} \cdot \mathbf{r}) + \mathbf{E_2} \cos(2\omega t - \mathbf{k_2} \cdot \mathbf{r}) \tag{19}$$

where $\mathbf{E_1}$ and $\mathbf{E_2}$ are generic elliptically polarized unit vectors in polarization planes transverse to $\mathbf{k_1}$ and $\mathbf{k_2}$ axes, respectively, and where the beam propagation axes have an opening angle between them. If the field in Eq. (19) is evaluated at the origin ($\mathbf{r} = 0$), one obtains the following local form:

$$\begin{aligned} E_x(t) = E_0 \cos(\alpha) \{ &\cos(\beta_1) \cos(\omega t + \eta) - \varepsilon_1 \sin(\beta_1) \sin(\omega t + \eta) \\ &+ A[\cos(\beta_2) \cos(2\omega t) + \varepsilon_2 \sin(\beta_2) \sin(2\omega t)] \}, \end{aligned} \tag{20}$$

$$E_y(t) = E_0\{\sin(\beta_1)\cos(\omega t + \eta) + \varepsilon_1\cos(\beta_1)\sin(\omega t + \eta)$$
$$+ A[\sin(\beta_2)\cos(2\omega t) - \varepsilon_2\cos(\beta_2)\sin(2\omega t)]\},$$
$$E_z(t) = E_0\sin(\alpha)\{\cos(\beta_1)\cos(\omega t + \eta) - \varepsilon_1\sin(\beta_1)\sin(\omega t + \eta)$$
$$- A[\cos(\beta_2)\cos(2\omega t) + \varepsilon_2\sin(\beta_2)\sin(2\omega t)]\}$$

where the different free parameters represent the various DOF in the two plane waves that were absorbed in $\mathbf{E_{1,2}}$ and $\mathbf{k_{1,2}}$ in Eq. (19): $\alpha$ denotes half of the opening angle between the beams, $\beta_{1,2}$ denote the angles of the major elliptical axis of each beam with respect to the $xz$ plane, $\varepsilon_{1,2}$ denotes the ellipticity-helicity of each beam, $\eta$ is the relative phase between the beams, and $A$ is their amplitude ratio. $E_0$ is the $\omega$ field's amplitude, and $\omega$ is its fundamental frequency, both of which do not affect whether or not the field is locally-chiral as they do not affect its symmetry group. On the other hand, the EM field in Eq. (20) has seven nontrivial DOF that do affect its potential symmetry structure. For any generic choice of parameters Eq. (20) yields locally-chiral light, but for instance for $\alpha = 0$ where the fields are colinear, one can see that $E_z(t) = 0$, such that the field is contained in a plane and would be locally-achiral. In fact, as long as the electric field is not monochromatic (i.e., $A \neq 0$), it is physically much easier to break symmetries then impose symmetric light states, e.g., in order to have locally-achiral light one needs to specifically engineer particular polarization states, beam opening angles, or frequency components in Eq. (20).

### 2.3.3.2 Degree of chirality

Having established that light can be chiral even in a single point in space, the next important stage is to describe/measure its chirality, as well as address other questions such as the stability of local-chirality under propagation, and its physical meaning for light-matter responses. Unfortunately, the group theory formalism only determines whether or not a field is locally-chiral, i.e. a go or no-go condition. However, for applications we are interested in understanding the so-called degree of chirality of the EM field, and how to maximize it – we would for instance not care if a field is locally-chiral when its chirality is miniscule and negligible. There are currently two separate approaches that have been developed successfully to characterize the degree of local-chirality (DOC) of EM fields. The first relies on symmetry arguments, and is therefore the first we present here, while the second relies on chiral correlation functions, which will be introduced subsequently.

The DOC of light can be constructed by a spatio-temporal overlap measure [250]. In essence, we mathematically formulate a normalized pseudoscalar that estimates the extent to which an EM field breaks improper-rotational symmetries. This approach is adopted again from the molecular chirality community [250], and implemented here to time-dependent vector fields instead of molecular potentials or orbitals. Given the field $\mathbf{E}(t)$, the DOC can be formulated as:

$$|\text{DOC}| = \frac{\min\limits_{\theta,\phi,\psi,\Delta t}\left\{\int dt\left|\hat{R}_z(\psi)\cdot\hat{R}_x(\phi)\cdot\hat{R}_z(\theta)\cdot\mathbf{E}(t + \Delta t) + \mathbf{E}(t)\right|\right\}}{\int dt\left|\mathbf{E}(t)\right|} \tag{21}$$

where the angles $\psi$, $\phi$, and $\theta$, are three Euler angles, $\hat{R}_j$ denotes a rotation operator about the $j$'th axis that acts on the field's polarization space, and $\Delta t$ is a temporal shift. The operator $\min\{\}$ is a minimization procedure over the three Euler angles and $\Delta t$. The numerator in Eq. (21) evaluates the overlap between $\mathbf{E}(t)$, and its inverted mirror twin, $-\mathbf{E}(t)$ (where the subtraction of the mirror field $-\mathbf{E}(t)$ has resulted in the plus sign), while allowing $\mathbf{E}(t)$ to freely rotate in polarization space through the three Euler rotations, as well as allowing it to 'rotate' along the time-axis through temporal translations of $\Delta t$. If $\mathbf{E}(t)$ is locally-achiral, then the minimization procedure will return a set of Euler angles and time-delay such that the numerator vanishes. If on the other hand $\mathbf{E}(t)$ is locally-chiral, it means it cannot be superposed onto its mirror image, and the minimization procedure would minimize the overlap of the two fields but still yield a nonzero scalar (see illustration in Fig. 5(d,f). That scalar is the DOC, since it estimates the extent to which mirror symmetries are broken within $\mathbf{E}(t)$.

Let us highlight some noteworthy points. First, due to the nature of the minimization procedure in Eq. (21), the DOC is given as an absolute valued number. Hence, it does not fulfill the desired properties of a pseudoscalar that should be odd under parity. It therefore must be artificially multiplied by a 'handedness', denoted as $h_{DOC}$ (which takes values either ±1,0):

$$\text{DOC} = h_{DOC} |\text{DOC}| \tag{22}$$

$h_{DOC}$ needs to be determined through a separate calculation, as we will shortly describe. Second, we note that the DOC is a normalized quantity ranging from -2 to 2, just like chiral dichroism [27]. Notably, even if the DOC is bounded from above by 2, it is not clear what is the maximal physically-attainable DOC is for realistic EM fields, and the current known maximal DOC is ~0.69 [241]. Third, for practical applications we should note that the minimization over $\Delta t$ should be performed over one optical cycle if the field is time-periodic, or over the full pulse duration if it is not. Fourth, the minimization problem in the DOC definition generally cannot be solved analytically, and therefore realistic calculations of the DOC in EM fields require numerical implementations. Finally, we note that the DOC constitutes a highly non-trivial measure of chirality that in many respects defies the standard intuition we are accustomed to from EM theory. This is because local-chirality evaluates the chirality of the EM field's time-dependent polarization, rather than helical-like structures in the beam. For instance, in the case of the two non-colinear $\omega$-$2\omega$ beams that are elliptically polarized described in Eq. (20), the overall highest DOC is obtained when both individual beams exhibit a generic intermediate ellipticity, rather than a circularly-polarized state [241].

### 2.3.3.3 Handedness

Next, we address the handedness of locally-chiral light, $h_{DOC}$. In order to understand how to consistently assign handedness to a time-dependent vector, it is helpful to discuss the role of handedness in molecular chirality (where locally-chiral light conceptually originated), and the role of the triple product. Let us review the definition of the handedness of chiral molecules as given by the International Union of Pure and Applied Chemistry (IUPAC) [251]. We follow the discussion in ref. [243] and take an example chiral molecule, bromochlorofluoromethane (CBrClFH) as a test case (see Fig. 5(g)). CBrClFH is perhaps the simplest example of a stable chiral molecule with just one chiral center at its origin (the carbon atom is bonded to four distinct atoms forming a stereocenter), making its analysis simple. Fig. 5(g) (left panel) illustrates the protocol for determining the molecular handedness, as taught in undergraduate organic chemistry [252]: (i) number the substituents around the chiral center according to their mass with the lightest elements taking the lowest priority in numbering, (ii) Orient the molecule such that the lowest priority substituent points towards the back of the viewer, (iii) draw a circular arrow across the remaining in-plane substituents directed form the top priority group to the lowest one, (iv) the handedness is determined by the directionality of the resulting arrow – either "R" for clockwise rotation, or "S" for counter-clockwise.

It's worth contemplating the logic behind this procedure, as well as its potential pitfalls. One thing that stands out is its arbitrariness. For instance, if heavy constituents were given the lowest priority numbers, all labels of "R" and "S" of chiral molecules would flip. Infinite other conventions could easily be established. The most important point here is not therefore not that the process is arbitrary, but that it is consistent and rigorous for all chemical species. This means all chemists agree on the labeling of molecular handedness if they all have the same protocol. Another important point regards the math behind this procedure. Indeed, choosing the sign according to the clockwise/counter-clockwise rotation of an arrow is in fact equivalent to taking the sign of a vectorial triple product. Fig. 5(g) illustrates the concept – taking a triple product of vectors emerging from the chiral center onto several constituents provides the handedness. The beauty is that a vectorial triple product is inherently a pseudoscalar that is odd under parity, which is exactly what one requires from handedness. Thus, triple products play a crucial role in the study of chiral phenomena, both in matter and in light.

In order to construct handedness for locally-chiral light, we can mimic the molecular procedure, but where instead of chemical substituent weights, the different moments in time for the electric field are used to construct vectors used in a triple product (see Fig. 5(g) (right panel) for illustration). That is, we find three unique moments in time, $t_1$, $t_2$, $t_3$, from which to obtain three vectors, $\mathbf{v_1} = \mathbf{E}(t_1), \mathbf{v_2} = \mathbf{E}(t_2), \mathbf{v_3} = \mathbf{E}(t_3)$. The handedness is then given as $h_{DOC} = \frac{(\mathbf{v_1} \times \mathbf{v_2}) \cdot \mathbf{v_3}}{|(\mathbf{v_1} \times \mathbf{v_2}) \cdot \mathbf{v_3}|}$. Such an approach is of course similarly arbitrary, but as long as the method for determining the vectors is unique and can be applied consistently to all vector fields, we do not care. Ref. [243] outlined one possibility for obtaining the three unique moments in time to construct the triple product, where the choice relied on moments in time that maximize either the field's instantaneous power, or its derivatives. There are of course many potential pitfalls here and subcases that one has to consider. But the conceptual reasoning is the most crucial part – there exists a unique way to define handedness for locally-chiral fields, and that allows for consistent labeling for applications, both for linear and nonlinear optics, as well as for general EM theory.

### 2.3.3.4 Chiral-correlation functions

Having outlined the DOC and its analogy to molecular chirality, we present a complimentary technique that follows slightly different reasoning. The DOC outlined above was constructed as a unique measure that describes the EM field itself and its intrinsic properties. On the other hand, one could construct chiral pseudoscalars that rely on the field's interaction with chiral matter. This idea is also adopted from the molecular chemistry community, where chirality and handedness can equivalently be defined solely based on a molecule's optical activity – that is, its interaction with light. In this manner, one determines a consensus from experimental results – optically active molecules are labeled as chiral, optically non-active molecules are labeled achiral, and the handedness of the molecule is defined by the direction of optical activity, i.e., if light's polarization axis rotates rightwards, or leftwards. Such a definition is completely legitimate, unique, and consistent, but it requires performing experiments to determine a molecule's handedness. Similarly, we could define light's handedness based on its interaction with matter in an equivalent approach.

From the perspective of locally-chiral light, we could imagine determining its degree of chirality by performing experiments where light interacts with known reference chiral molecules, and its nonlinear response is assessed. The size and sign of the chiral signal would correspond to the DOC. Mathematically, the connection between the EM chirality and generated chiral signals of various nonlinear orders can be expressed by chiral correlation functions [243] [242]. The lowest order chiral correlation function is third order, and takes the following form for a time-periodic field:

$$H^{(3)}(\tau_1, \tau_2) = \int_0^T dt \mathbf{E}(t) \cdot [\mathbf{E}(t + \tau_1) \times \mathbf{E}(t + \tau_2)] \tag{23}$$

$H^{(3)}$ is a triple product comprised of the electric field evaluated at different moments in time, much like $h_{DOC}$, but where the triple product is further averaged over one full laser cycle. The moments in time are not uniquely determined as in the calculation of $h_{DOC}$, but rather kept as free parameters that characterize the field's structure in time and polarization space. In that respect, $H^{(3)}$ not only characterizes whether the EM field breaks some mirror symmetry, or even the extent to which a symmetry is broken, but also corresponds to the typical magnitude of a time-averaged chiral response of chiral matter interacting with locally-chiral light. For instance, circular dichroism (CD) in two-photon absorption form locally-chiral light is expected to be proportional to $H^{(3)}$ [242] [23], much in the same manner that one-photon absorption standard CD from circularly-polarized light is proportional to the OC [218]. Importantly, such predictions still await experimental confirmation.

Chiral correlation functions are more easily analyzed in the frequency domain after Fourier transforming the $\tau_1, \tau_2$ dimensions:

$$h^{(3)}(\omega_1, \omega_2) = \tilde{\mathbf{E}}(\omega_0) \cdot [\tilde{\mathbf{E}}(\omega_1) \times \tilde{\mathbf{E}}(\omega_2)] \tag{24}$$

where $\omega_0$ is not a free parameter, but rather determined by a photon energy conservation condition for the nonlinear interaction, with $\omega_0 = -\sum_{i=1}^n \omega_n$. For the 3$^{\text{rd}}$ order case in Eq. (24) we would have $\omega_0 = -(\omega_1 + \omega_2)$. The advantage of the Fourier transform is that most standard EM fields have a finite bandwidth, such that the different possible frequency arguments in Eq. (24) are finite. For instance, for the locally-chiral field in Eq. (20) one could only take $\omega_{1,2} = \omega, 2\omega$, because any other choice would yield zero. Moreover, all of the calculations are analytical, avoiding issues associated with multi-dimensional minimization problems as in DOC overlap measures. Overall, this methodology presents some clear attractive features – one can analytically calculate a specific quantity, which in itself is a pseudoscalar that embeds both light's degree of chirality, and its handedness.

On the other hand, there are some slightly less attractive features of the chiral correlation functions compared to the DOC symmetry-based definition, which make the two quite complimentary and ideal to employ in tandem. The first issue is that since chiral correlation functions describe a temporal average of triple products, one could have a locally-chiral field that yields $H^{(n)} = 0$, up to some order of $n$. For instance, in the synthetic locally-chiral light presented in ref. [242], $H^{(3)}$ identically vanishes. This is the case for the locally-chiral EM field in Eq. (20) regardless of any choice of the field's parameters, simply because it only comprises two frequency components. One must then evaluate higher order correlation functions (e.g., the first nonzero order correlation function for Eq. (20) is 5'th). The $n$'th order correlation function takes the form:

$$H^{(n)}(\tau_1, \dots \tau_{n-1}) = \int_0^T dt \mathbf{E}(t) \cdot [\mathbf{E}(t+\tau_1) \times \mathbf{E}(t+\tau_2)] \cdots [\mathbf{E}(t+\tau_{n-2}) \cdot \mathbf{E}(t+\tau_{n-1})] \tag{25}$$

where $n$ is odd. Note that in the general form in Eq. (25) there is only one triple product, multiplied by scalars comprised of the values of the EM field in different moments in time. For general $n$'th order cases it becomes cumbersome to perform analytical calculations. Moreover, one is left with ambiguity in determining light's DOC, because the choice of which correlation function to use is arbitrary. Even if this is solved by deciding on a consensus (just as was chosen above for $h_{DOC}$), e.g., using the first nonzero $H^{(n)}$, there is the issue of evaluating this function in different times (or different frequency components) and averaging it in some manner over those arguments in order to obtain a single scalar rather than a multi-dimensional function. It is also unclear how to compare the degree of local-chirality of different fields, for instance, which field has a higher degree of local-chirality – a field with nonzero but small $H^{(3)}$, or a field with very large $H^{(5)}$ but where $H^{(3)} = 0$? There is no clear prescription for the answer in the formalism of chiral correlation functions. Lastly, an important point for technical applications is that in the frequency domain $h^{(n)}$ are complex. All of these are remaining open challenges in this active field of research, especially concerning the fundamental formalism of local-chirality in EM theory. Nevertheless, chiral correlation functions can be extremely useful in many cases; especially for bi-chromatic fields where the choice of frequency arguments to employ in $h^{(n)}$ is obvious, and if the interest is in analyzing a particular order of nonlinear response rather than the fundamental handedness of a given laser field.

Let us close this section by contemplating the relationship between the descriptions of locally-chiral light either with the DOC based on overlap measures, or with chiral correlation functions. Crucially, both mathematical formulations describe the same physical phenomena.

Indeed, there is a clear connection between the two approaches. Besides the similar mathematical form of the equations, it has been proven in ref. [242] that for locally-achiral fields (i.e., with DOC=0), all orders of chiral correlation functions vanish. On the other hand, some differences can arise in practical cases, and for instance the functional behavior for different EM fields can be distinct (even qualitatively) if evaluated based on correlation functions, or using the DOC. This can become important when including the EM field's spatial-dependence in calculations as discussed in the next sub-section. It represents the manifestation of the arbitrariness of the definition of handedness in chiral phenomena. Nevertheless, all physical observables should be independent of such arbitrariness, meaning that potential physical definitions for the degree of local-chirality of light should eventually prevail as those that agree with experimental and numerical results. At the current stage of the field, it remains unclear if such a definition is one of those presented here, or an altogether different approach.

### 2.3.4 Global-chirality

Having discussed the main physical attributes of locally-chiral light in a single point in space, it is interesting to now allow the EM field to be spatially-dependent once again. That is, allow the vector field to vary in space such that $\mathbf{E} = \mathbf{E}(\mathbf{r}, t)$. However, we ignore any potential chiral structures induced by OAM, and assume that the dominant chiral nature of such an EM field would be an average of its local-chirality over different regions in space. Besides the pure fundamental interest in such chiral light forms, the spatial dependence is crucial for applications employing locally-chiral light. This is because it has been established that unless specific actions are taken to prevent it, the EM field's local-chirality can rapidly oscillate in space on length scales smaller than the wavelength. For instance, the paradigmatic example of non-colinear $\omega - 2\omega$ fields in Fig. 5(d) was shown in ref. [243] to flip the field's handedness every $\lambda/2$. Physically, this behavior is not very surprising given the functional form in Eq. (20) – the relative two-color phase plays a crucial role in determining the fields' handedness, but the spatial position of the plane wave acts the role of shifting the two-color relative phase between different positions in space.

Consequently, when such fields are employed for imaging chiral matter, the response within the beam's interaction region would contain many handedness-flipped zones, whose contributions to the chiral signal would average to zero [242]. It is therefore essential to address both the theoretical description of the field's spatial chirality, as well as to devise methodologies that prevent such handedness flipped regions averaging out, generating so-called globally-chiral light fields – fields that are locally-chiral in at least a single point in space, and their handedness is constant [242] (see example in Fig. 5(e)). Alternatively, one would require at least that if the field's handedness flips in different points in space, that those points have varying DOC such that the overall average chiral response would not vanish. Both definitions of globally-chiral light have been employed in numerical calculations and predicted to yield a nonzero far-field chiral signals of varying natures [242] [253] [254] [255] [256].

At the heart of these techniques, one relies on either calculating the DOC of light at different spatial regions [243], or, its lowest order chiral correlation function [242], and analyzing its functional structure. As a point of contact with the multi-scale theory for DS presented in section 2.1.3, the weaker condition for global-chirality – demanding that even if light's handedness flips in different regions in space, those regions have varying DOCs that wouldn't average-out – can be described by multi-scale DS [36]. Essentially, if one restricts the dynamical group of the EM field in (3+3+1)D to exclude improper-rotational elements, this would guarantee that the spatial averaging procedure for the chiral nonlinear responses of different regions in space would not average to zero [36]. It remains currently an open question if the stronger condition – demanding a field's handedness is constant everywhere – can be formulated just based on multi-scale DS considerations.

### 2.4 Dynamical symmetries in synthetic dimensions

In this subsection we expand the concept of DSs for employing light's DOF as synthetic dimension, in which additional symmetry relations can be defined. Synthetic dimensions have initially emerged in the fields of condensed matter physics and quantum simulation, offering a unique framework to engineer, explore, and understand novel physical phenomena [257] [258] [259]. By introducing additional DOF, synthetic dimensions enable investigating physics in higher-dimensional spaces without physically extending the underlying experimental apparatus. Most often, the term synthetic dimension represents non-spatial DOF, which can be either discrete or continuous, e.g. a phase or other parameter that can be tuned.

In the context of photonics, synthetic dimensions generally fall into one of three categories [257]: (i) "parametric" synthetic dimensions, i.e., parameters of the Hamiltonian that lack a kinetic term, (ii) "eigenstate ladder" synthetic dimensions that have a kinetic term and therefore support transport, and (iii) "time-bin" synthetic dimensions, typically corresponding to a sequence of optical pulses. The idea of DS in synthetic dimensions of light is to make use of these parametric DOF for additional complex symmetry relations, that may also lead to novel selection rules (as will be discussed in later sections). Dynamical symmetries in synthetic dimensions often arise when one EM field imposes a DS, that is subsequently broken by another EM field. By treating the polarization DOF of the symmetry breaking field as synthetic dimensions, it is often feasible to devise a symmetry operation of the complete system, which is a composition of the broken symmetry operation and an operation in the synthetic dimensions. These symmetries are also denoted as "real-synthetic symmetries" because they operate on both real and synthetic dimensions concurrently, possessing a composite structure.

To illustrate this concept, we consider a concrete example: a bi-chromatic laser beam of frequencies $\omega = 2\pi/T$ and $3\omega$, polarized along the $\hat{x}$ and $\hat{y}$ axes, respectively. The electric field of this beam may be written as $\boldsymbol{F}(t) = \sin(\omega t)\,\hat{x} + \sin(3\omega t + \pi/7)\,\hat{y}$, which exhibits the symmetry relation $\boldsymbol{F}(t) = -\boldsymbol{F}(t + T/2)$, corresponding to the symmetry operation $\hat{r}_2\hat{t}_2$ by the notation of Floquet groups [35]. Adding a symmetry breaking field of frequency $4\omega$ and amplitude $\lambda$ to the system, the total field takes the form:

$$\mathbf{F}(t, \lambda) = (\sin(\omega t) + \lambda\cos(4\omega t))\hat{x} + \sin(3\omega t + \pi/7)\,\hat{y} \tag{26}$$

For $\lambda \neq 0$, the driving field does not exhibit $\hat{r}_2\hat{t}_2$ DS because the term $\lambda cos(4\omega t)\hat{x}$ changes sign under the operation. However, by treating $\lambda$ as a parameter for a synthetic dimension, we may find that total electric field exhibits the relation $\boldsymbol{F}(t, \lambda) = -\boldsymbol{F}(t + T/2, -\lambda)$, corresponding to the symmetry operation $\hat{r}_2\hat{t}_2 \cdot \hat{\zeta}$ where $\hat{\zeta}$ takes $\lambda \to -\lambda$ (i.e. a mirror operation in $\lambda$ space). As stated above, the symmetry $\hat{r}_2\hat{t}_2 \cdot \hat{\zeta}$ belongs to the class of symmetry operations termed real-synthetic symmetries, because $\hat{r}_2\hat{t}_2$ and $\hat{\zeta}$ operate on real (space and time) and synthetic ($\lambda$) dimensions, respectively. Similar symmetries can be obtained for relative phase shifts in poly-chromatic fields [260] [261] [262].

Generally, we could consider a Floquet system that exhibit a generic DS, $\hat{X}$, which is subsequently broken by a 2nd EM field term, monochromatic with a frequency $\Omega$ and polarization vector $\boldsymbol{Q}$, given by $W(t) = \boldsymbol{Q}e^{i\Omega t} + c.c.$, where $\boldsymbol{Q} = (q_x, q_y, q_z)$ is a complex valued polarization vector. It is then possible to formulate a symmetry operation of the complete symmetry broken field $\hat{X} \cdot \hat{\zeta}$, where $\hat{\zeta}$ operates on the synthetic space spanned by the components of $\boldsymbol{Q}$. The operation $\hat{\zeta}$ can derived by solving the equation $W(t) = \hat{X} \cdot \hat{\zeta}W(t)$ (see further discussion in refs. [41] [42]). Such a DS is always guaranteed when considering DS breaking in the dipole approximation [42]. We note that similarly to standard DS, DS in synthetic dimensions result in unique kind of selection rules termed 'selection rules for breaking selection rules' (which will be addressed in section 3.3).

## 3. Selection rules in nonlinear optics

### 3.1 Selection rules by symmetries of nonlinear media

Before we embark on the journey to understand how DS of EM fields lead to selection rules in NLO, it is worthwhile to start from the historic perspective and analyze NLO selection rules as they are usually obtained – from the symmetries of the medium. Here we closely follow ref. [2].

Consider a nonlinear interaction of light in a material with dispersion and loss. The nonlinear susceptibility is a complex quantity, relating the complex amplitudes of the electric field and polarization density. We expand the electric field and polarization terms in Fourier components as:

$$\widetilde{\boldsymbol{E}}(r,t) = \sum_n \boldsymbol{E}(\omega_n)e^{-i\omega_n t} + c.c. \tag{27}$$

$$\widetilde{\boldsymbol{P}}(\boldsymbol{r},t) = \Sigma_n \boldsymbol{P}(\omega_n)e^{-i\omega_n t}$$

where $\boldsymbol{E}(\omega_n)$ is the amplitude of the $\omega_n$ frequency carrier component (after assuming the dipole approximation), and the sum in principle runs over multiple potential components (including both positive and negative frequencies). We may now define the 2$^{nd}$ order nonlinear susceptibility tensor $\chi_{ijk}^{(2)}$ as:

$$P_i(\omega_n + \omega_m) = \epsilon_0 \sum_{jk} \sum_{(nm)} \chi_{ijk}^{(2)}(\omega_n + \omega_m, \omega_n, \omega_m)E_j(\omega_n)E_k(\omega_m) \tag{28}$$

where $ijk$ are induces for the cartesian components of the fields, and the summation $(nm)$ is restricted to fixed $\omega_n + \omega_m$ values. The quantity $\chi_{ijk}^{(2)}(\omega_n + \omega_m, \omega_n, \omega_m)$ is sometimes denoted as $\chi_{ijk}^{(2)}(\omega_n + \omega_m; \omega_n, \omega_m)$ or $\chi_{ijk}^{(2)}(\omega_3; \omega_2, \omega_1)$, where $\omega_3 = \omega_1 + \omega_2$, and refers to 2$^{nd}$ order nonlinear optical processes. For instance, if $\omega_1 = \omega_2 = \omega$, then $\omega_3 = 2\omega$ and we retrieve SHG. If $\omega_2 = -\omega$ instead, then we obtain optical rectification, etc.

In an equivalent manner, the 3$^{rd}$ order nonlinear response can be given as a function of the 3$^{rd}$-order susceptibility tensor, $\chi_{ijkl}^{(3)}$:

$$\begin{aligned} P_i(\omega_o + \omega_n + \omega_m) \\ = \epsilon_0 \sum_{jkl} \sum_{(mno)} \chi_{ijkl}^{(3)}(\omega_o + \omega_n \\ + \omega_m, \omega_0, \omega_n, \omega_m)E_j(\omega_o)E_k(\omega_n)E_l(\omega_m) \end{aligned} \tag{29}$$

, in which $jkl$ and $(mno)$ are summation indices, analogously to the 2$^{nd}$-order case.

The susceptibility tensors, up to arbitrary order, encode the nonlinear response of a material to external EM fields. They are essentially the coefficients in a series expansion of the polarization density, **P**, in the applied field, **E**. Symmetries of the interacting medium impose constraints on the nonlinear susceptibility tensor. These constraints in turn lead to NLO selection rules. That is, if the various elements in $\chi^{(n)}$ exhibit symmetry-induced constraints (e.g. perhaps some terms vanish, or are degenerate), then this imposes constraints on the polarization through eqs. (28) and (29). In the following, we shall briefly outline example symmetry properties of $\chi_{ijk}^{(2)}$ without proof (which can be found in ref. [2]). This discussion can be extended to symmetry properties of $\chi^{(3)}$ and in principle also any higher order processes described by $\chi^{(n)}$, but for brevity we restrain to low-order processes.

Firstly, independently of the symmetry of the material, the requirement that the EM fields and induced polarization are real leads to:

$$\chi_{ijk}^{(2)}(-\omega_n - \omega_m, -\omega_n, -\omega_m) = \chi_{ijk}^{(2)}(\omega_n + \omega_m, \omega_n, \omega_m)^* \tag{30}$$

Additionally, $\chi_{ijk}^{(2)}$ adheres to an intrinsic permutation symmetry of the indices $j$ and $k$:

$$\chi_{ijk}^{(2)}(\omega_n + \omega_m, \omega_n, \omega_m) = \chi_{ikj}^{(2)}(\omega_n + \omega_m, \omega_m, \omega_n) \tag{31}$$

In a lossless medium all components of $\chi_{ijk}^{(2)}(\omega_n + \omega_m, \omega_n, \omega_m)$ are real; hence, a full permutation symmetry implies:

$$\chi_{ijk}^{(2)}(\omega_3 = \omega_1 + \omega_2) = \chi_{jki}^{(2)}(-\omega_1 = \omega_2 - \omega_3) = \chi_{jki}^{(2)}(\omega_1 = -\omega_2 + \omega_3)$$
$$= \chi_{kij}^{(2)}(\omega_2 = \omega_3 - \omega_1) \tag{32}$$

In the absence of dispersion, Kleinman symmetry applies, which enforces:

$$\chi_{ijk}^{(2)}(\omega_3 = \omega_1 + \omega_2) = \chi_{jki}^{(2)}(\omega_3 = \omega_1 + \omega_2) = \chi_{kij}^{(2)}(\omega_3 = \omega_1 + \omega_2) =$$
$$= \chi_{ikj}^{(2)}(\omega_3 = \omega_1 + \omega_2) = \chi_{jik}^{(2)}(\omega_3 = \omega_1 + \omega_2) \tag{33}$$
$$= \chi_{kji}^{(2)}(\omega_3 = \omega_1 + \omega_2)$$

When Kleinman's symmetry applies, it is typical to use contracted notation and replace the $\chi_{ijk}^{(2)}$ tensor with $d_{ijk} \equiv \frac{1}{2}\chi_{ijk}^{(2)}$. Exploiting the permutation symmetry of the last two indices, we may use contracted notation and write $d_{ijk}$ as a 3×6 tensor with 18 potential independent elements:

$$d_{il} \equiv \begin{bmatrix} d_{11} & d_{12} & d_{13} & d_{14} & d_{15} & d_{16} \\ d_{21} & d_{22} & d_{23} & d_{24} & d_{25} & d_{26} \\ d_{31} & d_{32} & d_{33} & d_{34} & d_{35} & d_{36} \end{bmatrix} \tag{34}$$

, and

$$\begin{bmatrix} P_x(\omega_3) \\ P_y(\omega_3) \\ P_z(\omega_3) \end{bmatrix} = 4\epsilon_0 \begin{bmatrix} d_{11} & d_{12} & d_{13} & d_{14} & d_{15} & d_{16} \\ d_{21} & d_{22} & d_{23} & d_{24} & d_{25} & d_{26} \\ d_{31} & d_{32} & d_{33} & d_{34} & d_{35} & d_{36} \end{bmatrix}$$
$$\times \begin{bmatrix} E_x(\omega_1)E_x(\omega_2) \\ E_y(\omega_1)E_y(\omega_2) \\ E_z(\omega_1)E_z(\omega_2) \\ E_y(\omega_1)E_z(\omega_2) + E_z(\omega_1)E_y(\omega_2) \\ E_x(\omega_1)E_z(\omega_2) + E_z(\omega_1)E_x(\omega_2) \\ E_x(\omega_1)E_y(\omega_2) + E_y(\omega_1)E_x(\omega_2) \end{bmatrix} \tag{35}$$

The Kleinman condition imposes that $d_{ijk}$ has additional constrains with only 10 independent components and can be rewritten as:

$$d_{il} \equiv \begin{bmatrix} d_{11} & d_{12} & d_{13} & d_{14} & d_{15} & d_{16} \\ d_{16} & d_{22} & d_{23} & d_{24} & d_{14} & d_{12} \\ d_{15} & d_{24} & d_{33} & d_{23} & d_{13} & d_{14} \end{bmatrix}$$

For each of the 32 crystal classes, there are additional constrains on $\chi$ and $d_{ijk}$. For example, for a Triclinic $C_1$ family, all elements of $\chi_{ijk}^{(2)}$ are independent and nonzero, whereas for the Triclinic $S_2$ family (with inversion), all elements vanish, i.e. there is no 2nd-order nonlinear response.

Overall, this approach can be used to uncover NLO selection rules. For instance, it can be trivially shown that in media with inversion symmetry all even order of the nonlinear tensor vanish (i.e. no even harmonic generation is allowed to arbitrary orders). Similarly, one may note that in rotationally invariant media , driving with circularly-polarized lasers will lead to constraints on the allowed emitted polarization such that it only permits certain orders of response, and those emitted harmonics must be circularly-polarized with a specific handedness [3]. However, mathematically this approach can be somewhat limiting because it is cumbersome to describe highly nonlinear phenomena. Imagine attempting to understand if the 30th harmonic order is allowed or not for a given driving condition – one must sum not only the polarization coming from $\chi^{(30)}$, but also every other sequential cascaded process allowed, such as $\chi^{(15)}\chi^{(15)}$, $\chi^{(16)}\chi^{(24)}$, etc. This series is formally infinite as arbitrarily high order responses can contribute with cancellation of terms (creating and annihilating photons). This situation is even more complex if multiple field polarization and carrier waves are involved, making the summation hard to follow even in perturbative orders, and meaning selection rules are not necessarily apparent. Clearly, extracting selection rules for arbitrary situations is very challenging in this approach. Moreover, in HHG and other highly nonlinear phenomena the perturbation series should inherently fail, because the process is explicitly non-perturbative and cannot be described by a series expansion in the EM field (tunnel ionization, the first step in HHG, is inherently exponential). Lastly, there are other considerations which are simply not taken into account here, such as symmetries involving time-reversal operations of the EM field, or various phase relations between fields (i.e. synthetic DS). Such symmetries can in principle lead to NLO selection rules, and while they could be derived for specific cases using ad-hoc expansions, it is impossible to mathematically derive them from the typical NLO approach.

All of these shortcomings motivate a different approach to be employed, especially in the ever evolving field of tailored light [183] [185] [33] [6] where very complex polarization and spatially tailored waves can be employed as discussed in previous sections. In the following, we will employ the DS approach of deriving general selection rules instead.

### 3.2 Selection rules by symmetries of light-matter systems

### 3.2.1   Nonlinear optics in the dipole approximation

The approach described above, which has dominated the field of NLO since its creation, has several shortcomings as discussed above. Recapping, first, on a conceptual level, it does not treat the EM field and the driven medium on equal footing. At face value, this is not an issue, but rather an aesthetic problem that becomes ever present when employing complex tailored light fields with varying symmetries, as were discussed in the previous section. Second, it is formulated in the form of a perturbative series expansion. From a theoretical correctness level, this is again not necessarily a deal-breaker for the theory, as even in non-perturbative nonlinear processes such as HHG one can take the limit of an infinite sum and draw conclusions on processes to arbitrary order. However, it seems a somewhat unnatural approach towards deriving selection rules, and there is no guarantee that one could always draw conclusions to arbitrarily high orders of nonlinear processes. These two points are already indicative that a more 'polished' version for a theory of selection rules is desired. Lastly, we would point out that by not explicitly treating the complex symmetries of the EM field, certain selection rules can be completely overlooked. For instance, those involving time-reversal operations. This last point certainly justifies replacing the standard approach for selection rules with a more general one.

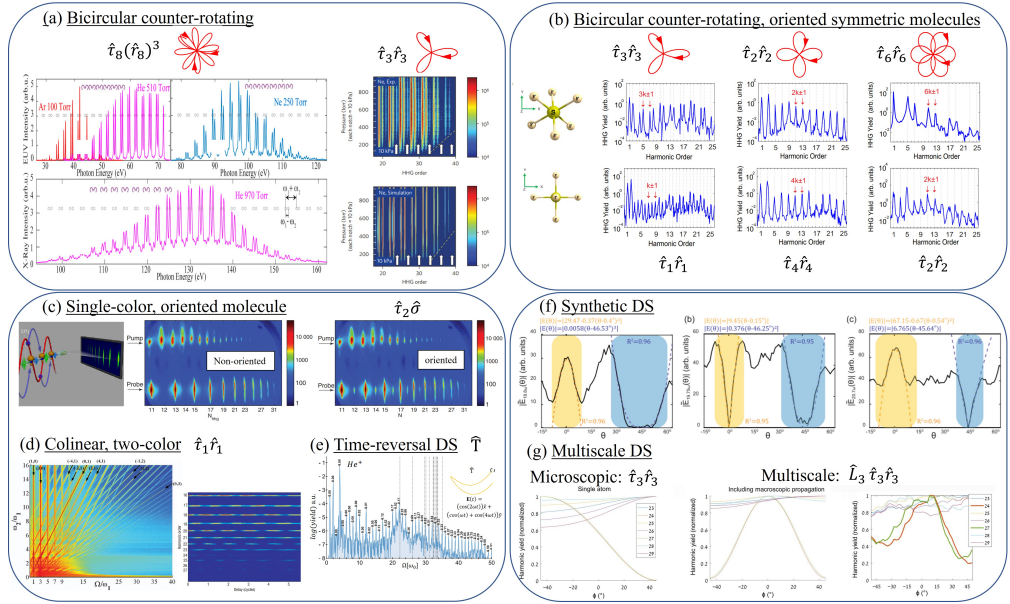

**Fig. 6. Dynamical symmetry induced selection rules in highly nonlinear optics.** In isotropic (atomic) media, the selection rules are dictated by the DS of the EM field alone. In (a) a two-color, bi-circular counter-rotating field of wavelengths (1.3µm, 0.79µm) gives rise to an extended rotational DS (figures repurposed from ref. [263] [88]). (b) In non-isotropic media, such as ensemble of oriented molecules, the order of DS is reduced (figure repurposed from ref. [264]). Oriented $SF_6$ molecules exhibit a four-fold rotational symmetry transverse to the $xy$ plane. A bicircular $\omega$-$m\omega$ EM field exhibits a rotational DS (of order $m+1$, 3-fold (left), 4-fold (middle), 6-fold (right)). Their jointed DS is only of order two, eliminating the forbidden harmonics selection rules. (c) Oriented heteronuclear gaseous media of point group $C_{\infty v}$ under linearly-polarized monochromatic driving yields both odd and even harmonics due to the existence of a mirror DS (figures repurposed from refs. [61] [265]). (d) Similarly, isotropic media driven by a collinearly-polarized two-color EM field yields similar allowed harmonics, but possesses a different DS (figures repurposed from refs. [87] [266]). (e) Time-reversal DS (Figure repurposed from ref. [35]). (f) Synthetic DS (figure repurposed from ref. [42]). (g) Multi-scale DS (Figure repurposed from ref. [36]). The relevant DS operators and Lissajous curves are indicated near corresponding plots.

We will now review the DS description of selection rules in NLO, focusing on HHG (both in the perturbative and non-perturbative regimes). We will start fully within the electric dipole approximation where the spatial DOF of the EM field are neglected, as well as interactions with light's magnetic components. We will assume that the electronic response plays the dominant role in the light-matter interaction, which is usually a good approximation for ultrafast and highly nonlinear physics. However, we emphasize that other entities might also play a role, e.g. phononic (or Raman) responses in certain regimes, spin-orbit interactions, Beyond-dipole terms, etc. Regardless of the origin of the main light-matter response, we will demonstrate below that the selection rules arise from an analysis of the symmetry group of the light-matter system's Hamiltonian, such that equivalent theoretical derivations can be systematically performed on other Hamiltonian terms to complement and extend the selection rules.

With these considerations in mind, let's assume the most general form of a system of quantum mechanical electrons (neglecting spin, as well as assuming frozen nuclei) within a molecular or solid potential, interacting with classical light within the dipole approximation. The many-body time-dependent Schrödinger equation (TDSE) governing the electronic wave function is given in atomic units and in the length-gauge [32] by:

$$i\frac{\partial}{\partial t}|\Psi(t)\rangle = \left[\sum_j\left(-\frac{\nabla_j^2}{2} + V(\mathbf{r}_j) + \mathbf{E}(t)\cdot\mathbf{r}_j\right) + \frac{1}{2}\sum_{j\neq i}\frac{1}{|\mathbf{r}_i - \mathbf{r}_j|}\right]\cdot|\Psi(t)\rangle \tag{36}$$

where $|\Psi(t)\rangle$ is a many-body electronic state, $\mathbf{r}_i$ is the coordinate of the $i$'th electron, $\nabla_j^2$ is the Laplacian with respect to $\mathbf{r}_j$, $V(\mathbf{r})$ is any potential energy term describing the field-free system, and we have included the full many-body electron-electron interaction term. Eq. (36) describes most harmonic generation and nonlinear optical phenomena generated by electronic responses of matter to light's electric field. In particular, let us emphasize that all phases of matter are captured by Eq. (36) with choosing an appropriate potential energy term, and that by allowing Eq. (36) to couple to Maxwell's equation for the classical light field (e.g. as an electronic current source term [2]), one captures all coherent stimulated light emission and absorption channels in the electronic sub-system. As such, by solving Eq. (36) one can determine which nonlinear optical effects should be observed.

Of course, this is a purely conceptual discussion, since in realistic systems Eq. (36) can hardly be solved, perhaps only for very low-dimensional systems and with no more than a few electrons. Approximated versions where various terms are neglected can of course be solved, but overall, the numeric paradigm of simply solving the dynamics directly is not effective for deriving general selection rules in NLO. Nonetheless, quantum mechanics allows us to infer a great deal of information about the solutions by simply analyzing the symmetry group of the underlying Hamiltonian. In the standard approach of NLO, this procedure is essentially performed for the field-free part, which we denote as $H_0$ (while the full Hamiltonian is denoted $H(t)$). It's noteworthy that the form of $H_0$ is universal up to the choice of potential $V(\mathbf{r})$. Since both the kinetic energy and electron-electron interaction terms are spherically symmetric, the resulting symmetry group of $H_0$ is determined by the symmetries of $V(\mathbf{r})$. We also highlight that having fixed the nuclei positions, $H_0$ is time-independent, making it symmetric under both continuous time-translations and time-reversal (having neglected spin). The light-matter term on the other hand is time-dependent and non-universal. It's symmetries can be altered by considering different forms of laser pulses as illustrated in previous sections.

Overall, the underlying symmetry group of the time-dependent Hamiltonian can be determined by separately analyzing the symmetries of $V(\mathbf{r})$ (as accomplished in standard approaches), analyzing the symmetries of the light-matter term, and performing an intersection. From a practical perspective, since only the light-matter term is time-dependent, the intersection of symmetry groups will reduce the symmetry elements in the group to those with spatial operations matching between the group of $H_0$ and the dynamical group of the light-matter term. For instance, if $V(\mathbf{r})$ describes a 6-fold symmetric Benzene molecule, and the laser is a circularly-polarized field with continuous rotational DS $\hat{r}_\infty \hat{t}_\infty$, then the intersection would reduce the symmetry to $\hat{r}_6 \hat{t}_6$ (as well as 2-fold and 3-fold elements that are subset and generated by powers of the generator $\hat{r}_6 \hat{t}_6$, see some illustrations of symmetry reduction in Fig. 6(b)). Let us further highlight a very crucial point – in the length gauge, the symmetries of the electric field's time-dependent polarization (analyzed in section 2.1.2) are directly mapped to spatial point-group symmetries of the light-matter term. This is a consequence of the dot product between the field polarization and the electronic position operator that couples each of the field's individual polarization components to the corresponding spatial position operator. Thus, analyzing symmetries of the time-dependent polarization of $\mathbf{E}(t)$ can be done on equal footing to point-group symmetries of $V(\mathbf{r})$. The implication of this mathematical relation is that the Floquet group description in section 2.1.2 applied in the polarization space of EM fields is directly applicable to light-matter Hamiltonians in the dipole approximation (this result holds in different gauges).

Having outlined the procedure by which the symmetry group of $H(t)$ is determined, we turn to the practical derivation of NLO selection rules. The derivation is based on the pioneering works in refs. [73] [74] in the context of Floquet theory [267], which has been extended in ref. [35] to include the full symmetry group considerations in the dipole approximation. At its heart, this derivation formally uncovers the symmetry-induced properties of the solutions to the TDSE and utilizes them to derive constraints in observable quantities such as light emission.

Let us begin by assuming that $\boldsymbol{E}(t)$ in Eq. (36) is time-periodic with a period $T$, in accordance with section 2.1.2. $H(t)$ is then also time-periodic with $H(t) = H(t + T)$, where $T$ is the minimal period. Consequently, the system can be described with Floquet theory, and we define the Floquet Hamiltonian:

$$\mathcal{H}_F = \left(H(t) - i\frac{\partial}{\partial t}\right) \tag{37}$$

The eigenstates of $\mathcal{H}_F$ are $T$-periodic functions, $|u_n(t)\rangle$, with corresponding Floquet quasi-energies, $\varepsilon_n$. Using these eigenstates and quasi-energies, a solution to the TDSE can be expanded in the form of:

$$|\Psi(t)\rangle = \sum_n a_n \exp(i\varepsilon_n t)|u_n(t)\rangle = \sum_n a_n |\phi_n(t)\rangle \tag{38}$$

where we can identify $|\phi_n(t)\rangle$ as Floquet states with amplitudes $a_n$. It is straightforward to substitute Eq. (38) back into Eq. (36) and show that it's a solution. We further assume for simplicity that $|\phi_n(t)\rangle$ are non-degenerate.

At this stage, we categorize the symmetry group of $H(t)$ to one of the classes of Floquet groups discussed in section 2.1.2, according to the procedure described above, which we denote as $G$. Let's assume that the DS $\hat{X}$ is an operator in $G$ such that $\hat{X} \in G$. That means it commutes with the Floquet Hamiltonian, $[\hat{X}, \mathcal{H}_F] = 0$. The resulting Floquet functions must therefore (due to inherent quantum mechanical considerations [29]) be simultaneous eigenmodes of $\mathcal{H}_F$ and $\hat{X}$. Given that $\hat{X}$ is either unitary or anti-unitary (if it involves time-reversal), its eigenvalues are roots of unity: $\hat{X}|u_n(t)\rangle = \exp(i\theta_n)|u_n(t)\rangle$, where $\theta_n$ is real. Let us now further make a crucial assumption which is that the system's state can be described by a single Floquet function, i.e. that for some index state $n = m$ we have $a_m = 1$, and for all other states we have $a_{n \neq m} = 0$. This physical condition can be obtained by a slow adiabatic ramping of the electric field that avoids population of Floquet superpositions. It is a typical (though not all-inclusive) condition in NLO. Within this assumption, the electronic wave function evolves as:

$$|\Psi(t)\rangle = \exp(i\varepsilon_m t)|u_m(t)\rangle \tag{39}$$

Let us then calculate an observable expectation value for a property governing NLO phenomena – the time-dependent induced electronic polarization in the medium, $\mathbf{P}(t)$:

$$\mathbf{P}(t) = \langle\Psi(t)|\hat{\mathbf{r}}|\Psi(t)\rangle = \langle u_m(t)|\hat{\mathbf{r}}|u_m(t)\rangle \tag{40}$$

In other words, since the total wave function is a pure Floquet state, its phase cancels out, leaving only the time-periodic part that is an eigenstate of $\hat{X}$ involved in the expression of $\mathbf{P}(t)$. Therefore, we can derive the resulting symmetries of the observable:

$$\mathbf{P}(t) = \langle u_m(t)|\hat{\mathbf{r}}|u_m(t)\rangle = \left\langle u_m(t)\left|\hat{X}^\dagger\hat{X}\hat{\mathbf{r}}\hat{X}^\dagger\hat{X}\right|u_m(t)\right\rangle = \hat{X}\cdot\mathbf{P}(t) \tag{41}$$

Note that in Eq. (41) we have interchangeably used the quantum mechanical notation of $\hat{X}$ that acts on operators (transforming as $\hat{X}\hat{\mathbf{r}}\hat{X}^\dagger$), to the vector notation compatible with the DS analysis of vectorial entities such as EM fields, or $\mathbf{P}(t)$, (which acts as a dot product on the vector). The main result is that $\mathbf{P}(t)$ upholds the symmetry $\hat{X}$, and that relation holds for any $\hat{X} \in G$, since we did not make any assumptions. Since $\mathbf{P}(t)$ is the induced polarization in the medium, any light emission due to acceleration of dipoles should be proportional to it.

We will soon show that the fact that $\hat{X}$ is a symmetry of $\mathbf{P}(t)$ directly associates to certain degeneracies (or constraints) that $\mathbf{P}(t)$ must uphold in the spectral domain, i.e. for the Fourier transformed function $\bar{\mathbf{P}}(\Omega)$. These are interpreted as selection rules in NLO. But before that, it's worth going back and reassessing the various assumptions made in the above derivation to understand its validity domains. Some assumptions are quite straightforward. For instance, we had assumed a particular form for the Hamiltonian, and that the dominant NLO response is electronic in nature. Those assumptions can in general be removed by simply performing a more extensive symmetry analysis that includes whatever Hamiltonian interaction terms are relevant for the particular process of interest. For instance, in the presence of coherent phonons one might imagine that the potential term becomes time-dependent, $V(\mathbf{r}) \rightarrow V(\mathbf{r}, t)$, which might reduce the symmetry group of the Hamiltonian [7] [268]. In such a case, a similar symmetry analysis holds, as long as the correct symmetry group of the full light-matter Hamiltonian is identified. The same approach can be applied to any other term of interest, e.g. interactions with magnetic field, electric quadrupole terms, spin-orbit couplings, etc. This is a strength of the formalism – it is quite simple to include analysis of other interaction terms.

At the same time, it's important to enumerate the less-straightforward assumptions that led to the result in Eq. (41). The most major assumption we have made is that Eq. (39) holds, that is, that the dynamics can be described with a single Floquet state. From the point of view of the driving field, this assumption is valid when the laser pulses are sufficiently long [269] (in the context of their minimal period, which can be different from the period defined by their carrier frequencies, e.g. when detuned frequencies or pulse trains are involved [86] [248] [249] [270]), and when the turn-on is sufficiently slow [269] (though in some other cases it has been argued Floquet phenomena can be turned on already within one laser cycle [271]). Otherwise, we cannot derive DS constraints on the induced polarization. At the same time, certain constraints might still apply if they arise from a purely spatial symmetry in the Floquet group, e.g. a mirror plane. Therefore, mirror symmetries that are not 'dynamical' induce selection rules also in the very short few-cycle regime [256] [272], and in certain pump-probe geometries where the total electric field might appear to not have any periodic behavior (which generally causes DS selection rules to break [11] [273]). From the point of view of the system, there are also considerations that go beyond the envelope of the driving field (that can also induce non-Floquet states [57]). For instance, occupations of multiple Floquet states can also arise near resonances of the system [274] [275] [276] [277], most dominantly near the ionization threshold in atomic and molecular systems (see supplementary information (SI) in ref. [35] for additional discussion). Moreover, the condition of describing the dynamics with a single Floquet state inherently breaks when that state is degenerate. This could be the case when the field-free Hamiltonian is initiated in a superposition of quantum states [278] [279] [280] [281], or even if the initial quantum state lacks certain symmetries of the Hamiltonian (e.g. atomic $p$-states that might not be fully isotropic). Degeneracy may also arise due to other considerations such as if the symmetry group $G$ is non-abelian (having generators that do not commute). In such a case, one cannot choose $|u_m(t)\rangle$ to be a simultaneous eigenstate of all of the generators in $G$, and only some of the symmetries of $G$ will translate to symmetries in $\mathbf{P}(t)$. Degeneracy also arises in groups that involve time-reversal as a consequence of Kramers theorem [282]. We should also note that phenomena such as HHG and photoemission are inherently described in open quantum systems that are non-Hermitian (since electrons effectively exit the system as they ionize), and therefore there is a certain degree of time-translation and time-reversal symmetry breaking by construction (because more and more electrons get emitted over time, meaning the dynamics slightly differ from cycle to cycle and the Floquet assumption is not exact). Similar effects arise even in solids which are formally closed, since the electronic excitation dynamics is asymmetric (excitation timescales are much faster than timescales for relaxation). This issue is especially prominent for time-reversal based DS, which are very prone to symmetry breaking (see further discussion in ref. [35]).

All of these issues might not necessarily lead to a breaking of associated selection rules if an isotropic ensemble of states is considered, as one typically has in a gas or liquid; but each case needs to be analyzed individually. Another implicit assumption we have made was to calculate specifically the observable $\mathbf{P}(t)$, which arises as the expectation value of the position operator. Generally, a similar derivation can be obtained for any generic observable $\hat{O}$, but the type of symmetry constraints that develop would depend on the structure of that operator.

Essentially, because $\hat{\mathbf{r}}$ is linear in position, it is parity-odd, and it transforms like the point-group symmetries of $\mathcal{H}_F$. Hence, $\mathbf{P}(t)$ directly complies with the symmetries in the group $G$. However, if we consider a parity-even operator such as $\hat{r}^2$ that statement changes, since $\hat{r}^2$ transforms differently under point group operations [29]. In fact, in this particular example $\hat{r}^2$ is spherically symmetric, meaning it is invariant under any point-group operations, and the operator would only transform under the temporal part (we can separate the spatial and temporal parts of a DS as: $\hat{X} = \hat{X}_r \hat{X}_t$, and denote $\hat{X}_t$ as the temporal part and $\hat{X}_r$ as the spatial part associated with the point-group operation). Then we should have $\hat{X}_t \cdot \langle r^2(t) \rangle = \langle r^2(t) \rangle$. Such an analysis can be performed for all observables of interest and possible symmetry operators.

With this in mind, we are in a good position to derive the selection rules associated with symmetries of $\mathbf{P}(t)$. Let us decompose $\mathbf{P}(t)$ into a Fourier series and obtain an eigenvalue problem for its polarization components. Here we assume time-periodicity in $\mathbf{P}(t)$, but this is anyways assumed for the Floquet approach we used to formulate the DS relations $\mathbf{P}(t) = \hat{X} \cdot \mathbf{P}(t)$. The only symmetries that do not require this assumption are the pure spatial point group operations – these can be derived more generally with a full Fourier transform, but practically, that approach would lead to the same result as in the periodic case. Overall, in the spectral domain we can expand the induced polarization:

$$\mathbf{P}(t) = \sum_n \mathbf{F}_n \exp\left(in\frac{2\pi}{T}\right) \tag{42}$$

where $\mathbf{F}_n$ are the amplitudes of the different time-harmonic modes that comprise $\mathbf{P}(t)$. Note that since $\mathbf{P}(t)$ is real (as is any Hermitian physical observables of $|\Psi(t)\rangle$), its Fourier components uphold the inherent constraint $\mathbf{F}_n = \mathbf{F}_{-n}^*$. At the next stage we can re-write the symmetry relation in the Fourier domain:

$$\hat{X} \cdot \sum_n \mathbf{F}_n \exp\left(in\frac{2\pi}{T}t\right) = \sum_n \mathbf{F}_n \exp\left(in\frac{2\pi}{T}t\right) \tag{43}$$

, and separate the spatial and temporal parts of the operator $\hat{X}$:

$$\sum_n \hat{X}_r \cdot \mathbf{F}_n \cdot \hat{X}_t \cdot \exp\left(in\frac{2\pi}{T}t\right) = \sum_n \mathbf{F}_n \exp\left(in\frac{2\pi}{T}t\right) \tag{44}$$

Since the temporal and spatial parts of $\hat{X}$ commute, the temporal part can directly act on the oscillatory exponent, and the spatial part on the vector in polarization space. In order for Eq. (51) to be upheld, every single Fourier coefficient on both sides of the equation must equal individually. As the temporal operator cannot change the mode's frequency (the number $n$ in the exponents in Eq. (44)), we can directly write an eigenvalue problem for $\mathbf{F}_n$, where $n$ is the index of the $n$'th harmonic mode:

$$\hat{X}_r \cdot \mathbf{F}_n \cdot \hat{X}_t \cdot \exp\left(in\frac{2\pi}{T}t\right) = \mathbf{F}_n \exp\left(in\frac{2\pi}{T}t\right) \tag{45}$$

At this point, we should note that if $\hat{X}_t$ involves only temporal translations, it can only lead to a phase shift in the exponent such that the main time-dependent oscillation will cancel out from both sides of the equation, leaving an eigenvalue problem for $\mathbf{F}_n$ of the type:

$$\hat{X}_r \cdot \mathbf{F}_n = \exp(i\phi_n)\mathbf{F}_n \tag{46}$$

where the phase $\phi_n$ arises from the action of $\hat{X}_t$.

Let us consider a concrete example, e.g. that $\hat{X} = \hat{r}_3 \hat{t}_3$ is a 3-fold rotational DS as discussed in the previous sections and expressed in Eq. (6) (see examples in Fig. 6(a,b)). In this case the point-group part $\hat{X}_r$ is a 3-fold rotation, and the phases $\phi_n$ are 3rd order roots of unity that arise from $\hat{t}_3$. Then, the only non-trivial (i.e. nonzero) amplitudes $\mathbf{F}_n$ that physically solve

Eq. (46) are modes with indices $3n \pm 1$ for integer $n$ for the components transverse to the polarization plane, and those modes are eigenstates of left/right CP, respectively (because those are eigenstates of the rotational operator). In other words, if one attempts to solve the equation for instance with $n = 6$, which is a multiple of 3, vectors with nonzero amplitudes in the plane transverse to the rotation axis that can solve the equation simply do not exist. Thus, emission of $3n$ harmonic orders is forbidden in HHG (Fig. 6(a), right pannel). This is the well-known selection rule that led to the emission of circularly-polarized harmonics driven by $\omega$-$2\omega$ fields [85] [84] [85] [86], and it also appears in some other cases [74] [82] [283]. Along the axis of the rotational symmetry however (denoted here for simplicity as the $j$'th axis), one can show that a trivial condition arises of the type $F_{n,j} = F_{n,j}$, for which any choice is a solution, which we practically interpret as a situation without selection rules for the induced polarization along the rotational DS axis.

On the other hand, if $\hat{X}_t$ also involves a time-reversal operation, then that operation flips the arrow of time on the left-hand side of Eq. (45). Practically, that takes $\mathbf{F}_n \rightarrow \mathbf{F}_{-n}$ on the left-hand side, which can be combined with the general property $\mathbf{F}_n = \mathbf{F}_{-n}^*$ that is anyways upheld. Thus, DS with time-reversal can lead to complex generalized eigenvalue equations that also involve complex conjugation. Let us again consider an example where the DS is a pure time-reversal operator: $\hat{X} = \hat{T}$. Then Eq. (45) reduces to $\mathbf{F}_n = \mathbf{F}_{-n} = \mathbf{F}_n^*$. In other words, the vector must be its own conjugate, making it real. This may not sound like a big deal, but it's still a symmetry-induced selection rule. Let's attempt to further analyze its physical consequences, keeping in mind that $\mathbf{F}_n$ is already a Fourier transform of a purely real quantity, and there is no particular reason that it must be real. Indeed, that would inherently set the absolute phase of all three of its polarization components to zero. Of course, we remember there could be an overall absolute phase factor on top of the current selection rule, such that this constraint doesn't in fact set the phase of $\mathbf{P}(t)$ to zero. On the other hand, it does dictate the relative phase between all of its polarization components and all of its harmonics to zero, meaning that every harmonic order emitted in HHG should be in phase with all of the others, for every polarization direction. This is already a measurable property, because in-phase polarization components are linearly-polarized. Note though that having time-reversal symmetry in $\mathbf{E}(t)$ does not assume pure linear polarization, and so we have obtained a slightly more interesting selection rule – even if the driving field has polarization components spanning multiple axes in space, as long as it upholds time-reversal symmetry, emitted harmonics should be linearly-polarized (see Fig. 6(e) showing linear harmonic emission from a field polarized in 2D space exhibiting $\hat{T}$). Selection rules arising from time-reversal-related operators are especially intriguing, because they generally cannot be derived from the perturbative theory described in section 3.1.

At this point, we could go on and systematically derive all of the different selection rules associated with every possible DS. Table 2 summarizes these results (for derivation of each selection rule we refer readers to ref. [35]). These symmetries can be practically used either for controlling the spectral properties of harmonics, or for spectroscopic purposes, as will be later expanded on in detail.

**Table 2.** DSs and their associated selection rules in harmonic generation (from gas, liquid and solid media) in general three polarization dimensions and within the dipole approximation.

| Row # | Symmetry | Order | Harmonic generation selection rules |
|---|---|---|---|
| 1 | $\hat{r}_n$ | $n$ | Only allowed harmonics polarized along the rotational symmetry axis (this is a purely spatial symmetry that can only arise in freely propagating fields with only one nonzero polarization axis) |
| 2 | $\hat{\sigma}$ | 2 | Harmonics transverse to the mirror plane are forbidden (this is a purely spatial symmetry that can only arise in freely propagating fields with up to two nonzero polarization axes) |
| 3 | $\hat{T}\hat{\sigma}, \hat{T}\hat{t}_2\hat{\sigma}$ | 2 | The polarization ellipsoid has a major/minor axis normal to the reflection plane. |

| 4 | $\widehat{T}\hat{r}_2,\ \widehat{T}\hat{t}_2\hat{r}_2$ | 2 | The rotation axis is a major/minor axis of the polarization ellipsoid. |
|---|---|---|---|
| 5 | $\hat{t}_2\hat{\sigma}$ | 2 | Odd harmonics are polarized linearly and orthogonally to the reflection plane, only even harmonics allowed polarized within the reflection plane |
| 6 | $\hat{t}_2\hat{r}_2$ | 2 | Odd-only harmonics in any polarization are allowed polarized in the plane orthogonal to rotation axis, any harmonic emission is allowed polarized parallel to the rotation axis. |
| 7 | $\widehat{T},\ \widehat{T}\hat{\imath},\ \widehat{T}\hat{t}_2\hat{\imath}$ | 2 | Linearly polarized harmonics only. |
| 8 | $\hat{t}_2\hat{\imath}$ | 2 | Odd-only harmonics in any polarization. |
| 9 | $\hat{t}_n(\hat{r}_n)^m$ | $n>2$ | ($\pm$) circularly polarized ($nq \mp m$) harmonics, $q \in \mathbb{N}$, within the plane orthogonal to the rotation axis. Linearly polarized $nq$ harmonics are also allowed, but polarized parallel to the rotation axis. |
| 10 | $\hat{t}_n(\hat{e}_n)^m$ | $n>2$ | ($\pm$) elliptically polarized ($nq \mp m$) harmonics, $q \in \mathbb{N}$, with an ellipticity $b$ within the plane orthogonal to rotation axis. Linearly polarized $nq$ harmonics are also allowed, but polarized parallel to the rotation axis. |
| 11 | $\hat{t}_{2n}\hat{\sigma}_h(\hat{r}_{2n})^m$ | $2n>2$ | ($\pm$) circularly polarized ($2nq \mp m$) harmonics, $q \in \mathbb{N}$, within the plane orthogonal to the improper rotation axis ('$h$' denotes that plane). $n(2q+1)$ harmonics are also allowed, but polarized parallel to improper rotation axis. |
| 12 | $\hat{t}_{2(n+1)}\hat{\sigma}_h(\hat{r}_{2n+1})^m$ | $2(2n+1)>2$ | ($\pm$) circularly polarized $2q(2n+1) \mp 2m$ harmonics, $q \in \mathbb{N}$, with an ellipticity $b$ within the plane orthogonal to the improper rotation axis ('$h$' denotes that plane). $(2n+1)(2q+1)$ harmonics are also allowed, but polarized parallel to improper rotation axis. |
| 13 | $\hat{t}_{2n}\hat{\sigma}_h(\hat{e}_{2n})^m$ | $2n>2$ | ($\pm$) elliptically polarized ($2nq \mp m$) harmonics, $q \in \mathbb{N}$, with an ellipticity $b$ within the plane orthogonal to the improper rotation axis ('$h$' denotes that plane). $n(2q+1)$ harmonics are also allowed, but polarized parallel to improper rotation axis. |
| 14 | $\hat{t}_{2(n+1)}\hat{\sigma}_h(\hat{e}_{2n+1})^m$ | $2(2n+1)>2$ | ($\pm$) elliptically polarized $2q(2n+1) \mp 2m$ harmonics, $q \in \mathbb{N}$, with an ellipticity $b$ within the plane orthogonal to the improper rotation axis ('$h$' denotes that plane). $(2n+1)(2q+1)$ harmonics are also allowed, but polarized parallel to improper rotation axis. |

The next order of business would be to ask what happens if the light-matter system exhibits multiple DS, i.e. the Hamiltonian belongs to a group with several symmetries. This situation is quite common in NLO, because light waves tend to be highly symmetric. In this case, one can prove (see the SI in ref. [35]) that it is sufficient to only derive selection rules arising from the generators of the symmetry group of $\mathcal{H}_F$.

Lastly, let us go back for a moment and analyze an exemplary observable that transforms differently than $\hat{\mathbf{r}}$, such as $\hat{r}^2$ discussed above. As we have shown, in that case only the temporal part of $\hat{X}$ matters, and we should have the following eigenvalue problem:

$$\hat{X}_t \cdot F_n = F_n \tag{47}$$

Let's consider one of the most ubiquitous symmetries, $\hat{r}_2\hat{t}_2$, which is a 2-fold rotational DS that any linearly-polarized monochromatic wave exhibits. In this case $\hat{X}_t$ is a temporal translation by $T/2$. Substituting in Eq. (47) leads to $F_n = -F_n$ for odd values of $n$, but to $F_n = F_n$ for even values of $n$. That is, only even harmonic components are allowed to develop in $\langle r^2(t)\rangle$ due to the 2-fold rotational DS in the Hamiltonian. This is the exact opposite of the result obtained for the induced polarization that leads to odd-only harmonics (see Table 2). The discrepancy is due to the different transformation natures of the operators. Note that in this case since $\hat{r}^2$ is a scalar, the equations do not involve vector operations and are inherently simpler. Such derivations can be quite useful, for instance for deriving selection rules for other quantities of interest in nonlinear optics such as the system's time-dependent ionization rate, quadrupole emission [20], or its magnetization [48].

We should also address another important point, which is selection rules for ensembles. Consider for example a gas of randomly oriented water molecules (each of a $C_{2v}$ point group [29], exhibiting mirror symmetries). Even though each molecule belongs to $C_{2v}$ group, the gas as a whole is of a higher symmetry, because it consists of an ensemble average of all randomly oriented molecular states (another way to put it is that the system is described by a quantum mechanical partition function, rather than a single microscopic Hamiltonian). In such a case one need only consider the symmetry group of the ensemble as a whole, and the individual constituent symmetry is washed out (although the properties of the molecules will still be expressed in the NLO response). This situation is very common in gas-phase experiments, because pre-oriented molecules require pump-probe laser geometries [61] (see illustration in Fig. 6(c)). It is also the typical situation in liquids [284] [285] [286] [287] [288] [289] [290] [291], and in amorphous solids [292] [293]. Single crystals however, are a more prevalent example where the symmetries of the medium always play a major role in NLO selection rules [4] [272] [294].

### 3.2.2 Above-threshold ionization

To complement the discussion above relating to HHG selection rules, similar selection rules for other NLO observables can be obtained. One such interesting case is ionization phenomena, leading to above threshold ionization (ATI) spectra and photoelectron momentum-resolved spectra (PES). At the rigorous level of a full group theory, these selection rules have yet to be derived. However, several ad-hoc cases have been discussed, for instance for fields with bi-chromatic components [295] [296] [297] [298] [299] [300] [301], for double-ionization in Helium [302], for backscattered photoelectrons [303], and for photoelectron circular dichroism (PECD) where several ad-hoc cases have been analyzed [216] [304] [305] (see Fig. 7(c-e)).

Among these various cases, one often obtains pronounced symmetric features in photoemission spectra that are inherited from the laser-field DS, or the molecular symmetry (see illustrations in Fig. 7(b)). These features can be used for applications such as shaping electron vortex beams (Fig. 7(a)), as well as for analyzing symmetry breaking components for ultrafast spectroscopy.

We would highlight that as the field evolves, we expect rigorous symmetry theories for ATI and PES driven by tailored-light to arise, which takes into account the full Floquet group of the light-matter system (just as for HHG discussed above).

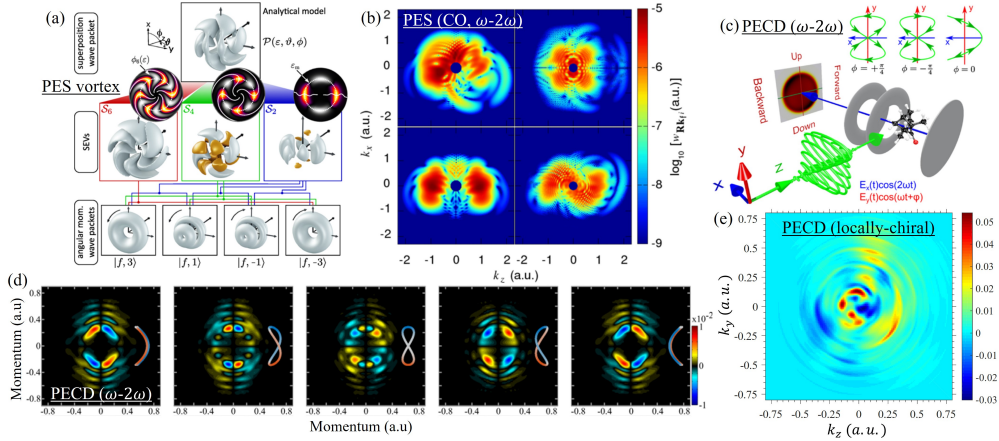

**Fig. 7. Symmetries and selection rules arising in photoemission spectra (PES) and photoelectron circular dichroism (PECD).** (a) Generation of photoelectron vortices using light with rotational dynamical symmetries, which imprints rotational symmetries onto the electron vortices. Figure repurposed from ref. [301]. (b) Rotational and mirror symmetries (selection rules) arising in PES form CO molecules driven by tailored $\omega$-$2\omega$ bi-chromatic fields. Figure repurposed from ref. [297]. (c) Schematic of PECD driving with $\omega$-$2\omega$ light exhibiting mirror dynamical symmetry, generating up/down asymmetries in PECD shown in (d) vs. the driving pulse relative phase. Figures repurposed from refs. [304] and [216], respectively. (e) PECD driven by locally-chiral light, showing a breaking of all PECD selection rules (e.g. lack of forwards/backwards asymmetry). Figure repurposed from ref. [305].

### 3.2.3 Multi-scale DS

To complement section 3.2.1, a full selection rules derivation can also be obtained in the multi-scale DS case. This analysis is a little more mathematically cumbersome, which is a result of the high dimensionality of the theory. However, the logic is the same as in the microscopic analysis. We briefly explain the procedures for the derivation with some exemplary cases, while readers are referred to ref. [36] (including the SI there) for a full formulation.

In order to understand what constraints multi-scale DS lead to in NLO responses, we need to follow a procedure that is similar to the microscopic analysis, but also involves spatial DOF. The induced polarization is expanded into a series of harmonics in both space and time, and the symmetry structure of the field is assumed to be inherited by the induced polarization. In the microscopic DS case this assumption was proven and validated for cases where a single Floquet state is excited adiabatically, allowing the derivation of selection rules (in section 3.2). Naturally, the same constraints are required here, since microscopic operations are part of the multi-scale DS manifold. However, due to the spatial structure of multi-scale DS we must further assume that the driving field and emitted fields are phase matched in order for the symmetry of the drive to transfer onto harmonics. This is immediately upheld in thin optical media (as well as in vacuum), but is not the general case. Nonetheless, this assumption is crucial for any interesting spatially-symmetric structure in the field – if the HHG emission and field are not phase-matched then the harmonics being propagated 'see' a field with broken symmetry, since the phase-mismatch will cause it to propagate in a different velocity and experience time- and space-shifted driving field.

Upon employing these assumptions, general tables connecting multi-scale DS to selection rules can be derived (see ref. [36]). Let us discuss a few noteworthy examples. To start, in ref. [36] it was shown that $3n$ harmonics (for integer $n$) are forbidden in a case where the driving field exhibited a $\hat{\tau}_3\hat{r}_3\hat{L}_{z,3}$ multi-scale symmetry (generated by a combination of a gaussian bi-chromatic $\omega$ - $2\omega$ beam and a Bessel beam, see Fig. 6(g)). Generically, measurements of HHG driven by beams with OAM can also be analyzed with the multi-scale DS approach, which will yield the OAM conservation laws that have been measured [152] [153] [154] [155] [156] [157] [158] [37] [39] [159] [38], including conservation of so-called torus-knot angular momentum [39] [38] (see Fig. 2(e)). Some more interesting cases are HHG driven by vector beams that couple spatial and polarization DOF [43] [306] [104] [307] [308] [179]. Notably, in materials that are not isotropic (e.g. in solids [104]) the multi-scale symmetry theory can be ideally employed to analyze results also without invoking conservation laws, see Fig. 11(g)).

We will end by stating that we expect the application of multi-scale DS and induced selection rules to become more prominent as a theoretical technique for analyzing experiments as the field evolves, and more complex light forms are employed in ultrafast light-matter interactions.

### 3.2.4  Reverse engineering fields with desired symmetry

One potentially applicative use of the selection rules theory (apart from spectroscopic applications that will be discussion in section 4) is its use for engineering light fields with a desired symmetry group. To put differently, up until this stage with have immersed ourselves in the various symmetry and asymmetry properties of EM fields, and how those can affect NLO responses; but we have not stopped to ask how one can generate such symmetry properties to begin with. From a conceptual point of view the problem is substantial – imagine having arbitrary access to a laser source with fully tunable wavelength, polarization, and spatial properties – how would you choose parameters to generate a specific symmetry in the EM wave?

In this section we will show that by inversely applying the selection rules derived above, at least for the microscopic theory within the dipole approximation, one can uncover the various constraints that the different laser parameters must uphold in order to exhibit a particular symmetry.

Let us consider the benchmark example of the 3-fold rotational DS that we previously discussed (Eq. (6)). The selection rules for HHG driven by a field with such symmetry in atomic isotropic media permit only the generation of $3n \pm 1$ harmonics (for integer $n$), where the harmonics are circularly polarized with alternating helicities (see row 9, Table 2). It's important to note here that since the medium is fully isotropic, the selection rule is essentially determined by the DS of the laser field. Imagine now attempting to construct a poly-chromatic field that upholds such a DS – It must comply to its own selection rules, because the medium does not contribute any significant symmetry information. Thus, any combination of carrier waves with frequencies $(3n \pm 1)\omega_0$ which have the correct polarization and helicity will suffice. Of course, the $\omega$-$2\omega$ counter-rotating circularly-polarized fields comply to this relation, with only two nonzero modes. However, other choices also can be made. For instance, a monochromatic circularly-polarized laser field (i.e. with only one nonzero mode) trivially exhibits a 3-fold DS. This is because such a field exhibits a continuous rotational DS $\hat{r}_\infty \hat{t}_\infty$, such that $\hat{r}_3 \hat{t}_3$ is also a part of the symmetry group. On the other end, one could make things more complicated, e.g. by adding a $4\omega_0$ on top of the usual $\omega_0$-$2\omega_0$ counter-rotating configuration. In that case the $4\omega_0$ must be circularly-polarized and co-rotating with the $\omega_0$ mode. Note that the relative phase between the modes is not necessarily fixed to any unique value, because that's not a selection rule arising from $\hat{r}_3 \hat{t}_3$. Indeed, let's consider an example where the $\omega$-$2\omega$ waves are in phase, but the $4\omega$ mode is phase shifted by $\pi/2$:

$$\mathbf{E}(t) = E_0 \mathrm{Re}\{e^{i\omega t}\hat{\mathbf{e}}_\mathbf{R} + e^{2i\omega t}\hat{\mathbf{e}}_\mathbf{L} + e^{4i\omega t + i\pi/2}\hat{\mathbf{e}}_\mathbf{R}\} \tag{48}$$

where $\hat{\mathbf{e}}_{\mathbf{R/L}}$ are right/left CPL eigenmodes. While it may not be initially clear from Eq. (48) that this electric field exhibits a 3-fold DS, one can show this analytically and also by illustrating the field's time-dependent polarization (its Lissajous curve, see Fig. 2(a)). From the reverse engineering perspective, it is enough to note that the spectral content of the field exactly follows the desired selection rule – it contains only $(3n \pm 1)\omega_0$ modes with alternating circular components. One can repeat this approach for any desired symmetry.

It is worth noting is that with this minimal technique the resulting EM field is not guaranteed to contain only the desired symmetry operator. Indeed, the symmetry group might contain other operations (like the example of the monochromatic CPL above). It is however possible to follow a similar approach in order to forbid the appearance of certain symmetries in the EM field. Taking the benchmark case above of $\hat{r}_3\hat{t}_3$, one could ask what is the particular choice of laser frequencies and polarization that must be employed in order to have a symmetry group with $\hat{r}_3\hat{t}_3$ as the single generator. In that case we should go over the full Table 2 and make sure that our chosen polarizations and frequencies only uphold the selection rule from the symmetry $\hat{r}_3\hat{t}_3$, while all other selection rules are not upheld. For example, the tri-circular field in Eq. (48) does not uphold DS $\hat{\partial}\hat{T}$, which is part of the DS group of the $\omega$-$2\omega$ counter-rotating circularly-polarized fields. That is, by adding an additional $4\omega$ we made sure that $\hat{\partial}\hat{T}$ is no longer part of the field's symmetry group. This is because the additional frequency component in this case was chosen to be phase shifted from those at $\omega$-$2\omega$, making sure that the desired selection rule of $\hat{\partial}\hat{T}$ is broken. Indeed, by setting the relative phase to zero, $\hat{\partial}\hat{T}$ DS can be restored. In that respect, the phase shift in this tri-circular case plays a crucial role in controlling the symmetry of the time-dependent polarization. This approach is exactly the one that was employed in section 2.3.3 to derive the conditions in which one can have locally-chiral light [242], i.e. light for which the symmetry group fully excludes operators with improper-rotations, reflections, and inversion.

As a whole, the technique is highly useful, because it guides to a direct course of action for engineering the symmetry properties of EM fields, even in very complex cases. It also makes clear which DOF need tuning in order to smoothly alter an EM field's symmetry group.

### 3.3 Selection rules for synthetic dynamical symmetries

In this subsection, we review the concept of selection rules for breaking selection rules [41] (i.e. arising from DS in synthetic dimensions). As we discussed above, synthetic DS typically arise when a standard DS is broken by an additional perturbation. If the DS is broken, deviations from the selection rules that it enforces are anticipated, and the deviation is expected to be proportional to the perturbation strength. For instance, if the medium undergoes a transition from isotropic to a non-isotropic state, it might not uphold inversion symmetry, leading to generation of even harmonics.

This can also occur if the perturbation is itself a time-dependent field, e.g. if HHG is driven by a bright field of frequency $\omega$, superimposed with a perturbative field of frequency $2\omega$, both of which are linearly polarized along the same axis. The $2\omega$ field breaks the $2^{\text{nd}}$ order rotational DS that usually leads to the generation of even harmonics. However, if the perturbation is sufficiently weak, the broken DS is still expected to hold a significance in the properties of the emitted light. Indeed, upon the application of an external symmetry breaking perturbation, the broken DSs may impose "selection rules for breaking selection rules". Selection rules for breaking selection rules are propensity rules that restrict the scaling of such deviations with the strength (amplitude) of the symmetry breaking perturbation. They may be imposed jointly by the broken symmetry, and the symmetry breaking perturbation, through a synthetic DS (section 2.4).

We consider the general situation where a periodically driven system with Hamiltonian $H(t) = H(t + T)$ exhibits a DS $\hat{X}$ due to interaction with a pump optical field with fundamental frequency $\omega$. Consequently, the emission of the system is subject to selection rules. When $H(t)$ is perturbed by a perturbation $\lambda W(t) = \lambda W(t + T)$, the optical emission exhibits selection rule deviations. For example, $H(t)$ is perturbed by an electric field of amplitude $\lambda$, polarization $\boldsymbol{p} = (p_x, p_y)$ and frequency $s\omega = 2\pi s/T$ ($s$ being a rational number), the perturbation is given by

$$\lambda W(t) = \lambda \Re\{\mathbf{p} \cdot \hat{\boldsymbol{\mu}} e^{is\omega t}\} \tag{49}$$

in which $\hat{\boldsymbol{\mu}}$ is the electric dipole moment operator such that:

$$\mathbf{p} \cdot \hat{\boldsymbol{\mu}} = p_x x + p_y y \tag{50}$$

where, $x, y$ are cartesian spatial coordinates. The $\Omega$ frequency component of the emission can be expanded by Floquet perturbation theory [309]:

$$\tilde{\mathbf{E}}(\Omega, \lambda) = \tilde{\mathbf{E}}_0(\Omega) + \lambda \tilde{\mathbf{E}}_1(\Omega) + \lambda^2 \tilde{\mathbf{E}}_2(\Omega) \qquad (51)$$

The term $\tilde{\mathbf{E}}_0(\Omega)$ complies with any selection rules imposed by the DS of the unperturbed system, which are discussed above. The new contributions $\tilde{\mathbf{E}}_1(\Omega), \tilde{\mathbf{E}}_2(\Omega)$ are a result of the perturbation, inducing selection rule deviations that scale linearly and quadratically with the perturbation strength, respectively. Analyzing the explicit structure of $\tilde{\mathbf{E}}_1(\Omega)$ and $\tilde{\mathbf{E}}_2(\Omega)$ obtained by Floquet perturbation theory, one finds that $\tilde{\mathbf{E}}_1(\Omega)$ and $\tilde{\mathbf{E}}_2(\Omega)$ exhibit selection rules that are imposed by $\hat{X}$ (the broken DS). They are tabulated in Table 3. These selection rules are termed 'selection rules for breaking selection rules' because they constrain the scaling of selection rule deviations as a DS breaks by an external perturbation.

**Table 3.** Selection rules for breaking selection rules by a linearly polarized monochromatic laser. The broken symmetry imposes universal selection rules on $\tilde{\mathbf{E}}_{0,1,2}(n\omega)$, which also depend on the polarization (**p**) and frequency ($s\omega$) of the perturbation. The selection rules for $\tilde{\mathbf{E}}_0(n\omega)$ were tabulated in ref. [35] and the selection rules for $\tilde{\mathbf{E}}_{1,2}$ are derived in ref. [41].

| Symmetry | Selection rules & selection rules for breaking selection rules | | |
|---|---|---|---|
| | *Time-reversal symmetries* | | |
| $\hat{T}$ | $\tilde{\mathbf{E}}_0(n\omega) \in \mathbb{R}^2$ | $\tilde{\mathbf{E}}_1(n\omega) \in \mathbb{R}^2$ | $\tilde{\mathbf{E}}_2(n\omega) \in \mathbb{R}^2$ |
| $\hat{t}_2\hat{r}_2$ | $\tilde{\mathbf{E}}_0(n\omega) \in i\mathbb{R}^2$ | $\tilde{\mathbf{E}}_1(n\omega) \in \mathbb{R}^2$ | $\tilde{\mathbf{E}}_2(n\omega) \in i\mathbb{R}^2$ |
| $\hat{T}\hat{t}_2\hat{r}_2$ | $\tilde{\mathbf{E}}_0(n\omega) \in i^{n+1}\mathbb{R}^2$ | $\tilde{\mathbf{E}}_1(n\omega) \in i^{n+s}\mathbb{R}^2$ | $\tilde{\mathbf{E}}_2(n\omega) \in i^{n+1}\mathbb{R}^2$ |
| | *Dynamical reflection symmetries* | | |
| $\hat{T}\hat{\sigma}_y$ | $\tilde{E}_{0x}, i\tilde{E}_{0y} \in i\mathbb{R}$ | $\tilde{E}_{1x}, i\tilde{E}_{1y}(n\omega) \in p_x\mathbb{R}$ $+ ip_y\mathbb{R}$ | $\tilde{E}_{2x}(n\omega), i\tilde{E}_{2y}(n\omega) \in ip_x^2\mathbb{R} + ip_y^2\mathbb{R}$ $+ p_xp_y\mathbb{R}$ |
| $\hat{T}\hat{t}_2\hat{\sigma}_y$ | $\tilde{E}_{0x}(n\omega), i\tilde{E}_{0y}(n\omega)$ $\in i^{n+1}\mathbb{R}$ | $\tilde{E}_{1x}(n\omega), i\tilde{E}_{1y}(n\omega)$ $\in i^{n+s}p_x\mathbb{R} + i^{n+s+1}p_y\mathbb{R}$ | $\tilde{E}_{2x}(n\omega), i\tilde{E}_{2y}(n\omega) \in i^{n+1}p_x^2\mathbb{R}$ $+ i^{n+1}p_y^2\mathbb{R}$ $+ i^n p_xp_y\mathbb{R}$ |
| $\hat{t}_2\hat{\sigma}_y$ | $\tilde{\mathbf{E}}_0(n\omega)$ $\in \begin{pmatrix} 0 & 1 \\ 1 & 0 \end{pmatrix}^n \begin{pmatrix} 1 & 0 \\ 0 & 0 \end{pmatrix}\mathbb{C}^2$ | $\tilde{\mathbf{E}}_1(n\omega)$ $\in \begin{pmatrix} 0 & 1 \\ 1 & 0 \end{pmatrix}^{n+s} \begin{pmatrix} p_x & 0 \\ 0 & p_y \end{pmatrix}\mathbb{C}^2$ | $\tilde{\mathbf{E}}_2(n\omega) \in$ $\begin{pmatrix} 0 & 1 \\ 1 & 0 \end{pmatrix}^n \begin{pmatrix} p_xp_y & 0 \\ 0 & p_x^2 + \alpha p_y^2 \end{pmatrix}\mathbb{C}^2$ ; $\alpha \in \mathbb{C}$ |
| | *Rotational & elliptical dynamical symmetries* | | |
| $\hat{\tau}_n(\hat{r}_n)^m$ | $\tilde{\mathbf{E}}_0(n\omega) \neq 0$ for | $\tilde{\mathbf{E}}_1(n\omega) \neq 0$ for | $\tilde{\mathbf{E}}_2(n\omega) \neq 0$ for |
| $\hat{\tau}_n(\hat{e}_n)^m$ | $n = Nm \pm M$ | $n = Nm \pm (1 \pm 1)M \pm s$ | $n = Nm \pm (2 \pm 1)M \pm (1 \pm 1)s$ |

We note that selection rules for breaking selection rules are imposed by the real-synthetic symmetries outlined in section 2.4. Hence, the anticipation of selection rules for breaking selection rules arises whenever a real-synthetic symmetry can be identified within a DS-broken system at hand. It is important to also mention that while Table 3 outlines these selection rules up to $2^{nd}$ order in the strength of the perturbation ($\lambda$), real-synthetic symmetries enforce selection rules for breaking selection rules to all orders in the strength of the perturbation. This concept is further elaborated in reference [42], which also provides experimental evidence for this phenomenon (see Fig. 6(f)). In general terms, synthetic DS can be viewed as imposing scaling laws for symmetry breaking, which are fundamentally determined by internal symmetry relations between the original and perturbed Hamiltonians. They outline a clear path for analyzing experiments in various situations, and engineering controlled symmetry breaking.

### 3.4 Photonic conservation laws in parametric NLO

In the sections above we took the path of symmetry-induced selection rules as the main tool to analyze constraints on allowed/forbidden light-matter interaction pathways. However, there is a complementary point of view where instead conservation laws are employed. As expected, for the most part, these two views overlap and can give two physical pictures explaining the same phenomena. This is obvious if one recalls that mathematically conservation laws in fact arise from continuous symmetries through Noether's theorem. However, at times the two pictures can give different perspectives that do not necessarily agree, in which case, as we will argue below, the symmetry picture is somewhat more general.

Let us begin by explaining in more detail the conservation law perspective, and how it can be used to explain several well-known results in nonlinear optics. For parametric nonlinear optical phenomena, one assumes that the medium ends up in the same state as it started in prior to light irradiation (e.g. electronic, vibronic, etc.). In HHG for instance, the electron is assumed to start in the ground state, to which it also returns after a process of recombination. It follows that for parametric phenomena there is no exchange of conserved quantities with the medium [34] [2]. That is, one can 'count' conserved quantities in-between only the involved photons. In this manner we can employ various conservation laws on the photonic side and formulate what constraints they produce for NLO. As an example, let's consider one of the more straightforward constraints in HHG spectra – the appearance of only integer harmonic peaks (in units of the fundamental frequency) in the emission spectrum. This result can be understood in the conservation laws picture as arising from photon energy counting – since the driving laser contains only photons of a given frequency, $\omega$, the emitted photon must carry integer units of this fundamental quantity $n\omega$ (where $n$ is any integer), because there cannot be an energy exchange with the medium (since the process is assumed parametric). Conceptually this is very obvious, as $n$ photons must come together to create a single photon of frequency $n\omega$, and there cannot be halves (or fractions) of photon energies involved. To connect with the symmetry picture, we note that this energy conservation law leads to equivalent constraints to those resulting from the pure time-translation symmetry ($\tau_1$) that we enforced in all Floquet systems. This makes complete sense recalling that by assuming the system is 'Floquet' in nature, we immediately enforce that it is also parametric – there cannot be any energy transfer if the dynamics are time-periodic, only steady-states.

Moving on, we can also analyze the well-known odd-only harmonic selection rule in inversion-symmetric systems. In the DS picture, this arises from dynamical inversion symmetry (or a two-fold dynamical rotation in 2D) [35] [73]. However, in the photonic conservation law picture one could deduce the same result from parity conservation. Indeed, parity conservation means that only an even number of photons can be involved in a given nonlinear parametric process (because otherwise the parity of the photonic system cannot be conserved). If only an even number of photons is allowed to interact, and one photon is always reserved for the outgoing state, this means only an odd number of photons are permitted to mix together and annihilate, yielding odd-only HHG selection rules. In this unique case the symmetry and conservation law perspectives inherently have the same origin.

Looking at the polarization state of the NLO emission, it is common to consider conservation of SAM. In this approach for every single photonic pathway (in which $n$ photons annihilate to emit a single high energy photon at energy $n\omega$), the outgoing photon can only have a SAM of $\pm 1$. Circularly-polarized photons have SAM of $\pm 1$ depending on their handedness, while linearly-polarized photons can be described as superpositions in the circular-photon basis set. We can apply this principle to the well-known case of the counter-rotating bi-circular driven HHG with frequencies $\omega$ - $2\omega$ , which has been analyzed from multiple perspective both using the DS approach and conservation laws [86] [171] [278] [310] [311] [263] [59] [74] [35]. Here the incoming photons (being annihilated) have either SAM=1 for the $\omega$ component, or SAM=-1 for the $2\omega$ component. As an example, we consider the process of third harmonic generation where $n = 3$ for the outgoing photon. It can be generated by several pathways, but dominantly either mixing three photons from the $\omega$ beam, or one from each $\omega$ and $2\omega$ beams. For the former case the outgoing photon should have SAM of 3, which is forbidden, while for the latter case the outgoing photon should have a SAM of 0, which is also forbidden. It turns out also all other pathways involving more photons lead to similar forbidden SAM on the outgoing photon, meaning that the emission of $3\omega$ light in this configuration is forbidden. In a similar manner one can derive that emission of any photon with a frequency multiple of $3\omega$ is forbidden, i.e. $3n\omega$ components in the HHG spectrum are not allowed for integer $n$. We should also analyze the polarization components of the allowed emission, taking the $4\omega$ photon as a test case. Due to SAM conservation, we can derive that only cases where there is one more photon annihilated from one of the beams compared to the other will be allowed in general (or the outgoing photon will not have a SAM of $\pm1$). For the $4\omega$ photon this leaves a channel taking two $\omega$ photons plus one $2\omega$ photon, yielding a SAM of 2-1=1, such that the $4\omega$ photon in this cannel would have a SAM similar to the $\omega$ beam. Generalizing, as has been shown in refs. [86] [171], this analysis is upheld for all other allowed channels, and also all outgoing photons with energies $(3m-2)\omega$. For the $(3m-1)\omega$ photons one finds in a similar manner that only -1 SAM (same as the $2\omega$ beam) is allowed. Thus, we obtain for this case the exact same selection rules for HHG as can be derived from the 3-fold rotational DS. In fact, the analogy is completely natural as SAM conservation can only arise in rotationally-invariant media, and the DS selection rule is a result of a form of rotational symmetry. Similar considerations can be used when one of the fields has slightly detuned frequency components, e.g. the $2\omega$ field relaced with $1.95\omega$, which can shed light on the various photonic channels involved in the coherent sum (see Fig. 8(c))

In a similar approach OAM conservation can be applied for helical laser beams [38] [155] [156] [159] (see Fig. 8(a)), which we will not follow here, but merely point out that such constraints can be derived in the multi-scale DS theory approach discussed in section 3.2.3.

Now, let us focus on where these two approaches differ. First, there are DS-induced selection rules that do not appear to have any conservation law analogue. This includes selection rules invoked by time-reversal symmetry, and dynamical mirror planes, which lead to linearly-polarized harmonics [35]. Such selection rules are not directly connected with the conservation of a particular known quantity. This leads to a potential gap in the conservation law perspective, which seems not general enough to incorporate all potential spectral constraints. Second, the conservation law picture cannot easily incorporate the medium in the analysis. For instance, for the $\omega$-$2\omega$ case the DS picture predicts without difficulty that as long as the medium preserves 3-fold rotational symmetry, the same selection rules will be obtained. This includes molecules that are aligned and 3-fold symmetric [35] [79] [283] [264]. In the conservation law picture however, this complexifies the analysis since SAM is then allowed to exchange with the molecule. Moreover, the molecule breaks the continuous rotational symmetry itself such that SAM conservation should not be expected. Of course, one can in principle incorporate potential exchanges of quanta of SAM with the molecule into the framework (as for instance is common for atomic states with nonzero angular momentum [278]), but that is not a very natural approach and cannot be applied ad-hoc for each system. Lastly, we would highlight that the selection rules derived in the DS picture arise from discrete symmetries, rather than continuous ones. This implies that the points of view of conservation laws and symmetry-induced selection rules for parametric photonic processes might not overlap, because conservation laws should only be expected for continuous symmetries from Noether's theorem. It is true that in certain cases continuous DS also exist [312], which might lead to interesting constraints, but generally discrete DS usually dominate spectral signatures in experiments as thoroughly presented throughout this review. This last point connects to a trickier issue in the conservation law picture, which is that there are no photons to begin with, since both irradiating and outgoing fields are fully classical! In that respect, when we define 'photons' in photonic channels for conservation law analysis, we are in fact only analyzing classical fields that are periodically oscillating and assuming the Floquet picture is valid. All of these points lead us to conclude that generally, at least for parametric NLO, DS-induced selection rules are a more general approach than photonic conservation laws.

As a final crucial point in this discussion, we highlight ref. [40], which make substantial headway in further connecting the two perspectives. Here the DS selection rule picture was invoked by artificially increasing the dimensionality of the light-matter Hamiltonian (as is commonly employed for quasi-crystal systems [313]). The higher dimensional theory then leads to continuous DS, which can formally be used to derive conservation laws through Noether's theorem (unlike discrete DS). This approach was shown to reconstruct known conservation laws in HHG as introduced above, but also yielded two new laws for conservation of new forms of parity (see Fig. 8(b)). The constraints lead to linear-polarization constraints on HHG spectra that otherwise cannot be derived purely from SAM conservation, as well as phase constraints. Both of these types of constraints can arise in DS-induced selection rules, but do not associate with known conservation laws. Thus, the formalism in ref. [40] could pave the way to a reorganization of the way that we currently think of conservation laws and symmetry-induced selection rules in NLO, and truly hybridize the two concepts, even in the absence of quantized photonic states.

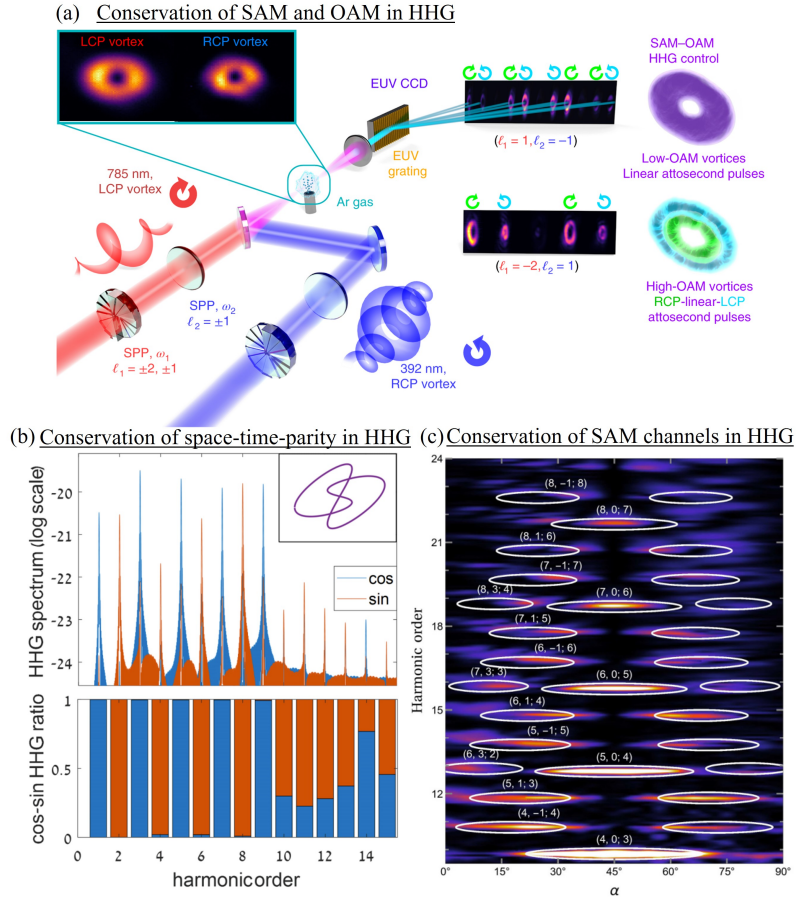

**Fig. 8. Photonic conservation law perspective in HHG.** (a) simultaneous conservation of SAM and OAM in HHG in atomic media driven by bi-chromatic driving fields. Figure repurposed from ref. [156]. (b) Conservation of a novel type of parity in atomic HHG, which leads to a phase-specific and polarization specific response in HHG. Figure repurposed from ref. [40]. (c) Theoretical analysis of SAM conservation in HHG driven by slightly detuned bi-circular b-chromatic fields, analyzing the experiment in ref. [86]. Figure repurposed from ref. [171].

## 4.  Applications of selection rules in nonlinear optics

### *4.1 Generation of light with desired properties*

The symmetries of a physical system, as exhibited in its Hamiltonian, can shape its response to external driving, also affecting light emitted as a result of interactions such as scattered or generated light. This leads to an obvious yet highly rewarding application of symmetry-tailoring – tuning the spectral, temporal, polarization, spatial, or essentially any other property (e.g. coherence), of new light sources that are generated from light-matter interactions. In this section we limit our scope to processes driven by well-defined driving fields, and to the new frequencies produced in parametric NLO interactions.

### 4.1.1    Stable instrumentation

The selection rules and the accompanying suppression of specific harmonic orders are particularly useful for applications of the extreme-UV and X-ray radiation generated (since HHG naturally emits very broad spectra up to the X-ray range, and since controlling polarization states of coherent X-rays is generally challenging). Their polarization can be inferred from the spectrum and can be flipped by a mirror-symmetry operation, implemented for example, by flipping or rotating a waveplate by 90°. A practical development for applications of circularly polarized high harmonics was the in-line apparatuses, such as the MAZEL-TOV (MAch-ZEhnder-Less Trifold Optical Virginia-spiderwort) [59]. The term was coined as such since upon the insertion of this device in path of a focusing beam it creates at the focal point a three-fold DS field (discussed thoroughly in examples above), with a Lissajous curve resembling three-leafed spiderwort flower, without the need to split the two colors to different arms of a Mach-Zehnder-like system. The device comprises several elements: a nonlinear crystal for type-I phase matching generation of the second harmonic, that is, partly converting a linearly polarized fundamental into a perpendicularly linearly polarized second harmonic. The Perpendicularly polarized two colors then pass in birefringent crystals for controlling their relative time delays and a single quarter-wave retardation plate that converts their polarization to counter-rotating circular. Since the device is placed in a focusing beam with solely transmissive optics, the two colors co-focus to the same geometrical spot regardless of any vibrations of the optical elements. By eliminating transverse shifts between the two colors, the beam is concentric and uniform throughout the length of the HHG process. Thus, the resulting high-order harmonics beam is extremely stable, of a high-quality profile, and it maintains experimental alignment. For these reasons it achieved challenging applications such as nanoscopic imaging of weakly scattering chiral samples, such as spin-polarized domains in magnetic materials [314], even with multi-frame dynamics and subwavelength resolution [315]. Fig. 9(a) illustrates the experimental setup using the MAZEL-TOV inline apparatus for high-resolution imaging.

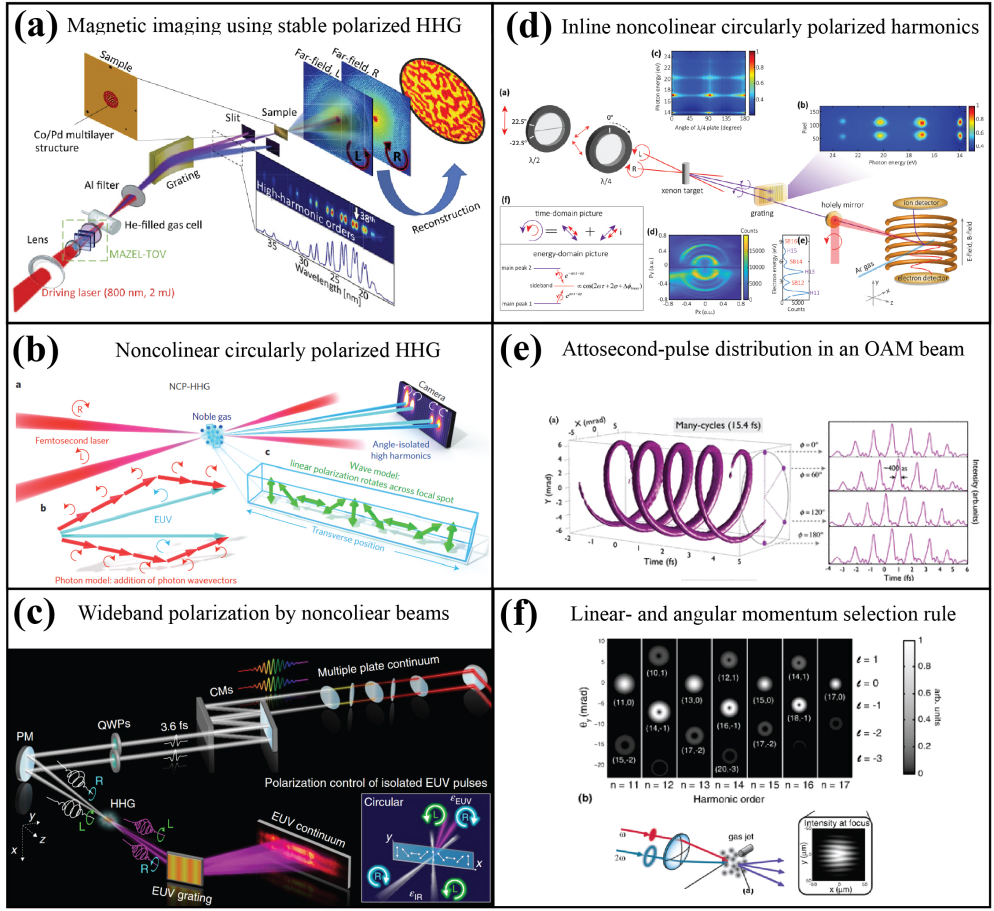

**Fig. 9. Controlling light properties through dynamical symmetries in high harmonic generation.** (a) Stable instrumentation, such as the inline MAZEL-TOV apparatus enables photon demanding applications, such as coherent diffractive imaging of chiral features (here magnetic domains imaged on the M-edge of cobalt). Figure repurposed from ref. [314]. (b) Noncolinear generation of circularly polarized high harmonics, illustrated as the photon spin-conservation or as a rotation structure $\hat{L}_{x,n_1}\hat{r}_{n_2}$. Figure repurposed from ref. [316]. (c) Driving circularly polarized HHG with noncolinear single-cycle laser results in a circularly polarized continuum. Figure repurposed from ref. [317]. (d) Noncolinear generation of circularly polarized HHG beams links the beams angle and helicity, where one helicity is chosen spatially to drive helically-selective photoionization. Figure repurposed from ref. [318]. (e) Simulation of the energy density of HHG from a linearly-polarized vortex (OAM) beam, showing attosecond OAM beams. Figure repurposed from ref. [153]. (f) Noncolinear beams with different angle, frequency, and OAM allows angularly separating vortex beams at every harmonic order according to their winding number $\ell$. Figure repurposed from ref. [154].

### 4.1.2 Controlling the Helicity of high harmonic beams

The helicity of the HHG emerges from the Hamiltonian symmetry, thus, applying a mirror symmetry to the Hamiltonian flips the radiation's handedness. For helical HHG in achiral medium, the helicity originates from the input light fields, for which a mirroring on a plane parallel to the beam axis is a polarization change. For bi-circular fields having an opposite handedness for frequency components $\omega_1$ and $\omega_2$, the emission at $m(\omega_1 + \omega_2) + \omega_i$, where $i = 1, 2$, is polarized as the $\omega_i$ photon (which has been derived above from both the DS and the conservation law perspective). Replacing the input polarization helicities changes the helicity of each of the harmonic orders. The same logic partly applies for non-colinear circularly polarized harmonic generation [316] – flipping the input polarization changes the output HHG accordingly. However, the symmetry argument is a bit more delicate since the field of the non-colinear system locally achiral (and also globally). The field's polarization is locally linear, with a translation-dependent orientation, $\hat{L}_{x,n_1}\hat{n}_2$ (see Fig. 9(b)). Since the symmetry is continuous, $n_1, n_2 \to \infty$, albeit with a ratio of, $n_2 = n_1(\Lambda/\lambda)$, such that a full polarization rotation, $n_2 = 1$, is equivalent to the translation of a one spatial cycle $\Lambda$, that is $n_1(\Lambda/\lambda) = n_2 = 1$, corresponding to the definition of $\hat{L}_{x,n_1}$ with $\lambda/n_1 = \Lambda$. The far-field of such a symmetry at the focus is two replicas of the laser with opposing circular polarization and with a spatial separation determined by $\Lambda$ and the radiation wavelength. In other words, the angle of the two incoming (and outgoing) beams of the fundamental laser is $2\theta_1 = \pi\frac{\lambda}{\Lambda}$, such that each of them is $\pm\theta_1$ with respect to the axis. Globally, the output radiation is achiral since for each left-going photon there is a right-going one with an opposite helicity, flying in an opposite angle. Helicity is added by post-selecting a collection angle $\theta_q$ corresponding to harmonic order $q$, and blocking emission to angle $-\theta_q$. Thus, the symmetry of the left- and right-going HHG (and driving beams) is broken [316].

As a side note, we would mention that there are schemes where the polarization relays on coherently combining two orthogonal linearly polarized HHG sources, where the helicity can be set by their relative temporal delay [319]. There is no overarching symmetry in such fields, and hence the circular polarization is a tailored result rather than a natural one since the time delay operator produces a harmonic-order dependent phase $\hat{\tau}_n = e^{iq\omega_0 T/n}$. For example, choosing timing $t_0$, between the horizontal- and vertical-polarized components, say $\hat{x}, \hat{y}$ respectively, the polarization dependent harmonic is $1/\sqrt{2}(\hat{x} + \hat{y}e^{qi\omega_0 t_0})$. Therefore, if one harmonic order is circularly polarized, the order-dependent phase shearing makes the polarization of the other harmonics elliptical and even with an opposite helicity.

### 4.1.3 Helical HHG with a tunable frequency

Since the discrete harmonic spectrum comprises integer multiples of the driving frequencies, its tunability relays on controlling the driver fields. For monochromatic driving, the frequency $\omega$ is the single influencing the system's DOF. When it is tuned, the harmonics rescale to its integer multiples, $q\omega$, e.g., as done for linearly polarized HHG from an optical parametric chirped-pulsed amplifier (OPCPA) source [320] [321]. For bi-chromatic sources, one can scale both of the spectral components; however, it is more efficient to tune just one of them. Conveniently, circularly polarized harmonics driven by bi-circular fields are efficient even if intensities differ and their relative phase fluctuates freely. Thus, one driver can stay fixed at the fundamental frequency (e.g., 800 nm for Ti:Sapphire lasers) while the other can be tuned freely using an OPA [263]. The relative phase between two circularly polarized fields defines a spatial axis, such that downstream circularly polarized HHG is not affected. However, the phase matters for HHG in non-isotropic media, such as crystals, as we elaborate on in section 4.2.2.1 below.

In the abovementioned tunability mechanism, we have omitted macroscopic mechanisms and post-selection controllability. Post-selection can be accomplished by, for example, a bandwidth-selective slit positioned at the image-plane downstream, after a dispersive and focusing optics, as well phase-matching constraints [322] [323] [324]. For supercontinuum generation, the slit position or a shift in the pre-slit optics defines the central frequency of the transmission, or the transmitted harmonic order. The macroscopic process of the HHG involves nonlinear phenomena due to the field-induced ionization accompanied by a time-varying plasma dispersion [325] [326]. In practice, these small nonlinear frequency shifts are optimized per experimental realization to produce the brightest source, whereas tunability can be achieved by tuning the driver.

### 4.1.4 Helical Attosecond-pulse generation

The circular polarization of HHG does not warranty the circular polarization or even the helicity of the attosecond pulses that comprise such an emission. For instance, a repeating set of three linearly-polarized (achiral) pulses that are rotated by 120° is the radiation pattern underlying the circularly polarized high HHG from a bi-chromatic laser driver [327]. The global helicity is conserved here since the harmonics are both left- and right-rotating, and hence the radiation is overall non-helical. For the attosecond pulses to be helical in the colinear scheme of bi-chromatic fields the yield of the co- and counter-rotating harmonic orders should be imbalanced. Temporally helical emission can be formed by a chiral generation medium [328] [329] [10] [330] [331], by tuning other DOF such as the intensity ratios between the driving fields [217] [223], or the addition of other frequency components [217] [223], or by breaking the symmetry in the field directly and using elliptically polarized driving fields [332]. The non-colinear geometry for circularly polarized HHG is especially suitable for generation of circularly polarized attosecond pulses since the two helicities are separable spatially [316]. Isolated circularly polarized attosecond pulses can be generated with few-cycle pulses that extend the spectrum of the non-colinear harmonic generation to a supercontinuum [318] [317], relaying on the same mechanisms of linearly-polarized attopulse formation. These are relevant since the driving field is locally linearly-polarized, albeit with the periodic polarization grating characteristic of non-colinear generation of circularly polarized HHG [316].

Motivated by the generation of circularly polarized attosecond pulses from non-colinear beam [317], Han *et al.*, utilized the concept of in-line transmissive geometry for a stable generation of single-attosecond pulses [318]. The in-line apparatus forms the necessary counter-rotating pair of beams by simply inserting a segmented half-wave retardation plate into the beam path, followed by a quarter waveplate. The crystalline axis of the two segments is tilted by ±22.5° such that half of the outgoing beam is diagonally-polarized, and the other half is anti-diagonally-polarized, and upon passage through a quarter waveplate the two parts are counter-rotating circularly-polarized. In this method the counter-rotating segments are spatially separated, thus, it is mirror-symmetric. The output radiation is mirror symmetric as well, where each side is counter-rotating. One part can be focused onto the sample while the other is blocked for ultrafast applications. The helicity of the selected beamlet can be flipped by rotating the quarter waveplate. The optimal waveplate orientations for the two helicities here are selected by the maximum of the HHG yield, rather than the symmetry-based suppression of the bi-circular schemes.

### 4.1.5 Compounded spectro-polarization symmetries

As mentioned above, in the case of rotational DS, the selection rules forbid the appearance of certain harmonic orders, allowing nonzero HHG for discrete harmonic multiples of the fundamental frequency only for some orders which must be circularly-polarized. When the driving fields are $\omega$ and $2\omega$ with counter-rotating circular polarizations, as in Eq. (6), the selection rule is $3n \pm 1$ for integer $n$. Importantly, the selection rule imposes the spectral constraint compounded with the radiating polarization. All the allowed harmonics are circularly polarized, where harmonic orders $n = 3m + 1$ co-rotate with the fundamental and the orders $n = 3m - 1$ are counter rotating, that is, co-rotating with the second harmonic. That can be inferred by solving the eigenvalue described in detail above [80] [74] [329]. This scheme was originally used to break the paradigm that the polarization of high-order harmonic generation from lasers is limited to linear polarization, or perturbations around linear. Driving HHG with these so-called bi-circular lasers demonstrated for the first time circularly polarized high harmonics [86], and in addition, controlled the helicity of the circular polarization for chiral and/or magnetic spectroscopy [88]. When the DS is not fully imposed, the forbidden $3m$ orders revive. Hence, when the selection rule is observable experimentally, with absent $3m$ orders, the polarization is guaranteed to be generated as circularly polarization with a known helicity, relying on the symmetry theory. Polarizing elements in the beam path, such as gratings and grazing-reflections off either focusing or flat mirrors should be accounted for in order to determine the final polarization of the beam.

Interestingly, the rotational DS does not require to be an integer or even rational integer to produce circularly polarized high-frequency radiation. Any bi-circular field comprising of fields at frequencies $\omega_1 = \omega(1 + \delta)$ and $\omega_2 = \omega(1 - \delta)$ will generate fields at $\omega_x = x\omega$, where x conforms to the selection rule $x = (2m - 1) \pm \delta$. If Delta is irrational, x is irrational as well. In other words, the symmetry-protected emission does not necessarily emerge as harmonics of some fundamental frequency. It is the suppression of the harmonic orders $q = 2m$ that guarantees the symmetry of the driving fields and the co- or counter-rotating helicity of the spectral components that correspond to the "+" and "–" of the selection rules. This is an intriguing case, because for irrational driving frequencies we obtain a breaking of the essential time-periodic condition in the Hamiltonian in Floquet theory, showing that in some cases DS-based selection rules for polarization control are even more general than the derivation above (i.e. the Floquet condition can be lifted). For comparison, in the $\omega$-$2\omega$ scheme the mean frequency is $\frac{3}{2}\omega$, $\delta = 1/3$, hence we can rewrite the driving carrier waves as $\omega_{1,2} = \omega_{+,-} = \frac{3}{2}\omega(1 + 1/3)$. The suppression and selection rules for outgoing modes is in this language for $q\omega = 2m\frac{3}{2}\omega = 3m\omega$, and the allowed harmonics are $x\frac{3}{2}\omega = 2\frac{3}{2}\omega\left(m \pm \frac{1}{3}\right) = 3m\omega \pm \omega$, Indeed, when the input fields are set to an extreme frequency ratio, $\delta \ll \omega$, the radiation is separated by $2\delta$ exhibit opposing polarization [247]. Thus, this scheme allows for circular polarization control over continuously tunable upconverted frequencies, matchable to particular spectroscopic needs.

### 4.1.6 Spatial selectivity and optical angular momentum

The application of nonlinear optics with GL beams can be viewed as an expansion of local polarization-related symmetries alongside closing any spatial variation to a loop, as described in section 2.2.1. The main order parameter in GL beams is the phase winding number, covering multiples of $2\pi$ in a closed loop around the beam's center. It is represented as a continuous symmetry $\hat{R}_n\hat{t}_n$ or $\hat{R}_n\hat{L}_{z,n}$ in the scalar beam equation of the GL beam (Eq. (12)), with $n \to \infty$. In the case of linearly polarized beams, the polarization is traced out, given definition of $\hat{R}_n$ as not affecting local polarization. It is worth mentioning that circular polarization links phase with direction, hence, circularly polarized GL beams are indeed eigenmodes of the operation of time translation and a complete coordinate-rotation, which in our notation, is $(\hat{R}_n\hat{n}_n)\hat{t}_n$. However, circularly polarized beams produce no harmonics in isotropic, and hence, are less of an interest here. The simplest form of GL beams is linearly polarized. Harmonics generated in isotropic media will be linearly polarized as well, hence the polarization can be traced out entirely, and the radiation physics is nicely represented by the scalar form of the GL beam (Eq. (12)). Fig. 9(e) utilizes the simplicity of the scalar form, showing that the intensity of high harmonics in the extreme UV regime forms a screw pattern at a finite radius [153]. When driven with a long (quasi-continuous) driving pulse, the radiation intensity forms a screw pattern with a visible rotation-translation relation, $\hat{R}_n\hat{L}_{z,n}$. Since the nonlinear radiation inherits the topological nature of the parent GL beam, albeit with a winding number multiplied by the harmonic order $q$. Operating $\hat{R}_n\hat{t}_n$ for $n$ times, $n\hat{R}_n = \hat{I}$ and of the d $n\hat{t}_n = e^{i\omega_q T} = e^{2\pi i q}$, where T is the optical cycle associated with the driver's angular frequency $\omega_0$ and $\omega_q = q\omega_0$. This is the conservation of OAM, where the creation of a photon with an OAM of $\ell = q$ originates from the annihilation of $q$ photons with an OAM of $\ell = 1$. Although the winding number is not conserved, the node at the center is. Thus, applications that use the node can in principle be replicated to the harmonics, though to our knowledge this has not been applied yet. However, it also implies that one cannot produce a harmonic beam with an OAM winding of $\ell = 1$, as the fundamental. Gariepy *et al.*, used the concept of conservation of angular momentum alongside the conservation of linear momentum by driving harmonic generation with two non-colinear beams of different photon energy and OAM, including $\ell = 1$, as illustrated in Fig. 9(f) [154]. The outgoing beams differ in their energy, momentum (i.e., deflection angle), and OAM. Every specific harmonic order is emitted at several angles, where adjacent angles represent an exchange of two photons with $(\omega, k, \ell) = (\omega_0, k_1, 0)$ by one photon of $(\omega, k, \ell) = (2\omega_0, -2k_1, 1)$. Thus, forming a link where the OAM rises with the emission angle. This demonstrates the simplicity and intuitiveness of the approach of photonic decomposition, whereas the symmetry operation is more complicated, involving translation, rotation and a different operator for both spectral components. We note that the interference in the subfigure is instantaneous, since it is formed by mixing two colors, the intensity peaks oscillate at $2\omega_0 - \omega_0 = \omega_0$.

### 4.2 Nonlinear spectroscopy

Perhaps the most obvious utilization of symmetry theories in electromagnetism and NLO is their application towards probing properties of matter, either dynamically evolving, or static. In this paradigm one can employ multiple schemes from engineering light with desired symmetries, light with desired asymmetries, scanning multiple available DOF in tailored light, or tuning the material properties themselves (e.g. by doping, orienting, changing temperature, etc.) while fixing the laser field properties. All of these ideally should allow extracting meaningful information about the system. Due to the inherently good temporal resolution of highly nonlinear processes such as HHG, this effort is often geared towards addressing attosecond dynamics of electrons in various systems. We will now focus on NLO spectroscopy from the perspective of symmetries in the EM field and light-matter system, i.e. in correspondence with the general theme of this review. The discussion is divided largely based on the system type, where we separately address atomic and molecular systems in the gas phase, solids, and chiral systems. Of course, we note that this separation is somewhat artificial, and from the symmetry perspective there is no fundamental motive for it. Nonetheless, since the intrinsic properties of interest in these systems tend to be very different from one another, we believe such a separation can be useful.

#### 4.2.1 Atomic and molecular systems

The topic of NLO spectroscopy in atomic and molecular systems is extremely wide reaching. Even when considering only the symmetry perspective of this field, there exist a multitude of works and important results. In order to focus our discussion, we divide it based on several classes of employing symmetry theories for NLO probing of atoms and molecules. First, we will discuss the most obvious application, which is probing the symmetries (i.e. point group) of the atomic and molecular media directly. Second, we will discuss probing induced symmetry breaking in the medium, e.g. by excitation of electronic modes, vibrational modes, or other means. Lastly, we will discuss symmetry breaking applications induced directly by the laser field, whereby breaking a symmetry in a controlled manner allows probing interesting dynamical phenomena. To be clear, this will by no means be a complete list of recent achievements, but it should give a taste of the novel approaches that have been employed over the last two decades, and how they connect to symmetries and asymmetries of light-matter interactions.

#### 4.2.1.1 Probing symmetries of atomic and molecular media

As an initial discussion, consider the following – in section 3.2 we derived the NLO selection rules considering the full light-matter Hamiltonian, i.e. that including both the medium and the driving laser fields. We saw that both terms equally contribute to selection rules. Therefore, it follows that by observing a certain symmetry-based selection rule in nonlinear spectra, or its absence, we might infer the symmetry of the medium. For instance, consider an example – if driving a molecular gas with a simple monochromatic intense laser field we observe odd-only harmonics being emitted, we can directly conclude that the medium is inversion symmetric (because we know that the laser field exhibits dynamical inversion symmetry by construction). At face value, this might seem like a trivial or even bad idea – why should we take the trouble of measuring the symmetry group of a medium through performing nonlinear optical measurements? It is much simpler to follow the standard spectroscopy schemes such as X-ray scattering or other techniques. However, this is a good point in time to recall that since intense laser pulses are often also ultrashort; hence, we could track symmetries in time with such a scheme, which encodes a great deal of non-trivial information.

In the early days of HHG [333] [333] [334], measurements were taken from noble gases, which are of the highest SO(3) symmetry group and exhibit all potential mirror and rotational symmetries. This means any observed HHG selection rules such as odd-only HHG [73], or absence of HHG when driving with circular driving [335], were in fact a direct result of the DS group of the laser field itself [35]. Soon, it became clear similar results are obtained from gas phase ensembles of randomly oriented molecules as well. Here the logic is the same – since the molecular ensemble is orientation averaged, it upholds all possible SO(3) operations and only the symmetries of the laser field play a role. An exception here is chiral media that break mirror symmetries (complying only to an O(3) point group) but might still be rotationally invariant and isotropic, which will be separately and elaborately discussed in section 4.2.3.

However, not long after, first experiments were possible with aligned molecular media. In this scheme one pre-excites the molecular gas with an aligning laser pulse, e.g. resonant with rotational energy levels, which induces preferential orientation in the ensemble. On average, this breaks the medium's SO(3) symmetry and reduces it to the molecular point group instead (depending on the level of alignment). Such orientating pulses were explored in various approach from THz [336] to optical centrifuge [337] [245] and more [338] [61] [339] [340] [265], and are today a relatively widespread technique. By using a delayed probe pulse that generates HHG (which is potentially polarization-tailored as well), one can sample NLO selection rules in emerging spectra. In ref. [10], Baykusheva *et al.* generated HHG using a counter-rotating bi-circular field that should lead to forbidden $3n$ harmonics (for integer $n$), as long as the medium exhibits 3-fold rotational symmetry, same as the DS of the field in these conditions (see section 3.2). They showed that in the absence of alignment, gases of $N_2$ and $SF_6$ molecules indeed exhibit this selection rule. However, when aligning the media, $3n$ harmonics appear as a result of rotational symmetry breaking (see Fig. 10(a)). Thus, symmetry breaking spectroscopy can be directly applied to probe the molecular structure. Similar symmetry-broken even harmonics are obtained from aligned asymmetric systems such as the OCS molecule, with their yield indicating the rotational period [341][xx].

In ref. [342] Levesque *et al.* showed that molecular alignment also breaks the symmetries that lead to linearly-polarized harmonics (see Fig. 10(b)), i.e. the harmonic ellipticity can be tuned by the degree of symmetry breaking (which can even be employed for generating helical few-cycle pulses [343]). Others have shown similar effects [344] [345]. We should note that there have been numerous theory works predicting a variety of spectral signatures due to aligned molecular symmetries in various cases [76] [264] [82] [311], including in semi-periodic systems such as nano-tubes [79] [283]. These all fall under the category of the more general group theory discussed throughout the review, and can be used for pinning down the molecular point group. In principle, this technique can also be utilized to extract additional information such as the molecular orbitals involved and their symmetry if the HHG yield is measured with respect to the molecular alignment [346] [347] [348] [349] [350].

Returning to atomic media, one might initially expect that no interesting symmetry breaking can be probed in noble gases. However, theory predictions have shown that if the medium is initiated in a state with non-zero angular momentum (e.g. $p$-shells of hydrogenic atoms), then typical mirror symmetries can be broken. In ref. [351] it was shown numerically that this effect should lead to elliptical harmonic emission from linearly-polarized driving, which would generally lead to only linearly-polarized harmonics due to a static mirror symmetry. This effect can be considered the atomic analogue of the above molecular alignment case, because inducing an asymmetric electronic distribution in orbitals carrying energy levels requires pre-pumping the media, which is a form of alignment in itself. It should also arise in a system initiated in a super-position of states (which can carry nonzero currents that might break a symmetry). We should note that similar phenomena were measured in photoemission spectroscopies [352], though the type of symmetry breaking there is a little different and induces an asymmetry in the photoemission spectrum that otherwise would not be present in it (see Fig. 16).

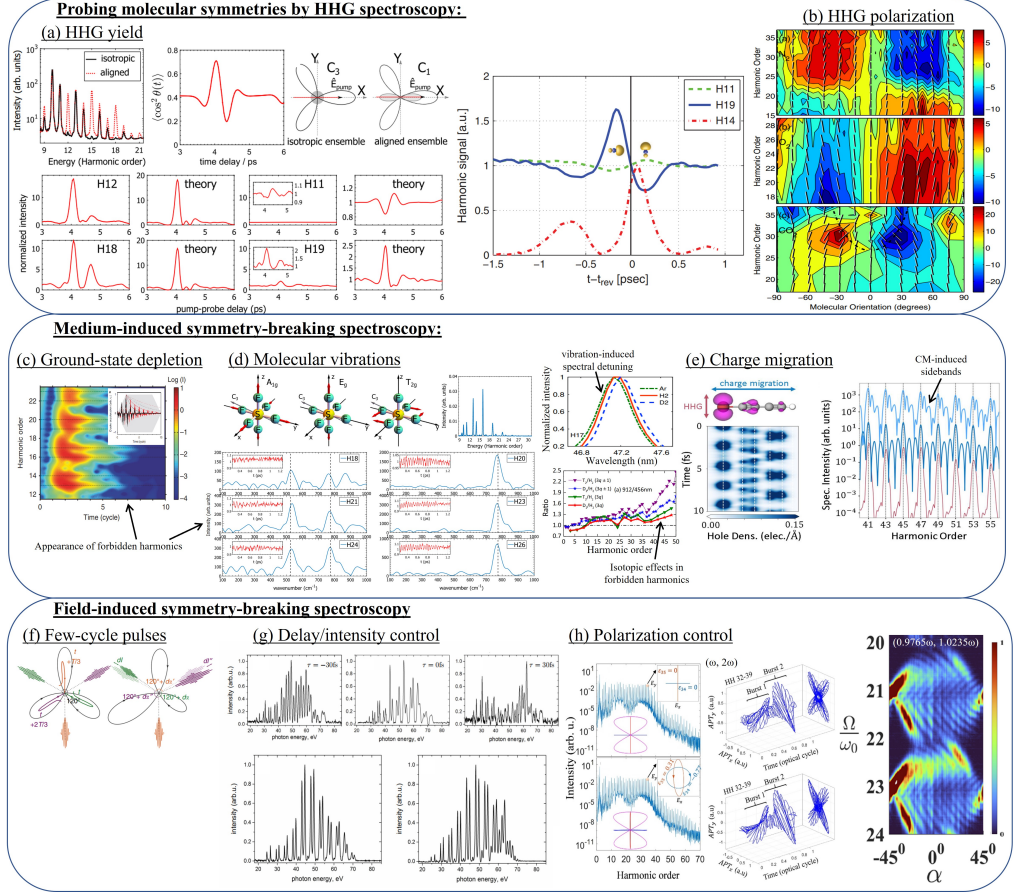

**Fig. 10. AMO spectroscopies by symmetry and symmetry-breaking.** Upper row: Revealing the symmetry of the medium through HHG spectroscopy; Middle row: medium-induced symmetry-breaking; Lower row: Optical field-induced symmetry-breaking. (a) Tracking the temporal evolution of the medium's symmetry forbidden harmonics: Orientational-rotational wavepacket in CO molecule tracked through the appearance of even-harmonics (Figure repurposed from ref. [265]), and alignment-rotational wavepacket in $N_2$ molecule by appearance of forbidden $3n$ harmonics (Figure repurposed from ref. [10]). (b) Tracking the alignment rotational wavepackets in $N_2$, $O_2$, and $CO_2$ molecules through the HHG polarization (Figure repurposed from ref. [342]). (c) Symmetry-breaking in the medium could result from any processes causing coherent temporal variations in the electronic structure. For instance, ground-state depletion (Figure repurposed from refs. [353]), or (d) molecular vibrations and isotopic features (Figures repurposed from refs. [10] [354]) leading to otherwise forbidden harmonics or red/blue-shifting of harmonic peaks (Figure repurposed from ref. [355]). (e) Medium symmetry-breaking could result also from charge migration, manifested as harmonic side-peaks (Figure repurposed from ref. [356]). (f) Field-induced symmetry breaking in the bicircular $\omega$-$2\omega$ case by shortening the pulse duration (Figure repurposed from ref. [353]), tweaking the relative delay or amplitudes of the two colors (Figure repurposed from ref. [11]), or tweaking the driving laser polarizations (Figure repurposed from refs. [332] [357]), including during spectral detuning of the driving frequencies away from the $\omega$-$2\omega$ case.

### 4.2.1.2 Probing induced symmetry breaking phenomena

The next natural step is to employ an analogous approach which is time-resolved. This can be accomplished either *via* pump-probe spectroscopy, where the pulse generating the NLO response is the probe for a given symmetry breaking signal, or *via* high harmonic spectroscopy in certain limiting cases, where temporal information can be extracted from the HHG spectrum directly through models or reconstruction algorithms (e.g. as in the case of orbital tomography [346], ionization [358], or other). Let us briefly highlight several noteworthy dynamical molecular phenomena that can be probed through symmetry breaking spectroscopy, demonstrating the general principle.

First, perhaps the most obvious physical effect that can be probed is molecular vibrations (or the quantum nuclei nature of the molecule). For instance, in $SF_6$ forbidden harmonic yields were measured to oscillate (vanish and then revive) with respect to pump-probe delay, indicating the contributions of symmetry-breaking vibrational modes [10] (see Fig. 10(d)). Similar effects can appear in diatomics driven by rotationally symmetric fields [354]. More generally, vibrational and quantum nuclei effects were explored in several molecular systems with HHG, typically causing slight shifting in the harmonic frequency [355] [359] (see Fig. 10(d)). This effect can be analyzed thoroughly with various numerical approaches. From a symmetry perspective it can be understood as a failure of photonic energy conservation (or time translational symmetry) – otherwise only pure harmonic modes should be emitted. Practically, the vibrating molecule breaks the temporal periodicity of the laser because the eigenfrequencies of vibrations are usually not an integer multiple of the laser frequency.

Another interesting dynamical phenomena is the presence of ring-currents, or helical motion of charges within molecules [360] [168] [361] [362]. This can be conceptually understood in similar terms to the symmetry breaking induced by atomic states carrying nonzero angular momentum discussed above; however, in complex molecules the charge motion can have interesting features that differ from the atomic case such as a non-uniform distribution of the current in space and time. In ref. [363] it was shown that like in the atomic case, this ring current density can be probed through symmetry breaking in HHG in various schemes. By tracking the symmetry breaking signal in time, the impact of dynamically evolving electron-electron interactions (on attosecond timescales) could also be isolated. More generally, helical ring currents are a sub-group of the more general phenomena of charge migration [364] [365] [366] [367] [368] [369] [370] [371] [372]. Depending on the system at hand, charge migration can cause similar symmetry breaking effects that can be probed by HHG in a pump-probe configuration. Indeed, it was recently suggested that the HHG spectrum would contain information on charge migration through the appearance of HHG side-bands at energy spacing associated with the characteristic temporal motion of the hole [356] [373]. From a DS perspective this is similar to the case of molecular vibrational modes, where instead of atomic motion the hole motion directly breaks time translation symmetry (see Fig. 10(e)). Equivalent phenomena can occur due to inherent temporal asymmetries in electron populations that build up excitations over time [353] (see Fig. 10(c))

Lastly, let us also highlight another intrinsic effect that can break symmetries and cause selection rule deviations – atomic and molecular resonances. Recall in the derivation of light-matter selection rules from dynamical group theory in section 3.2, we assumed that a single Floquet state dominates the dynamics, or selection rules would void. Naturally, this condition can is broken by using short few cycle pulses [269], or engineered pulses with varying temporal envelopes and frequency content [270] [374]. However, it can also arise if the molecule/atom has an intrinsic resonance that breaks the single Floquet state condition. Such resonances often cause exceptional features in spectra such as minima [375,376] [377], or enhanced emission [274] [275,378] [379] [378]. This can naturally be accompanied by various types of NLO selection rule breaking. For instance, HHG enhanced emission near resonances in atoms causes a broad-range spectral emission rather than a sharp harmonic peak, indicating the breaking of time-translation symmetry breaking as the resonance is populated (though here care must be taken because at high energies it is not often clear if the absence of sharp harmonic peaks is a result of symmetry breaking, or poor spectral resolution). In principle, other selection rules can be broken near resonances, though to our knowledge harmonic ellipticities were not yet measured in such conditions.

### 4.2.1.3 Symmetry-tailored laser pulses for ultrafast spectroscopy

As a final topic in this section, let us consider nonlinear spectroscopy in atoms and molecules which is enabled by engineering a desired symmetry breaking in the driving EM field itself. Such schemes can be used to probe intrinsic properties of matter such as electron dynamics and orbital structure, but also properties of the light-matter interaction, e.g. conservation laws or contributions from non-dipole terms.

One of the clearest examples of inducing symmetry breaking in a laser field to help probe an inherent property of atomic and molecular systems is utilizing short few-cycle pulses instead of longer time-periodic pulses. For few-cycle pulses the selection rule derivations discussed throughout obviously break, which can be understood in the symmetry picture as an absence of time-periodicity, and from a conservation law perspective as a result of the broad spectral range of the driving field. For instance, by employing such fields which are also circularly-polarized (or highly elliptical), one can probe the process of tunnel ionization through the attoclock configuration [380], and even effects of long-range Coulomb interactions [381] [382] [383] [384]. In this configuration it is made possible since a single peak in the electric field dominates the photoelectron response, precluding contributions from other moments in time. Therefore, a single photoelectron peak can be connected to the tunnel ionization process near the field peak. Short pulses that move away from the adiabatic regime required for Floquet DS selection rules naturally induce also symmetry-broken harmonics [349] (see Fig. 10(f)).

More generally, one can supplement the typical dominating electric-dipolar interaction in the Hamiltonian by additional terms that might break symmetries. These can be beyond-dipole terms originating from interactions with magnetic fields [385] [386] [387] [388] [389] [390] [391] [392], which allow studying the light-matter interaction itself with ultrafast time resolution.

In the bi-chromatic HHG configuration there have been many works with similar motives. For instance, by spectrally detuning the frequency of one of the driving field components it was shown in ref. [86] that SAM conservation laws can be probed, also prompting theoretical analysis [171] (see Fig. 8(c)). By detuning the polarization of the driving components away from the circular configuration instead one can probe the medium's response to symmetry breaking, which can be indicative for electronic correlations [393], or the nature of the atomic orbitals [394], and is also accompanied by spectral and polarization changes to the spectrum (see Fig. 10(h)). This is in principle not a direct result of a symmetry or asymmetry in the light-matter Hamiltonian, but instead arises naturally due to the high nonlinearity of HHG and its strong sensitivity to small changes in the Hamiltonian. Consequently, such small changes can easily be mapped onto changes in harmonic deviations from selection rules, e.g. polarization states [395] [247] [248] [249] [295] [396] [397] [357]. Similarly, by slightly detuning the relative phase of the two frequency components the tunnel ionization process can be probed [266].

For the bi-circular counter-rotating configuration, deviations from DS selection rules were measured as a result of detuning the delay between the two frequency components. From the symmetry point of view this is clear, as time translation symmetry breaks. However, the physical mechanism accompanying this type of symmetry breaking was shown to involve excitations of Rydberg states in atoms, allowing probing their dynamical occupations on ultrafast times [11] (see Fig. 10(g)).

Lastly, with a two-color frequency detuned configuration it was shown that synthetic symmetries of the Hamiltonian can also be probed in terms of HHG selection rules [42]. In principle, this approach can connect to a variety of potential synthetic-space symmetries that can be explored in multi-dimensional fields or Hamiltonians with many DOF, such as phase-space symmetries [398] [262] [399]. More generally, it can be employed in the context of multi-dimensional spectroscopy such as for carrier-envelope-phase effects, and more.

### 4.2.2 Condensed-matter systems

The crystal structure of solids inherently breaks the typical isotropy associated with randomly-oriented gases and liquids. While this means that the symmetry group describing the material medium is usually of a lower order than that of randomly oriented media (no longer a fully translationally and rotationally invariant), it is also an opportunity for interesting symmetries to play a more dominant role in the light-matter interaction. One clear example are even order nonlinear optical responses that become allowed in inversion-asymmetric crystals [2] [400] [401] [402] [403] [294] [404], which are not observed in gas phase unless it is pre-oriented by additional laser pulses. While reviewing the role of symmetries in nonlinear light-matter interactions with condensed-matter, we consider two main topics of interest: (i) Highly nonlinear HHG, light-induced magnetization, and photocurrent generation, in which the driving field initiates electron and/or spin dynamics within and in-between the bands that are weakly perturbed by the laser. We will review efforts in this regime for both utilizing nonlinear optics driven by unique symmetries for probing material properties, probing ultrafast charge dynamics (electronic or nuclear), and coherently controlling dynamical phenomena. (ii) A regime where the intense lasers optically dress bands resulting in so-called Floquet engineering of the energy-momentum landscape, i.e. whereby driving the system with unique forms of light the band structure can be tuned along with transient material properties.

### 4.2.2.1 Light-driven electron dynamics in solids

The physics of HHG in solids bears many similarities to that of HHG from atomic gases. In particular, the concept of electron trajectories can often be used to understand the emission profiles and analyze the light-driven dynamics. Instead of an electron trajectory starting at the atom's bound state and quivering through the continuum until it ends at a recombination with the hole that remained localized at the parent ion [405] [406], in solids both the electron and the hole are mobile in their respective bands (in $k$-space), where the spatiotemporal intersection of their trajectories results in a recombination and a photon emission [195] [402] [407] [408] [409] [194] [410] [411] [412] [413]. We should note, however, that this picture can be somewhat limited in its success. Thus, the laser polarization and amplitude, or time-dependent polarization in the generalized case of a multi-color field, governs the trajectories. These can be confined to a single band [400] [414] or span multiple bands [408] [415] [416] of the crystalline potential, enabling band structure reconstruction using HHG [417]. Indeed, the crystalline anisotropy of HHG under linear polarization [400] [418] [419] [420] [412] was of the first observations in solids (see Fig. 11(a) below), and has recently been proposed as a probe for electronic dephasing [421] and the transition to non-perturbative regime [422]. Complementarily, the discrete translation symmetry enables a circularly polarized driving field to emit harmonics [412] [415] (see Fig. 11(d), showing DS-induced circular harmonic selection rules in solid HHG), which are forbidden in the spatially isotropic phases due to continuous rotational DS. The absence of certain crystalline symmetries were shown to be imprinted on harmonic sideband generation [418]. The selection rules can indicate the passage of electron trajectories near high symmetry-points in $k$-space, such as band extrema [407] [423]. Further, the symmetry of the crystal can also affect the phase of the HHG with respect to the intensity envelope (carrier envelope phase, CEP) [419], and the appearance of generic phenomena such as HHG circular or elliptical dichroism [424] [415] [425](see Fig. 11(c)), as well as multi-scale DS and their associated angular momentum conservation [426] (see Fig. 11(g)).

One highly intriguing topic of interest is using HHG for probing topological properties of matter on ultrafast timescales. Here, several research directions proposed to use various generic properties of nonlinear optical responses such as circular dichroism [427–429], ellipticity-dependence [424,430,431], elliptical dichroism [424], CEP-sensitivity [432], forbidden harmonic emission [433], time-delays [434], and HHG enhancement [435], as signature fingerprints of topology. These types of universal features are often attributed to a crystal space group and the DS of the driving laser field, defining the dynamical Floquet group [35] that determines the selection rules for HHG and allowed/forbidden processes as discussed thoroughly in this review. The difficulty however lies in that topology is inherently defined by a different symmetry class (e.g. that of $Z_2$ or other topological invariants, connected to how the gap evolves under continuous deformation of the Hamiltonian [436] [437] [438] [439]) compared to the dynamical Floquet group that determines the selection rules for HHG and allowed/forbidden NLO processes. This poses an inherent challenge for theory and experiment to attempt to uncover a universal probe for material topology in ultrafast nonlinearities, which remains an open question [425] [440].

Besides topology, HHG and inherent symmetry breaking was utilized for probing Berry phases (intra and interband), which are generic properties of inversion-broken solids [22] [441] [294] [442] (see example in Fig. 11(b)), and chiral excitation states, e.g. emerging in the chiral quartz and absent from cubic MgO [442]. Instantaneous breaking of lattice symmetries, as induced by coherent phonons, was also utilized as a method of analyzing time-resolved harmonic yields, polarization, and selection rules, as a novel method of resolving phonon dynamics and phase transitions in real time [443] [268] [444] [445] [446] (see illustration in Fig. 11(e)). Even topological beams carrying OAM were employed for exploring material structure in symmetry-induced selection rules [104]. Generally speaking, by utilizing fields that have inherently broken or engineered symmetries one can gain access to a plethora of potential condensed matter phenomena of interest, but doing so with the inherent atto- to femtosecond temporal resolution provided by HHG.

Lastly, we would indicate complementary fields where highly-nonlinear phenomena also gain from a DS analysis, including nonlinear photocurrent generation [44] [447], and ultrafast magnetism [48] [49] [448]. For bulk photo-galvanic photocurrents, it was shown that fields with inherently broken symmetry can induce shift current responses that are highly tunable even in inversion-symmetric systems (where they are commonly forbidden) [162] [44] [163] [164] [449], and which can be used for probing various properties of matter such as evolving dynamical correlations [44] through symmetry-breaking spectroscopy. A symmetry theory connecting the Floquet group of the light-matter system to allowed photocurrent responses and their properties was recently developed [44], and is a trivial extension of the theory for HHG given that photocurrents constitute a zeroth-order nonlinear response (i.e. even order). For the field of ultrafast magnetism, a recent analysis developed the symmetry-induced selection rules for the allowed frequency components with which magnetism can evolve in Floquet systems [48] (see illustration in Fig. 11(f), showing spin and excitation dynamics also adhere to DS selection rules), also affecting the speed of the magnetization process (which was shown to be as fast as several hundreds of attoseconds).

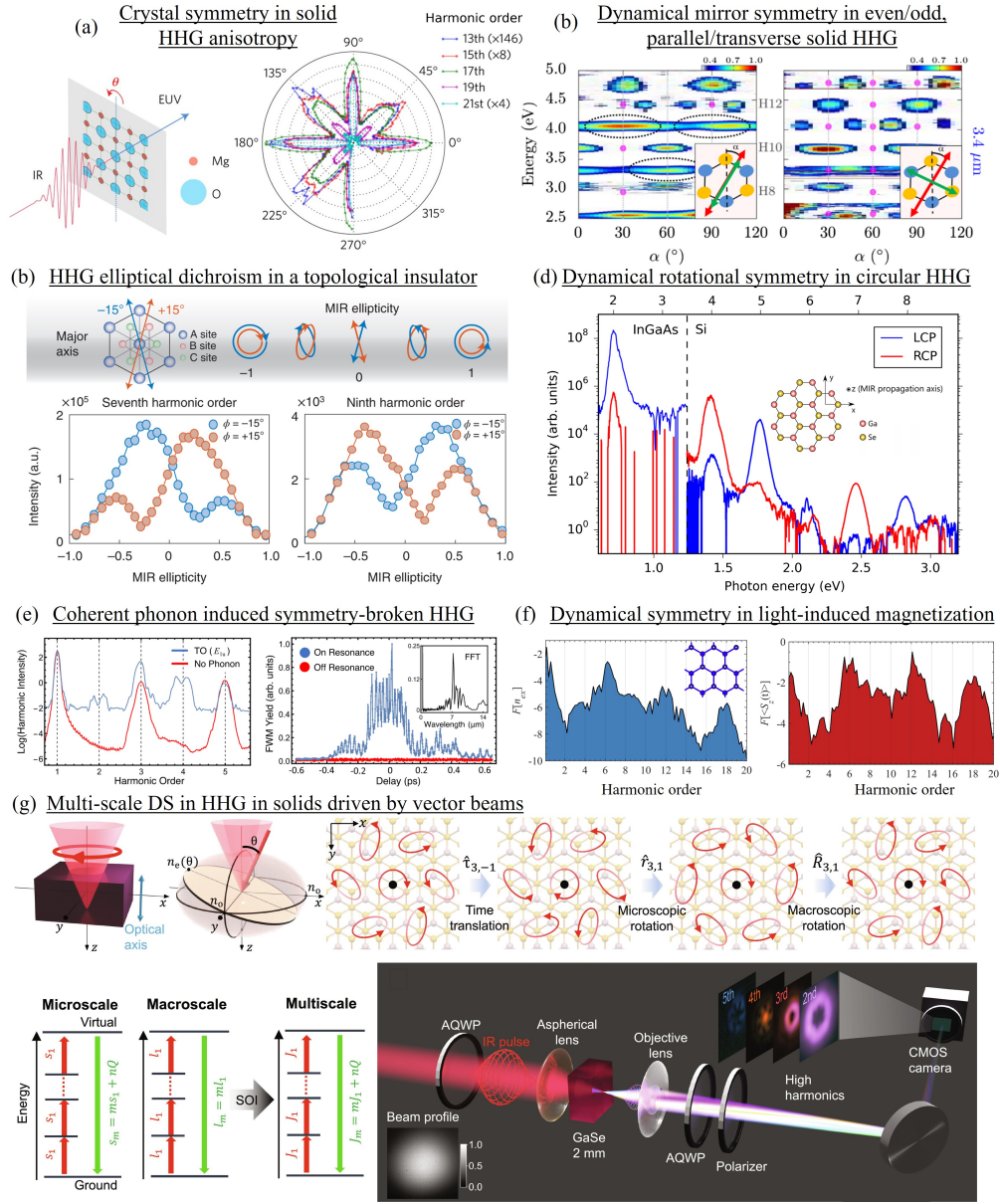

**Fig. 11. Symmetries and selection rules arising in HHG and strong-field phenomena in solids driven by intense laser pulses.** (a) Appearance of crystal symmetry (rotational axes and mirror planes) in HHG anisotropy driven by linearly-polarized lasers with respect to the crystal plane and polarization axis. Figure repurposed from ref. [412]. (b) Dynamical mirror symmetries and selection rules arising in HHG selection rules from $MoS_2$, dictating the polarization and allowed orders of various harmonic components. The symmetry considerations here arise regardless of the mechanism of the emission (interband, intraband, anomalous, etc.). Figure repurposed form ref. [404]. (c) HHG elliptical dichroism arising from material structure in the three-dimensional topological insulator $Bi_2Se_3$, which has been proposed as a method to probe material topology. Figure repurposed from ref. [424]. (d) Appearance of dynamical symmetry induced circular harmonics selection rules in a rotationally-symmetric solid driven by circularly-polarized lasers. Figure repurposed from ref. [4]. (e) Forbidden even harmonics emitted in HHG from hexagonal-boron-nitride (inversion symmetric) as a result of coherent phonon excitations (breaking the Hamiltonian's dynamical symmetry). Figure repurposed from ref. [7]. (f) Dynamical symmetry selection rules arising in strong-field HHG from monolayer bismuth in the electronic excitation probability over time (left), and light-induced magnetization (right) – only $6n$ harmonic frequencies are allowed in those dynamics due to the six-fold dynamical symmetry of the Hamiltonian (for integer $n$). Figure repurposed from ref. [48]. (g) Role of multi-scale DS in HHG driven in solids by vector beams, indicating exchange and conservation of angular momentum of light and matter. Figure repurposed from ref. [426].

### 4.2.2.2 Floquet-engineering and light-dressing

From the point of view of Floquet physics, the use of light for inducing or breaking symmetries in solids has been at the forefront of the field since its conception. The field of Floquet engineering gained a lot of attention around 2010 with several seminal works showing that hexagonal two-dimensional solids can turn topological upon driving with circularly-polarized lasers [450] [451]. Even at that level, there is a fundamental requirement for the driving field to break time-reversal symmetry in order to induce a topological state, hinting towards the importance of symmetry and symmetry breaking from the EM field.

The physics behind the effect relies on Floquet phases of matter, whereby the light-matter system enters into a steady-state upon continuous time-periodic driving that can be described with Floquet theory [267]. From the theory standpoint, the tools employed are very similar to those utilized for deriving the HHG selection rules in section 3.2 – we invoked symmetries of the Floquet Hamiltonian in order to formulate selection rules for observables, while here the structure of the Floquet states and quasi-energy eigenvalues themselves are being investigated. In solids, this takes on additional meaning when looking at the quasi-energy band dispersion in momentum space, allowing the band structure itself (together with material properties such as topology, surface states, transport, etc.) to evolve and be manipulated by laser irradiation (see several examples in Fig. 12).

We should note that the field of Floquet band engineering is very well established. Floquet band replicas and light-induced gaps have been measured directly with time- and angle-resolved photoelectron spectroscopy (Tr-ARPES) [452] [453] [454] [455], indirectly through transport [456], and in analogous systems such as photonic crystals [200] and cold atoms [457]. They have been thoroughly analyzed theoretically in various settings [458]. Even though the typical physical effects have been explored within a weak field driving regime and towards shorter wavelengths (for reasons having to do mostly with thermalization), the field has begone to merge with NLO phenomena in recent years, and even towards the strong-field regime. In particular, Floquet band dressing has not only been explored in the strong-field limit and shown to persist [56] [459], but also been suggested to yield unique properties on attosecond timescales [55] [271] [460] [57] [461], as well as being employed in higher-order wave mixing [462] [463]. What then is its connection to the symmetry and asymmetry properties of light and matter? We will argue here by reviewing several seminal works that the two are inherently connected, and it is in fact the ability of light to break additional crystal symmetries that allows further control knobs for Floquet band engineering.

Let us start by briefly reviewing the Floquet topological insulator out-of-equilibrium states predicted in graphene driven by circularly-polarized lasers. Since graphene is a Dirac semimetal with linear band touching points at K and K', perturbations breaking inversion, time-reversal, and rotational symmetries of the system, inherently open the gap in those points, even vanishingly small perturbations. If this perturbation explicitly breaks time-reversal symmetry (e.g. circularly-polarized light), then a topological Floquet phase can be formed [458]. From a practical viewpoint this means that the Floquet states, which are eigenstates of the symmetry-broken Floquet Hamiltonian with eigenvalues being the Floquet quasi-energy bands, exhibit a non-vanishing topological index such as a Chern number [458]. Another way to look at it is that the hybrid out-of-equilibrium light-matter system now exhibits the broken symmetry of the Floquet Hamiltonian itself (just as we discussed for optical selection rules), meaning that a Berry curvature is allowed (whereas in the field-free system it was symmetry-forbidden).

Nonetheless, this is not the full story for Floquet topology, since topology does not merely require the Hamiltonian to have a broken symmetry with respect to Floquet dynamical groups, but also with respect to other types of deformations and symmetry operations. In that respect, breaking time-reversal symmetry is a necessary but insufficient condition for obtaining a topological state in most systems. Instead, the system must pass a gap closing/opening event, which is also controlled by various chemical and other considerations [439]. Still, the fact that light's symmetries play a crucial role is evidence for the great potential it has in such phenomena.

More recently, it was understood that by employing poly-chromatic laser beams one can gain further control of topological states, e.g. flipping between Dirac and Weyl semimetal states [54], and changing parity in superconducting order parameters [58]. Indeed, the simultaneous breaking of multiple symmetries by the driving field in the Floquet Hamiltonian accomplishes several goals. First, it allows additional fundamental DOF for manipulating the electronic structure. By fundamental, we mean that every additional broken symmetry inherently changes the system's behavior as a result of decoupling degeneracies in the Floquet Bloch states. For instance, when time-reversal symmetry is broken, Floquet-Bloch states are no longer degenerate upon taking $\mathbf{k} \rightarrow -\mathbf{k}$ (see example of this effect in Fig. 12(a)). By breaking inversion symmetry, one can induce nonzero shift vectors [434] [464], allowing also even-order nonlinear optical responses from the out-of-equilibrium system such as photocurrents [447] [465]. By breaking rotational or mirror symmetries, one alters the irreducible BZ, modifying gaps and tuning the bands, and so on and so forth (e.g. see example in Fig. 12(d)). This provides tunability that otherwise could only be obtained by manipulating the crystal structure through strain, defects, etc., whereas here the control is all-optical and occurs on potentially femtosecond timescales. Second, driving the system with polychromatic fields introduces additional control knobs for Floquet engineering of the bands. The more tunable parameters are introduced into the Hamiltonian, e.g. in cases of poly-chromatic fields relative intensities, relative phases, polarization states, etc., the more freedom one has to achieve a desired target band structure [14] [53] [57] [466] [467]. Lastly, we also note that by introducing fields with multi-scale broken symmetries one can also break symmetries in novel ways that could lead to similar physics such as Floquet topological insulators induced by spatially-broken symmetries [468] (see Fig. 12(e)), but also new phenomena yet to be discovered. This is especially true for the emerging field of vortex and structured light beams that exhibit multi-scale structures as discussed in sections 2.1.3 and 2.2.

Moving towards the strong-field regime where highly nonlinear phenomena arise, it was recently shown that similar Floquet states of matter are not 'melted away' by the intense laser, and in fact persist during strong-field driving [56] [460]. It was experimentally shown that such states are established very quickly within one or two driving cycles [271], in accordance with our expectation for the required pulse durations for generating nonlinear optical selection rules [269] (and potentially even faster, see Fig. 12(b)). Indeed, the two physical phenomena are connected by the applicability of Floquet theory, even though very different observables are being probed in each case. On the one hand, these states are currently being explored for their fundamental involvement in nonlinear optical phenomena such as HHG or photocurrent generation, which is an interesting and emerging field of research in its own. On the other hand, strongly-driven Floquet states can be utilized for altering material properties on femtosecond timescales. Recently, such strong-field Floquet dressing was proposed for tuning valley DOF in hexagonal materials [51] [460] [14] [447] (e.g. in Fig. 12(a)). Moreover, it was shown to also alter properties of excited states such as excitons [469]. The intricate attosecond dynamics within the Floquet states is also a topic of recent studies that connects with HHG and attosecond science [57] [461]. Besides the standard notion of a time-averaged effective Hamiltonian, recent experiments propose novel dressed states that arise on timescales below a single laser cycle, allowing gaps to close within few hundred attoseconds [55] (see Fig. 12(f)), as well as non-Floquet states that can emerge during the laser pulse turn-on and turn-off period [57]. It currently remains unclear to what extent this phenomenon is general, as it might depend on the system properties or laser regime. Using symmetry adapted bi-chromatic fields was also recently employed as a novel approach to study photocurrent excitation in Floquet topological systems [447] (see Fig. 12(c)). What it is clear is that by utilizing light with dynamical symmetries in the study of Floquet or other light-dressed phases of matter, one gains additional control over the dynamics, and potentially new physical effects can be uncovered. In our opinion, as the field will evolve in coming years, also the application of light fields with multi-scale symmetries and asymmetries will be applied in Floquet band engineering, paving the way to novel phenomena at the intersection of condensed matter and nonlinear optics.

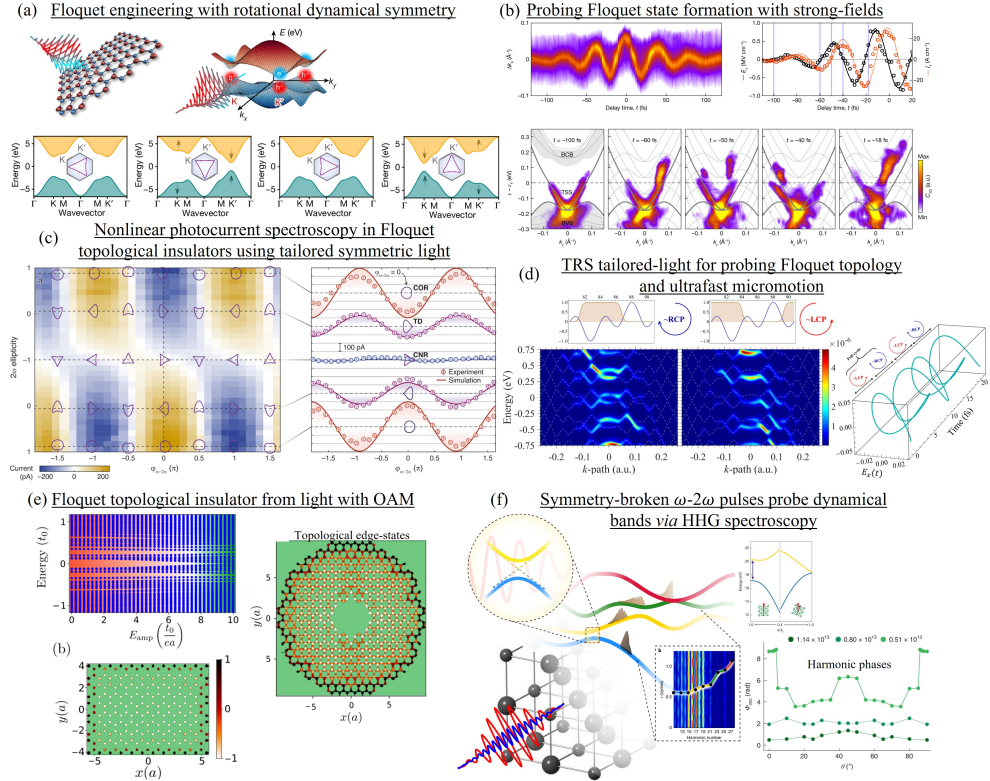

**Fig. 12. Utilizing symmetric and asymmetric light forms for Floquet band engineering and ultrafast spectroscopy of Floquet physics.** (a) Bi-chromatic bi-circular ω-2ω fields used for Floquet band dressing in hexagonal-boron-nitride monolayer, permitting breaking of time-reversal symmetry and tuning band edge positions. Figure repurposed from ref. [14]. (b) Few-cycle THz pulses employed for probing Floquet state formation via time-and angle- resolved photoelectron spectroscopy (Tr-ARPES), probing the adiabatic assumption in Floquet theory. Figure repurposed from ref. [271]. (c) Utilizing dynamical mirror symmetric bi-chromatic pulses to induce photocurrents in Floquet topological insulators, and probe the Floquet state dynamical phases on sub-cycle timescales. Figure reproposed from ref. [447]. (d) Employing three-color ω-2ω-3ω pulses that are individually linearly-polarized, and overall exhibit time-reversal symmetry, in order to probe topological Floquet state formation and fundamental responses. In this case by tuning the relative color phases and amplitudes one can engineer a field that over a full cycle is time-reversal symmetric, but in each half-cycle is approximately circularly-polarized and flips its handedness between half-cycles. Combined with Tr-ARPES this allows probing potential sub-cycle Floquet state formation and non-Floquet physics. Figure repurposed from ref. [57]. (e) Generation of Floquet topological insulators driven by linearly-polarized light with OAM. Figure repurposed form ref. [468]. (f) Probing dynamical sub-cycle changes in electronic bands (dressed states) through two-color (broken symmetry) solid HHG spectroscopy. Figure repurposed from ref. [55].

### 4.2.3 Chiral systems

In this subsection we review the particular use of tailored symmetric and asymmetric forms of light for nonlinear spectroscopy of chirality, focusing mainly on molecular chirality in randomly-oriented media. As presented in chapter 2.3, chirality is inherently a form of asymmetry that can only be probed by an additional form of asymmetry. In that respect, an object's chirality only comes in to play when it interacts with another chiral object (be it light or matter). Consequently, tailoring the spatio-temporal properties of light plays a crucial role in the study of chirality as it allows generating physical situations with enhanced chiral signatures by controlling the asymmetries of EM fields (in the polarization domain, the spatial domain, or both).

Before starting this section, it is worthwhile to discuss the motivation for exploring chiral systems to begin with. This is a vast field of research encompassing many scientific disciplines, and it can therefore be difficult for new researchers to comprehend its 'big picture' importance. Let us highlight some main physical aspects that motivate the study of chiral phenomena, and especially its manifestation in material and chemical systems. First, we should impress the fact that all biological life as we know it is chiral. In that respect, chirality manifests on all biological length scales ranging from macroscopic objects such as hands, anatomy, and other structures (e.g. snail shells [470]), down to the micro-scale of cell configurations, and all the way down to the nano-scale in the chirality of DNA, amino-acids, or other functional chemical groups [471] [472]. A direct consequence of this is that chirality plays a basic role in drug-development and molecular biology [473] [252]. For instance, a drug might either trigger a desired effect such as symptom relief for one handedness, or in extreme cases be poisonous in its mirror twin [474]. Since chirality (and asymmetry) is so prevalent in nature, most drugs involve chiral molecules in one form or another. Hence, chiral characterization and enantio-separation is perquisite for drug development, and is a main challenge in industry, especially as moving towards bigger bio-molecules with multiple active chirality centers [475]. On a more fundamental level, the origin of this chiral symmetry breaking in nature is not understood. That is, the question of why life has chosen one particular handedness for certain structures remains largely unanswered, especially since two pairs of chiral molecules are essentially energy-degenerate [476]. This is an intriguing field with immense consequences in the life sciences, and it is very active. Interestingly, a set of very recent works suggest that life's homochirality could have originated due to cooperative feedback between bio-molecules and magnetic surfaces [477] [478] [479] in connection with chirality-induced spin selectivity [480] [481]. The biological aspects connect to a second avenue of importance in the field of physics. Here chirality plays an essential role in the standard model of particles [482]. For instance, in interactions involving the strong and weak nuclear forces, one form of chirality for a particle might interact more strongly than another, which is an empirically-observed phenomenon that implies the universe has preference for a certain handedness. Such effects are usually described as a result of a spontaneous symmetry breaking, and their origin remains elusive. It is not yet clear how this type of chiral anomaly connects with the one observed in chemistry and biology, though this is also an active area of research [483] [484] [485] [486]. It would be especially interesting if one could link the origin of chiral phenomena ranging across scales, e.g. from fundamental particles, up to scales of the cosmos [487]. However, such connections are not clear, as there could be multiple sources of chiral symmetry breaking in different physical systems. Lastly, we should mention that there is a currently evolving field that studies ultrafast and dynamically evolving chirality [23] [488]. Here there are fundamental questions such as how do chiral molecules switch handedness during chemical reactions, how can the chirality of matter be controlled in order to generate desired states and tune material properties, and what are the fundamental mechanisms that allow such phenomena. This very brief overview is meant to paint the general context for the existence of the field, and why it is important and gaining more traction in recent years.

Even though the field is old and well-studied, it has been partially revolutionized in the last two decades with an outflux of novel methodologies for detecting and manipulating chirality based on tailored light and highly nonlinear light-matter interactions. We attempt to review here the latest progress in the context of the established older methodologies, and in the language of symmetries and asymmetries. Of course, we cannot include all possible efforts, and refer interested readers to more focused recent reviews and books [23] [488] [489] [27] [27].

We will start the section by describing standard established methodologies such as circular dichroism in absorption and optical rotation, and move on to more recent progress in methods either relying on weak beyond-dipole light-matter interactions, or more intense purely electric-dipole light-matter interactions.

### 4.2.3.1 Chiral spectroscopy beyond the electric-dipole

#### 4.2.3.1.1 Absorption circular-dichroism

The first and most abundantly employed methodologies for chiral spectroscopy rely on the interactions of chiral media and EM fields within the linear-optics domain. Indeed, these methodologies are as old as the original discovery of optical activity in quartz [490], and are commonly employed in the drug industry for characterization of chiral molecules, and even in the sugar industry for measuring the sugar concentration of various products. The fundamental physical mechanism in operation here involves either electric-quadropole interactions, or, a combination of electric- and magnetic-dipole interactions [27], as well as contributions from higher order beyond electric-dipole terms. Electric-dipole-based contributions end up not contributing to the chiral signal [27], which can be understood directly from the formula for OC in Eq. (15) – the OC vanishes within the dipole approximation. Electric-quadropole interactions also vanish in unoriented media due to symmetry considerations. Put differently, if one removes all spatial dependency in a CP EM field, what is left is just a rotating polarization in a plane, in which case the molecule can't 'distinguish' which direction the light is coming from – is it right-CP coming head on, or left-CP propagating coming from behind? That immediately means there cannot be chiral signals that are purely electric-dipole-based within the dipole approximation, unless there is some other manner the molecule can distinguish the handedness of light.

With this in mind, let's review the standard mechanism for chiroptical spectroscopy. Here chiral media is irradiated by CPL and its absorption (or reflection) is measured. One then performs a second experiment in the same conditions by reversing light's handedness, in which case if the medium is chiral, a different absorption coefficient is observed. The normalized difference in absorption is referred to as circular dichroism (CD):

$$CD = 2\frac{I_+ - I_-}{I_+ + I_-} \tag{52}$$

where here $I_{+/-}$ denotes to the absorption under left/right-CPL, and the factor 2 arises from a standard notation in the community, bounding the CD from -200 to 200%. Importantly, this experiment is fully equivalent to one performed with a fixed handedness for light, but where the handedness of the chiral media is inverted. From the CD one can infer not only if the medium is chiral, but also its 'degree of chirality', or in other words, the relative excess of one handedness of molecules over the other. This is denoted as enantiomeric-excess (EE), since the two copies of a chiral molecules connected by mirror symmetry are called enantiomers. Detecting EE is essential for drug development and control verification.

An alternative technique that relies on the same physical principle is optical rotation [27]. Here one shines linearly-polarized light onto material samples instead of CPL, and measures the rotation of light's polarization axis instead of the CD. The physical mechanism is equivalent, because one can decompose linearly-polarized light into a sum of left- and right-CP components, where if one component is preferentially absorbed the overall polarization axis rotates.

Such techniques are extremely useful due to their simplicity and robustness, but also have some underlying weaknesses that should be clarified. First, the resulting chiral signal (or CD) is inherently very small, owing to its beyond-electric-dipole origin (with terms proportional to one over the speed of light in the Hamiltonian making the main contributions). From a mathematical perspective of the equations of motion, this is also clear because the typical electric field of optical pulses is much smaller than atomic-scale values (even in the strong-field regime), meaning that higher order terms are suppressed. Indeed, the signal is typically on the order of ~$10^{-4}$-$10^{-6}$, requiring high-sensitivity and expensive instruments for accurate detection. Another way to understand the weak origin of these CD signals is to analyze the length scales for chirality – molecules are typically chiral on angstrom to nanometer length scales, which is the length scales upon which inversion symmetry is broken (on the order of the bond length). CPL on the other hand shows a helical screw-like structure on length scales of the wavelength, which is in the optical domain few hundred nanometers, and orders of magnitude larger. One then needs to imagine a molecule 'looking' at the structure of incoming light and trying to understand if its chiral or not and how it should react – it would be almost analogous to a human being trying to establish that the earth is round just by looking at the horizon – the effect is very weak (see illustration in Fig. 5(a)). Expectedly though, if the wavelength of the light is reduced to the X-ray range (e.g. as in magnetic circular dichroism [491] [492]), then chiral signals substantially increase, but then one can run into other issues connected with enhanced absorption and poor optics. In the other direction one could also look at the chirality of larger nanoparticles [493] [494], which would also enhance chiral signals, but often is not the fundamental interest of chemists. A second weakness is that the method is partially limited in its capabilities. For instance, if a molecule has more than one chirality center it is not trivial to analyze the signal's source, or how to differentiate in measurements between enantiomers (stereoisomers), or diastereomers of all types (molecules with the same chemical formula but with different bond arrangements around the chiral centers that are not connected by mirror symmetry). Indeed, in such situations one might get a vanishing CD from a sample with non-equal concentrations of two diastereomers that have opposite contributions to the optical activity. Overall, for complex molecules a combination of several experimental techniques is then required. Thirdly, CD signals are inherently quite weak for gas-phase, and even liquid systems. Indeed, most realizations rely on pre-crystallization of molecular samples into molecular-solid phases, which cannot always be performed. Furthermore, for fundamental physical investigations one often wishes to analyze the cleanest possible physical set-up, where the pure chiral response of the molecules is separated from other factors such as interactions with solvents. This response typically arises in the gas phase. Lastly, we wish to point out that CD can generally arise also in non-chiral systems, meaning one should take care in interpreting signals in certain experimental conditions. If the medium is randomly oriented or amorphous, then indeed CD inherently means the sample is overall chiral and lacks inversion symmetry. On the other hand, CD also arises in samples of oriented matter, either oriented molecules in the gas phase, or single crystal solids. It can also arise in randomly-oriented achiral media if that sample is pre-excited. Therefore, a crucial point in all chiral signal analyses is to separate the origin of chiral signals arising from matter's intrinsic chirality and other sources.

We should also mention that such methodologies have been quite successfully extended to the infrared and near-infrared regions, yielding vibrational-CD spectroscopy [495]. Here the molecular vibrational modes play an additional role as in Raman spectroscopy, resulting in similar types of signals. This methodology can also be applied for measuring ultrafast chiral responses in the timescales of picoseconds to few hundreds of femtoseconds [496]. Notably, the challenge of obtaining ultrafast time-resolved chiral signals, e.g. for probing chemical reactions or other dynamically evolving chiral physics, is even greater than that of resolving EE in the ground state of material and chemical systems. The main challenges here are that: (i) the dynamical part of the chiral signals are typically smaller and on order of few precents of the ground state CDs [496], (ii) the experimental set-ups are much more involved, requiring pump-probe configurations and a stability of the laser and material sample over a long time duration. Another noteworthy extension is the use of cavities for enhancing CD signals, which rely on polaritonic and quantum electrodynamics considerations [497]. So far, and to our knowledge, cavity-enhanced chiral sensing has only been predicted [498] [499] [500], and experimental efforts are ongoing. Moreover, similar in origin signals can be obtained in enantio-selective orientation and rotational dynamics of chiral molecules driven by helical pulses [501] [245] [502] [503].

#### 4.2.3.1.2 Chiral HHG

At this stage, we move the discussion towards more modern methodologies for probing chirality with enhanced resolution, which are based on tailoring the symmetries and asymmetries of light fields. We still restrict this subsection to beyond-electric-dipole-based interactions. There are two main techniques we wish to discuss. The first relies on HHG employing light with nonzero OC [504]. In 2015, a pioneering experiment by Cireasa *et al.* generated high harmonics by driving a gas of the chiral molecule epoxypropane with elliptically-polarized light (see schematic in Fig. 13(a)). The main innovation here is that the light field was purposefully restricted to relatively low ellipticity states (up to ~±0.3), which deviates from the typical conditions where one expects the strongest chiral signals to appear (usually in circular driving conditions). The reason for the small ellipticity in the driving field is connected with the fundamental physical mechanism for HHG in the gas phase – recollisions of ionized electrons with the parent molecular ion [348] [335]. Such recollisions simply do not occur for circular or high-ellipticity driving.

In their experiment, Cireasa *et al.* showed that one can obtain quite strong chiral signals on the order of ~1% from HHG. The chiral signal is practically defined in accordance with the CD in Eq. (52):

$$\text{ED}(n) = 2 \frac{I_{HHG_+}(n) - I_{HHG_-}(n)}{I_{HHG_+}(n) + I_{HHG_-}(n)} \tag{53}$$

where ED is in this case the elliptical dichroism, $n$ is the particular harmonic order (since one can have separate signals for each separate harmonic), $I_{HHG_\pm}(n)$ is the yield of the $n$'th harmonic driven by left/right-handed elliptical light, and exchanging light's helicity in Eq. (53) is equivalent to an enantiomeric exchange. One might even further renormalize these values to the ellipticity of the drive (as often done for elliptical dichroism (ED) in the linear-response regime), yielding effective numbers on the order of ~10%; however, that is a matter of terminology that is not necessarily effective here (since there would not be any harmonics generated for CP driving). Regardless, and no matter how the numbers are expressed, it is clear that these figures are at least an order of magnitude larger than standard CDs obtained in optical absorption from CPL. The main reason, as was explained in refs. [505] [506] [507], is attributed to the highly nonlinear nature of the HHG process, where the multi-photon contribution to the light-matter interaction appears to enhance the relative contributions of dichroic terms.

Another advantage of the experiment in ref. [504] is that it benefits from the inherent temporal resolution of HHG – with pump-probe geometries signals of resolutions of few femtoseconds are typically obtained, and through high harmonic spectroscopy (HHS) one can even improve such resolutions to the sub-cycle attosecond-resolved domain. In that respect, one needs to consider the wealth of information in a given harmonic spectra – since the chiral signal is resolved over a broadband (often more than an octave) spectral domain, it contains much more data than an equivalent linear-response CD for a given driving wavelength. Through the fundamental physics of gas-phase HHG, each harmonic order can be connected with a typical time of ionization and recombination [32], providing potential for following chiral dynamics with sub-cycle resolutions, directly in the gas-phase, and with table-top light sources.

This paper led to a larger follow up of theory works attempting to understand the origin of the enhanced response, its potential, as well as possible directions to further enhance it. One noteworthy avenue that was first proposed through theory predictions, is the use of $\omega$-$2\omega$ bi-circular fields for driving HHG from chiral media [505] [506] [507]. This represents a natural extension of the concept of ref. [504], as bi-circular driving was shown to allow generation of circular harmonics from gases [86] [88] [59], overcoming issues connecting to recombination cross-sections (see section 4.1). Moreover, it was argued that such light carries enhanced optical chirality that can be tuned by the field's intensity ratios [223] [217]. Consequently, ref. [505] made the first predictions for chiral signals driven in HHG with bi-circular light to be as high as ~80% (although later more accurate calculations found signals should be on the scale of ~5% [507]). Note here that the physical mechanism responsible for these signals remarkably still relies on an interplay of electric- and magnetic-dipole interactions.

Following the prediction in ref. [505], several groups set-out to perform experiments. The predictions seemed to be somewhat too optimistic, but nevertheless extremely stable signals on the order of ~10% were measured in ETH in ref. [508]. Slightly larger (although also somewhat noisier) signals were reported in ref. [509] with a similar methodology. In a continuation study, the group at ETH studied the dissociation of a chiral molecule in a pump-probe setup using this configuration (see Fig. 13(c)), shedding light onto the mechanism by which a molecule's chirality decays and even changes sign as it dissociates, with a temporal resolution of ~50 fs [510]. Thus, the potentials of this approach for studying ultrafast chirality were validated.

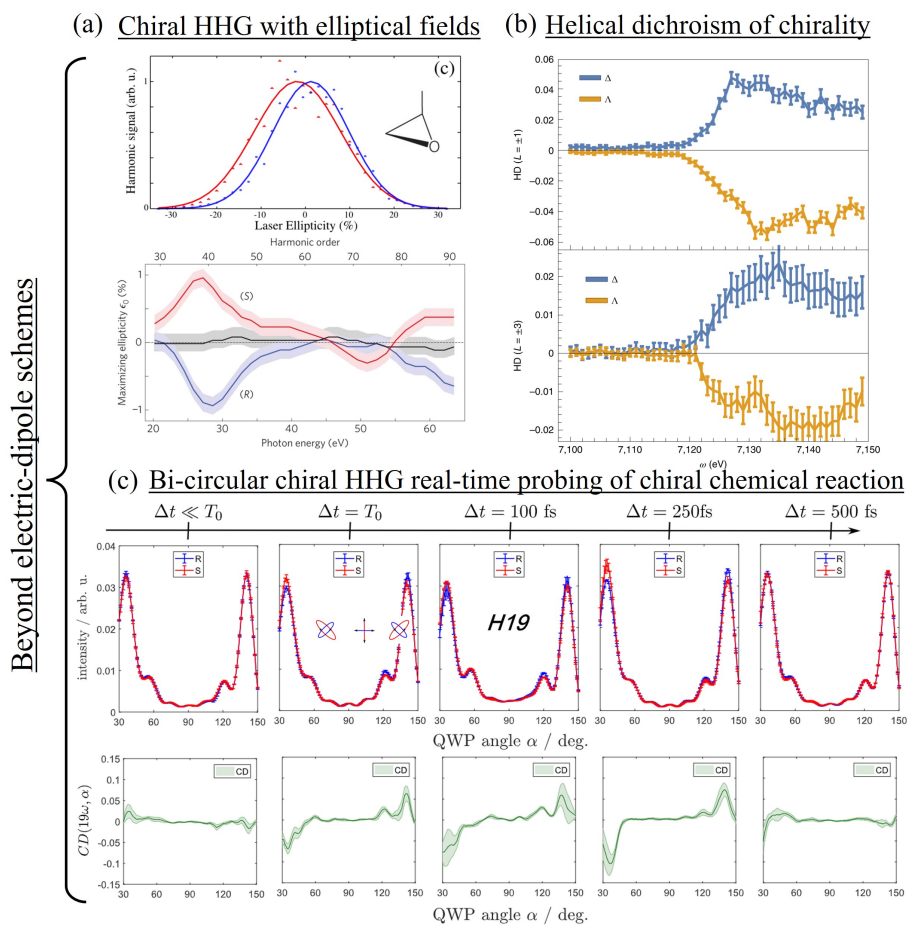

**Fig. 13**. **Spectroscopy of chirality using symmetric/asymmetric forms of light in various configurations with methods requiring an interplay of magnetic- and electric-dipolar interactions.** (a) Chirality spectroscopy via HHG employing elliptically-polarized laser fields with low ellipticity. The chiral signal here still originates from an interplay of electric and magnetic dipole interactions, but the extreme nonlinearity makes it much larger on order of ~1%. The signal arises in the maximizing ellipticity of the HHG yield. Figure repurposed from ref. [504]. (b) Helical dichroism using light with OAM employed for probing molecular chirality. The signal is still based on an interplay of electric and magnetic dipoles, but arises even when the driving beam is linearly-polarized due to an asymmetry in the evolving phase of vortex beams. Figure repurposed from Ref. [511]. (c) ω-2ω bicircular light probes molecular chirality during a chemical reaction with a resolution of few tens of femtoseconds. This chiral signal originates similarly to the mechanism in (a), but the bichromatic field allows for non-vanishing HHG signals even with circular driving components, inducing larger signals on order of ~10%. The signal is tracked in a pump-probe experiment as a molecule dissociates, showing handedness flipping. Figure repurposed from Ref. [510].

### 4.2.3.1.3 Chiral-spectroscopy with OAM

The second noteworthy approach for probing chirality through beyond-electric-dipole interactions with chiral matter utilizes topological light. Here, the light still carries nonzero OC, but the OC arises due to the helical spatial wave-front of the beam, i.e. due to the OAM-associated term in Eq. (17). Moreover, one could consider light carrying a combination of both OAM and polarization-associated OC (SAM), which makes the analysis more complex. Let us first review recent advances along these lines while focusing on the linear-optical regime, and then discuss directions moving towards NLO.

Brullot *et al.* demonstrated the use of light with OAM for chiral discrimination [512]. They employed GL beams that were linearly-polarized (i.e. not carrying SAM or polarization-associated OC), carrying an OAM of $m = \pm 1$ (see Eq. (12)), and measured the transmission of light at 800nm carrier wavelengths from a sample of surface adsorbed chiral molecules. Upon flipping the handedness of light's OAM, a dichroism of a form similar to Eq. (52) is obtained:

$$\text{HD} = 2\frac{I_{0,+} - I_{0,-}}{I_{0,+} + I_{0,-}} \tag{54}$$

where HD stands for 'helical dichroism', a dichroism equivalent to CD but arising not from light's polarization, but rather its OAM, and we have labeled $I_{0,\pm}$ to represent the absorbance (or transmission) of a light beam carrying a SAM of zero (i.e. linearly-polarized, denoted by the first sub-index), and nonzero OAM (denoted by the second sub-index, $\pm$). Brullot *et al.* obtained substantial HD signals of up to ~0.5%, though note that the samples were not in the gas phase, where signals are typically weaker. For HD it turns out magnetic-dipole interactions are the dominant source for chiral signals (with electric-quadrupole interactions possibly contributing in oriented media [513] [514]).

An important point here is that despite the mechanism and physical setup being different than the well-established methodologies of CD and optical rotation, the chiral signals are typically on similar scales. Indeed, the screw-like shape formed by the GL beam's wavefront evolves on the length scale of the wavelength, just as for the SAM case. Thus, one typically does not expect HD to be substantially more efficient than CD, though it remains to be seen what will happen in the highly-nonlinear regime of light-matter interactions. Besides, there are some additional complexities in the measurements of HD that should be reflected upon. These might either be considered a source for interesting physical phenomena, or disadvantages for deriving simple and robust spectroscopies of chirality. First, OAM can convert to SAM (and vice versa) upon interactions with matter [135] [130]. Thus, its propagation might not be sufficiently stable for high-resolution measurements. This is especially true for beams with very high OAM that tend to separate through dispersion and other macroscopic processes to lower-order OAM [515] [516]. Second, the beams unique donut-like spatial structure means that the signal is not collected from the center of the beam, but rather from the ring-like shape where light's intensity is maximal. This effect can be useful for high-resolution imaging, but might also result in anisotropies of the sample becoming more prevalent in HD signals, masking the chiral contribution to HD. Another issue is that the additional complexity in the light state inherently means more control is required in order to obtain well defined chiral signals. Consider for instance a case where the beam's polarization state is not perfectly linear, but with a small ellipticity due to polarization imperfections. In that case one could replace the HD formula in Eq. (54) with the form:

$$\text{HD} = 2\frac{I_{\varepsilon,+} - I_{\varepsilon,-}}{I_{\varepsilon,+} + I_{\varepsilon,-}} \tag{55}$$

with $\varepsilon$ representing the beam's ellipticity. The dichroism here would not perfectly reflect the system's chirality, since the driving conditions for $I_{\varepsilon,+}$ and $I_{\varepsilon,-}$ are not perfect mirror images of one another. Similarly, imperfections in the beams' OAM between positive and negative handedness states could be imprinted onto the HD. Such matters require careful experimental examination of the light source. Similar issues of course in principle occur also for CD, but the status of polarization control and measurement devices is much more advanced than similar methodologies for controlling OAM. Nonetheless, we should impress that the methodology brings forward potential advantages, simply because it enhances the tunability one has in the asymmetry of the light source. Recent predictions [517] and measurements [511] in the X-ray domain (where such signals are much stronger) showed that it can be quite useful leading to signals of ~10%, and even with beams with angular momentum as high as ±3 (see Fig. 13(b)). Though, contrary to the intuitive expectations beams with higher OAM did not lead to a stronger HD signal.

More recently, results in similar spirit were demonstrated in the liquid phase but through a multiphoton nonlinear absorption mechanism [518]. The unique thing about the results in ref. [518], apart from measuring dichroic singles of up to ~5%, is that they showed obtain helical-circular dichroism can also be observed in achiral media. That is, if one defines a dichroic signal by flipping only the handedness of light's SAM, but fixing the handedness of the OAM, a differential absorption should be obtained also if the medium is achiral. Such a signal could be defined as helical-circular-dichroism (HCD) [517]:

$$\text{HCD} = 2\frac{I_{s,m} - I_{-s,m}}{I_{s,m} + I_{-s,m}} \tag{56}$$

where the indices $s$ and $m$ represent the driving beams SAM and OAM, respectively. One could similarly flip the OAM handedness but fix the SAM:

$$\text{HCD} = 2\frac{I_{s,m} - I_{s,-m}}{I_{s,m} + I_{s,-m}} \tag{57}$$

Nonzero HCD could arise simply because in these conditions the optical beams driving the two measured absorption spectra are not connected by a mirror symmetry. Indeed, the true chiral signal could only be obtained by a proper mirror operation on the light field, yielding:

$$\text{HCD} = 2\frac{I_{s,m} - I_{-s,-m}}{I_{s,m} + I_{-s,-m}} \tag{58}$$

, which generalizes the formulas for HD in Eq. (55) and CD in Eq. (52).

From this discussion it is clear that light carrying complex polarization and spatial asymmetries has a great deal of potential for chiral recognition. It also requires developing theoretical methodologies for properly describing the light-matter interactions, and for analyzing dichroic signals. We also note that similar in nature extensions were employed to Raman optical activity [519] [520], though that is beyond our scope.

We will end this subsection by noting that the current state of the art in the field of HHG indeed has been advanced towards the use of light beams exhibiting complex orbital and polarization states [157] [153] [159] [156] [38] [36] [156] [43]. Even though most research has focused on atomic media and on light-source-based applications, we believe that the application of such intense beams of light for probing chirality is immanent, and holds great potential to combine the best of both worlds. This would be highly relevant even if only considering light-matter interactions beyond the electric dipole, because highly nonlinear interactions can cause an enhancement of the chiral signal and provide superb temporal resolution. We should also mention a recent theoretical suggestion for employing a light beam that is both topological, and locally-chiral, for probing chirality [521], which is predicted to lead to other unique signatures of topology imprinted in the harmonics' polarization and angular states. Such an approach has the potential to advance helical-circular dichroic signals into the realm of the electric-dipole intense interactions. It connects with our next sub-section that discusses novel nonlinear chiroptical techniques that rely on purely electric-dipole-based interactions.

### 4.2.3.2 Modern electric-dipole-based chiral spectroscopy

To contrast with the varying approaches described above (that rely on non-electric-dipole light-matter interactions), the recent decades have seen a rapid development in modern methodologies capable of resolving chirality within the electric-dipole [23]. These rely on a rather diverse paradigm of either uniquely tailoring the asymmetry in the laser field, measuring much more complex multi-dimensional observables, or combinations thereof. Naturally, the great advantage here is that the typical scale of the chiral signals are orders of magnitude larger than in other techniques. In certain cases, they are even predicted to reach the ultimate limit of 200% chiral discrimination, and have been shown to distinguish the absolute configuration of chiral molecules – meaning to experimentally resolve the molecular handedness without comparison to theory or prior reference data. We will now review these developments while focusing on the symmetry (asymmetry) perspective, and on the technical aspect of the different types of observables from which chiral signals are constructed. Note that in choosing this path the various methodologies are not chronologically ordered.

### 4.2.3.2.1 Circularly-polarized light and photoemission

The first type of novel methodologies we review relies on using driving fields with nonzero OC, that is, light that is chiral only outside of the dipole approximation (see section 2.3.1). However, in this case the main observables of interest are not optical fields (e.g. reflected/transmitted light, high harmonics, etc.), but rather photoemitted particles. These could be photoelectrons, photoions, dissociated molecular fragments, or combinations thereof. When such photoemission is resolved along specific axes, it provides chiral signals that arise from purely electric-dipole based interactions. The origin of this enhanced chiral interaction has been analyzed in different ways in the literature, and in recent years has been reformulated in connection with chiral triple-products [522] [523] [524] [525]. Essentially, the ability to resolve a particular observable along different orientations in space allows distilling the electric-dipole based chiral signals. Intuitively, this is not so surprising if we recall that the purely electric-dipolar interactions are always presents and dominant for any molecular species, but simply average out to zero when integrating the chiral signal in the orientation averaged ensemble of molecules. The projection operation allows this averaging procedure to not fully cancel out all of the chiral electric-dipolar response. Another way to think about it is that the projection axis is a way for the molecules to 'know' which direction the CPL is coming from without 'knowing' the fields spatial dependence, since the observer (or detector) can tell that by measuring along specific axes. For the rigorous mathematical analysis, we refer interested readers to a recent perspective by Ayuso *et al.* [23].

With this in mind, let us analyze several different set-ups and their main chiral signals. First, let us discuss photoelectron circular dichroism (PECD), which is a very popular and successful technique (see ref. [526] for extensive discussion). Here an ensemble of gas-phase molecules is driven by CPL (or generally light with polarization-associated OC), and the angle-resolved photoemission is detected in a reaction COLTRIMS microscope [527] [528]. From this measurement one resolves (potentially in full 3D) the photoelectron *vs.* the electron momenta, $P_{\pm}(k_x, k_y, k_z)$, where the index $\pm$ indicates the handedness of the driving light. In practice, one of the momentum axes is usually integrated over, and an effective 2D distribution is obtained where one of the resolving axes must be the light's propagation axis (the axis along which some angular momentum and mirror symmetry breaking is present, denoted here as $z$). By performing the measurements with flipped handedness for the light field one can obtain the standard form of PECD in photoemission:

$$\text{PECD}(k_x, k_z) = 2 \frac{P_+(k_x, k_z) - P_-(k_x, k_z)}{P_{\text{norm}}(k_x, k_z)} \tag{59}$$

Note that there are several different methodologies for normalizing the denominator in Eq. (59) such that there are no erroneous signals in regions of the spectra with very low photoelectron yields. Typically, $P_{\text{norm}}$ is either taken as the total photoelectron yield, or as the maximal yield for a given momenta, though there are some other conventions, and these change the absolute size of the chiral signals.

PECD presents a two-dimensional function that is enantio-sensitive. Depending on the symmetry properties of the driving light, the PECD spectra respects certain selection rules. For instance, if the driving beam is CP and propagating along the z-axis, then one necessarily has that $-\text{PECD}(k_x, -k_z) = \text{PECD}(k_x, k_z)$, which arises from the mirror plane in the driving field within the dipole approximation (connected with the selection rules derived in section 3.2). This is the main chiral signal denoted as a forward-backwards asymmetry. Essentially, the photoemission process tends to prefer ionizing electrons with either positive or negative momenta along $k_z$, depending on the handedness of the molecule (or on the handedness of the light, the interchanging of which is fully equivalent). This is a purely chiral effect that arises in isotropic randomly-oriented ensembles of chiral molecules (similar asymmetry arises also in aligned systems, but is dominated by orientation rather than intrinsic chirality). PECD is therefore formally odd (antisymmetric) with respect to a mirror plane along $k_z$. Moreover, in the limit of long duration CPL pulses (where CEP effects do not play a role) there is also an up-down symmetry that arises from the field's 2-fold rotational DS in the $xy$ plane: $\text{PECD}(k_x, k_z) = \text{PECD}(-k_x, k_z)$. The field's continuous rotational symmetry also means that $\text{PECD}(k_x, k_z) = \text{PECD}(k_y, k_z)$. Let us clarify though that these are not absolute symmetries of PECD, because they arise from the remaining symmetries of the driving light within the dipole approximation, and can therefore be tailored by using other forms of light or if other types of interactions become relevant. For instance, in refs. [529] [216] [304] PECD was explored using $\omega$-$2\omega$ fields breaking the up-down symmetry and allowed to temporally-resolve the moment of ionization (see Fig. Fig. 7(d)). In ref. [305] PECD was explored using locally-chiral light fields, which generated a fully asymmetric PECD function (we will address this particular technique in a later section). We should note that no experiment or theory yet has considered using light's non-instantaneous OC for driving PECD, although it is not clear if this would have any benefit.

PECD contains a great richness of information about the medium's chirality, since much like in HHG the spectra is energy resolved. Different photoelectron energies can refer to varying molecular orbitals, and/or multiphoton processes. Usually, PECD is decomposed into a Legendre polynomial expansion [530]. Even-order coefficients, $b_{2n}$, then do not survive orientation averaging, or in other words vanish when fitting the experimental PECD spectra. Odd-order Legendre coefficients, $b_{2n+1}$, are backwards-forward anti-symmetric, as expected, and form the chiral signal. There are various ways of averaging the signals in different energies in order to obtain a single scalar the quantifies the enantiomeric excess of the media. Notably, these signals are typically very large. In the single photon ionization regime where the technique was originally pioneered [531] [532] [533], signals can be as high as ~30% in the gas phase. PECD has since then been extended also to the multi-photon regime where photoemission is performed either with few perturbative photon absorption [534] [535], or in the strong-field regime due to tunneling ionization [536] [537], where it can also provide time-resolved chiral signals with resolutions down to the attosecond domain [538] [539] [540] [541] [542] (see Fig. 14(c) for exemplary use of PECD to probe ultrafast electron dynamics within a chiral molecule with a perturbative NLO process). Typically, NLO chiral signals have a similar scale to those in the single photon case, usually reaching ~10%. Ref. [543] also used the approach to explore a time-resolved dissociation of a chiral molecule, complimenting results that have been obtained with HHG spectroscopy in ref. [510].

An important point of connection for PECD with the symmetry perspective is that without the $k_z$-resolving axis (the axis of light propagation), the signal vanishes. This is clear because of the asymmetry of PECD with respect to mirror planes along $k_z$, meaning that $\int dk_z PECD(k_x, k_y, k_z) = 0$. As a sub-case of this integration, let us highlight two common observables in NLO. The first is the total ionization rate that fully integrates the photoemission rates. The total ionization rates from left- or right-handed chiral molecules driven by CPL are therefore equal and lack a chiral signal. The origin of this cancellation is the mirror symmetry in the PECD. Similarly, ATI spectra (PES integrated angularly, but resolved in photoelectron energy) from enantiomeric partners are identical. In order to break these connections, one must break all mirror, inversion, and improper-rotational symmetries in the light, as provided by locally-chiral light (see section 2.3.3, and discussion in later section 4.2.3.2.3, see Fig. 7).

At this stage we would like to discuss some extensions and permutations of PECD. The first is photoexcitation-induced photoelectron circular dichroism (PEXCD) [544] [545]. Here the photoemission process is driven by a pair of coherent pulses (a pump and a probe), one of which is linear, and the other circular. A CD can be defined in the forward-backwards part of the PECD upon changing handedness in the circular driving pulse. Notably, in this case it would not matter which pulse is the pump and which is the probe – a PECD would arise anyways just because the mirror symmetry is broken. The properties of the PECD on the other hand would depend on exchanging the pump and probe pulses, and some excitation schemes are more favorable. Due to the inherent pump-probe structure of the method it is optimally suited for resolving ultrafast chiral dynamics. This methodology was pioneered in ref. [545] and shown to resolve chiral photoexcitation dynamics on the timescale of ~100fs, with signals of ~2%.

Another option is measuring photo-ion yields upon molecular dissociation, proving photo-ion circular dichroism (PICD) [546] [547] [548]. PICD is practically very similar in origin to PECD, and provides roughly similarly-sized chiral signals. It can potentially be measured in coincidence with PECD [548], which can shed light on electron-nuclear correlations [549].

To our knowledge, these techniques have thus far only been implemented using light with polarization-associated OC, though there is no fundamental aspect preventing its utilization also with light carrying OAM. Indeed, we foresee this becoming another prevalent technique in coming years, where photoelectron (or photo-ion) helical dichroism (PEHD, PIHD) and combination thereof with PECD are also explored.

Lastly, let us review the technique of Coulomb explosion imaging as used for chiral detection [550] [551]. Here a strong laser photoionizes a molecule, potentially with multiple photons that generates a highly charged molecular cation. The resulting cation dissociates into fragments due to intense electrostatic repulsion forces within the molecule. By measuring the fragments' charge, mass, momenta, etc., one can reconstruct the molecular geometry prior to the explosion process. This technique is quite general and can handle practically any molecular species, also achiral. For chiral molecules however, it presents a very unique ability – reconstructing the exact molecular geometry means that the measurement directly provides the molecular handedness. Note that in all other techniques we have presented, and will present, this is not the case – even if the chiral signal contains handedness specific information, one can never infer from it if the molecular sample has an excess of the (R) or the (S) enantiomer. The only way to standardly achieve this is by comparing to theory, or to known reference data measured by other techniques such as X-ray spectroscopy for condensed phases, or chemical techniques. One way to exemplify this struggle is through an age-old organic chemistry undergraduate-type test question: student X was working in the lab and accidently erased the labels (S) and (R) from two enantiomerically-pure chiral samples. What kind of measurements can X perform to know which is which? The answer is that with almost all known techniques he cannot reconstruct the labels. For instance, standard CD absorption signals will result in one of the samples absorbing CPL more strongly, but not give any information towards which molecular handedness causes the stronger absorption. In Coulomb explosion imaging, this information is a direct observable, making it truly unique. On the other hand, it is a very experimentally-challenging approach that is not a first choice for chiral detection.

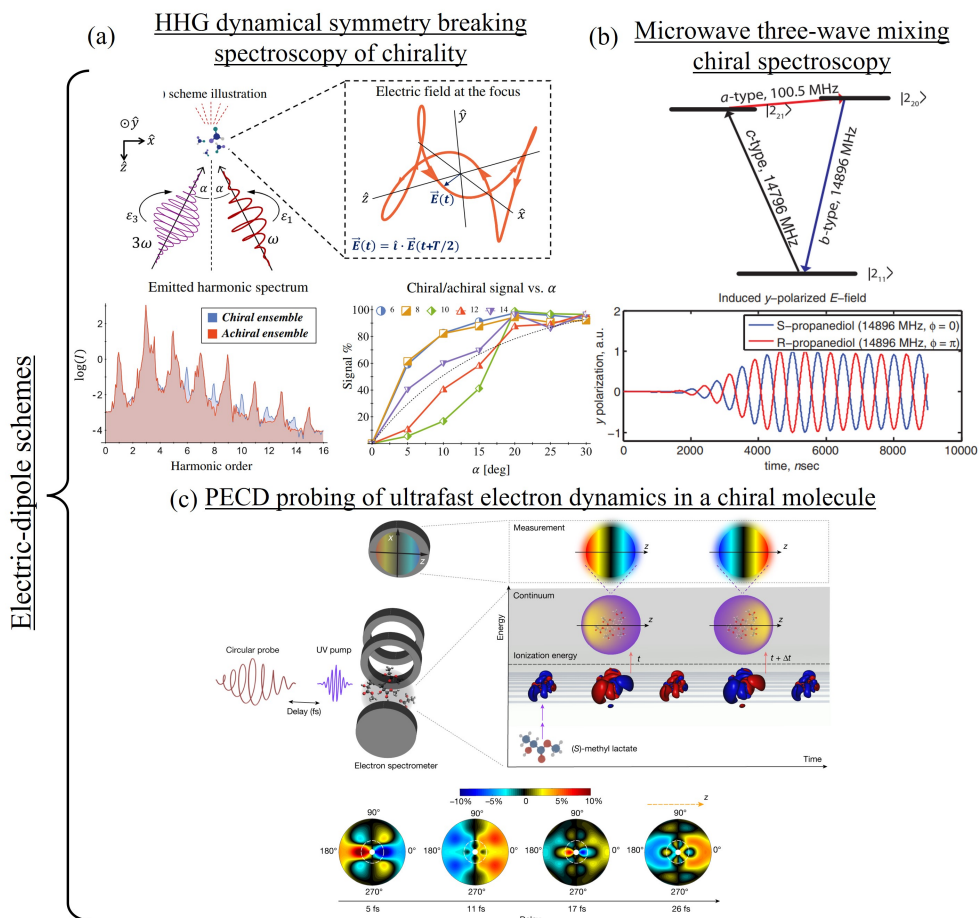

**Fig. 14**. **Spectroscopy of chirality using symmetric/asymmetric forms of light in various configurations with modern methods that are purely electric-dipole based.** (a) HHG symmetry breaking spectroscopy with engineered dynamical symmetries used to probe molecular chirality. Here the chiral signal arises in the electric dipole approximation as a background free signal, allowing reconstruction of the enantiomeric excess. Figure repurposed from ref. [13]. (b) microwave three-wave mixing spectroscopy employed for tracking chiral signals. The chiral signal here is obtained from amplitude and phase of the nonlinear mixing, which can be measured directly in microwaves, creating a background free signal. Figure repurposed from ref. [552]. (c) PECD used for probing ultrafast electronic currents within a chiral molecule. Figure repurposed from ref. [542].

### 4.2.3.2.2    All-optical symmetry-breaking spectroscopy

Along the lines of PECD, one can obtain electric-dipole-based chiral signals also from all-optical measurements even if the driving fields are locally-achiral, but carry nonzero OC. In this paradigm, instead of resolving the photoemission along particular axes in space like in PECD, one resolves the optical response along various axes in space. The potential caveat is that optical responses are prone to macroscopic and phase matching issues in certain geometries, and it is not always possible to fully resolve the optical fields in space. Indeed, if one drives a chiral ensemble of molecules with CPL propagating along the z-axis, a chiral current arises along the z-axis, which is enantio-sensitive and flips sign upon flipping the light or molecular handedness [27]. This is exactly the current that contributes to the PECD signal. It inevitably can also emit an optical near field that is chiral-sensitive; which unfortunately, cannot be practically measured because it does not propagate to the far due to phase matching conditions (there is no way to conserve photon momenta along light's propagation axis). However, by constructing unique optical set-ups where the polarization of the driving beams are not necessarily transverse to the main optical axis, one can engineer conditions where these chiral currents phase match.

From a symmetry perspective we denote these techniques as symmetry-breaking methodologies. The idea being that a particular mirror or inversion symmetric state of the beam is chosen which is locally-achiral, but that symmetry is broken by the chiral medium, causing the expected symmetry-based selection rule to be broken. The simplest example is perturbative nonlinear optical $\chi(2)$ methodologies [489] [553] [554] [555] [9]. For instance, consider driving a randomly oriented molecular ensemble by two carrier wavelengths, $\omega_1$ and $\omega_2$. From the standard NLO perspective, if a medium is inversion-symmetric one has that $\chi(2) = 0$, and a sum frequency generation (SFG) signal will not be observed. On the other hand, chiral media breaks inversion symmetry, even if it is isotropic. Consequently, one could measure the sum frequency signal at $\omega_1 + \omega_2$. The limiting factor would be to choose the frequencies, driving polarization, and beam propagation axes, such that the SFG signal could practically be observed. Evidently, due to symmetry constraints one must have that the two beams are non-collinear (where some optical angle exists that maximizes both the microscopic response and the phase matching conditions) [554]. This methodology has been quite widely used, mostly for exploring chiral surfaces, but also in the gas phase. The beauty of the approach is that the signal is essentially background-free in the sense that the SFG signal is identically zero if the medium is achiral. The yield can then be used to measure the absolute value of the EE. However, the sign of the EE is more difficult to infer. In order to obtain it, one needs to measure the phase of the generated $\chi(2)$ response, which flips sign upon flipping the molecular handedness. It has been shown in the microwave (see demonstration in Fig. 14(b)), where phase measurements are technologically easier, that this is a very good methodology for chiral spectroscopy [8] [556] [552]. Due to the background free nature of it, it essentially leads to the maximal 200% chiral discrimination. Its shortcomings are generally in the microwave implementation, and in relatively weak signals in the optical domain and gas phases. Other extensions based on higher order nonlinear processes have also been explored such as chiral coherent anti-stokes Raman scattering (CARS) [557], and third harmonic Rayleigh scattering in liquids [558].

This methodology was recently extended to HHG. In ref. [13], the concept of symmetry-breaking spectroscopy for chirality was introduced based on the dynamical-symmetry perspective for the process. The general paradigm based on symmetry is then that instead of looking for nonzero $\chi(2)$ responses in chiral media, one can design a great wealth of mirror-based, inversion-based, or improper-rotational-based symmetries in the driving beam that generates HHG. These symmetries lead to selection rules for the high harmonics (as derived in section 3.2), either in the allowed harmonic orders or in their polarization states. The deviations from the selection rules in chiral media can then be inferred as background-free chiral signals from which the EE, including its sign, can be reconstructed. For instance, it was predicted that for high harmonics the phase of the chiral signal (the sign of the EE) can be resolved by measuring the polarization state (including handedness) of emitted harmonics, suffice that the driving beam conditions are engineered in order to carry such signal to the far field (see examples in Fig. 14(a)). Extensions have also considered these signals in few-cycle pulses [559].

The methodology can also be applied to explore chiral phenomena in non-chiral molecules. For instance, ref. [12] explored a pump-probe geometry in achiral molecules and atomic systems. If a circular pump field excites a helical ring current in the molecular system, then a second probe pulse that generates harmonics should be able to sense that chiral current as deviations from selection rules that are upheld in the non-pumped system. This concept therefore provides a background-free approach towards measuring ring currents that is all-optical (avoiding the need for measuring photoelectron emission in coincidence [352]). The inherent pump-probe nature also means dynamical phenomena can be explored on the scale of attoseconds, e.g. electronic-correlations [12]. We will further discuss these possibilities in section 4.2.3.3.

### 4.2.3.2.3    Locally-chiral light

The next route towards electric-dipole based chiral signals is employing locally-chiral light, which is chiral within the dipole approximation and carries a nonzero chirality density in its temporally evolving polarization state (see section 2.3.3). This approach has not yet been validated experimentally, but several groups are currently working towards realizing experiments. Nevertheless, there have been quite a few theory predictions for utilizing locally-chiral light in enhanced chiral spectroscopies of various methodologies. From a paradigmatic point of view, every previous optical approach for probing chirality that works with CPL can be rederived in conditions where locally-chiral light is employed instead, producing a wide range of novel possibilities (actually, this statement is correct even for spectroscopies of physical and chemical properties other than chirality, many of which also employ CPL and CD, e.g. spin, topology, valley physics, etc.). Let us briefly review here the proposals set forward in recent years.

The first proposal for employing locally-chiral light for chirality spectroscopy involved replacing elliptical/CPL light in chiral HHG with locally-chiral fields [242]. Practically, this means that HHG should be generated by a non-collinear two-beam geometry, which is also non-monochromatic (although some other geometries have been suggested employing vector beams that have an intense field component along the beam propagation axis [256] [254]). The main simulated geometry in ref. [242] suggested a bi-chromatic set-up where all beams are linearly-polarized with varying polarization axes and relative phases, such that the resulting field overall has a nonzero DOC (see section 2.3.3.2). The simulated HHG spectra predicted a chiral signal reaching the maximal 200% discrimination for certain harmonic orders. Let us emphasize here a peculiar nature of that chiral signal compared to standard CD measurements – since the light is not necessarily comprised of CP components, and since the origin of its chiral density is different than the OC in CPL, the chiral signal is not uniquely defined by the discrimination signal between two experiments with different light handedness. Rather, the chiral-dichroism and circular-dichroism become separate entities, and the chiral dichroism carries the main chiral signal. It is denoted here as X$D$ in order to avoid ambiguities, and is defined according to its molecular origin in HHG:

$$\mathrm{XD}(n) = 2 \frac{I_{HHG_R}(n) - I_{HHG_S}(n)}{I_{HHG_R}(n) + I_{HHG_S}(n)} \qquad (60)$$

where we explicitly use the molecular handedness notation ('R' or 'S') to clarify that the two experiments require switching the handedness of the medium in order to obtain the dichroism. Alternatively, one can fix the medium's handedness between both set-ups, but mirror the geometry of the driving beams. The mirror operation acting on the light field flips it's handedness in the locally-chiral sense, but also in a spatial manner. Notably, it is not the same as replacing all left rotating frequency components with right rotating ones and vice versa. This is another aspect making experiments slightly more challenging than standard CD geometries.

A main issue analyzed in ref. [242] was the contribution of so-called globally-chiral light to the XD, i.e. where the handedness of light is fixed throughout the interaction region (see section 2.3.4, and Fig. 5). It was shown that by generating such conditions, the strong chiral signals obtained in the local near field should propagate to the far field. At the same time, we should note that ref. [242] only considered the two-dimensional spatial structure of the driving beams, and did not perform full three-dimensional propagation calculations of Maxwell's equations coupled to the quantum-electron dynamics driven in the chiral medium. This is a main concern for locally-chiral light, as its bi-chromatic and non-collinear nature often mean that the relative phase between the field components evolves in space also along the laser propagation direction, even if light's handedness is fixed. Thus, the effect of full three-dimensional propagation and phase matching effects on chiral signals generated by locally-chiral fields still remains to be explored. Recently, it was suggested that an additional third carrier beam in a different frequency should be added in order to phase match similar phenomena of electronic chiral state transfers [560]. It is possible a similar constraint is needed for locally/globally-chiral light, though this is currently an open question.

The next suggestions also relied on HHG, and focused on employing locally-chiral light in slightly different spatial conditions [253] [243] [561]. As it turns out, it is enough for light's DOC to be asymmetric across the interaction region rather than having a fixed handedness everywhere in space. In other words, even if light's local handedness does change sign in several places in space, strong chiral signals should still propagate to the far field if these sign changes happen asymmetrically. This connects with the symmetry perspective of locally- and globally-chiral light, which states that it is enough to demand the driving laser field not respect reflection, inversion, or improper rotational multi-scale DS.

Subsequent theory work showed that HHG with locally-chiral light could also be employed in the few-cycle regime with monochromatic fields, where signals can arise in the harmonics polarizations [256]; for probing chiral nuclear dynamics [255]; for probing chirality of multi-chiral systems (including accurately sensing the concentrations of stereo-isomers in mixtures) [562]; and in combination with topological light [521]. It was also shown in ref. [562] that the HHG chiral signal is proportional to the DOC of the driving beam, painting a clear connection between the chiral observables and the main physical origin for the signal. This connection is crucial, because it allows one to think of locally-chiral light's DOC as a direct chiral density in analogy with other EM-derived properties such as OC.

Beyond the realm of harmonic generation, locally-chiral light was also predicted useful for chiral sensing through free-induction decay signals [563], as well as for PECD (denoted as PE*X*D when driven by locally-chiral light) [305] [564]. We now briefly focus on the PEXD case, as it connects with previous discussions on the symmetry-origins of the chiral signals. For PE*X*D driven by locally-chiral light, all of the standard symmetries of PECD can be lifted, leading to photoelectron spectra that are not up-down, left-right, or forward-backwards symmetric (or anti-symmetric). Thus, a chiral signal can generically arise in the full three-dimensional photoelectron phase space:

$$\text{PEXD}(k_x, k_y, k_z) = 2 \frac{P_R(k_x, k_y, k_z) - P_S(k_x, k_y, k_z)}{P_{\text{norm}}(k_x, k_y, k_z)} \qquad (61)$$

Potentially, this could greatly increase PECD signals, especially if specific points in momentum space are taken into consideration. It also provides more freedom in choosing the dimensions over which one averages the photoelectron momenta. Still, in ref. [305] chiral signals were shown to be on the same scale as those typically obtained in PECD with CPL (see Fig. 7(e)). On the other hand, there is one major physical consequence for this additional symmetry breaking – the ATI and total molecular ionization rates now also constitute chiral signals within the electric-dipole approximation. Thus, one obtains a X*D* in ATI:

$$ATI_{XD}(\epsilon) = 2 \frac{ATI_R(\epsilon) - ATI_S(\epsilon)}{ATI_{norm}(\epsilon)} \qquad (62)$$

, and in the total molecular photoionization rates (MPI):

$$MPI_{XD} = 2 \frac{MPI_R - MPI_S}{MPI_R + MPI_S} \qquad (63)$$

where $\epsilon$ is the photoelectron energy, and the denominator in Eq. (62) can be determined in several ways to normalize the data (which we do not discuss here). Mathematically, it is worthwhile to remember that Eq. (62) represents the angle-integrated data from Eq. (61), and Eq. (63) the energy-integrated data from Eq. (62). Typically, each additional integration of DOF causes the chiral signal to vanish, but this is not the case for locally-chiral light driven processes. Indeed, in ref. [305] ATI chiral signals of up to ~10% were predicted, and up to ~5% in the molecular ionization rates. Those signals were also shown to correlate with the DOC of the driving field, just like in the HHG case. One main physical consequence of this effect is that locally-chiral light can be employed for all-optical enantio-separation through selective ionization. In other words, since a given molecular handedness can be photoionized in a large excess (faster) compared to the other, a continuous excitation of a racemic mixture (with proper filtration of the dissociated molecular fragments) should eventually lead to an enantiomerically pure sample. Thus, locally-chiral light could end up solving one of the holy grails of optical chiral manipulation (which has otherwise only been suggested in refs. [565] [566] [567] [568] [569] [570], but has never been experimentally realized, possibly due to ensemble averaging issues [571]). One possible connected solution could come from using chiral light sources (such as locally-chiral light or others) for optical tweezers, to selectively tweeze a given handedness of chiral moelcules [572] [573].

Let us end this discussion by considering another methodology that might not seem related at first glance, but which on the asymmetry level of the driving field shares common attributes with locally-chiral light. This is enantio-specific state transfer, which in the last decade was pioneered as a novel and highly sensitive approach for chiral spectroscopy [556] [574] [575] [576] [577]. Here a set of on-resonance three separate laser pulses with orthogonal polarization components are employed. The frequencies and polarization components of each pulse are selected for the specific molecule in question in order to excite some resonant transitions. After interaction with this complex three-pulse set-up, chiral molecules with left (right) handedness are found to occupy with different amplitudes certain rotational levels of the system. Thus, a chiral signal is born in the rotational level occupations. The discrepancy has been predicted in models to reach as high as 100%, and measurements of ~60% were reported [574] (see Fig. 15), with further improvement reaching up to 92% enrichment [578].

The method has also been suggested in a cavity geometry [579] [580], and with probing other signals such as Stark shifts [581]. Since very large chiral signals that are electric-dipole based are generated, and since the method inherently incorporate also ro-vibrational levels, similar set-ups have also been suggested for probing weak nuclear force chiral symmetry breaking and parity violations [582] [485], as well as for enantio-separation schemes [583] [584], potentially offering similar capabilities to those predicted with locally-chiral light.

While at first glance this methodology may seem disconnected from those employing locally-chiral light, we would like to emphasize in the context of this review the similarities on the asymmetry level – it has been shown that enantioselective state transfer requires the polarization components of the pulses to span the full three-dimensional polarization space, and also carry more than a single frequency component [585]. These are also necessary conditions for obtaining locally-chiral light. The methodology was also suggested for probing enantiomeric excess *via* light deflection [586], which is quite similar to the proposals using locally-chiral light [253]. Thus, it seems quite likely that the two set-ups could eventually be connected by a more extensive chiral theory that includes also resonant responses and ro-vibrational couplings. Moreover, there has been recent work suggesting an onset of chiral forces arising in similar conditions [587], and we expect many more chiral phenomena to appear and be discovered in these geometries.

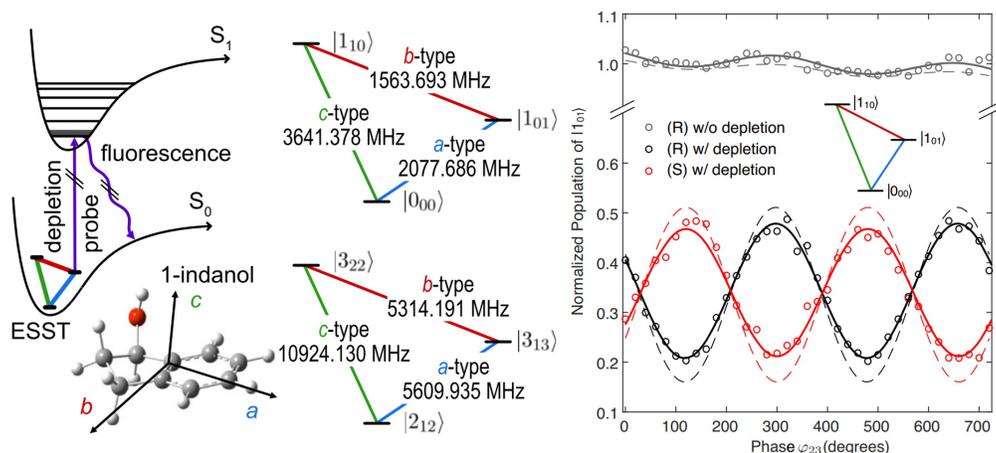

**Fig. 15. Enantiospecific state transfer (ESST) demonstrated in various systems and conditions, utilizing three independent polarization axes and laser frequencies that control transfer between two- or three-level systems.** (left) Illustration of the energy diagram and axes polarization with respect to the molecule. (right) Resultant high enrichment signals vs. phase (right). Figure repurposed from ref. [574].

#### 4.2.3.3 Probing chirality in non-molecular systems

Let us end this section by contemplating the utilization of these techniques also for probing chiral phenomena besides molecular chirality. Naturally, this includes chiral solids [588] [253], as well as aligned chiral molecules [589] [590] [591], which have already been suggested to be accessible within HHG and nonlinear optics methodologies. But, in the broad sense, the term chiral effects could refer to any physical phenomena breaking parity, which is much more widespread than molecular systems. This of course involves many different scientific fields and is too wide a scope for our review to fully contain. Nonetheless, we would like to highlight some example cases that connect with the arguments and methodologies discussed here.

First, we would mention the use of symmetry-breaking based methodologies for probing atomic and molecular ring currents [360] [168] [592] [361] [593] [594]. Such currents can be generated by interactions with intense magnetic pulses, or *via* strong-field ionization. They comprise a multi-electron wave packet that propagates helically throughout space and time, breaking any mirror symmetry in the non-excited system. The symmetry breaking is unique, because it arises from the excited electron wave function rather than from direct geometrical arrangement of the nuclei positions.

In the symmetry-breaking perspective, such currents were experimentally probed using photoemission spectroscopies [352], and suggested to be probed through HHG [352] [12]. In the photoemission case, the emitted photoelectron possessed nonzero initial momenta causing the expected symmetries of the photoelectron spectra to break (essentially resulting in photoelectron spectra that is not mirror symmetric even though the neutral atoms do exhibit such symmetry, see Fig. 16(a)). In HHG, the ring currents were shown to cause a shift in the maximizing ellipticity for the HHG yield [351], as well as directly breaking the HHG selection rules causing the emission of symmetry forbidden harmonics and forbidden polarization states [12] (see Fig. 16(b)). The selection rules deviation were analytically shown to scale parabolically with the current density, providing an analogy with enantiomeric-excess detection through similar methodologies [13].

(a) <u>Probing chiral ring currents in atoms with photoemission</u>

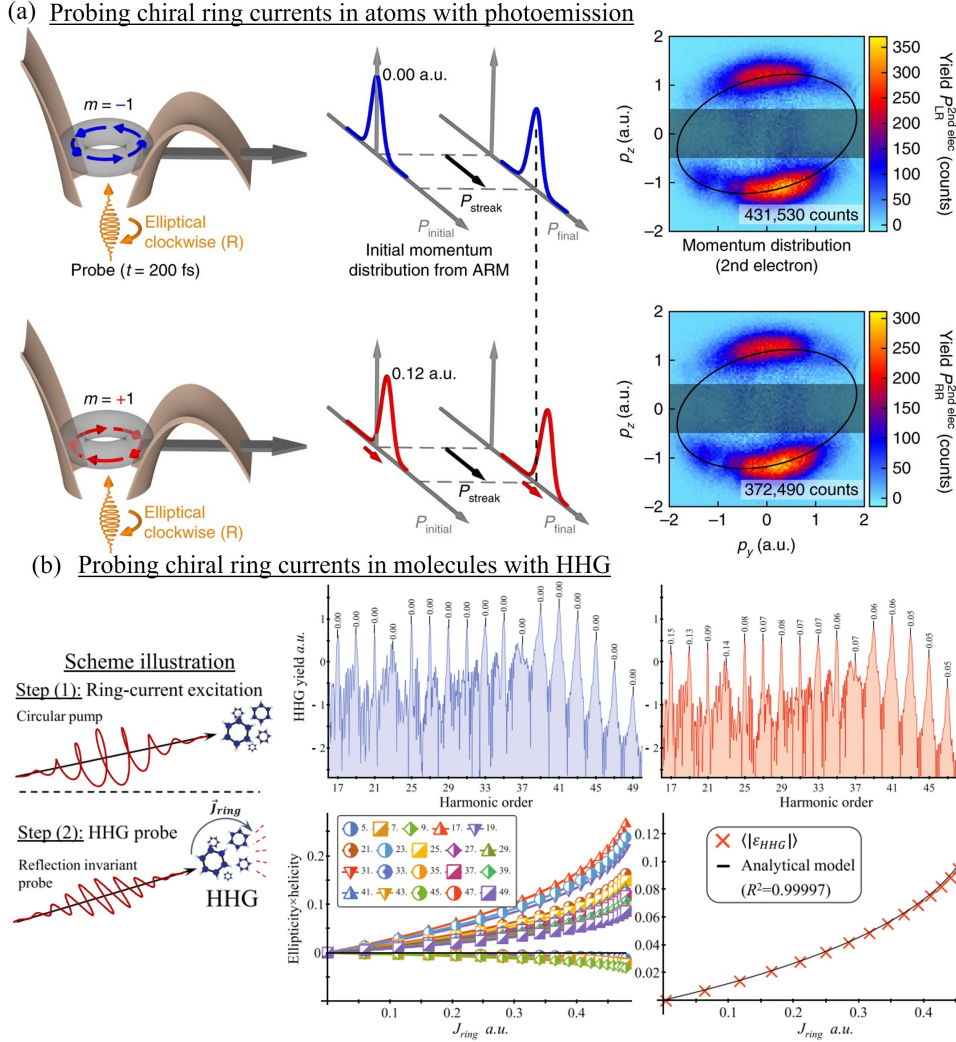

(b) <u>Probing chiral ring currents in molecules with HHG</u>

**Fig. 16. Probing laser-induced ring currents through symmetry breaking spectroscopy.** (a) Pump-probe photoemission measurements reveal broken mirror symmetry in the spectra arising from ring-currents. Figure repurposed from ref. [352]. (b) HHG symmetry breaking spectroscopy predicts nonlinear elliptical polarization (indicated over individual harmonics in the plot) as a result of a pumped ring current, driven by linearly-polarized light. Figure repurposed from ref. [12].

Second, we note that chiral light sources can also be used to imprint chirality onto non-chiral systems such as atoms and solids [458] [595] (as well as aligned achiral molecules [596]), which can be especially prevalent using locally-chiral light [597]. All of the methodologies discussed above could be employed for studying light-imprinted chiral phenomena. This includes also extensions of CD, elliptical dichroism, and generally asymmetric light forms, that are currently being employed to probe topological systems with nonlinear optics [424] [428] [427] [425], as well as induce and probe valleytronics (that generates valley asymmetric populations that break the material symmetry) [464] [598] [51] [50], and induce and probe ultrafast magnetism [599] [314] [315] [48].

Third, we would comment that there is a vast and growing field of generating chiral metasurfaces in the hopes of enhancing local OC fields, and potentially utilizing these fields for chiral sensing (see e.g. refs. [600] [601] [602] [603] [604] [605] [606] [607] [608] [609] and therein for a partial picture of the field), also recently including beams carrying

OAM [610]. It remains at this stage unclear how the two fields connect, but it is reasonable to expect a mutual fertilization of concepts in the coming years.

Lastly, we would comment on the connection to some unique chiral phenomena in condensed systems (also partially discussed in section 4.2.2). These are Weyl semimetals which can support chiral Fermion states. Probing the chirality of these states can be accomplished using perturbative NLO [611] [612] [613] [614] [615] [616]. However, they have also been found to lead to a unique effect coined 'topological frequency conversion' [462] [463] [617] [618]. Here the system is driven by a pair of non-mutually-polarized bi-chromatic fields, and the power can convert between the fields in a manner similar to wave mixing phenomena. Contrarily though, the yield and efficiency of the process scales differently than standard wave mixing effects due to the topological nature of the physical mechanism [463]. We mention this particular case, because the driving beam geometry utilized for topological frequency conversion in fact employs locally-chiral light, drawing a clear connection between the systems. Other connections to chiral systems have also been drawn out [463]. Therefore, there is likely a deep, and yet not fully known, physical connection between chiral molecular sensing and topological phenomena in certain condensed systems.

## 5. Summary, outlook, and establishing a database

In summary, we reviewed how hybrid symmetries of light and matter govern selection rules in nonlinear optical processes. We first discussed traditional approaches based on the symmetry of the nonlinear medium and photonic conservation laws, then shifted our focus to a more general framework that explores the dynamical symmetries of the light-matter system – the symmetries of its Hamiltonian – in both real and synthetic dimensions. This formalism is particularly important for the emerging field of nonlinear optics driven by structured and tailored light, where conventional methods fail to capture the full range of selection rules.

We then moved on to discuss how structured light with engineered symmetries and asymmetries can be leveraged for a wide range of spectroscopic applications across atomic, molecular, and solid-state systems, including chiral systems. This includes tailoring laser fields in polarization, space, and time to either enforce or break specific symmetry relations, as well as cases where symmetries emerge or are disrupted by the nonlinear medium itself. Within this broad scope, we aimed to unify results obtained over the past years into a comprehensive framework that highlights the fundamental role of dynamical symmetries in nonlinear spectroscopy and electromagnetic field control. This perspective deepens our understanding of light-matter interactions and provides a foundation for designing novel spectroscopic techniques and optimizing ultrafast control strategies.

We believe that the general light-matter dynamical symmetry approach for analyzing selection rules will play a growing role in advancing nonlinear optics and ultrafast science, both in symmetric and symmetry-broken systems. Given its recent success in HHG, its extension to other nonlinear processes is highly promising. One particularly exciting avenue is the application of this approach to strong-field angularly-resolved photoelectron spectroscopy [295] [296] [297] [298] [299] [300] [301] [216] [304] [305]. Similar to HHG, the measured photoelectron distributions in such experiments are highly sensitive to the dynamical symmetries of the light-matter system. However, this system offers even richer physics, as the selection rules also depend on the momentum of the emitted electrons. Mapping these selection rules could enhance our ability to probe ultrafast dynamics in molecules and condensed matter systems.

There are numerous other promising extensions of this symmetry-based framework. For instance, it should enable the derivation of selection rules for non-electric dipole harmonics in solids [619], molecules [385] [620] [621], and nanostructures [622] [623], as well as for non-relativistic high harmonics [624], nonlinear generation of photo-currents [625] [626] [44] [627] [50] [465], magnetism [628] [599] [48] [629], Floquet physics [450] [451] [456] [457] [458] [630], and more. We also expect extensions of the

theorems could be derived for systems occupying multiple Floquet states through superposition (beyond the adiabatic limit), which are currently not included in the framework.

Another particularly intriguing direction is the interplay between dynamical symmetries and quantum optical effects in high harmonic generation [631] [20] [632] [633] [634] [635] [636]. This raises a fundamental question: What is the general theoretical framework that describes the interplay between classical dynamical symmetries and quantum properties of light and matter, such as entanglement and squeezing, in nonlinear optical processes? Developing such a theory would not only deepen our understanding of light-matter interactions at the quantum level but also provide a guiding principle for engineering structured light and matter to generate intricate photonic quantum states. Furthermore, it could offer new ways to induce and control quantum phase transitions in complex materials, opening novel pathways for quantum technologies.

It is likely that artificial intelligence (AI) will be used in the future for uncovering selection rules in nonlinear optics by identifying hidden symmetries and correlations in complex optical interactions. AI-driven approaches can complement traditional group-theoretical methods by learning patterns from vast datasets, potentially revealing selection rules that might be challenging to derive analytically. Furthermore, AI can aid in classifying nonlinear optical responses based on symmetry constraints and experimental conditions, enabling predictive modeling of novel light-matter interactions. AI will also benefit data analysis in nonlinear optics experiments [637], offering enhanced signal reconstruction, noise reduction, and feature extraction [562] [638] [639] [640] [641].

Finally, we invite you, the reader of this review, to contribute to building a comprehensive database of dynamical symmetries and their associated selection rules for nonlinear optical processes. This effort will be a valuable service to the scientific community, much like the work of researchers in the 1960s and 1970s who systematically derived and tabulated selection rules based on the symmetries of the nonlinear medium. To this end, when investigating selection rules for a nonlinear process, we encourage you to go beyond the specifics of the problem at hand. Instead, adopt a general perspective by formulating selection rules in terms of the dynamical symmetries of the light-matter system and expressing them in the language of group theory. Furthermore, we invite you to share the dynamical symmetries, selection rules, and corresponding citations of your work with us (following the guidelines at [642]), so we can systematically organize and tabulate this data for the benefit of the community.

Symmetry is one of the most profound and elegant principles in physics, underlying our understanding of fundamental interactions across disciplines. As nonlinear optics continues to evolve, systematically mapping dynamical symmetries and their associated selection rules will be essential for both enhancing theoretical insights and expanding practical applications. We hope that this comprehensive review, together with our initiative to develop a continually expanding database of dynamical symmetries and their associate selection rules, will serve as a lasting resource – fostering new discoveries and propelling the field forward.

**Acknowledgments.** This research was funded by Israel Science Foundation (ISF) grant number 2626/23. O.K gratefully acknowledges the support of The Israel Science Foundation (grants No. 2992/24 and 1021/22) and the young faculty award from the National Quantum Science and Technology program of the Israeli Planning and Budgeting Committee. M.E.T. gratefully acknowledges the support of the Council for Higher Education scholarship for excellence in quantum science and technology. M.E.T. and O.C. thank the Helen Diller Quantum Center for partial financial support.

**Disclosures.** The authors declare no conflicts of interest.

**Data availability statement.** All data underlying the results presented in this paper is available within the body of text directly.

## References

1. P. A. Franken, A. E. Hill, C. W. Peters, and G. Weinreich, "Generation of Optical Harmonics," Phys. Rev. Lett. **7**, 118–119 (1961).
2. R. W. Boyd, "Nonlinear Optics," in 3rd ed. (Academic Press, 2008).
3. C. L. Tang and H. Rabin, "Selection Rules for Circularly Polarized Waves in Nonlinear Optics," Phys. Rev. B **3**, 4025–4034 (1971).
4. N. Saito, P. Xia, F. Lu, T. Kanai, J. Itatani, and N. Ishii, "Observation of selection rules for circularly polarized fields in high-harmonic generation from a crystalline solid," Optica **4**, 1333–1336 (2017).
5. Y. R. Shen, "Surface Second Harmonic Generation: A New Technique for Surface Studies," Annu. Rev. Mater. Res. **16**, 69–86 (1986).
6. D. Habibović, K. R. Hamilton, O. Neufeld, and L. Rego, "Emerging tailored light sources for studying chirality and symmetry," Nat. Rev. Phys. (2024).
7. J. S. Ginsberg, M. M. Jadidi, J. Zhang, C. Y. Chen, N. Tancogne-Dejean, S. H. Chae, G. N. Patwardhan, L. Xian, K. Watanabe, T. Taniguchi, J. Hone, A. Rubio, and A. L. Gaeta, "Phonon-enhanced nonlinearities in hexagonal boron nitride," Nat. Commun. **14**, 7685 (2023).
8. D. Patterson, M. Schnell, and J. M. Doyle, "Enantiomer-specific detection of chiral molecules via microwave spectroscopy," Nature **497**, 475 (2013).
9. P. Fischer, "Nonlinear Optical Spectroscopy of Chiral Molecules," Chirality **17**, 421–437 (2005).
10. D. Baykusheva, M. S. Ahsan, N. Lin, and H. J. Wörner, "Bicircular High-Harmonic Spectroscopy Reveals Dynamical Symmetries of Atoms and Molecules," Phys. Rev. Lett. **116**, 123001 (2016).
11. Á. Jiménez-Galán, N. Zhavoronkov, M. Schloz, F. Morales, and M. Ivanov, "Time-resolved high harmonic spectroscopy of dynamical symmetry breaking in bi-circular laser fields: the role of Rydberg states," Opt. Express **25**, 22880–22896 (2017).
12. O. Neufeld and O. Cohen, "Background-Free Measurement of Ring Currents by Symmetry-Breaking High-Harmonic Spectroscopy," Phys. Rev. Lett. **123**, 103302 (2019).
13. O. Neufeld, D. Ayuso, P. Decleva, M. Y. Ivanov, O. Smirnova, and O. Cohen, "Ultrasensitive Chiral Spectroscopy by Dynamical Symmetry Breaking in High Harmonic Generation," Phys. Rev. X **9**, 031002 (2019).
14. S. Mitra, Á. Jiménez-Galán, M. Aulich, M. Neuhaus, R. E. F. Silva, V. Pervak, M. F. Kling, and S. Biswas, "Light-wave-controlled Haldane model in monolayer hexagonal boron nitride," Nature **628**, 752–757 (2024).
15. K. Uchida, G. Mattoni, S. Yonezawa, F. Nakamura, Y. Maeno, and K. Tanaka, "High-Order Harmonic Generation and Its Unconventional Scaling Law in the Mott-Insulating Ca2RuO4," Phys. Rev. Lett. **128**, 127401 (2022).
16. R. E. F. Silva, I. V Blinov, A. N. Rubtsov, O. Smirnova, and M. Ivanov, "High-harmonic spectroscopy of ultrafast many-body dynamics in strongly correlated systems," Nat. Photonics **12**, 266–270 (2018).
17. J. Alcalà, U. Bhattacharya, J. Biegert, M. Ciappina, U. Elu, T. Graß, P. T. Grochowski, M. Lewenstein, A. Palau, T. P. H. Sidiropoulos, T. Steinle, and I. Tyulnev, "High-harmonic spectroscopy of quantum phase transitions in a high-Tc superconductor," Proc. Natl. Acad. Sci. **119**, e2207766119 (2022).
18. S. Ghimire, G. Ndabashimiye, A. D. DiChiara, E. Sistrunk, M. I. Stockman, P. Agostini, L. F. DiMauro, and D. A. Reis, "Strong-field and attosecond physics in solids," J. Phys. B At. Mol. Opt. Phys. **47**, 204030 (2014).
19. J. Freudenstein, M Borsch, M. Meierhofer, D. Afanasiev, C. P. Schmid, F. Sandner, M. Liebich, A. Girnghuber, M. Knorr, M. Kira, and R. Huber, "Attosecond clocking of correlations between Bloch electrons," Nature **610**, 290–295 (2022).
20. A. Gorlach, O. Neufeld, N. Rivera, O. Cohen, and I. Kaminer, "The quantum-optical nature of high harmonic generation," Nat. Commun. **11**, 4598 (2020).
21. A. Gorlach, M. E. Tzur, M. Birk, M. Krüger, N. Rivera, O. Cohen, and I. Kaminer, "High-harmonic generation driven by quantum light," Nat. Phys. (2023).
22. A. J. Uzan-Narovlansky, L. Faeyrman, G. G. Brown, S. Shames, V. Narovlansky, J. Xiao, T. Arusi-Parpar, O. Kneller, B. D. Bruner, O. Smirnova, R. E. F. Silva, B. Yan, Á. Jiménez-Galán, M. Ivanov, and N. Dudovich, "Observation of interband Berry phase in laser-driven crystals," Nature **626**, 66–71 (2024).
23. D. Ayuso, A. F. Ordonez, and O. Smirnova, "Ultrafast chirality: the road to efficient chiral measurements," Phys. Chem. Chem. Phys. **24**, 26962–26991 (2022).
24. F. Krausz and M. Ivanov, "Attosecond physics," Rev. Mod. Phys. **81**, 163–234 (2009).
25. M. F. Ciappina, J. A. Pérez-Hernández, A. S. Landsman, W. A. Okell, S. Zherebtsov, B. Förg, J. Schötz, L. Seiffert, T. Fennel, T. Shaaran, T. Zimmermann, A. Chacón, R. Guichard, A. Zaïr, J. W. G. Tisch, J. P. Marangos, T. Witting, A. Braun, S. A. Maier, L. Roso, M. Krüger, P. Hommelhoff, M. F. Kling, F. Krausz, and M. Lewenstein, "Attosecond physics at the nanoscale," Reports Prog. Phys. **80**, 54401 (2017).
26. A. Yariv and P. Yeh, *Photonics: Optical Electronics in Modern Communications* (Oxford University Press, 2007).
27. N. Berova, P. L. Polavarapu, K. Nakanishi, and R. W. Woody, *Comprehensive Chiroptical Spectroscopy* (Wiley, 2013), Vol. 2.
28. C. Li, "Nonlinear optics," Princ. Appl. (2017).
29. D. M. Bishop, *Group Theory and Chemistry* (Dover Publications, 2012).
30. D. C. Harris and M. D. Bertolucci, *Symmetry and Spectroscopy: An Introduction to Vibrational and Electronic Spectroscopy* (Dover Publications, 1978).

31.  C. Kittel, P. McEuen, and P. McEuen, *Introduction to Solid State Physics* (Wiley New York, 1996), Vol. 8.

32.  T. Schultz and M. Vrakking, *Attosecond and XUV Physics: Ultrafast Dynamics and Spectroscopy* (Wiley, 2014).

33.  H. Rubinsztein-Dunlop, A. Forbes, M. V Berry, M. R. Dennis, D. L. Andrews, M. Mansuripur, C. Denz, C. Alpmann, P. Banzer, T. Bauer, E. Karimi, L. Marrucci, M. Padgett, M. Ritsch-Marte, N. M. Litchinitser, N. P. Bigelow, C. Rosales-Guzmán, A. Belmonte, J. P. Torres, T. W. Neely, M. Baker, R. Gordon, A. B. Stilgoe, J. Romero, A. G. White, R. Fickler, A. E. Willner, G. Xie, B. McMorran, and A. M. Weiner, "Roadmap on structured light," J. Opt. **19**, 13001 (2017).

34.  N. Bloembergen, "Conservation laws in nonlinear optics*," J. Opt. Soc. Am. **70**, 1429–1436 (1980).

35.  O. Neufeld, D. Podolsky, and O. Cohen, "Floquet group theory and its application to selection rules in harmonic generation," Nat. Commun. **10**, 405 (2019).

36.  G. Lerner, O. Neufeld, L. Hareli, G. Shoulga, E. Bordo, A. Fleischer, D. Podolsky, A. Bahabad, and O. Cohen, "Multiscale dynamical symmetries and selection rules in nonlinear optics," Sci. Adv. **9**, eade0953 (2023).

37.  E. Pisanty, G. J. Machado, V. Vicuña-Hernández, A. Picón, A. Celi, J. P. Torres, and M. Lewenstein, "Knotting fractional-order knots with the polarization state of light," Nat. Photonics **13**, 569–574 (2019).

38.  M. Luttmann, M. Vimal, M. Guer, J.-F. Hergott, A. Z. Khoury, C. Hernández-García, E. Pisanty, and T. Ruchon, "Nonlinear up-conversion of a polarization Möbius strip with half-integer optical angular momentum," Sci. Adv. **9**, eadf3486 (2023).

39.  E. Pisanty, L. Rego, J. San Román, A. Picón, K. M. Dorney, H. C. Kapteyn, M. M. Murnane, L. Plaja, M. Lewenstein, and C. Hernández-Garcia, "Conservation of Torus-knot Angular Momentum in High-order Harmonic Generation," Phys. Rev. Lett. **122**, 203201 (2019).

40.  G. Lerner, M. Even Tzur, O. Neufeld, A. Fleischer, and O. Cohen, "Reflection parity and space-time parity photonic conservation laws in parametric nonlinear optics," Phys. Rev. Res. **6**, L042034 (2024).

41.  M. Even Tzur, O. Neufeld, A. Fleischer, and O. Cohen, "Selection rules for breaking selection rules," New J. Phys. **23**, 103039 (2021).

42.  M. E. Tzur, O. Neufeld, E. Bordo, A. Fleischer, and O. Cohen, "Selection rules in symmetry-broken systems by symmetries in synthetic dimensions," Nat. Commun. **13**, 1312 (2022).

43.  S. Sederberg, F. Kong, F. Hufnagel, C. Zhang, E. Karimi, and P. B. Corkum, "Vectorized optoelectronic control and metrology in a semiconductor," Nat. Photonics **14**, 680–685 (2020).

44.  O. Neufeld, N. Tancogne-Dejean, U. De Giovannini, H. Hübener, and A. Rubio, "Light-Driven Extremely Nonlinear Bulk Photogalvanic Currents," Phys. Rev. Lett. **127**, 126601 (2021).

45.  Y. Ikeda, S. Kitamura, and T. Morimoto, "Photocurrent Induced by a Bicircular Light Drive in Centrosymmetric Systems," Phys. Rev. Lett. **131**, 96301 (2023).

46.  C. Heide, T. Boolakee, T. Eckstein, and P. Hommelhoff, "Optical current generation in graphene: CEP control vs. ω + 2ω control," **10**, 3701–3707 (2021).

47.  A. Bharti and G. Dixit, "Photocurrent generation in solids via linearly polarized laser," Phys. Rev. B **109**, 104309 (2024).

48.  O. Neufeld, N. Tancogne-Dejean, U. De Giovannini, H. Hübener, and A. Rubio, "Attosecond magnetization dynamics in non-magnetic materials driven by intense femtosecond lasers," npj Comput. Mater. **9**, 39 (2023).

49.  B. S. Mendoza, N. Arzate-Plata, N. Tancogne-Dejean, and B. M. Fregoso, "Nonlinear photomagnetization in insulators," Phys. Rev. B **110**, 224412 (2024).

50.  S. Sharma, P. Elliott, and S. Shallcross, "THz induced giant spin and valley currents," Sci. Adv. **9**, eadf3673 (2023).

51.  Á. Jiménez-Galán, R. E. F. Silva, O. Smirnova, and M. Ivanov, "Lightwave control of topological properties in 2D materials for sub-cycle and non-resonant valley manipulation," Nat. Photonics **14**, 728–732 (2020).

52.  I. Tyulnev, Á. Jiménez-Galán, J. Poborska, L. Vamos, P. S. J. Russell, F. Tani, O. Smirnova, M. Ivanov, R. E. F. Silva, and J. Biegert, "Valleytronics in bulk MoS2 with a topologic optical field," Nature **628**, 746–751 (2024).

53.  Y. Wang, A.-S. Walter, G. Jotzu, and K. Viebahn, "Topological Floquet engineering using two frequencies in two dimensions," Phys. Rev. A **107**, 043309 (2023).

54.  T. V Trevisan, P. V. Arribi, O. Heinonen, R.-J. Slager, and P. P. Orth, "Bicircular Light Floquet Engineering of Magnetic Symmetry and Topology and Its Application to the Dirac Semimetal Cd3As2," Phys. Rev. Lett. **128**, 66602 (2022).

55.  A. Uzan-Narovlansky, A. Jimenez-Galan, G. Orenstein, R. Silva, T. Arusi-Parpar, S. Shames, B. Bruner, B. Yan, O. Smirnova, M. (Mikhail) Ivanov, and N. Dudovich, "Observation of light-driven band structure via multiband high-harmonic spectroscopy," Nat. Photonics **16**, 428–432 (2022).

56.  O. Neufeld, W. Mao, H. Hübener, N. Tancogne-Dejean, S. A. Sato, U. De Giovannini, and A. Rubio, "Time- and angle-resolved photoelectron spectroscopy of strong-field light-dressed solids: Prevalence of the adiabatic band picture," Phys. Rev. Res. **4**, 033101 (2022).

57.  O. Neufeld, H. Hübener, U. De Giovannini, and A. Rubio, "Tracking electron motion within and outside of Floquet bands from attosecond pulse trains in time-resolved ARPES," J. Phys. Condens. Matter **36**, 225401 (2024).

58.  S. Gassner, C. S. Weber, and M. Claassen, "Light-induced switching between singlet and triplet superconducting states," Nat. Commun. **15**, 1776 (2024).

59.  O. Kfir, E. Bordo, G. Ilan Haham, O. Lahav, A. Fleischer, and O. Cohen, "In-line production of a bi-circular field for generation of helically polarized high-order harmonics," Appl. Phys. Lett. **108**, 211106 (2016).

60.  J. B. Bertrand, H. J. Wörner, H.-C. Bandulet, É. Bisson, M. Spanner, J.-C. Kieffer, D. M. Villeneuve, and P. B.

Corkum, "Ultrahigh-Order Wave Mixing in Noncollinear High Harmonic Generation," Phys. Rev. Lett. **106**, 023001 (2011).

61. E. Frumker, N. Kajumba, J. B. Bertrand, H. J. Wörner, C. T. Hebeisen, P. Hockett, M. Spanner, S. Patchkovskii, G. G. Paulus, D. M. Villeneuve, A. Naumov, P. B. Corkum, E. Frumker, N. Kajumba, J. B. Bertrand, and H. J. Wo, "Probing polar molecules with high harmonic spectroscopy," Phys. Rev. Lett. **109**, 233904 (2012).

62. G. Vampa, T. J. Hammond, M. Taucer, X. Ding, X. Ropagnol, T. Ozaki, S. Delprat, M. Chaker, N. Thiré, B. E. Schmidt, F. Légaré, D. D. Klug, A. Y. Naumov, D. M. Villeneuve, A. Staudte, and P. B. Corkum, "Strong-field optoelectronics in solids," Nat. Photonics **12**, 465–468 (2018).

63. J. D. Jackson, *Classical Electrodynamics*, 3rd ed. (John Wiley & Sons, 2007).

64. W. I. Fushchich and A. G. Nikitin, *Symmetries of Maxwell's Equations* (Springer Science & Business Media, 1987), Vol. 8.

65. M. G. Calkin, "An Invariance Property of the Free Electromagnetic Field," Am. J. Phys. **33**, 958–960 (1965).

66. E. Noether and M. A. Tavel, "Invariant Variation Problems," arXiv Prepr. physics/0503066 (n.d.).

67. I. Fernandez-Corbaton, X. Zambrana-Puyalto, N. Tischler, X. Vidal, M. L. Juan, and G. Molina-Terriza, "Electromagnetic Duality Symmetry and Helicity Conservation for the Macroscopic Maxwell's Equations," Phys. Rev. Lett. **111**, 60401 (2013).

68. I. Fernandez-Corbaton, M. Fruhnert, and C. Rockstuhl, "Objects of maximum electromagnetic chirality," Phys. Rev. X **6**, 031013 (2016).

69. R. P. Cameron, S. M. Barnett, and A. M. Yao, "Optical helicity, optical spin and related quantities in electromagnetic theory," New J. Phys. **14**, 53050 (2012).

70. I. Fernandez-Corbaton and G. Molina-Terriza, "Role of duality symmetry in transformation optics," Phys. Rev. B **88**, 85111 (2013).

71. G. Y. Rainich, "Electrodynamics in the General Relativity Theory," Trans. Am. Math. Soc. **27**, 106–136 (1925).

72. C. E. Rüter, K. G. Makris, R. El-Ganainy, D. N. Christodoulides, M. Segev, and D. Kip, "Observation of parity–time symmetry in optics," Nat. Phys. **6**, 192–195 (2010).

73. N. Ben-Tal, N. Moiseyev, and A. Beswick, "The effect of Hamiltonian symmetry on generation of odd and even harmonics," J. Phys. B At. Mol. Opt. Phys. **26**, 3017 (1993).

74. O. E. Alon, V. Averbukh, and N. Moiseyev, "Selection Rules for the High Harmonic Generation Spectra," Phys. Rev. Lett. **80**, 3743 (1998).

75. V. Averbukh, O. Alon, and N. Moiseyev, "Crossed-beam experiment:   High-order harmonic generation and dynamical symmetry," Phys. Rev. A **60**, 2585–2586 (1999).

76. F. Ceccherini, D. Bauer, and F. Cornolti, "Dynamical symmetries and harmonic generation," J. Phys. B **34**, 5017–5029 (2001).

77. V. Averbukh, O. Alon, and N. Moiseyev, "Stability and instability of dipole selection rules for atomic high-order-harmonic-generation spectra in two-beam setups," Phys. Rev. A **65**, 063402 (2002).

78. V. S. Liu, B. K. VanLeeuwen, J. M. Munro, H. Padmanabhan, I. Dabo, V. Gopalan, and D. B. Litvin, "Spatio-temporal symmetry - crystallographic point groups with time translations and time inversion," Acta Crystallogr. Sect. A **74**, 399–402 (2018).

79. O. E. Alon, V. Averbukh, and N. Moiseyev, "Atoms, Molecules, Crystals and Nanotubes in Laser Fields: From Dynamical Symmetry to Selective High-Order Harmonic Generation of Soft X-Rays," Adv. Quantum Chem. **47**, 393–421 (2004).

80. Ofir E. Alon, "Dynamical symmetries of time-periodic Hamiltonians," Phys. Rev. A **66**, 013414 (2002).

81. H. M. N. and L. B. M. and J. P. Hansen, "On selection rules for atoms in laser fields and high harmonic generation," J. Phys. B At. Mol. Opt. Phys. **35**, L403 (2002).

82. V. Averbukh, O. E. Alon, and N. Moiseyev, "High-order harmonic generation by molecules of discrete rotational symmetry interacting with circularly polarized laser field," Phys. Rev. A **64**, 033411 (2001).

83. S. Long, W. Becker, and J. K. McIver, "Model calculations of polarization-dependent two-color high-harmonic generation," Phys. Rev. A **52**, 2262–2278 (1995).

84. D. B. Milošević, W. Becker, and R. Kopold, "Generation of circularly polarized high-order harmonics by two-color coplanar field mixing," Phys. Rev. A **61**, 063403 (2000).

85. D. B. Milošević, W. Becker, R. Kopold, and S. W., "High-harmonic generation by a bichromatic bicircular laser field," Laser Phys. **11**, 165–168 (2001).

86. A. Fleischer, O. Kfir, T. Diskin, P. Sidorenko, and O. Cohen, "Spin angular momentum and tunable polarization in high-harmonic generation," Nat. Photonics **8**, 543–549 (2014).

87. A. Fleischer and N. Moiseyev, "Attosecond laser pulse synthesis using bichromatic high-order harmonic generation," Phys. Rev. A **74**, 53806 (2006).

88. O. Kfir, P. Grychtol, E. Turgut, R. Knut, D. Zusin, D. Popmintchev, T. Popmintchev, H. Nembach, J. M. Shaw, A. Fleischer, H. Kapteyn, M. Murnane, and O. Cohen, "Generation of bright phase-matched circularly-polarized extreme ultraviolet high harmonics," Nat. Photonics **9**, 99–105 (2015).

89. S. Kerbstadt, L. Englert, T. Bayer, and M. Wollenhaupt, "Ultrashort polarization-tailored bichromatic fields," J. Mod. Opt. **64**, 1010–1025 (2017).

90. S. Kerbstadt, K. Eickhoff, T. Bayer, and M. Wollenhaupt, "Control of free electron wave packets by polarization-tailored ultrashort bichromatic laser fields," Adv. Phys. X **4**, 1672583 (2019).

91. K. Eickhoff, L. Feld, D. Köhnke, L. Englert, T. Bayer, and M. Wollenhaupt, "Coherent control mechanisms in bichromatic multiphoton ionization," J. Phys. B At. Mol. Opt. Phys. **54**, 164002 (2021).

92. K. Eickhoff, L. Englert, T. Bayer, and M. Wollenhaupt, "Multichromatic Polarization-Controlled Pulse Sequences for Coherent Control of Multiphoton Ionization," Front. Phys. **9**, (2021).

93. J. M. Ngoko Djiokap, S. X. Hu, L. B. Madsen, N. L. Manakov, A. V. Meremianin, and A. F. Starace, "Electron Vortices in Photoionization by Circularly Polarized Attosecond Pulses," Phys. Rev. Lett. **115**, 113004 (2015).

94. H. Weyl, *The Theory of Groups and Quantum Mechanics* (Martino Fine Books, 1950).

95. R. Mirman, *Point Groups, Space Groups, Crystals, Molecules* (WORLD SCIENTIFIC, 1999).

96. T. Janssen, A. Janner, and E. Ascher, "Crystallographic groups in space and time," Physica **41**, 541–565 (1969).

97. H. Padmanabhan, M. L. Kingsland, J. M. Munro, D. B. Litvin, and V. Gopalan, "Spatio-Temporal Symmetry—Point Groups with Time Translations," Symmetry (Basel). **9**, 187 (2017).

98. S. Xu and C. Wu, "Space-Time Crystal and Space-Time Group," Phys. Rev. Lett. **120**, 96401 (2018).

99. M. Damnjanović and I. Milošević, "Introduction," in *Line Groups in Physics* (Springer, 2010).

100. M. Damnjanovic and M. Vujicic, "Magnetic line groups," **25**, (1982).

101. M. Damnjanovic and I. Milosevic, *Line Groups in Physics: Theory and Applications to Nanotubes and Polymers (Lecture Notes in Physics Vol 801)* (Berlin: Springer, 2010).

102. M. Vujicic, "Construction of the symmetry groups of polymer molecules," J. Phys. A **10**, 1271–1279 (1977).

103. L. Allen, M. W. Beijersbergen, R. J. C. Spreeuw, and J. P. Woerdman, "Orbital angular momentum of light and the transformation of Laguerre-Gaussian laser modes," Phys. Rev. A **45**, 8185–8189 (1992).

104. A. García-Cabrera, R. Boyero-García, Ó. Zurrón-Cifuentes, J. Serrano, J. S. Román, L. Plaja, and C. Hernández-García, "Topological high-harmonic spectroscopy," Commun. Phys. **7**, 28 (2024).

105. S. M. Barnett, L. Allen, R. P. Cameron, C. R. Gilson, M. J. Padgett, F. C. Speirits, and A. M. Yao, "On the natures of the spin and orbital parts of optical angular momentum in electromagnetic theory," J. Opt. **18**, 064004 (2016).

106. M V Berry, "Optical vortices evolving from helicoidal integer and fractional phase steps," J. Opt. A Pure Appl. Opt. **6**, 259 (2004).

107. G. Milione, S. Evans, D. A. Nolan, and R. R. Alfano, "Higher Order Pancharatnam-Berry Phase and the Angular Momentum of Light," Phys. Rev. Lett. **108**, 190401 (2012).

108. M. V Berry and W. Liu, "No general relation between phase vortices and orbital angular momentum," J. Phys. A Math. Theor. **55**, 374001 (2022).

109. T. Senthil, "Symmetry-Protected Topological Phases of Quantum Matter," Annu. Rev. Condens. Matter Phys. **6**, 299–324 (2015).

110. B. Yan and C. Felser, "Topological Materials: Weyl Semimetals," Annu. Rev. Condens. Matter Phys. **8**, 337–354 (2017).

111. J. Wang and S.-C. Zhang, "Topological states of condensed matter," Nat. Mater. **16**, 1062–1067 (2017).

112. M. J. Gilbert, "Topological electronics," Commun. Phys. **4**, 70 (2021).

113. M. A. Bandres and J. C. Gutiérrez-Vega, "Circular beams," Opt. Lett. **33**, 177–179 (2008).

114. T. Doster and A. T. Watnik, "Laguerre–Gauss and Bessel–Gauss beams propagation through turbulence: analysis of channel efficiency," Appl. Opt. **55**, 10239–10246 (2016).

115. K Volke-Sepulveda, V Garcés-Chávez, S Chávez-Cerda, J Arlt, K Dholakia, K. Volke-Sepulveda, V. Garcés-Chávez, S. Chávez-Cerda, J. Arlt, and K. Dholakia, "Orbital angular momentum of a high-order Bessel light beam," J. Opt. B Quantum Semiclassical Opt. **4**, S82 (2002).

116. R. Kadlimatti and P. V Parimi, "Millimeter-Wave Nondiffracting Circular Airy OAM Beams," IEEE Trans. Antennas Propag. **67**, 260–269 (2019).

117. V. V Kotlyar, A. A. Kovalev, and A. P. Porfirev, "Vortex Hermite-Gaussian laser beams," Opt. Lett. **40**, 701–704 (2015).

118. M. A. Bandres and J. C. Gutiérrez-Vega, "Ince–Gaussian beams," Opt. Lett. **29**, 144–146 (2004).

119. A. Chong, W. H. Renninger, D. N. Christodoulides, and F. W. Wise, "Airy–Bessel wave packets as versatile linear light bullets," Nat. Photonics **4**, 103–106 (2010).

120. Y. Shen, S. Pidishety, I. Nape, and A. Dudley, "Self-healing of structured light: a review," J. Opt. **24**, 103001 (2022).

121. J. D. Ring, J. Lindberg, A. Mourka, M. Mazilu, K. Dholakia, and M. R. Dennis, "Auto-focusing and self-healing of Pearcey beams," Opt. Express **20**, 18955–18966 (2012).

122. A. Forbes, A. Dudley, and M. McLaren, "Creation and detection of optical modes with spatial light modulators," Adv. Opt. Photonics **8**, 200–227 (2016).

123. J. Jin, J. Luo, X. Zhang, H. Gao, X. Li, M. Pu, P. Gao, Z. Zhao, and X. Luo, "Generation and detection of orbital angular momentum via metasurface," Sci. Rep. **6**, 24286 (2016).

124. S. Zheng and J. Wang, "Measuring Orbital Angular Momentum (OAM) States of Vortex Beams with Annular Gratings," Sci. Rep. **7**, 40781 (2017).

125. G. Spektor, D. Kilbane, A. K. Mahro, B. Frank, S. Ristok, L. Gal, P. Kahl, D. Podbiel, S. Mathias, H. Giessen, F.-J. Meyer zu Heringdorf, M. Orenstein, and M. Aeschlimann, "Revealing the subfemtosecond dynamics of orbital angular momentum in nanoplasmonic vortices," Science **355**, 1187–1191 (2017).

126. N. V. Bloch, K. Shemer, A. Shapira, R. Shiloh, I. Juwiler, and A. Arie, "Twisting Light by Nonlinear Photonic Crystals," Phys. Rev. Lett. **108**, 233902 (2012).

127. C. Maurer, A. Jesacher, S. Fürhapter, S. Bernet, and M. Ritsch-Marte, "Tailoring of arbitrary optical vector fields," New J. Phys. **9**, 78 (2007).

128. K. S. Youngworth and T. G. Brown, "Focusing of high numerical aperture cylindrical-vector beams," Opt.

Express **7**, 77–87 (2000).

129. Q. Zhan, "Cylindrical vector beams: from mathematical concepts to          applications," Adv. Opt. Photonics **1**, 1–57 (2009).

130. L. Marrucci, E. Karimi, S. Slussarenko, B. Piccirillo, E. Santamato, E. Nagali, and F. Sciarrino, "Spin-to-orbital conversion of the angular momentum of light and its classical and quantum applications," J. Opt. **13**, 64001 (2011).

131. K. Y. Bliokh, F. J. Rodríguez-Fortuño, F. Nori, and A. V Zayats, "Spin–orbit interactions of light," Nat. Photonics **9**, 796–808 (2015).

132. F. Bouchard, I. De Leon, S. A. Schulz, J. Upham, E. Karimi, and R. W. Boyd, "Optical spin-to-orbital angular momentum conversion in ultra-thin metasurfaces with arbitrary topological charges," Appl. Phys. Lett. **105**, 101905 (2014).

133. R. C. Devlin, A. Ambrosio, D. Wintz, S. L. Oscurato, A. Y. Zhu, M. Khorasaninejad, J. Oh, P. Maddalena, and F. Capasso, "Spin-to-orbital angular momentum conversion in dielectric metasurfaces," Opt. Express **25**, 377–393 (2017).

134. R. C. Devlin, A. Ambrosio, N. A. Rubin, J. P. B. Mueller, and F. Capasso, "Arbitrary spin-to-orbital angular momentum conversion of light," Science **358**, 896–901 (2017).

135. Y. Zhao, J. S. Edgar, G. D. M. Jeffries, D. McGloin, and D. T. Chiu, "Spin-to-Orbital Angular Momentum Conversion in a Strongly Focused Optical Beam," Phys. Rev. Lett. **99**, 73901 (2007).

136. O. Yesharim, S. Pearl, J. Foley-Comer, I. Juwiler, and A. Arie, "Direct generation of spatially entangled qudits using quantum nonlinear optical holography," Sci. Adv. **9**, eade7968 (2023).

137. E. Rozenberg, A. Karnieli, O. Yesharim, J. Foley-Comer, S. Trajtenberg-Mills, D. Freedman, A. M. Bronstein, and A. Arie, "Inverse design of spontaneous parametric downconversion for generation of high-dimensional qudits," Optica **9**, 602–615 (2022).

138. Q. Yang, Z. Xie, M. Zhang, X. Ouyang, Y. Xu, Y. Cao, S. Wang, L. Zhu, and X. Li, "Ultra-secure optical encryption based on tightly focused perfect optical vortex beams," **11**, 1063–1070 (2022).

139. R. Fickler, R. Lapkiewicz, W. N. Plick, M. Krenn, C. Schaeff, S. Ramelow, and A. Zeilinger, "Quantum Entanglement of High Angular Momenta," Science **338**, 640–643 (2012).

140. H. Zhou, B. Sain, Y. Wang, C. Schlickriede, R. Zhao, X. Zhang, Q. Wei, X. Li, L. Huang, and T. Zentgraf, "Polarization-Encrypted Orbital Angular Momentum Multiplexed Metasurface Holography," ACS Nano **14**, 5553–5559 (2020).

141. T. Stav, A. Faerman, E. Maguid, D. Oren, V. Kleiner, E. Hasman, and M. Segev, "Quantum entanglement of the spin and orbital angular momentum of photons using metamaterials," Science **361**, 1101–1104 (2018).

142. P. S. Tan, X.-C. Yuan, J. Lin, Q. Wang, T. Mei, R. E. Burge, and G. G. Mu, "Surface plasmon polaritons generated by optical vortex beams," Appl. Phys. Lett. **92**, 111108 (2008).

143. G. Spektor, E. Prinz, M. Hartelt, A.-K. Mahro, M. Aeschlimann, and M. Orenstein, "Orbital angular momentum multiplication in plasmonic vortex cavities," Sci. Adv. **7**, eabg5571 (2021).

144. H. Ren, G. Briere, X. Fang, P. Ni, R. Sawant, S. Héron, S. Chenot, S. Vézian, B. Damilano, V. Brändli, S. A. Maier, and P. Genevet, "Metasurface orbital angular momentum holography," Nat. Commun. **10**, 2986 (2019).

145. L. Li and F. Li, "Beating the Rayleigh limit: Orbital-angular-momentum-based super-resolution diffraction tomography," Phys. Rev. E **88**, 33205 (2013).

146. M. Krenn and A. Zeilinger, "On small beams with large topological charge: II. Photons, electrons and gravitational waves," New J. Phys. **20**, 63006 (2018).

147. B. Wang, N. J. Brooks, P. Johnsen, N. W. Jenkins, Y. Esashi, I. Binnie, M. Tanksalvala, H. C. Kapteyn, and M. M. Murnane, "High-fidelity ptychographic imaging of highly periodic structures enabled by vortex high harmonic beams," Optica **10**, 1245–1252 (2023).

148. T. A. Klar and S. W. Hell, "Subdiffraction resolution in far-field fluorescence microscopy," Opt. Lett. **24**, 954–956 (1999).

149. S. W. Hell, "Far-Field Optical Nanoscopy," Science **316**, 1153–1158 (2007).

150. G. Li, S. Zhang, L. Isenhower, K. Maller, and M. Saffman, "Crossed vortex bottle beam trap for single-atom qubits," Opt. Lett. **37**, 851–853 (2012).

151. L. Isenhower, W. Williams, A. Dally, and M. Saffman, "Atom trapping in an interferometrically generated bottle beam trap," Opt. Lett. **34**, 1159–1161 (2009).

152. M. Zürch, C. Kern, P. Hansinger, A. Dreischuh, and C. Spielmann, "Strong-field physics with singular light beams," Nat. Phys. **8**, 743–746 (2012).

153. C. Hernández-García, A. Picón, J. San Román, and L. Plaja, "Attosecond Extreme Ultraviolet Vortices from High-Order Harmonic Generation," Phys. Rev. Lett. **111**, 83602 (2013).

154. G. Gariepy, J. Leach, K. T. Kim, T. J. Hammond, E. Frumker, R. W. Boyd, and P. B. Corkum, "Creating high-harmonic beams with controlled orbital angular momentum," Phys. Rev. Lett. **113**, 153901 (2014).

155. D. Gauthier, P. R. Ribič, G. Adhikary, A. Camper, C. Chappuis, R. Cucini, L. F. DiMauro, G. Dovillaire, F. Frassetto, R. Géneaux, P. Miotti, L. Poletto, B. Ressel, C. Spezzani, M. Stupar, T. Ruchon, and G. De Ninno, "Tunable orbital angular momentum in high-harmonic generation," Nat. Commun. **8**, 14971 (2017).

156. K. M. Dorney, L. Rego, N. J. Brooks, J. San Román, C.-T. Liao, J. L. Ellis, D. Zusin, C. Gentry, Q. L. Nguyen, J. M. Shaw, A. Picón, L. Plaja, H. C. Kapteyn, M. M. Murnane, and C. Hernández-García, "Controlling the polarization and vortex charge of attosecond high-harmonic beams via simultaneous spin–orbit momentum conservation," Nat. Photonics **13**, 123–130 (2019).

157. L. Rego, K. M. Dorney, N. J. Brooks, Q. L. Nguyen, C.-T. Liao, J. San Román, D. E. Couch, A. Liu, E. Pisanty, M. Lewenstein, L. Plaja, H. C. Kapteyn, M. M. Murnane, and C. Hernández-Garcia, "Generation of extreme-ultraviolet beams with time-varying orbital angular momentum," Science **364**, eaaw9486 (2019).

158. L. Rego, J. S. Román, A. Picón, L. Plaja, and C. Hernández-García, "Nonperturbative Twist in the Generation of Extreme-Ultraviolet Vortex Beams," Phys. Rev. Lett. **117**, 163202 (2016).

159. A. de las Heras, A. K. Pandey, J. San Román, J. Serrano, E. Baynard, G. Dovillaire, M. Pittman, C. G. Durfee, L. Plaja, S. Kazamias, O. Guilbaud, and C. Hernández-García, "Extreme-ultraviolet vector-vortex beams from high harmonic generation," Optica **9**, 71–79 (2022).

160. Z. Liu, Y. Liu, Y. Ke, Y. Liu, W. Shu, H. Luo, and S. Wen, "Generation of arbitrary vector vortex beams on hybrid-order Poincare sphere," Photonics Res. **5**, 15–21 (2017).

161. K. Jana, Y. Mi, S. H. Møller, D. H. Ko, S. Gholam-Mirzaei, D. Abdollahpour, S. Sederberg, and P. B. Corkum, "Quantum control of flying doughnut terahertz pulses," Sci. Adv. **10**, eadl1803 (2025).

162. R. Atanasov, A. Haché, J. L. P. Hughes, H. M. van Driel, and J. E. Sipe, "Coherent Control of Photocurrent Generation in Bulk Semiconductors," Phys. Rev. Lett. **76**, 1703–1706 (1996).

163. A. Haché, Y. Kostoulas, R. Atanasov, J. L. P. Hughes, J. E. Sipe, and H. M. van Driel, "Observation of Coherently Controlled Photocurrent in Unbiased, Bulk GaAs," Phys. Rev. Lett. **78**, 306–309 (1997).

164. R. D. R. Bhat and J. E. Sipe, "Optically Injected Spin Currents in Semiconductors," Phys. Rev. Lett. **85**, 5432–5435 (2000).

165. C. Heide, T. Eckstein, T. Boolakee, C. Gerner, H. B. Weber, I. Franco, and P. Hommelhoff, "Electronic Coherence and Coherent Dephasing in the Optical Control of Electrons in Graphene," Nano Lett. **21**, 9403–9409 (2021).

166. O. D. Jefimenko, "Comment on '"On the equivalence of the laws of Biot–Savart and Ampere,"' by T. A. Weber and D. J. Macomb [Am. J. Phys. 57, 57–59 (1989)]," Am. J. Phys. **58**, 505 (1990).

167. K. J. Yuan and A. D. Bandrauk, "Attosecond-magnetic-field-pulse generation by electronic currents in bichromatic circularly polarized UV laser fields," Phys. Rev. A **92**, 063401 (2015).

168. I. Barth and J. Manz, "Electric ring currents in atomic orbitals and magnetic fields induced by short intense circularly polarized π laser pulses," Phys. Rev. A **75**, 012510 (2007).

169. K. Jana, K. R. Herperger, F. Kong, Y. Mi, C. Zhang, P. B. Corkum, and S. Sederberg, "Reconfigurable electronic circuits for magnetic fields controlled by structured light," Nat. Photonics **15**, 622–626 (2021).

170. K. E. Ballantine, J. F. Donegan, and P. R. Eastham, "There are many ways to spin a photon: Half-quantization of a total optical angular momentum," Sci. Adv. **2**, e1501748 (2023).

171. E. Pisanty, S. Sukiasyan, and M. Ivanov, "Spin conservation in high-order-harmonic generation using bicircular fields," Phys. Rev. A **90**, 1–7 (2014).

172. A. Zdagkas, C. McDonnell, J. Deng, Y. Shen, G. Li, T. Ellenbogen, N. Papasimakis, and N. I. Zheludev, "Observation of toroidal pulses of light," Nat. Photonics **16**, 523–528 (2022).

173. C. Wan, Q. Cao, J. Chen, A. Chong, and Q. Zhan, "Toroidal vortices of light," Nat. Photonics **16**, 519–522 (2022).

174. N. Jhajj, I. Larkin, E. W. Rosenthal, S. Zahedpour, J. K. Wahlstrand, and H. M. Milchberg, "Spatiotemporal Optical Vortices," Phys. Rev. X **6**, 031037 (2016).

175. S. W. Hancock, S. Zahedpour, A. Goffin, and H. M. Milchberg, "Free-space propagation of spatiotemporal optical vortices," Optica **6**, 1547–1553 (2019).

176. N. Papasimakis, T. Raybould, V. A. Fedotov, D. P. Tsai, I. Youngs, and N. I. Zheludev, "Pulse generation scheme for flying electromagnetic doughnuts," Phys. Rev. B **97**, 201409 (2018).

177. S. Feng, H. G. Winful, and R. W. Hellwarth, "Spatiotemporal evolution of focused single-cycle electromagnetic pulses," Phys. Rev. E **59**, 4630–4649 (1999).

178. S. Feng, H. G. Winful, and R. W. Hellwarth, "Gouy shift and temporal reshaping of focused single-cycle electromagnetic pulses," Opt. Lett. **23**, 385–387 (1998).

179. Y. Fang, S. Lu, and Y. Liu, "Controlling Photon Transverse Orbital Angular Momentum in High Harmonic Generation," Phys. Rev. Lett. **127**, 273901 (2021).

180. Z. Lyu, Y. Fang, and Y. Liu, "Formation and Controlling of Optical Hopfions in High Harmonic Generation," Phys. Rev. Lett. **133**, 133801 (2024).

181. A. Forbes, "Structured Light from Lasers," Laser Photon. Rev. **13**, 1900140 (2019).

182. O. V Angelsky, A. Y. Bekshaev, S. G. Hanson, C. Y. Zenkova, I. I. Mokhun, and Z. Jun, "Structured Light: Ideas and Concepts   ," Front. Phys. **8**, (2020).

183. A. Forbes, M. de Oliveira, and M. R. Dennis, "Structured light," Nat. Photonics **15**, 253–262 (2021).

184. W. T. Buono and A. Forbes, "Nonlinear optics with structured light," Opto-Electronic Adv. **5**, 210119–210174 (2022).

185. C. He, Y. Shen, and A. Forbes, "Towards higher-dimensional structured light," Light Sci. Appl. **11**, 205 (2022).

186. S. Tsesses, E. Ostrovsky, K. Cohen, B. Gjonaj, N. H. Lindner, and G. Bartal, "Optical skyrmion lattice in evanescent electromagnetic fields," Science **361**, 993–996 (2018).

187. S. S., Y. X. Z., I. S., and T. Y., "Observation of Skyrmions in a Multiferroic Material," Science **336**, 198–201 (2012).

188. S. Das, Y. L. Tang, Z. Hong, M. A. P. Gonçalves, M. R. McCarter, C. Klewe, K. X. Nguyen, F. Gómez-Ortiz, P. Shafer, E. Arenholz, V. A. Stoica, S.-L. Hsu, B. Wang, C. Ophus, J. F. Liu, C. T. Nelson, S. Saremi, B. Prasad, A. B. Mei, D. G. Schlom, J. Íñiguez, P. García-Fernández, D. A. Muller, L. Q. Chen, J. Junquera, L. W. Martin, and R. Ramesh, "Observation of room-temperature polar skyrmions," Nature **568**, 368–372 (2019).

189. S. Woo, K. Litzius, B. Krüger, M.-Y. Im, L. Caretta, K. Richter, M. Mann, A. Krone, R. M. Reeve, M. Weigand, P. Agrawal, I. Lemesh, M.-A. Mawass, P. Fischer, M. Kläui, and G. S. D. Beach, "Observation of room-temperature magnetic skyrmions and their current-driven dynamics in ultrathin metallic ferromagnets," Nat. Mater. **15**, 501–506 (2016).

190. A. Karnieli, S. Tsesses, G. Bartal, and A. Arie, "Emulating spin transport with nonlinear optics, from high-order skyrmions to the topological Hall effect," Nat. Commun. **12**, 1092 (2021).

191. T. Ozawa, H. M. Price, A. Amo, N. Goldman, M. Hafezi, L. Lu, M. C. Rechtsman, D. Schuster, J. Simon, O. Zilberberg, and I. Carusotto, "Topological photonics," Rev. Mod. Phys. **91**, 15006 (2019).

192. L. Lu, J. D. Joannopoulos, and M. Soljačić, "Topological photonics," Nat. Photonics **8**, 821–829 (2014).

193. M. Segev and M. A. Bandres, "Topological photonics: Where do we go from here?," **10**, 425–434 (2021).

194. S. Ghimire and D. A. Reis, "High-harmonic generation from solids," Nat. Phys. **15**, 10–16 (2019).

195. L. Yue and M. B. Gaarde, "Introduction to theory of high-harmonic generation in solids: tutorial," J. Opt. Soc. Am. B **39**, 535–555 (2022).

196. A. F. Rañada, "A topological theory of the electromagnetic field," Lett. Math. Phys. **18**, 97–106 (1989).

197. A. F. Ranada, "Knotted solutions of the Maxwell equations in vacuum," J. Phys. A. Math. Gen. **23**, L815–L820 (1990).

198. W. T. M. Irvine and D. Bouwmeester, "Linked and knotted beams of light," Nat. Phys. **4**, 716–720 (2008).

199. T. Bauer, P. Banzer, E. Karimi, S. Orlov, A. Rubano, L. Marrucci, E. Santamato, R. W. Boyd, and G. Leuchs, "Observation of optical polarization Möbius strips," Science **347**, 964–966 (2015).

200. M. C. Rechtsman, J. M. Zeuner, Y. Plotnik, Y. Lumer, D. Podolsky, F. Dreisow, S. Nolte, M. Segev, and A. Szameit, "Photonic Floquet topological insulators," Nature **496**, 196–200 (2013).

201. M. A. Bandres, S. Wittek, G. Harari, M. Parto, J. Ren, M. Segev, D. N. Christodoulides, and M. Khajavikhan, "Topological insulator laser: Experiments," Science **359**, (2018).

202. G. Harari, M. A. Bandres, Y. Lumer, M. C. Rechtsman, Y. D. Chong, M. Khajavikhan, D. N. Christodoulides, and M. Segev, "Topological insulator laser: Theory," Science **359**, (2018).

203. W. T. M. Irvine, "Linked and knotted beams of light, conservation of helicity and the flow of null electromagnetic fields," J. Phys. A. Math. Theor. **43**, 385203 (2010).

204. H. Kedia, D. Foster, M. R. Dennis, and W. T. M. Irvine, "Weaving Knotted Vector Fields with Tunable Helicity," Phys. Rev. Lett. **117**, 274501 (2016).

205. H. Larocque, D. Sugic, D. Mortimer, A. J. Taylor, R. Fickler, R. W. Boyd, M. R. Dennis, and E. Karimi, "Reconstructing the topology of optical polarization knots," Nat. Phys. **14**, 1079–1082 (2018).

206. M. Arrayás, D. Bouwmeester, and J. L. Trueba, "Knots in electromagnetism," Phys. Rep. **667**, 1–61 (2017).

207. M. F. Ferrer-Garcia, A. D'Errico, H. Larocque, A. Sit, and E. Karimi, "Polychromatic electric field knots," Phys. Rev. Res. **3**, 33226 (2021).

208. H. Kedia, I. Bialynicki-Birula, D. Peralta-Salas, and W. T. M. Irvine, "Tying Knots in Light Fields," Phys. Rev. Lett. **111**, 150404 (2013).

209. L.-J. Kong, J. Zhang, F. Zhang, and X. Zhang, "Topological Holography and Storage with Optical Knots and Links," Laser Photon. Rev. **n/a**, 2300005 (2023).

210. J. Verbeeck, H. Tian, and P. Schattschneider, "Production and application of electron vortex beams," Nature **467**, 301–304 (2010).

211. B. J. McMorran, A. Agrawal, I. M. Anderson, A. A. Herzing, H. J. Lezec, J. J. McClelland, and J. Unguris, "Electron Vortex Beams with High Quanta of Orbital Angular Momentum," Science (80-. ). **331**, 192–195 (2011).

212. M. Kozák, "Electron Vortex Beam Generation via Chiral Light-Induced Inelastic Ponderomotive Scattering," ACS Photonics **8**, 431–435 (2021).

213. A. Luski, Y. Segev, R. David, O. Bitton, H. Nadler, A. R. Barnea, A. Gorlach, O. Cheshnovsky, I. Kaminer, and E. Narevicius, "Vortex beams of atoms and molecules," Science (80-. ). **373**, 1105–1109 (2021).

214. W. H. McMaster, "Polarization and the Stokes Parameters," Am. J. Phys. **22**, 351–362 (1954).

215. V. Strelkov, A. Zair, O. Tcherbakoff, R. López-Martens, E. Cormier, E. Mével, and E. Constant, "Generation of attosecond pulses with ellipticity-modulated fundamental," Appl. Phys. B **78**, 879–884 (2004).

216. S. Rozen, A. Comby, E. Bloch, S. Beauvarlet, D. Descamps, B. Fabre, S. Petit, V. Blanchet, B. Pons, N. Dudovich, and Y. Mairesse, "Controlling Subcycle Optical Chirality in the Photoionization of Chiral Molecules," Phys. Rev. X **9**, 031004 (2019).

217. K. M. Dorney, J. L. Ellis, C. Hernández-Garc'ia, D. D. Hickstein, C. A. Mancuso, N. Brooks, T. Fan, G. Fan, D. Zusin, C. Gentry, P. Grychtol, H. C. Kapteyn, and M. M. Murnane, "Helicity-Selective Enhancement and Polarization Control of Attosecond High Harmonic Waveforms Driven by Bichromatic Circularly Polarized Laser Fields," Phys. Rev. Lett. **119**, 63201 (2017).

218. Y. Tang and A. E. Cohen, "Optical chirality and its interaction with matter," Phys. Rev. Lett. **104**, 163901 (2010).

219. D. M. Lipkin, "Existence of a New Conservation Law in Electromagnetic Theory," J. Math. Phys. **5**, 696–700 (1964).

220. T. W. B. Kibble, "Conservation Laws for Free Fields," J. Math. Phys. **6**, 1022–1026 (2004).

221. D. J. Candlin, "Analysis of the new conservation law in electromagnetic theory," Nuovo Cim. **37**, 1390–1395 (1965).

222. G. Smith and P. Strange, "Lipkin's conservation law in vacuum electromagnetic fields," J. Phys. A Math. Theor.

**51**, 435204 (2018).

223. O. Neufeld and O. Cohen, "Optical Chirality in Nonlinear Optics: Application to High Harmonic Generation," Phys. Rev. Lett. **120**, 133206 (2018).

224. M. M. Coles and D. L. Andrews, "Photonic measures of helicity: optical vortices and circularly polarized reflection," Opt. Lett. **38**, 869–871 (2013).

225. P. Gutsche, L. V Poulikakos, M. Hammerschmidt, S. Burger, and F. Schmidt, "Time-harmonic optical chirality in inhomogeneous space," Proceeding SPIE **9756X**, (2016).

226. C. Rosales-guzmán, K. Volke-sepulveda, and J. P. Torres, "Light with enhanced optical chirality," Opt. Lett. **37**, 3486–3488 (2012).

227. Y. Tang and A. E. Cohen, "Enhanced Enantioselectivity in Excitation of Chiral Molecules by Superchiral Light," Science **332**, 333–336 (2011).

228. C. Kramer, M. Schäferling, T. Weiss, H. Giessen, and T. Brixner, "Analytic Optimization of Near-Field Optical Chirality Enhancement," ACS Photonics **4**, 396–406 (2017).

229. T. Kakkar, C. Keijzer, M. Rodier, T. Bukharova, A. J. Love, J. J. Milner, A. S. Karimullah, L. D. Barron, N. Gadegaard, A. J. Lapthorn, and M. Kadodwala, "Superchiral near fields detect virus structure," Light Sci. Appl. **9**, 195 (2020).

230. E. Mohammadi, A. Tittl, K. L. Tsakmakidis, T. V Raziman, and A. G. Curto, "Dual Nanoresonators for Ultrasensitive Chiral Detection," ACS Photonics **8**, 1754–1762 (2021).

231. T. J. Davis and E. Hendry, "Superchiral electromagnetic fields created by surface plasmons in nonchiral metallic nanostructures," Phys. Rev. B **87**, 85405 (2013).

232. M. M. Coles and D. L. Andrews, "Chirality and angular momentum in optical radiation," Phys. Rev. A **85**, 063810 (2012).

233. J. S. Choi and M. Cho, "Limitations of a superchiral field," Phys. Rev. A **86**, 63834 (2012).

234. Y. Liu, W. Zhao, Y. Ji, R.-Y. Wang, X. Wu, and X. D. Zhang, "Strong superchiral field in hot spots and its interaction with chiral molecules," Europhys. Lett. **110**, 17008 (2015).

235. T. G. Philbin, "Lipkin's conservation law, Noether's theorem, and the relation to optical helicity," Phys. Rev. A **87**, 043843 (2013).

236. S. M. Barnett, R. P. Cameron, and A. M. Yao, "Duplex symmetry and its relation to the conservation of optical helicity," Phys. Rev. A **86**, 013845 (2012).

237. D. L. Andrews and M. M. Coles, "Measures of chirality and angular momentum in the electromagnetic field," Opt. Lett. **37**, 3009–3011 (2012).

238. D. S. Bradshaw, J. M. Leeder, M. M. Coles, and D. L. Andrews, "Signatures of material and optical chirality: Origins and measures," Chem. Phys. Lett. **626**, 106–110 (2015).

239. S. Nechayev, J. S. Eismann, R. Alaee, E. Karimi, R. W. Boyd, and P. Banzer, "Kelvin's chirality of optical beams," Phys. Rev. A **103**, L031501 (2021).

240. A. Korobenko, "Control of molecular rotation with an optical centrifuge," J. Phys. B At. Mol. Opt. Phys. **51**, 203001 (2018).

241. O. Neufeld, M. Even Tzur, and O. Cohen, "Degree of chirality of electromagnetic fields and maximally chiral light," Phys. Rev. A **101**, 053831 (2020).

242. D. Ayuso, O. Neufeld, A. F. Ordonez, P. Decleva, G. Lerner, O. Cohen, M. Ivanov, and O. Smirnova, "Synthetic chiral light for efficient control of chiral light–matter interaction," Nat. Photonics **13**, 866–871 (2019).

243. O. Neufeld and O. Cohen, "Unambiguous definition of handedness for locally chiral light," Phys. Rev. A **105**, 23514 (2022).

244. J. Karczmarek, J. Wright, P. Corkum, and M. Ivanov, "Optical Centrifuge for Molecules," Phys. Rev. Lett. **82**, 3420–3423 (1999).

245. A. A. Milner, J. A. M. Fordyce, I. Macphail-bartley, W. Wasserman, I. Tutunnikov, I. S. Averbukh, and V. Milner, "Controlled enantioselective orientation of chiral molecules with an optical centrifuge," Phys. Rev. Lett. **122**, 223201 (2019).

246. D. M. Villeneuve, S. A. Aseyev, P. Dietrich, M. Spanner, M. Y. Ivanov, and P. B. Corkum, "Forced Molecular Rotation in an Optical Centrifuge," Phys. Rev. Lett. **85**, 542–545 (2000).

247. A. Fleischer, E. Bordo, O. Kfir, P. Sidorenko, and O. Cohen, "Polarization-fan high-order harmonics," J. Phys. B At. Mol. Opt. Phys. **50**, 034001 (2017).

248. O. Neufeld, E. Bordo, A. Fleischer, and O. Cohen, "High harmonic generation with fully tunable polarization by train of linearly-polarized pulses," New J. Phys. **19**, 023051 (2017).

249. O. Neufeld, E. Bordo, A. Fleischer, and O. Cohen, "High Harmonics with Controllable Polarization by a Burst of Linearly-Polarized Driver Pulses," Photonics **4**, 31 (2017).

250. T. P. E. Auf der heyde, A. B. Buda, and K. Mislow, "Desymmetrization and degree of chirality," J. Math. Chem. **6**, 255–265 (1991).

251. G. P. Moss, "Basic terminology of stereochemistry (IUPAC Recommendations 1996)," Pure Appl. Chem. **68**, 2193–2222 (1996).

252. P. Y. Bruice, *Organic Chemistry*, 5th ed. (Prentice Hall, 2006).

253. D. Ayuso, A. F. Ordonez, P. Decleva, M. Ivanov, and O. Smirnova, "Enantio-sensitive unidirectional light bending," Nat. Commun. **12**, 3951 (2021).

254. L. Rego, O. Smirnova, and D. Ayuso, "Tilting light's polarization plane to spatially separate the ultrafast nonlinear response of chiral molecules," Nanophotonics (2023).

255. D. Ayuso, "New opportunities for ultrafast and highly enantio-sensitive imaging of chiral nuclear dynamics enabled by synthetic chiral light," Phys. Chem. Chem. Phys. **24**, 10193–10200 (2022).

256. D. Ayuso, A. F. Ordonez, M. Ivanov, and O. Smirnova, "Ultrafast optical rotation in chiral molecules with ultrashort and tightly focused beams," Optica **8**, 1243–1246 (2021).

257. E. Lustig and M. Segev, "Topological photonics in synthetic dimensions," Adv. Opt. Photonics **13**, 426–461 (2021).

258. M. Ehrhardt, S. Weidemann, L. J. Maczewsky, M. Heinrich, and A. Szameit, "A Perspective on Synthetic Dimensions in Photonics," Laser Photon. Rev. **17**, 2200518 (2023).

259. T. Ozawa and H. M. Price, "Topological quantum matter in synthetic dimensions," Nat. Rev. Phys. **1**, 349–357 (2019).

260. S. Long, W. Becker, and J. K. McIver, "Model calculations of polarization-dependent two-color high-harmonic generation," Phys. Rev. A **52**, 2262–2278 (1995).

261. S. Bivona, R. Burlon, and C. Leone, "Symmetries in harmonic generation by a two-color field: an application to a simple model atom," J. Opt. Soc. Am. B **16**, 986–993 (1999).

262. A. Fleischer, V. Averbukh, and N. Moiseyev, "Non-Hermitian quantum mechanics versus the conventional quantum mechanics: Effect of the relative phase of bichromatic fields on high-order harmonic generation," Phys. Rev. A **69**, 43404 (2004).

263. T. Fan, P. Grychtol, R. Knut, C. Hernández-García, D. D. Hickstein, D. Zusin, C. Gentry, F. J. Dollar, C. A. Mancuso, C. W. Hogle, O. Kfir, D. Legut, K. Carva, J. L. Ellis, K. M. Dorney, C. Chen, O. G. Shpyrko, E. E. Fullerton, O. Cohen, P. M. Oppeneer, D. B. Milošević, A. Becker, A. A. Jaroń-Becker, T. Popmintchev, M. M. Murnane, and H. C. Kapteyn, "Bright circularly polarized soft X-ray high harmonics for X-ray magnetic circular dichroism," Proc. Natl. Acad. Sci. **112**, 14206–14211 (2015).

264. X. Liu, X. Zhu, L. Li, Y. Li, Q. Zhang, P. Lan, and P. Lu, "Selection rules of high-order-harmonic generation: Symmetries of molecules and laser fields," Phys. Rev. A **94**, 033410 (2016).

265. E. Frumker, C. T. Hebeisen, N. Kajumba, J. B. Bertrand, H. J. Wörner, M. Spanner, D. M. Villeneuve, A. Naumov, and P. B. Corkum, "Oriented Rotational Wave-Packet Dynamics Studies via High Harmonic Generation," Phys. Reivew Lett. **109**, 113901 (2012).

266. N. Dudovich, O. Smirnova, J. Levesque, Y. Mairesse, M. Y. Ivanov, D. M. Villeneuve, and P. B. Corkum, "Measuring and controlling the birth of attosecond XUV pulses," Nat Phys **2**, 781–786 (2006).

267. M. Holthaus, "Floquet engineering with quasienergy bands of periodically driven optical lattices," J. Phys. B At. Mol. Opt. Phys. **49**, 13001 (2016).

268. O. Neufeld, J. Zhang, U. De Giovannini, H. Hübener, and A. Rubio, "Probing phonon dynamics with multidimensional high harmonic carrier-envelope-phase spectroscopy," Proc. Natl. Acad. Sci. **119**, e2204219119 (2022).

269. A. Fleischer and N. Moiseyev, "Adiabatic theorem for non-Hermitian time-dependent open systems," Phys. Rev. A **72**, 032103 (2005).

270. O. Neufeld, A. Fleischer, and O. Cohen, "High-order harmonic generation of pulses with multiple timescales: selection rules, carrier envelope phase and cutoff energy," Mol. Phys. **117**, 1956–1963 (2019).

271. S. Ito, M. Schüler, M. Meierhofer, S. Schlauderer, J. Freudenstein, J. Reimann, D. Afanasiev, K. A. Kokh, O. E. Tereshchenko, J. Güdde, M. A. Sentef, U. Höfer, and R. Huber, "Build-up and dephasing of Floquet–Bloch bands on subcycle timescales," Nature (2023).

272. H. Liu, Y. Li, Y. S. You, S. Ghimire, T. F. Heinz, and D. A. Reis, "High-harmonic generation from an atomically thin semiconductor," Nat. Phys. **13**, 262–265 (2017).

273. G. Sansone, E. Benedetti, F. Calegari, C. Vozzi, L. Avaldi, R. Flammini, L. Poletto, P. Villoresi, C. Altucci, R. Velotta, S. Stagira, S. De Silvestri, and M. Nisoli, "Isolated Single-Cycle Attosecond Pulses," Science **314**, 443–446 (2006).

274. D. Faccialà, S. Pabst, B. D. Bruner, A. G. Ciriolo, S. De Silvestri, M. Devetta, M. Negro, H. Soifer, S. Stagira, N. Dudovich, and C. Vozzi, "Probe of Multielectron Dynamics in Xenon by Caustics in High-Order Harmonic Generation," Phys. Rev. Lett. **117**, (2016).

275. A. D. Shiner, B. E. Schmidt, C. Trallero-Herrero, H. J. Wörner, S. Patchkovskii, P. B. Corkum, J.-C. Kieffer, F. Légaré, and D. M. Villeneuve, "Probing collective multi-electron dynamics in xenon with high-harmonic spectroscopy," Nat. Phys. **7**, 464 (2011).

276. S. Pabst and R. Santra, "Strong-Field Many-Body Physics and the Giant Enhancement in the High-Harmonic Spectrum of Xenon," Phys. Rev. Lett. **111**, 233005 (2013).

277. P. M. Kraus, D. Baykusheva, and H. J. Wörner, "Two-Pulse Field-Free Orientation Reveals Anisotropy of Molecular Shape Resonance," Phys. Rev. Lett. **113**, 023001 (2014).

278. D. B. Milošević, "Circularly polarized high harmonics generated by a bicircular field from inert atomic gases in the p state: A tool for exploring chirality-sensitive processes," Phys. Rev. A **92**, 043827 (2015).

279. D. B. Milosevic, "Theoretical analysis of high-order harmonic generation from a coherent superposition of states," J. Opt. Soc. Am. B **23**, 308–317 (2006).

280. D. B. Milošević, B. Fetić, and P. Ranitovic, "High-order above-threshold ionization from a coherent superposition of states," Phys. Rev. A **106**, 13109 (2022).

281. S. Chelkowski, T. Bredtmann, and A. D. Bandrauk, "High-harmonic generation from a coherent superposition of electronic states: Controlling interference patterns via short and long quantum orbits," Phys. Rev. A **88**, 33423 (2013).

282. H. A. Kramers, "General theory of paramagnetic rotation in crystals," in *Proc. Acad. Sci. Amsterdam* (1930), Vol. 33, p. 959.

283. O. Alon, V. Averbukh, and N. Moiseyev, "High harmonic generation of soft X-rays by carbon nanotubes," Phys. Rev. Lett. **85**, 5218 (2000).

284. T. T. Luu, Z. Yin, A. Jain, T. Gaumnitz, Y. Pertot, J. Ma, and H. J. Wörner, "Extreme–ultraviolet high–harmonic generation in liquids," Nat. Commun. **9**, 3723 (2018).

285. A.-W. Zeng and X.-B. Bian, "Impact of Statistical Fluctuations on High Harmonic Generation in Liquids," Phys. Rev. Lett. **124**, 203901 (2020).

286. A. Mondal, O. Neufeld, Z. Yin, Z. Nourbakhsh, V. V. Svoboda, A. Rubio, N. Tancogne-Dejean, and H. J. Wörner, "High-harmonic spectroscopy of low-energy electron-scattering dynamics in liquids," Nat. Phys. **19**, 1813–1820 (2023).

287. A. Mondal, B. Waser, T. Balciunas, O. Neufeld, Z. Yin, N. Tancogne-Dejean, A. Rubio, and H. J. Wörner, "High-harmonic generation in liquids with few-cycle pulses: effect of laser-pulse duration on the cut-off energy," Opt. Express **31**, 34348–34361 (2023).

288. O. Neufeld, Z. Nourbakhsh, N. Tancogne-Dejean, and A. Rubio, "Ab Initio Cluster Approach for High Harmonic Generation in Liquids," J. Chem. Theory Comput. **18**, 4117–4126 (2022).

289. H. G. Kurz, M. Kretschmar, T. Binhammer, T. Nagy, D. Ristau, M. Lein, U. Morgner, and M. Kovačev, "Revealing the Microscopic Real-Space Excursion of a Laser-Driven Electron," Phys. Rev. X **6**, 31029 (2016).

290. O. Alexander, J. C. T. Barnard, E. W. Larsen, T. Avni, S. Jarosch, C. Ferchaud, A. Gregory, S. Parker, G. Galinis, A. Tofful, D. Garratt, M. R. Matthews, and J. P. Marangos, "Observation of recollision-based high-harmonic generation in liquid isopropanol and the role of electron scattering," Phys. Rev. Res. **5**, 43030 (2023).

291. Z.-W. Ding, Y.-B. Wang, Z.-L. Li, and X.-B. Bian, "High-order harmonic generation in liquids in bicircularly polarized laser fields," Phys. Rev. A **107**, 13503 (2023).

292. Y. S. You, Y. Yin, Y. Wu, A. Chew, X. Ren, F. Zhuang, S. Gholam-Mirzaei, M. Chini, Z. Chang, and S. Ghimire, "High-harmonic generation in amorphous solids," Nat. Commun. **8**, 724 (2017).

293. T. T. Luu and H. J. Wörner, "Observing broken inversion symmetry in solids using two-color high-order harmonic spectroscopy," Phys. Rev. A **98**, 41802 (2018).

294. T. T. Luu and H. J. Wörner, "Measurement of the Berry curvature of solids using high-harmonic spectroscopy," Nat. Commun. **9**, 916 (2018).

295. D. Habibović, W. Becker, and D. B. Milošević, "Symmetries and Selection Rules of the Spectra of Photoelectrons and High-Order Harmonics Generated by Field-Driven Atoms and Molecules," Symmetry (Basel). **13**, (2021).

296. D. Habibović, A. Gazibegović-Busuladžić, M. Busuladžić, A. Čerkić, and D. B. Milošević, "Strong-field ionization of homonuclear diatomic molecules using orthogonally polarized two-color laser fields," Phys. Rev. A **102**, 23111 (2020).

297. D. Habibović, A. Gazibegović-Busuladžić, M. Busuladžić, and D. B. Milošević, "Strong-field ionization of heteronuclear diatomic molecules using an orthogonally polarized two-color laser field," Phys. Rev. A **103**, 53101 (2021).

298. M. Busuladžić, A. Gazibegović-Busuladžić, and D. B. Milošević, "Strong-field ionization of homonuclear diatomic molecules by a bicircular laser field: Rotational and reflection symmetries," Phys. Rev. A **95**, 33411 (2017).

299. M. Busuladžić, A. Gazibegović-Busuladžić, A. Čerkić, and D. B. Milošević, "Signature of molecular symmetry in the plateau region of the photoelectron spectra: Above-threshold ionization of the C2 molecule," Phys. Scr. **95**, 75402 (2020).

300. S. Hashim, D. Habibović, and C. Faria, "Below threshold nonsequential double ionization with linearly polarized two-color fields I: symmetry and dominance," arXiv Prepr. arXiv2411.19658 (2024).

301. D. Köhnke, T. Bayer, and M. Wollenhaupt, "Shaped free-electron vortices," Phys. Rev. A **110**, 53109 (2024).

302. H. Ni, S. Chen, C. Ruiz, and A. Becker, "Selection rules in the few-photon double ionization of the helium atom," J. Phys. B At. Mol. Opt. Phys. **44**, 175601 (2011).

303. T. Rook, L. C. Rodriguez, and C. F. de M. Faria, "Influence of catastrophes and hidden dynamical symmetries on ultrafast backscattered photoelectrons," Phys. Rev. Res. **6**, 23329 (2024).

304. P. V Demekhin, A. N. Artemyev, A. Kastner, and T. Baumert, "Photoelectron Circular Dichroism with Two Overlapping Laser Pulses of Carrier Frequencies ω and 2ω Linearly Polarized in Two Mutually Orthogonal Directions," Phys. Rev. Lett. **121**, 253201 (2018).

305. O. Neufeld, H. Hübener, A. Rubio, and U. De Giovannini, "Strong chiral dichroism and enantiopurification in above-threshold ionization with locally chiral light," Phys. Rev. Res. **3**, L032006 (2021).

306. S. Sederberg, F. Kong, and P. B. Corkum, "Tesla-Scale Terahertz Magnetic Impulses," Phys. Rev. X **10**, 11063 (2020).

307. J. Wätzel and J. Berakdar, "Multipolar, polarization-shaped high-order harmonic generation by intense vector beams," Phys. Rev. A **101**, 43409 (2020).

308. L. Zhang, L. Ji, and B. Shen, "Intense harmonic generation driven by a relativistic spatiotemporal vortex beam," High Power Laser Sci. Eng. **10**, e46 (2022).

309. M. Rodriguez-Vega, M. Lentz, and B. Seradjeh, "Floquet perturbation theory: formalism and application to low-frequency limit," New J. Phys. **20**, 93022 (2018).

310. D. B. Milošević, "High-order harmonic generation by a bichromatic elliptically polarized field: conservation of

angular momentum," J. Phys. B At. Mol. Opt. Phys. **48**, 171001 (2015).

311. F. Mauger, A. D. Bandrauk, and T. Uzer, "Circularly polarized molecular high harmonic generation using a bicircular laser," J. Phys. B At. Mol. Opt. Phys. **49**, 10LT01 (2016).

312. Y. Kaneko and T. N. Ikeda, "Floquet systems with continuous dynamical symmetries: characterization, time-dependent noether charge, and solvability," Phys. Scr. **99**, 85231 (2024).

313. D. Shechtman, I. Blech, D. Gratias, and J. W. Cahn, "Metallic Phase with Long-Range Orientational Order and No Translational Symmetry," Phys. Rev. Lett. **53**, 1951–1953 (1984).

314. O. Kfir, S. Zayko, C. Nolte, M. Sivis, M. Möller, B. Hebler, S. S. P. K. Arekapudi, D. Steil, S. Schäfer, M. Albrecht, O. Cohen, S. Mathias, and C. Ropers, "Nanoscale magnetic imaging using circularly polarized high-harmonic radiation," Sci. Adv. **3**, eaao4641 (2017).

315. S. Zayko, O. Kfir, M. Heigl, M. Lohmann, M. Sivis, M. Albrecht, and C. Ropers, "Ultrafast high-harmonic nanoscopy of magnetization dynamics," Nat. Commun. **12**, 6337 (2021).

316. D. D. Hickstein, F. J. Dollar, P. Grychtol, J. L. Ellis, R. Knut, C. Hernández-García, D. Zusin, C. Gentry, J. M. Shaw, T. Fan, K. M. Dorney, A. Becker, A. Jarón-Becker, H. C. Kapteyn, M. M. Murnane, and C. G. Durfee, "Non-collinear generation of angularly isolated circularly polarized high harmonics," Nat. Photonics **9**, 743–750 (2015).

317. P.-C. Huang, C. Hernández-García, J.-T. Huang, P.-Y. Huang, C.-H. Lu, L. Rego, D. D. Hickstein, J. L. Ellis, A. Jaron-Becker, A. Becker, S.-D. Yang, C. G. Durfee, L. Plaja, H. C. Kapteyn, M. M. Murnane, A. H. Kung, and M.-C. Chen, "Polarization control of isolated high-harmonic pulses," Nat. Photonics **12**, 349–354 (2018).

318. M. Han, J.-B. Ji, K. Ueda, and H. J. Wörner, "Attosecond metrology in circular polarization," Optica **10**, 1044–1052 (2023).

319. D. Azoury, O. Kneller, M. Krüger, B. D. Bruner, O. Cohen, Y. Mairesse, and N. Dudovich, "Interferometric attosecond lock-in measurement of extreme-ultraviolet circular dichroism," Nat. Photonics **13**, 198–204 (2019).

320. T. Feng, A. Heilmann, M. Bock, L. Ehrentraut, T. Witting, H. Yu, H. Stiel, S. Eisebitt, and M. Schnürer, "27 W 2.1 μm OPCPA system for coherent soft X-ray generation operating at 10 kHz," Opt. Express **28**, 8724–8733 (2020).

321. P.-A. Chevreuil, F. Brunner, S. Hrisafov, J. Pupeikis, C. R. Phillips, U. Keller, and L. Gallmann, "Water-window high harmonic generation with 0.8-μm and 2.2-μm OPCPAs at 100 kHz," Opt. Express **29**, 32996–33008 (2021).

322. A. Bahabad, M. M. Murnane, and H. C. Kapteyn, "Quasi-phase-matching of momentum and energy in nonlinear optical processes," Nat. Photonics **4**, 571–575 (2010).

323. L. Hareli, L. Lobachinsky, G. Shoulga, Y. Eliezer, L. Michaeli, and A. Bahabad, "On-the-Fly Control of High-Harmonic Generation Using a Structured Pump Beam," Phys. Rev. Lett. **120**, 183902 (2018).

324. G. Lerner, T. Diskin, O. Neufeld, O. Kfir, and O. Cohen, "Selective suppression of high-order harmonics within phase-matched spectral regions," Opt. Lett. **42**, 1349–1352 (2017).

325. M.-C. Chen, C. Mancuso, C. Hernández-García, F. Dollar, B. Galloway, D. Popmintchev, P.-C. Huang, B. Walker, L. Plaja, A. A. Jarón-Becker, A. Becker, M. M. Murnane, H. C. Kapteyn, and T. Popmintchev, "Generation of bright isolated attosecond soft X-ray pulses driven by multicycle midinfrared lasers," Proc. Natl. Acad. Sci. **111**, E2361–E2367 (2014).

326. C. G. Durfee, A. R. Rundquist, S. Backus, C. Herne, M. M. Murnane, and H. C. Kapteyn, "Phase Matching of High-Order Harmonics in Hollow Waveguides," Phys. Reivew Lett. **83**, 2187–2190 (1999).

327. C. Chen, Z. Tao, C. Hernández-García, P. Matyba, A. Carr, R. Knut, O. Kfir, D. Zusin, C. Gentry, P. Grychtol, O. Cohen, L. Plaja, A. Becker, A. Jaron-Becker, H. Kapteyn, and M. Murnane, "Tomographic reconstruction of circularly polarized high-harmonic fields: 3D attosecond metrology.," Sci. Adv. **2**, e1501333 (2016).

328. L. Medišauskas, J. Wragg, H. Van Der Hart, and M. Y. Ivanov, "Generating Isolated Elliptically Polarized Attosecond Pulses Using Bichromatic Counterrotating Circularly Polarized Laser Fields," Phys. Rev. Lett. **115**, 153001 (2015).

329. O. Kfir, P. Grychtol, E. Turgut, R. Knut, D. Zusin, A. Fleischer, E. Bordo, T. Fan, D. Popmintchev, T. Popmintchev, H. Kapteyn, M. Murnane, and O. Cohen, "Helicity-selective phase-matching and quasi-phase matching of circularly polarized high-order harmonics: towards chiral attosecond pulses," J. Phys. B At. Mol. Opt. Phys. **49**, 123501 (2016).

330. K. M. Dorney, T. Fan, Q. L. D. Nguyen, J. L. Ellis, D. D. Hickstein, N. Brooks, D. Zusin, C. Gentry, C. Hernández-García, H. C. Kapteyn, and M. M. Murnane, "Bright, single helicity, high harmonics driven by mid-infrared bicircular laser fields," Opt. Express **29**, 38119–38128 (2021).

331. T. Heinrich, M. Taucer, O. Kfir, P. B. Corkum, A. Staudte, C. Ropers, and M. Sivis, "Chiral high-harmonic generation and spectroscopy on solid surfaces using polarization-tailored strong fields," Nat. Commun. **12**, 3723 (2021).

332. E. Bordo, O. Kfir, S. Zayko, O. Neufeld, A. Fleischer, C. Ropers, and O. Cohen, "Interlocked attosecond pulse trains in slightly bi-elliptical high harmonic generation," J. Phys. Photonics **2**, (2020).

333. M. Ferray, A. L'Huillier, X. F. Li, L. A. Lompre, G. Mainfray, and C. Manus, "Multiple-harmonic conversion of 1064 nm radiation in rare gases," J. Phys. B At. Mol. Opt. Phys. **21**, L31–L35 (1988).

334. X. F. Li, A. L'Huillier, M. Ferray, L. A. Lompré, and G. Mainfray, "Multiple-harmonic generation in rare gases at high laser intensity," Phys. Rev. A **39**, 5751–5761 (1989).

335. M. Möller, Y. Cheng, S. D. Khan, B. Zhao, K. Zhao, M. Chini, G. G. Paulus, and Z. Chang, "Dependence of high-order-harmonic-generation yield on driving-laser ellipticity," Phys. Rev. A **86**, 011401 (2012).

336. S. Fleischer, Y. Zhou, R. W. Field, and K. A. Nelson, "Molecular Orientation and Alignment by Intense Single-

Cycle THz Pulses," Phys. Rev. Lett. **107**, 163603 (2011).

337. A. A. Milner, A. Korobenko, and V. Milner, "Field-free long-lived alignment of molecules with a two-dimensional optical centrifuge," Phys. Rev. A **93**, 53408 (2016).

338. S. Fleischer, Y. Khodorkovsky, Y. Prior, and I. Sh Averbukh, "Controlling the sense of molecular rotation," New J. Phys. **11**, 105039 (2009).

339. R. Velotta, N. Hay, M. B. Mason, M. Castillejo, and J. P. Marangos, "High-Order Harmonic Generation in Aligned Molecules," Phys. Rev. Lett. **87**, 183901 (2001).

340. N. Hay, R. Velotta, M. Lein, R. de Nalda, E. Heesel, M. Castillejo, and J. P. Marangos, "High-order harmonic generation in laser-aligned molecules," Phys. Rev. A **65**, 53805 (2002).

341. P. M. Kraus, A. Rupenyan, and H. J. Wörner, "High-harmonic spectroscopy of oriented OCS molecules: emission of even and odd harmonics.," Phys. Rev. Lett. **109**, 233903 (2012).

342. J. Levesque, Y. Mairesse, N. Dudovich, H. Pépin, J.-C. Kieffer, P. B. Corkum, and D. M. Villeneuve, "Polarization State of High-Order Harmonic Emission from Aligned Molecules," Phys. Rev. Lett. **99**, 243001 (2007).

343. E. Skantzakis, S. Chatziathanasiou, P. A. Carpeggiani, G. Sansone, A. Nayak, D. Gray, P. Tzallas, D. Charalambidis, E. Hertz, and O. Faucher, "Polarization shaping of high-order harmonics in laser-aligned molecules," Sci. Rep. **6**, 39295 (2016).

344. X. Zhou, R. Lock, N. Wagner, W. Li, H. C. Kapteyn, and M. M. Murnane, "Elliptically polarized high-order harmonic emission from molecules in linearly polarized laser fields.," Phys. Rev. Lett. **102**, 073902 (2009).

345. Y. Mairesse, J. Higuet, N. Dudovich, D. Shafir, B. Fabre, E. Mével, E. Constant, S. Patchkovskii, Z. Walters, M. Y. Ivanov, and O. Smirnova, "High harmonic spectroscopy of multichannel dynamics in strong-field ionization.," Phys. Rev. Lett. **104**, 213601 (2010).

346. J. Itatani, J. Levesque, D. Zeidler, H. Niikura, H. Pepin, J. C. Kieffer, P. B. Corkum, and D. M. Villeneuve, "Tomographic imaging of molecular orbitals," Nature **432**, 867–871 (2004).

347. X. X. Zhou, X. M. Tong, Z. X. Zhao, and C. D. Lin, "Role of molecular orbital symmetry on the alignment dependence of high-order harmonic generation with molecules," Phys. Rev. A **71**, 61801 (2005).

348. T. Kanai, S. Minemoto, and H. Sakai, "Ellipticity dependence of high-order harmonic generation from aligned molecules.," Phys. Rev. Lett. **98**, 053002 (2007).

349. C. Vozzi, M. Negro, F. Calegari, G. Sansone, M. Nisoli, S. De Silvestri, and S. Stagira, "Generalized molecular orbital tomography," Nat. Phys. **7**, 822–826 (2011).

350. D. Shafir, Y. Mairesse, D. M. Villeneuve, P. B. Corkum, and N. Dudovich, "Atomic wavefunctions probed through strong-field light–matter interaction," Nat. Phys. **5**, 412–416 (2009).

351. X. Xie, A. Scrinzi, M. Wickenhauser, A. Baltuška, I. Barth, and M. Kitzler, "Internal Momentum State Mapping Using High Harmonic Radiation," Phys. Rev. Lett. **101**, 033901 (2008).

352. S. Eckart, M. Kunitski, M. Richter, A. Hartung, J. Rist, F. Trinter, K. Fehre, N. Schlott, K. Henrichs, L. P. H. Schmidt, T. Jahnke, M. Schöffler, K. Liu, I. Barth, J. Kaushal, F. Morales, M. Ivanov, O. Smirnova, and R. Dörner, "Ultrafast preparation and detection of ring currents in single atoms," Nat. Phys. **14**, 701–704 (2018).

353. L. Barreau, K. Veyrinas, V. Gruson, S. J. Weber, T. Auguste, J.-F. Hergott, F. Lepetit, B. Carré, J.-C. Houver, D. Dowek, and P. Salières, "Evidence of depolarization and ellipticity of high harmonics driven by ultrashort bichromatic circularly polarized fields," Nat. Commun. **9**, 4727 (2018).

354. S. Yue, S. Brennecke, H. Du, and M. Lein, "Probing dynamical symmetries by bicircular high-order harmonic spectroscopy beyond the Born-Oppenheimer approximation," Phys. Rev. A **101**, 53438 (2020).

355. L. He, Q. Zhang, P. Lan, W. Cao, X. Zhu, C. Zhai, F. Wang, W. Shi, M. Li, X.-B. Bian, P. Lu, and A. D. Bandrauk, "Monitoring ultrafast vibrational dynamics of isotopic molecules with frequency modulation of high-order harmonics," Nat. Commun. **9**, 1108 (2018).

356. K. A. Hamer, F. Mauger, A. S. Folorunso, K. Lopata, R. R. Jones, L. F. DiMauro, K. J. Schafer, and M. B. Gaarde, "Characterizing particle-like charge-migration dynamics with high-order harmonic sideband spectroscopy," Phys. Rev. A **106**, 13103 (2022).

357. E. Ragonis, E. Ben-Arosh, L. Merensky, and A. Fleischer, "Controlling the bandwidth of high harmonic emission peaks with the spectral polarization of the driver," Opt. Lett. **49**, 2741–2744 (2024).

358. D. Shafir, H. Soifer, B. D. Bruner, M. Dagan, Y. Mairesse, S. Patchkovskii, M. Y. Ivanov, O. Smirnova, and N. Dudovich, "Resolving the time when an electron exits a tunnelling barrier," Nature **485**, 343–346 (2012).

359. X.-B. Bian and A. D. Bandrauk, "Spectral Shifts of Nonadiabatic High-Order Harmonic Generation," Appl. Sci. **3**, 267–277 (2013).

360. I. Barth, J. Manz, Y. Shigeta, and K. Yagi, "Unidirectional Electronic Ring Current Driven by a Few Cycle Circularly Polarized Laser Pulse: Quantum Model Simulations for Mg - Porphyrin," J. Am. Chem. Soc. **128**, 7043–7049 (2006).

361. I. Barth and O. Smirnova, "Nonadiabatic tunneling in circularly polarized laser fields: Physical picture and calculations," Phys. Rev. A **84**, 063415 (2011).

362. A. de las Heras, F. P. Bonafé, C. Hernández-García, A. Rubio, and O. Neufeld, "Tunable Tesla-Scale Magnetic Attosecond Pulses through Ring-Current Gating," J. Phys. Chem. Lett. **14**, 11160–11167 (2023).

363. O. Neufeld and O. Cohen, "Background-Free Measurement of Ring Currents by Symmetry Breaking High-Harmonic Spectroscopy," Phys. Rev. Lett. **123**, 103202 (2019).

364. L. S. Cederbaum and J. Zobeley, "Ultrafast charge migration by electron correlation," Chem. Phys. Lett. **307**, 205–210 (1999).

365. A. I. Kuleff, S. Lünnemann, and L. S. Cederbaum, "Electron-correlation-driven charge migration in oligopeptides," Chem. Phys. **414**, 100–105 (2013).

366. F. Calegari, D. Ayuso, A. Trabattoni, L. Belshaw, S. De Camillis, S. Anumula, F. Frassetto, L. Poletto, A. Palacios, P. Decleva, J. B. Greenwood, F. Mart\\in, and M. Nisoli, "Ultrafast electron dynamics in phenylalanine initiated by attosecond pulses," Science **346**, 336–339 (2014).

367. F. Calegari, A. Trabattoni, A. Palacios, D. Ayuso, M. C. Castrovilli, J. B. Greenwood, P. Decleva, F. Martín, and M. Nisoli, "Charge migration induced by attosecond pulses in bio-relevant molecules," J. Phys. B At. Mol. Opt. Phys. **49**, 142001 (2016).

368. P. M. Kraus, B. Mignolet, D. Baykusheva, A. Rupenyan, L. Horný, E. F. Penka, G. Grassi, O. I. Tolstikhin, J. Schneider, F. Jensen, L. B. Madsen, A. D. Bandrauk, F. Remacle, H. J. Wörner, L. Horný, E. F. Penka, G. Grassi, O. I. Tolstikhin, J. Schneider, F. Jensen, L. B. Madsen, A. D. Bandrauk, F. Remacle, and H. J. Wörner, "Measurement and laser control of attosecond charge migration in ionized iodoacetylene," Science **350**, 790–795 (2015).

369. D. T. Matselyukh, V. Despré, N. V Golubev, A. I. Kuleff, and H. J. Wörner, "Decoherence and revival in attosecond charge migration driven by non-adiabatic dynamics," Nat. Phys. **18**, 1206–1213 (2022).

370. F. Mauger, A. S. Folorunso, K. A. Hamer, C. Chandre, M. B. Gaarde, K. Lopata, and K. J. Schafer, "Charge migration and attosecond solitons in conjugated organic molecules," Phys. Rev. Res. **4**, 13073 (2022).

371. L. He, S. Sun, P. Lan, Y. He, B. Wang, P. Wang, X. Zhu, L. Li, W. Cao, P. Lu, and C. D. Lin, "Filming movies of attosecond charge migration in single molecules with high harmonic spectroscopy," Nat. Commun. **13**, 4595 (2022).

372. M. Lara-Astiaso, D. Ayuso, I. Tavernelli, P. Decleva, A. Palacios, and F. Martin, "Decoherence, control and attosecond probing of XUV-induced charge migration in biomolecules. A theoretical outlook," Faraday Discuss. **194**, 41–59 (2016).

373. K. A. Hamer, A. S. Folorunso, K. Lopata, K. J. Schafer, M. B. Gaarde, and F. Mauger, "Tracking Charge Migration with Frequency-Matched Strobo-Spectroscopy," J. Phys. Chem. A **128**, 20–27 (2024).

374. E. I. Kiselev, M. S. Rudner, and N. H. Lindner, "Inducing exceptional points, enhancing plasmon quality and creating correlated plasmon states with modulated Floquet parametric driving," Nat. Commun. **15**, 9914 (2024).

375. Y. Mairesse, J. Levesque, N. Dudovich, P. B. Corkum, and D. M. Villeneuve, "High harmonic generation from aligned molecules–amplitude and polarization," J. Mod. Opt. **55**, 2591–2602 (2008).

376. H. J. Wörner, J. B. Bertrand, P. Hockett, P. B. Corkum, and D. M. Villeneuve, "Controlling the Interference of Multiple Molecular Orbitals in High-Harmonic Generation," Phys. Rev. Lett. **104**, 233904 (2010).

377. O. Neufeld, N. Tancogne-Dejean, and A. Rubio, "Benchmarking Functionals for Strong-Field Light-Matter Interactions in Adiabatic Time-Dependent Density Functional Theory," J. Phys. Chem. Lett. **7254**–7264 (2024).

378. N. Rosenthal and G. Marcus, "Discriminating between the Role of Phase Matching and that of the Single-Atom Response in Resonance Plasma-Plume High-Order Harmonic Generation," Phys. Rev. Lett. **115**, 133901 (2015).

379. M. A. Fareed, V. V Strelkov, M. Singh, N. Thiré, S. Mondal, B. E. Schmidt, F. Légaré, and T. Ozaki, "Harmonic Generation from Neutral Manganese Atoms in the Vicinity of the Giant Autoionization Resonance," Phys. Rev. Lett. **121**, 23201 (2018).

380. A. S. Kheifets, "The attoclock and the tunneling time debate," J. Phys. B At. Mol. Opt. Phys. **53**, 72001 (2020).

381. U. S. Sainadh, H. Xu, X. Wang, A. Atia-Tul-Noor, W. C. Wallace, N. Douguet, A. Bray, I. Ivanov, K. Bartschat, A. Kheifets, R. T. Sang, and I. V Litvinyuk, "Attosecond angular streaking and tunnelling time in atomic hydrogen," Nature **568**, 75–77 (2019).

382. V. P. Majety and A. Scrinzi, "Absence of electron correlation effects in the Helium attoclock setting," J. Mod. Opt. **64**, 1026–1030 (2017).

383. L. Torlina, F. Morales, J. Kaushal, I. Ivanov, A. Kheifets, A. Zielinski, A. Scrinzi, H. G. Muller, S. Sukiasyan, M. Ivanov, and O. Smirnova, "Interpreting attoclock measurements of tunnelling times," Nat. Phys. **11**, 503–508 (2015).

384. A. N. Pfeiffer, C. Cirelli, M. Smolarski, D. Dimitrovski, M. Abu-samha, L. B. Madsen, and U. Keller, "Attoclock reveals natural coordinates of the laser-induced tunnelling current flow in atoms," Nat. Phys. **8**, 76–80 (2011).

385. M. W. Walser, C. H. Keitel, A. Scrinzi, and T. Brabec, "High Harmonic Generation Beyond the Electric Dipole Approximation," Phys. Rev. Lett. **85**, 5082–5085 (2000).

386. A. D. Bandrauk and H. Lu, "Controlling harmonic generation in molecules with intense laser and static magnetic fields: Orientation effects," Phys. Rev. A **68**, 043408 (2003).

387. R. Martín-Hernández, H. Hu, A. Baltuska, L. Plaja, and C. Hernández-García, "Fourier-Limited Attosecond Pulse from High Harmonic Generation Assisted by Ultrafast Magnetic Fields," Ultrafast Sci. **3**, 36 (2024).

388. A. Ludwig, J. Maurer, B. W. Mayer, C. R. Phillips, L. Gallmann, and U. Keller, "Breakdown of the Dipole Approximation in Strong-Field Ionization," Phys. Rev. Lett. **113**, 243001 (2014).

389. A. S. Maxwell and L. B. Madsen, "Relativistic and spin-orbit dynamics at nonrelativistic intensities in strong-field ionization," Phys. Rev. A **110**, 33108 (2024).

390. M. C. Suster, J. Derlikiewicz, K. Krajewska, F. C. Vélez, and J. Z. Kamiński, "Nondipole signatures in ionization and high-order harmonic generation," Phys. Rev. A **107**, 53112 (2023).

391. A. Hartung, S. Brennecke, K. Lin, D. Trabert, K. Fehre, J. Rist, M. S. Schöffler, T. Jahnke, L. P. H. Schmidt, M. Kunitski, M. Lein, R. Dörner, and S. Eckart, "Electric Nondipole Effect in Strong-Field Ionization," Phys. Rev. Lett. **126**, 53202 (2021).

392. M. Schmidt, N. Melzer, M. Kircher, G. Kastirke, A. Pier, L. Kaiser, P. Daum, D. Tsitsonis, M. Astaschov, J.

Rist, N. Anders, P. Roth, K. Lin, J. Drnec, F. Trinter, M. S. Schöffler, L. P. H. Schmidt, N. M. Novikovskiy, P. V. Demekhin, T. Jahnke, and R. Dörner, "Role of the Binding Energy on Nondipole Effects in Single-Photon Ionization," Phys. Rev. Lett. **132**, 233002 (2024).

393. O. Neufeld and O. Cohen, "Probing ultrafast electron correlations in high harmonic generation," Phys. Rev. Res. **2**, 033037 (2020).

394. E. Bordo, O. Neufeld, O. Kfir, A. Fleischer, and O. Cohen, "Spectroscopy of atomic orbital sizes using bi-elliptical high-order harmonic generation," Phys. Rev. A **100**, 043419 (2019).

395. S. Zayko, O. Kfir, E. Bordo, A. Fleischer, O. Cohen, and C. Ropers, "A dynamical symmetry triad in high-harmonic generation revealed by attosecond recollision control," New J. Phys. **22**, 53017 (2020).

396. D. Habibović and D. B. Milošević, "Ellipticity of High-Order Harmonics Generated by Aligned Homonuclear Diatomic Molecules Exposed to an Orthogonal Two-Color Laser Field," Photonics **7**, (2020).

397. D. Habibović, W. Becker, and D. B. Milošević, "High-order harmonic generation by two linearly polarized laser fields with an arbitrary angle between their polarization axes," Phys. Rev. A **106**, 23119 (2022).

398. S. Long, W. Becker, and J. K. McIver, "Model calculations of polarization-dependent two-color high-harmonic generation," Phys. Rev. A **52**, 2262–2278 (1995).

399. A. Fleischer, A. K. Gupta, and N. Moiseyev, "Dynamical symmetry analysis of ionization and harmonic generation of atoms in bichromatic laser pulses," Int. J. Quantum Chem. **103**, 824–840 (2005).

400. S. Ghimire, A. D. Dichiara, E. Sistrunk, P. Agostini, L. F. DiMauro, and D. A. Reis, "Observation of high-order harmonic generation in a bulk crystal," Nat. Phys. **7**, 138–141 (2011).

401. X.-S. Kong, H. Liang, X.-Y. Wu, and L.-Y. Peng, "Symmetry analyses of high-order harmonic generation in monolayer hexagonal boron nitride," J. Phys. B At. Mol. Opt. Phys. **54**, 124004 (2021).

402. L. Yue and M. B. Gaarde, "Imperfect Recollisions in High-Harmonic Generation in Solids," Phys. Rev. Lett. **124**, 153204 (2020).

403. G. Le Breton, A. Rubio, and N. Tancogne-Dejean, "High-harmonic generation from few-layer hexagonal boron nitride: Evolution from monolayer to bulk response," Phys. Rev. B **98**, 165308 (2018).

404. L. Yue, R. Hollinger, C. B. Uzundal, B. Nebgen, Z. Gan, E. Najafidehaghani, A. George, C. Spielmann, D. Kartashov, A. Turchanin, D. Y. Qiu, M. B. Gaarde, and M. Zuerch, "Signatures of Multiband Effects in High-Harmonic Generation in Monolayer MoS2," Phys. Rev. Lett. **129**, 147401 (2022).

405. P. B. Corkum, "Plasma perspective on strong field multiphoton ionization," Phys. Rev. Lett. **71**, 1994–1997 (1993).

406. M. Lewenstein, P. Balcou, M. Y. Ivanov, A. L'Huillier, and P. B. Corkum, "Theory of high-harmonic generation by low-frequency laser fields," Phys. Rev. A **49**, 2117–2132 (1994).

407. Y. Sanari, T. Otobe, Y. Kanemitsu, and H. Hirori, "Modifying angular and polarization selection rules of high-order harmonics by controlling electron trajectories in k-space," Nat. Commun. **11**, 3069 (2020).

408. G. Vampa, C. R. McDonald, G. Orlando, D. D. Klug, P. B. Corkum, and T. Brabec, "Theoretical Analysis of High-Harmonic Generation in Solids," Phys. Rev. Lett. **113**, 073901 (2014).

409. L. Li, P. Lan, X. Zhu, and P. Lu, "Huygens-Fresnel Picture for High Harmonic Generation in Solids," Phys. Rev. Lett. **127**, 223201 (2021).

410. E. Goulielmakis and T. Brabec, "High harmonic generation in condensed matter," Nat. Photonics **16**, 411–421 (2022).

411. A. J. Uzan, G. Orenstein, Á. Jiménez-Galán, C. McDonald, R. E. F. Silva, B. D. Bruner, N. D. Klimkin, V. Blanchet, T. Arusi-Parpar, M. Krüger, A. N. Rubtsov, O. Smirnova, M. Ivanov, B. Yan, T. Brabec, and N. Dudovich, "Attosecond spectral singularities in solid-state high-harmonic generation," Nat. Photonics **14**, 183–187 (2020).

412. Y. S. You, D. A. A. Reis, and S. Ghimire, "Anisotropic high-harmonic generation in bulk crystals," Nat. Phys. **13**, 345–349 (2017).

413. E. N. Osika, A. Chacón, L. Ortmann, N. Suárez, J. A. Pérez-Hernández, B. Szafran, M. F. Ciappina, F. Sols, A. S. Landsman, and M. Lewenstein, "Wannier-bloch approach to localization in high-harmonics generation in solids," Phys. Rev. X **7**, 1–14 (2017).

414. O. Schubert, M. Hohenleutner, F. Langer, B. Urbanek, C. Lange, U. Huttner, D. Golde, T. Meier, M. Kira, S. W. Koch, and R. Huber, "Sub-cycle control of terahertz high-harmonic generation by dynamical Bloch oscillations," Nat. Photonics **8**, 119–123 (2014).

415. N. Tancogne-Dejean, O. D. Mücke, F. X. Kärtner, and A. Rubio, "Ellipticity dependence of high-harmonic generation in solids originating from coupled intraband and interband dynamics," Nat. Commun. **8**, 745 (2017).

416. N. Yoshikawa, K. Nagai, K. Uchida, Y. Takaguchi, S. Sasaki, Y. Miyata, and K. Tanaka, "Interband resonant high-harmonic generation by valley polarized electron–hole pairs," Nat. Commun. **10**, 3709 (2019).

417. G. Vampa, T. J. Hammond, N. Thiré, B. E. Schmidt, F. Légaré, C. R. McDonald, T. Brabec, D. D. Klug, and P. B. Corkum, "All-Optical Reconstruction of Crystal Band Structure," Phys. Rev. Lett. **115**, 193603 (2015).

418. F. Langer, M. Hohenleutner, C. P. Schmid, C. Poellmann, P. Nagler, T. Korn, C. Schüller, M. S. Sherwin, U. Huttner, J. T. Steiner, S. W. Koch, M. Kira, and R. Huber, "Lightwave-driven quasiparticle collisions on a subcycle timescale," Nature **533**, 225–229 (2016).

419. F. Langer, M. Hohenleutner, U. Huttner, S. W. Koch, M. Kira, and R. Huber, "Symmetry-controlled temporal structure of high-harmonic carrier fields from a bulk crystal," Nat. Photonics **11**, 227–231 (2017).

420. S. Han, L. Ortmann, H. Kim, Y. W. Kim, T. Oka, A. Chacon, B. Doran, M. Ciappina, M. Lewenstein, S.-W. Kim, S. Kim, and A. S. Landsman, "Extraction of higher-order nonlinear electronic response in solids using high harmonic generation," Nat. Commun. **10**, 3272 (2019).

421. Y. Kim, M. J. Kim, S. Cha, S. Choi, C.-J. Kim, B. J. Kim, M.-H. Jo, J. Kim, and J. Lee, "Dephasing Dynamics Accessed by High Harmonic Generation: Determination of Electron–Hole Decoherence of Dirac Fermions," Nano Lett. **24**, 1277−1283 (2024).

422. M. Kim, T. Kim, A. Galler, D. Kim, A. Chacon, X. Gong, Y. Yang, R. Fang, K. Watanabe, T. Taniguchi, B. J. Kim, S. H. Chae, M.-H. Jo, A. Rubio, O. Neufeld, and J. Kim, "Quantum interference and occupation control in high harmonic generation from monolayer WS2," arXiv Prepr. arXiv2503.04335 (2025).

423. P. Suthar, F. Trojánek, P. Malý, T. J.-Y. Derrien, and M. Kozák, "Role of Van Hove singularities and effective mass anisotropy in polarization-resolved high harmonic spectroscopy of silicon," Commun. Phys. **5**, 288 (2022).

424. C. Heide, Y. Kobayashi, D. R. Baykusheva, D. Jain, J. A. Sobota, M. Hashimoto, P. S. Kirchmann, S. Oh, T. F. Heinz, D. A. Reis, and S. Ghimire, "Probing topological phase transitions using high-harmonic generation," Nat. Photonics **16**, 620–624 (2022).

425. O. Neufeld, N. Tancogne-Dejean, H. Hübener, U. De Giovannini, and A. Rubio, "Are there universal signatures of topological phases in high harmonic generation? Probably not.," Phys. Rev. X **13**, 031011 (2023).

426. K. Nagai, T. Okamoto, Y. Shinohara, H. Sanada, and K. Oguri, "High-harmonic spin-orbit angular momentum generation in crystalline solids preserving multiscale dynamical symmetry," Sci. Adv. **10**, eado7315 (2021).

427. R. E. F. Silva, Á. Jiménez-Galán, B. Amorim, O. Smirnova, and M. Ivanov, "Topological strong-field physics on sub-laser-cycle timescale," Nat. Photonics **13**, 849–854 (2019).

428. A. Chacón, D. Kim, W. Zhu, S. P. Kelly, A. Dauphin, E. Pisanty, A. S. Maxwell, A. Picón, M. F. Ciappina, D. E. Kim, C. Ticknor, A. Saxena, and M. Lewenstein, "Circular dichroism in higher-order harmonic generation: Heralding topological phases and transitions in Chern insulators," Phys. Rev. B **102**, 134115 (2020).

429. C. Jürß and D. Bauer, "Helicity flip of high-order harmonic photons in Haldane nanoribbons," Phys. Rev. A **102**, 43105 (2020).

430. D. Baykusheva, A. Chacón, D. Kim, D. E. Kim, D. A. Reis, and S. Ghimire, "Strong-field physics in three-dimensional topological insulators," Phys. Rev. A **103**, 23101 (2021).

431. D. Baykusheva, A. Chacón, J. Lu, T. P. Bailey, J. A. Sobota, H. Soifer, P. S. Kirchmann, C. Rotundu, C. Uher, T. F. Heinz, D. A. Reis, and S. Ghimire, "All-Optical Probe of Three-Dimensional Topological Insulators Based on High-Harmonic Generation by Circularly Polarized Laser Fields," Nano Lett. **21**, 8970–8978 (2021).

432. C. P. Schmid, L. Weigl, P. Grössing, V. Junk, C. Gorini, S. Schlauderer, S. Ito, M. Meierhofer, N. Hofmann, D. Afanasiev, J. Crewse, K. A. Kokh, O. E. Tereshchenko, J. Güdde, F. Evers, J. Wilhelm, K. Richter, U. Höfer, and R. Huber, "Tunable non-integer high-harmonic generation in a topological insulator," Nature **593**, 385–390 (2021).

433. Y. Bai, F. Fei, S. Wang, N. Li, X. Li, F. Song, R. Li, Z. Xu, and P. Liu, "High-harmonic generation from topological surface states," Nat. Phys. **17**, 311–315 (2021).

434. C. Qian, C. Yu, S. Jiang, T. Zhang, J. Gao, S. Shi, H. Pi, H. Weng, and R. Lu, "Role of Shift Vector in High-Harmonic Generation from Noncentrosymmetric Topological Insulators under Strong Laser Fields," Phys. Rev. X **12**, 21030 (2022).

435. D. Bauer and K. K. Hansen, "High-Harmonic Generation in Solids with and without Topological Edge States," Phys. Rev. Lett. **120**, 177401 (2018).

436. A. P. Schnyder, S. Ryu, A. Furusaki, and A. W. W. Ludwig, "Classification of topological insulators and superconductors in three spatial dimensions," Phys. Rev. B **78**, 195125 (2008).

437. C.-K. Chiu, J. C. Y. Teo, A. P. Schnyder, and S. Ryu, "Classification of topological quantum matter with symmetries," Rev. Mod. Phys. **88**, 35005 (2016).

438. M. G. Vergniory, L. Elcoro, C. Felser, N. Regnault, B. A. Bernevig, and Z. Wang, "A complete catalogue of high-quality topological materials," Nature **566**, 480–485 (2019).

439. B. Bradlyn, L. Elcoro, J. Cano, M. G. Vergniory, Z. Wang, C. Felser, M. I. Aroyo, and B. A. Bernevig, "Topological quantum chemistry," Nature **547**, 298–305 (2017).

440. R. Qin and Z.-Y. Chen, "Angle-dependent high-order harmonic generation in a topological phase transition of monolayer black phosphorous," Phys. Rev. A **109**, 43102 (2024).

441. Y.-Y. Lv, J. Xu, S. Han, C. Zhang, Y. Han, J. Zhou, S.-H. Yao, X.-P. Liu, M.-H. Lu, H. Weng, Z. Xie, Y. B. Chen, J. Hu, Y.-F. Chen, and S. Zhu, "High-harmonic generation in Weyl semimetal β-WP2 crystals," Nat. Commun. **12**, 6437 (2021).

442. L. Yue and M. B. Gaarde, "Characterizing Anomalous High-Harmonic Generation in Solids," Phys. Rev. Lett. **130**, 166903 (2022).

443. M. R. Bionta, E. Haddad, A. Leblanc, V. Gruson, P. Lassonde, H. Ibrahim, J. Chaillou, N. Émond, M. R. Otto, Á. Jiménez-Galán, R. E. F. Silva, M. Ivanov, B. J. Siwick, M. Chaker, and F. Légaré, "Tracking ultrafast solid-state dynamics using high harmonic spectroscopy," Phys. Rev. Res. **3**, 023250 (2021).

444. N. Rana and G. Dixit, "Probing phonon-driven symmetry alterations in graphene via high-order-harmonic spectroscopy," Phys. Rev. A **106**, 53116 (2022).

445. J. Zhang, Z. Wang, F. Lengers, D. Wigger, D. E. Reiter, T. Kuhn, H. J. Wörner, and T. T. Luu, "High-harmonic spectroscopy probes lattice dynamics," Nat. Photonics (2024).

446. J. Zhang, O. Neufeld, N. Tancogne-Dejean, I.-T. Lu, H. Hübener, U. De Giovannini, and A. Rubio, "Enhanced high harmonic efficiency through phonon-assisted photodoping effect," npj Comput. Mater. **10**, 202 (2024).

447. T. Weitz, D. Lesko, S. Wittigschlager, W. Li, C. Heide, O. Neufeld, and P. Hommelhoff, "Lightwave-driven electrons in a Floquet topological insulator," arXiv Prepr. arXiv2407.17917 (2024).

448. S. Takayoshi, Y. Murakami, and P. Werner, "High-harmonic generation in quantum spin systems," Phys. Rev. B **99**, 184303 (2019).

449. D. A. Bas, K. Vargas-Velez, S. Babakiray, T. A. Johnson, P. Borisov, T. D. Stanescu, D. Lederman, and A. D. Bristow, "Coherent control of injection currents in high-quality films of Bi2Se3," Appl. Phys. Lett. **106**, 41109 (2015).

450. T. Oka and H. Aoki, "Photovoltaic Hall effect in graphene," Phys. Rev. B **79**, 81406 (2009).

451. N. H. Lindner, G. Refael, and V. Galitski, "Floquet Topological Insulator in Semiconductor Quantum Wells," Nat. Phys. **7**, 490–495 (2010).

452. W. Y. H., S. H., J.-H. P., G. N., Y. H. Wang, H. Steinberg, P. Jarillo-Herrero, and N. Gedik, "Observation of Floquet-Bloch States on the Surface of a Topological Insulator," Science **342**, 453 LP – 457 (2013).

453. S. S. S. Zhou, C. Bao, B. Fan, H. Zhou, Q. Gao, H. Zhong, T. Lin, H. Liu, P. Yu, P. Tang, S. Meng, W. Duan, and S. S. S. Zhou, "Pseudospin-selective Floquet band engineering in black phosphorus," Nature **614**, 75–80 (2023).

454. M. Merboldt, M. Schüler, D. Schmitt, J. P. Bange, W. Bennecke, K. Gadge, K. Pierz, H. W. Schumacher, D. Momeni, and D. Steil, "Observation of Floquet states in graphene," arXiv Prepr. arXiv2404.12791 (2024).

455. D. Choi, M. Mogi, U. De Giovannini, D. Azoury, B. Lv, Y. Su, H. Hübener, A. Rubio, and N. Gedik, "Direct observation of Floquet-Bloch states in monolayer graphene," arXiv Prepr. arXiv2404.14392 (2024).

456. J. W. McIver, B. Schulte, F.-U. Stein, T. Matsuyama, G. Jotzu, G. Meier, and A. Cavalleri, "Light-induced anomalous Hall effect in graphene," Nat. Phys. **16**, 38–41 (2020).

457. G. Jotzu, M. Messer, R. Desbuquois, M. Lebrat, T. Uehlinger, D. Greif, and T. Esslinger, "Experimental realization of the topological Haldane model with ultracold fermions," Nature **515**, 237–240 (2014).

458. M. S. Rudner and N. H. Lindner, "Band structure engineering and non-equilibrium dynamics in Floquet topological insulators," Nat. Rev. Phys. **2**, 229–244 (2020).

459. A. Galler, A. Rubio, and O. Neufeld, "Mapping light-dressed Floquet bands by highly nonlinear optical excitations and valley polarization," J. Phys. Chem. Lett. **14**, 11298–11304 (2023).

460. O. Neufeld, H. Hübener, G. Jotzu, U. De Giovannini, and A. Rubio, "Band nonlinearity-enabled manipulation of Dirac nodes, Weyl cones, and valleytronics with intense linearly polarized light," Nano Lett. **23**, 7568–7575 (2023).

461. M. Schüler and M. A. Sentef, "Theory of subcycle time-resolved photoemission: Application to terahertz photodressing in graphene," J. Electron Spectros. Relat. Phenomena **253**, 147121 (2021).

462. I. Martin, G. Refael, and B. Halperin, "Topological Frequency Conversion in Strongly Driven Quantum Systems," Phys. Rev. X **7**, 41008 (2017).

463. F. Nathan, I. Martin, and G. Refael, "Topological frequency conversion in Weyl semimetals," Phys. Rev. Res. **4**, 43060 (2022).

464. M. S. Mrudul, Á. Jiménez-Galán, M. Ivanov, and G. Dixit, "Light-induced valleytronics in pristine graphene," Optica **8**, 422–427 (2021).

465. M. W. Day, K. Kusyak, F. Sturm, J. I. Aranzadi, H. M. Bretscher, M. Fechner, T. Matsuyama, M. H. Michael, B. F. Schulte, and X. Li, "Nonperturbative Nonlinear Transport in a Floquet-Weyl Semimetal," arXiv Prepr. arXiv2409.04531 (2024).

466. A. Castro, U. De Giovannini, S. A. Sato, H. Hübener, and A. Rubio, "Floquet engineering the band structure of materials with optimal control theory," Phys. Rev. Res. **4**, 33213 (2022).

467. O. Neufeld, "Degree of Time-Reversal and Dynamical Symmetry Breaking in Electromagnetic Fields and Its Connection to Floquet Engineering," ACS Photonics (2025).

468. U. Bhattacharya, S. Chaudhary, T. Grass, A. S. Johnson, S. Wall, and M. Lewenstein, "Fermionic Chern insulator from twisted light with linear polarization," Phys. Rev. B **105**, L081406 (2022).

469. Y. Kobayashi, C. Heide, A. C. Johnson, V. Tiwari, F. Liu, D. A. Reis, T. F. Heinz, and S. Ghimire, "Floquet engineering of strongly driven excitons in monolayer tungsten disulfide," Nat. Phys. **19**, 171–176 (2023).

470. F. Maderspacher, "Snail Chirality: The Unwinding," Curr. Biol. **26**, R215–R217 (2016).

471. M. Kenji, "Bioactive natural products and chirality," Chirality **23**, 449–462 (2011).

472. U. J. Meierhenrich, "Amino Acids and the Asymmetry of Life," Eur. Rev. **21**, 190–199 (2013).

473. W. H. Brooks, W. C. Guida, K. G. Daniel, W. H Brooks, W. C Guida, and K. G Daniel, "The Significance of Chirality in Drug Design and Development," Curr. Top. Med. Chem. **11**, 760–770 (2011).

474. J. E. Ridings, "The thalidomide disaster, lessons from the past.," Methods Mol. Biol. **947**, 575–586 (2013).

475. H. X. Ngo and S. Garneau-Tsodikova, "What are the drugs of the future?," Medchemcomm **9**, 757–758 (2018).

476. R. R. Julian, S. Myung, and D. E. Clemmer, "Do Homochiral Aggregates Have an Entropic Advantage?," J. Phys. Chem. B **109**, 440–444 (2005).

477. S. F. Ozturk, Z. Liu, J. D. Sutherland, and D. D. Sasselov, "Origin of biological homochirality by crystallization of an RNA precursor on a magnetic surface," Sci. Adv. **9**, eadg8274 (2023).

478. S. F. Ozturk, D. K. Bhowmick, Y. Kapon, Y. Sang, A. Kumar, Y. Paltiel, R. Naaman, and D. D. Sasselov, "Chirality-induced avalanche magnetization of magnetite by an RNA precursor," Nat. Commun. **14**, 6351 (2023).

479. S. F. Ozturk, D. D. Sasselov, and J. D. Sutherland, "The central dogma of biological homochirality: How does chiral information propagate in a prebiotic network?," J. Chem. Phys. **159**, 61102 (2023).

480. F. Evers, A. Aharony, N. Bar-Gill, O. Entin-Wohlman, P. Hedegård, O. Hod, P. Jelinek, G. Kamieniarz, M. Lemeshko, K. Michaeli, V. Mujica, R. Naaman, Y. Paltiel, S. Refaely-Abramson, O. Tal, J. Thijssen, M. Thoss,

J. M. van Ruitenbeek, L. Venkataraman, D. H. Waldeck, B. Yan, and L. Kronik, "Theory of Chirality Induced Spin Selectivity: Progress and Challenges," Adv. Mater. **34**, 2106629 (2022).

481. B. P. Bloom, Y. Paltiel, R. Naaman, and D. H. Waldeck, "Chiral Induced Spin Selectivity," Chem. Rev. **124**, 1950–1991 (2024).

482. D. Griffiths, *Introduction to Elementary Particles* (John Wiley & Sons, 2008).

483. M. Quack, "Structure and Dynamics of Chiral Molecules," Angew. Chemie Int. Ed. English **28**, 571–586 (1989).

484. M. Quack, J. Stohner, and M. Willeke, "High-Resolution Spectroscopic Studies and Theory of Parity Violation in Chiral Molecules," Annu. Rev. Phys. Chem. **59**, 741–769 (2008).

485. I. Erez, E. R. Wallach, and Y. Shagam, "Simultaneous Enantiomer-Resolved Ramsey Spectroscopy Scheme for Chiral Molecules," Phys. Rev. X **13**, 41025 (2023).

486. A. Landau, Eduardus, D. Behar, E. R. Wallach, L. F. Pašteka, S. Faraji, A. Borschevsky, and Y. Shagam, "Chiral molecule candidates for trapped ion spectroscopy by ab initio calculations: From state preparation to parity violation," J. Chem. Phys. **159**, 114307 (2023).

487. D. K. Kondepudi and D. J. Durand, "Chiral asymmetry in spiral galaxies?," Chirality **13**, 351–356 (2001).

488. J. Meyer-Ilse, D. Akimov, and B. Dietzek, "Recent advances in ultrafast time-resolved chirality measurements: perspective and outlook," Laser Photon. Rev. **7**, 495–505 (2013).

489. G. J. Simpson, "Molecular Origins of the Remarkable Chiral Sensitivity of Second-Order Nonlinear Optics," ChemPhysChem **5**, 1301–1310 (2004).

490. T. Radhakrishnan, "The dispersion, briefringence and optical activity of quartz," in *Proceedings of the Indian Academy of Sciences-Section A* (Springer India New Delhi, 1947), Vol. 25, p. 260.

491. P. J. Stephens, "Theory of Magnetic Circular Dichroism," J. Chem. Phys. **52**, 3489–3516 (2003).

492. J. R. Rouxel and S. Mukamel, "Molecular Chirality and Its Monitoring by Ultrafast X-ray Pulses," Chem. Rev. **122**, 16802–16838 (2022).

493. A. Guerrero-Martínez, J. L. Alonso-Gómez, B. Auguié, M. M. Cid, and L. M. Liz-Marzán, "From individual to collective chirality in metal nanoparticles," Nano Today **6**, 381–400 (2011).

494. Z. Fan and A. O. Govorov, "Helical Metal Nanoparticle Assemblies with Defects: Plasmonic Chirality and Circular Dichroism," J. Phys. Chem. C **115**, 13254–13261 (2011).

495. T. A. Keiderling, "Vibrational Circular Dichroism," Appl. Spectrosc. Rev. **17**, 189–226 (1981).

496. H. Rhee, Y.-G. June, J.-S. Lee, K.-K. Lee, J.-H. Ha, Z. H. Kim, S.-J. Jeon, and M. Cho, "Femtosecond characterization of vibrational optical activity of chiral molecules," Nature **458**, 310–313 (2009).

497. H. Hübener, U. De Giovannini, C. Schäfer, J. Andberger, M. Ruggenthaler, J. Faist, and A. Rubio, "Engineering quantum materials with chiral optical cavities," Nat. Mater. **20**, 438–442 (2021).

498. P. Scott, X. Garcia-Santiago, D. Beutel, C. Rockstuhl, M. Wegener, and I. Fernandez-Corbaton, "On enhanced sensing of chiral molecules in optical cavities," Appl. Phys. Rev. **7**, 41413 (2020).

499. C. Schäfer and D. G. Baranov, "Chiral Polaritonics: Analytical Solutions, Intuition, and Use," J. Phys. Chem. Lett. **14**, 3777–3784 (2023).

500. R. R. Riso, L. Grazioli, E. Ronca, T. Giovannini, and H. Koch, "Strong Coupling in Chiral Cavities: Nonperturbative Framework for Enantiomer Discrimination," Phys. Rev. X **13**, 31002 (2023).

501. I. Tutunnikov, L. Xu, R. W. Field, K. A. Nelson, Y. Prior, and I. S. Averbukh, "Enantioselective orientation of chiral molecules induced by terahertz pulses with twisted polarization," Phys. Rev. Res. **3**, 13249 (2021).

502. I. Tutunnikov, E. Gershnabel, S. Gold, and I. S. Averbukh, "Selective Orientation of Chiral Molecules by Laser Fields with Twisted Polarization," J. Phys. Chem. Lett. **9**, 1105–1111 (2018).

503. I. Tutunnikov, J. Floβ, E. Gershnabel, P. Brumer, I. S. Averbukh, A. A. Milner, and V. Milner, "Observation of persistent orientation of chiral molecules by a laser field with twisted polarization," Phys. Rev. A **101**, 21403 (2020).

504. R. Cireasa, A. E. Boguslavskiy, B. Pons, M. C. H. Wong, D. Descamps, S. Petit, H. Ruf, N. Thiré, A. Ferré, J. Suarez, J. Higuet, B. E. Schmidt, A. F. Alharbi, F. Légaré, V. Blanchet, B. Fabre, S. Patchkovskii, O. Smirnova, Y. Mairesse, V. R. Bhardwaj, N. Thire, A. Ferre, J. Suarez, J. Higuet, B. E. Schmidt, A. F. Alharbi, F. Legare, V. Blanchet, B. Fabre, S. Patchkovskii, O. Smirnova, Y. Mairesse, and V. R. Bhardwaj, "Probing molecular chirality on a sub-femtosecond timescale," Nat. Phys. **11**, 654–658 (2015).

505. O. Smirnova, Y. Mairesse, and S. Patchkovskii, "Opportunities for chiral discrimination using high harmonic generation in tailored laser fields," J. Phys. B At. Mol. Opt. Phys. **48**, 234005 (2015).

506. D. Ayuso, P. Decleva, S. Patchkovskii, and O. Smirnova, "Strong-field control and enhancement of chiral response in bi-elliptical high-order harmonic generation: an analytical model," J. Phys. B At. Mol. Opt. Phys. **51**, 124002 (2018).

507. D. Ayuso, P. Decleva, S. Patchkovskii, and O. Smirnova, "Chiral dichroism in bi-elliptical high-order harmonic generation," J. Phys. B At. Mol. Opt. Phys. **51**, 06LT01 (2018).

508. D. Baykusheva and H. J. Wörner, "Chiral Discrimination through Bielliptical High-Harmonic Spectroscopy," Phys. Rev. X **8**, 031060 (2018).

509. Y. Harada, E. Haraguchi, K. Kaneshima, and T. Sekikawa, "Circular dichroism in high-order harmonic generation from chiral molecules," Phys. Rev. A **98**, 021401 (2018).

510. D. Baykusheva, D. Zindel, V. Svoboda, E. Bommeli, M. Ochsner, A. Tehlar, and H. J. Wörner, "Real-time probing of chirality during a chemical reaction," Proc. Natl. Acad. Sci. **116**, 23923–23929 (2019).

511. J. R. Rouxel, B. Rösner, D. Karpov, C. Bacellar, G. F. Mancini, F. Zinna, D. Kinschel, O. Cannelli, M. Oppermann, C. Svetina, A. Diaz, J. Lacour, C. David, and M. Chergui, "Hard X-ray helical dichroism of disordered molecular media," Nat. Photonics **16**, 570–574 (2022).

512. W. Brullot, M. K. Vanbel, T. Swusten, and T. Verbiest, "Resolving enantiomers using the optical angular momentum of twisted light," Sci. Adv. **2**, e1501349 (2023).

513. K. A. Forbes and D. L. Andrews, "Optical orbital angular momentum: twisted light and chirality," Opt. Lett. **43**, 435–438 (2018).

514. D. Green and K. A. Forbes, "Optical chirality of vortex beams at the nanoscale," Nanoscale **15**, 540–552 (2023).

515. F. Ricci, W. Löffler, and M. P. van Exter, "Instability of higher-order optical vortices analyzed with a multi-pinhole interferometer," Opt. Express **20**, 22961–22975 (2012).

516. Y. Egorov, M. Bretsko, Y. Akimova, and A. Volyar, "Instability of the OAM of higher-order optical vortices," in *2021 International Conference on Information Technology and Nanotechnology (ITNT)* (2021), pp. 1–5.

517. L. Ye, J. R. Rouxel, S. Asban, B. Rösner, and S. Mukamel, "Probing Molecular Chirality by Orbital-Angular-Momentum-Carrying X-ray Pulses," J. Chem. Theory Comput. **15**, 4180–4186 (2019).

518. J.-L. Bégin, A. Jain, A. Parks, F. Hufnagel, P. Corkum, E. Karimi, T. Brabec, and R. Bhardwaj, "Nonlinear helical dichroism in chiral and achiral molecules," Nat. Photonics **17**, 82–88 (2023).

519. K. A. Forbes, "Raman Optical Activity Using Twisted Photons," Phys. Rev. Lett. **122**, 103201 (2019).

520. K. A. Forbes and D. L. Andrews, "Enhanced optical activity using the orbital angular momentum of structured light," Phys. Rev. Res. **1**, 33080 (2019).

521. N. Mayer, D. Ayuso, P. Decleva, M. Khokhlova, E. Pisanty, M. Ivanov, and O. Smirnova, "Chiral topological light for detecting robust enantio-sensitive observables," Nat. Photonics **18**, 1155–1160 (2024).

522. A. F. Ordonez and O. Smirnova, "Generalized perspective on chiral measurements without magnetic interactions," Phys. Rev. A **98**, 063428 (2018).

523. G. A. Garcia, L. Nahon, S. Daly, and I. Powis, "Vibrationally induced inversion of photoelectron forward-backward asymmetry in chiral molecule photoionization by circularly polarized light," **4**, 2132 EP- (2013).

524. I. Powis, "Photoelectron circular dichroism in chiral molecules," Adv. Chem. Phys. **138**, 267–330 (2008).

525. I. Dreissigacker and M. Lein, "Photoelectron circular dichroism of chiral molecules studied with a continuum-state-corrected strong-field approximation," Phys. Rev. A **89**, 053406 (2014).

526. C. Sparling and D. Townsend, "Two decades of imaging photoelectron circular dichroism: from first principles to future perspectives," Phys. Chem. Chem. Phys. (2025).

527. J Ullrich, R Moshammer, A Dorn, R Dörner, L Ph H Schmidt, and H Schmidt-Böcking, "Recoil-ion and electron momentum spectroscopy: reaction-microscopes," Reports Prog. Phys. **66**, 1463 (2003).

528. R. Dörner, V. Mergel, O. Jagutzki, L. Spielberger, J. Ullrich, R. Moshammer, and H. Schmidt-Böcking, "Cold Target Recoil Ion Momentum Spectroscopy: a 'momentum microscope' to view atomic collision dynamics," Phys. Rep. **330**, 95–192 (2000).

529. P. V Demekhin, "Photoelectron circular dichroism with Lissajous-type bichromatic fields: One-photon versus two-photon ionization of chiral molecules," Phys. Rev. A **99**, 63406 (2019).

530. C. S. Lehmann, N. B. Ram, I. Powis, and M. H. M. Janssen, "Imaging photoelectron circular dichroism of chiral molecules by femtosecond multiphoton coincidence detection," J. Chem. Phys. **139**, 234307 (2013).

531. B. Ritchie, "Theory of the angular distribution of photoelectrons ejected from optically active molecules and molecular negative ions," Phys. Rev. A **13**, 1411–1415 (1976).

532. I. Powis, "Photoelectron Spectroscopy and Circular Dichroism in Chiral Biomolecules: l-Alanine," J. Phys. Chem. A **104**, 878–882 (2000).

533. N. A. Cherepkov, "Circular dichroism of molecules in the continuous absorption region," Chem. Phys. Lett. **87**, 344–348 (1982).

534. S. Beaulieu, A. Comby, D. Descamps, S. Petit, F. Légaré, B. Fabre, V. Blanchet, and Y. Mairesse, "Multiphoton photoelectron circular dichroism of limonene with independent polarization state control of the bound-bound and bound-continuum transitions," J. Chem. Phys. **149**, 134301 (2018).

535. A. Kastner, T. Ring, H. Braun, A. Senftleben, and T. Baumert, "Observation of Photoelectron Circular Dichroism Using a Nanosecond Laser," ChemPhysChem **20**, 1416–1419 (2019).

536. J. Miles, D. Fernandes, A. Young, C. M. M. Bond, S. W. Crane, O. Ghafur, D. Townsend, J. Sá, and J. B. Greenwood, "A new technique for probing chirality via photoelectron circular dichroism," Anal. Chim. Acta **984**, 134–139 (2017).

537. S. B. and A. F. and R. G. and R. C. and D. D. and B. F. and N. F. and F. L. and S. P. and T. R. and V. B. and Y. M. and B. Pons, "Universality of photoelectron circular dichroism in the photoionization of chiral molecules," New J. Phys. **18**, 102002 (2016).

538. S. Beaulieu, A. Comby, B. Fabre, D. Descamps, A. Ferre, G. Garcia, R. Geneaux, F. Legare, L. Nahon, S. Petit, T. Ruchon, B. Pons, V. Blanchet, and Y. Mairesse, "Probing ultrafast dynamics of chiral molecules using time-resolved photoelectron circular dichroism," Faraday Discuss. **194**, 325–348 (2016).

539. A. Comby, S. Beaulieu, M. Boggio-Pasqua, D. Descamps, F. Légaré, L. Nahon, S. Petit, B. Pons, B. Fabre, Y. Mairesse, and V. Blanchet, "Relaxation Dynamics in Photoexcited Chiral Molecules Studied by Time-Resolved Photoelectron Circular Dichroism: Toward Chiral Femtochemistry," J. Phys. Chem. Lett. **7**, 4514–4519 (2016).

540. V. Blanchet, D. Descamps, S. Petit, Y. Mairesse, B. Pons, and B. Fabre, "Ultrafast relaxation investigated by photoelectron circular dichroism: an isomeric comparison of camphor and fenchone," Phys. Chem. Chem. Phys. **23**, 25612–25628 (2021).

541. S. Beaulieu, A. Comby, A. Clergerie, J. Caillat, D. Descamps, N. Dudovich, B. Fabre, R. Géneaux, F. Légaré, S. Petit, B. Pons, G. Porat, T. Ruchon, R. Taïeb, V. Blanchet, and Y. Mairesse, "Attosecond-resolved photoionization of chiral molecules," Science **358**, 1288–1294 (2017).

542. V. Wanie, E. Bloch, E. P. Månsson, L. Colaizzi, S. Ryabchuk, K. Saraswathula, A. F. Ordonez, D. Ayuso, O. Smirnova, A. Trabattoni, V. Blanchet, N. Ben Amor, M.-C. Heitz, Y. Mairesse, B. Pons, and F. Calegari, "Capturing electron-driven chiral dynamics in UV-excited molecules," Nature **630**, 109–115 (2024).

543. V. Svoboda, N. B. Ram, D. Baykusheva, D. Zindel, M. D. J. Waters, B. Spenger, M. Ochsner, H. Herburger, J. Stohner, and H. J. Wörner, "Femtosecond photoelectron circular dichroism of chemical reactions," Sci. Adv. **8**, eabq2811 (2023).

544. A. G. Harvey, Z. Mašín, and O. Smirnova, "General theory of photoexcitation induced photoelectron circular dichroism," J. Chem. Phys. **149**, 064104 (2018).

545. S. Beaulieu, A. Comby, D. Descamps, B. Fabre, G. A. Garcia, R. Géneaux, A. G. Harvey, F. Légaré, Z. Mašín, and L. Nahon, "Photoexcitation circular dichroism in chiral molecules," Nat. Phys. **14**, 484–489 (2018).

546. K. Fehre, S. Eckart, M. Kunitski, C. Janke, D. Trabert, M. Hofmann, J. Rist, M. Weller, A. Hartung, L. P. H. Schmidt, T. Jahnke, H. Braun, T. Baumert, J. Stohner, P. V. Demekhin, M. S. Schöffler, and R. Dörner, "Strong Differential Photoion Circular Dichroism in Strong-Field Ionization of Chiral Molecules," Phys. Rev. Lett. **126**, 83201 (2021).

547. K. Fehre, S. Eckart, M. Kunitski, M. Pitzer, S. Zeller, C. Janke, D. Trabert, J. Rist, M. Weller, A. Hartung, L. P. H. Schmidt, T. Jahnke, R. Berger, R. Dörner, and M. S. Schöffler, "Enantioselective fragmentation of an achiral molecule in a strong laser field," Sci. Adv. **5**, eaau7923 (2023).

548. C. S. Lehmann and K.-M. Weitzel, "Coincident measurement of photo-ion circular dichroism and photo-electron circular dichroism," Phys. Chem. Chem. Phys. **22**, 13707–13712 (2020).

549. A. Geyer, O. Neufeld, D. Trabert, U. De Giovannini, M. Hofmann, N. Anders, L. Sarkadi, M. S. Schöffler, L. P. H. Schmidt, A. Rubio, T. Jahnke, M. Kunitski, and S. Eckart, "Quantum correlation of electron and ion energy in the dissociative strong-field ionization of H2," Phys. Rev. Res. **5**, 13123 (2023).

550. M. Pitzer, M. Kunitski, A. S. Johnson, T. Jahnke, H. Sann, F. Sturm, L. P. H. Schmidt, H. Schmidt-Böcking, R. Dörner, J. Stohner, J. Kiedrowski, M. Reggelin, S. Marquardt, A. Schießer, R. Berger, and M. S. Schöffler, "Direct Determination of Absolute Molecular Stereochemistry in Gas Phase by Coulomb Explosion Imaging," Science **341**, 1096–1100 (2013).

551. P. Herwig, K. Zawatzky, M. Grieser, O. Heber, B. Jordon-Thaden, C. Krantz, O. O. rich Novotný, R. Repnow, V. Schurig, D. Schwalm, Z. Vager, A. Wolf, O. Trapp, and H. Kreckel, "Imaging the Absolute Configuration of a Chiral Epoxide in the Gas Phase," Science **342**, 1084–1086 (2013).

552. D. Patterson and J. M. Doyle, "Sensitive Chiral Analysis via Microwave Three-Wave Mixing," Phys. Rev. Lett. **111**, 23008 (2013).

553. M. A. Belkin and Y. R. Shen, "Non-linear optical spectroscopy as a novel probe for molecular chirality," Int. Rev. Phys. Chem. **24**, 257–299 (2005).

554. P. Fischer, D. S. Wiersma, R. Righini, B. Champagne, and A. D. Buckingham, "Three-Wave Mixing in Chiral Liquids," Phys. Rev. Lett. **85**, 4253–4256 (2000).

555. L. M. Haupert and G. J. Simpson, "Chirality in Nonlinear Optics," (2009).

556. S. Eibenberger, J. Doyle, and D. Patterson, "Enantiomer-Specific State Transfer of Chiral Molecules," Phys. Rev. Lett. **118**, 123002 (2017).

557. K. Hiramatsu, H. Kano, and T. Nagata, "Raman optical activity by coherent anti-Stokes Raman scattering spectral interferometry," Opt. Express **21**, 13515–13521 (2013).

558. L. Ohnoutek, H.-H. Jeong, R. R. Jones, J. Sachs, B. J. Olohan, D.-M. Răsădean, G. D. Pantoş, D. L. Andrews, P. Fischer, and V. K. Valev, "Optical Activity in Third-Harmonic Rayleigh Scattering: A New Route for Measuring Chirality," Laser Photon. Rev. **15**, 2100235 (2021).

559. D. Ayuso, A. F. Ordonez, P. Decleva, M. Ivanov, and O. Smirnova, "Strong chiral response in non-collinear high harmonic generation driven by purely electric-dipole interactions," Opt. Express **30**, 4659–4667 (2022).

560. A. Ordóñez, P. Vindel-Zandbergen, and D. Ayuso, "Chiral coherent control of electronic population transfer: towards all-optical and highly enantioselective photochemistry," arXiv Prepr. arXiv2309.02392 (2023).

561. L. Rego and D. Ayuso, "Structuring the local handedness of synthetic chiral light: global chirality versus polarization of chirality," New J. Phys. **25**, 93005 (2023).

562. O. Neufeld, O. Wengrowicz, O. Peleg, A. Rubio, and O. Cohen, "Detecting multiple chiral centers in chiral molecules with high harmonic generation," Opt. Express **30**, 3729–3740 (2022).

563. M. Khokhlova, E. Pisanty, S. Patchkovskii, O. Smirnova, and M. Ivanov, "Enantiosensitive steering of free-induction decay," Sci. Adv. **8**, eabq1962 (2023).

564. G. P. Katsoulis, Z. Dube, P. B. Corkum, A. Staudte, and A. Emmanouilidou, "Momentum scalar triple product as a measure of chirality in electron ionization dynamics of strongly driven atoms," Phys. Rev. A **106**, 43109 (2022).

565. M. Shapiro, E. Frishman, and P. Brumer, "Coherently controlled asymmetric synthesis with achiral light.," Phys. Rev. Lett. **84**, 1669–1672 (2000).

566. E. Frishman, M. Shapiro, D. Gerbasi, and P. Brumer, "Enantiomeric purification of nonpolarized racemic mixtures using coherent light," J. Chem. Phys. **119**, 7237–7246 (2003).

567. P. Král, I. Thanopulos, M. Shapiro, and D. Cohen, "Two-Step Enantio-Selective Optical Switch," Phys. Rev. Lett. **90**, 033001 (2003).

568. D. Gerbasi, M. Shapiro, and P. Brumer, "Theory of "laser distillation" of enantiomers: Purification of a racemic mixture of randomly oriented dimethylallene in a collisional environment," J. Chem. Phys. **124**, 74315 (2006).

569. D. Gerbasi, P. Brumer, I. Thanopulos, P. Král, and M. Shapiro, "Theory of the two step enantiomeric purification of 1,3 dimethylallene," J. Chem. Phys. **120**, 11557–11563 (2004).

570. X. Li and M. Shapiro, "Theory of the optical spatial separation of racemic mixtures of chiral molecules," J. Chem. Phys. **132**, 194315 (2010).

571. M. Shapiro, E. Frishman, and P. Brumer, "Erratum: Coherently Controlled Asymmetric Synthesis with Achiral Light [Phys. Rev. Lett. 84, 001669 (2000)]," Phys. Rev. Lett. **91**, 129902 (2003).

572. K. A. Forbes and D. Green, "Enantioselective optical gradient forces using 3D structured vortex light," Opt. Commun. **515**, 128197 (2022).

573. R. Ali, F. A. Pinheiro, R. S. Dutra, F. S. S. Rosa, and P. A. Maia Neto, "Enantioselective manipulation of single chiral nanoparticles using optical tweezers," Nanoscale **12**, 5031–5037 (2020).

574. J. Lee, J. Bischoff, A. O. Hernandez-Castillo, B. Sartakov, G. Meijer, and S. Eibenberger-Arias, "Quantitative Study of Enantiomer-Specific State Transfer," Phys. Rev. Lett. **128**, 173001 (2022).

575. J.-L. Wu, Y. Wang, J.-X. Han, C. Wang, S.-L. Su, Y. Xia, Y. Jiang, and J. Song, "Two-Path Interference for Enantiomer-Selective State Transfer of Chiral Molecules," Phys. Rev. Appl. **13**, 44021 (2020).

576. C. Pérez, A. L. Steber, S. R. Domingos, A. Krin, D. Schmitz, and M. Schnell, "Coherent Enantiomer-Selective Population Enrichment Using Tailored Microwave Fields," Angew. Chemie Int. Ed. **56**, 12512–12517 (2017).

577. B. T. Torosov, M. Drewsen, and N. V Vitanov, "Efficient and robust chiral resolution by composite pulses," Phys. Rev. A **101**, 63401 (2020).

578. J. Lee, E. Abdiha, B. G. Sartakov, G. Meijer, and S. Eibenberger-Arias, "Near-complete chiral selection in rotational quantum states," Nat. Commun. **15**, 7441 (2024).

579. Y.-Y. Chen, J.-J. Cheng, C. Ye, and Y. Li, "Enantiodetection of cyclic three-level chiral molecules in a driven cavity," Phys. Rev. Res. **4**, 13100 (2022).

580. Y.-Y. Chen, C. Ye, and Y. Li, "Enantio-detection via cavity-assisted three-photon processes," Opt. Express **29**, 36132–36144 (2021).

581. C. Ye, Q. Zhang, Y.-Y. Chen, and Y. Li, "Determination of enantiomeric excess with chirality-dependent ac Stark effects in cyclic three-level models," Phys. Rev. A **100**, 33411 (2019).

582. B. A. Stickler, M. Diekmann, R. Berger, and D. Wang, "Enantiomer Superpositions from Matter-Wave Interference of Chiral Molecules," Phys. Rev. X **11**, 31056 (2021).

583. N. V Vitanov and M. Drewsen, "Highly Efficient Detection and Separation of Chiral Molecules through Shortcuts to Adiabaticity," Phys. Rev. Lett. **122**, 173202 (2019).

584. B. Liu, C. Ye, C. P. Sun, and Y. Li, "Spatial enantioseparation of gaseous chiral molecules," Phys. Rev. A **104**, 13113 (2021).

585. M. Leibscher, T. F. Giesen, and C. P. Koch, "Principles of enantio-selective excitation in three-wave mixing spectroscopy of chiral molecules," J. Chem. Phys. **151**, 14302 (2019).

586. Y.-Y. Chen, C. Ye, Q. Zhang, and Y. Li, "Enantio-discrimination via light deflection effect," J. Chem. Phys. **152**, 204305 (2020).

587. R. P. Cameron, D. McArthur, and A. M. Yao, "Strong chiral optical force for small chiral molecules based on electric-dipole interactions, inspired by the asymmetrical hydrozoan Velella velella," New J. Phys. **25**, 83006 (2023).

588. Z.-Y. Chen and R. Qin, "Probing structural chirality of crystals using high-order harmonic generation in solids," Phys. Rev. A **101**, 53423 (2020).

589. D. Wang, X. Zhu, X. Liu, L. Li, X. Zhang, P. Lan, and P. Lu, "High harmonic generation from axial chiral molecules," Opt. Express **25**, 23502 (2017).

590. S. Giri, A. M. Dudzinski, J. C. Tremblay, and G. Dixit, "Time-dependent electronic current densities in chiral molecules," Phys. Rev. A **102**, 63103 (2020).

591. S. Giri, J. C. Tremblay, and G. Dixit, "Imaging charge migration in chiral molecules using time-resolved x-ray diffraction," Phys. Rev. A **104**, 53115 (2021).

592. T. Heine, C. Corminboeuf, and G. Seifert, "The Magnetic Shielding Function of Molecules and Pi-Electron Delocalization," Chem. Rev. **105**, 3889–3910 (2005).

593. I. Barth and M. Lein, "Numerical verification of the theory of nonadiabatic tunnel ionization in strong circularly polarized laser fields," J. Phys. B At. Mol. Opt. Phys. **47**, 204016 (2014).

594. R. Gershoni-Poranne and A. Stanger, "The NICS-XY-Scan: Identification of Local and Global Ring Currents in Multi-Ring Systems," Chem. – A Eur. J. **20**, 5673–5688 (2014).

595. J. R. Rouxel, M. Kowalewski, and S. Mukamel, "Photoinduced molecular chirality probed by ultrafast resonant X-ray spectroscopy," Struct. Dyn. **4**, 44006 (2017).

596. T. Moitra, L. Konecny, M. Kadek, O. Neufeld, A. Rubio, and M. Repisky, "Light-induced persistent electronic chirality in achiral molecules probed with transient absorption circular dichroism spectroscopy," arXiv:2503.16986 (2025).

597. N. Mayer, S. Patchkovskii, F. Morales, M. Ivanov, and O. Smirnova, "Imprinting Chirality on Atoms Using Synthetic Chiral Light Fields," Phys. Rev. Lett. **129**, 243201 (2022).

598. Á. Jiménez-Galán, R. E. F. Silva, O. Smirnova, and M. Ivanov, "Sub-cycle valleytronics: control of valley polarization using few-cycle linearly polarized pulses," Optica **8**, 277–280 (2021).

599. F. Siegrist, J. A. Gessner, M. Ossiander, C. Denker, Y.-P. Chang, M. C. Schröder, A. Guggenmos, Y. Cui, J. Walowski, U. Martens, J. K. Dewhurst, U. Kleineberg, M. Münzenberg, S. Sharma, and M. Schultze, "Light-wave dynamic control of magnetism," Nature **571**, 240–244 (2019).

600. P. T. Probst, M. Mayer, V. Gupta, A. M. Steiner, Z. Zhou, G. K. Auernhammer, T. A. F. König, and A. Fery, "Mechano-tunable chiral metasurfaces via colloidal assembly," Nat. Mater. **20**, 1024–1028 (2021).

601. Z. Han, F. Wang, J. Sun, X. Wang, and Z. Tang, "Recent Advances in Ultrathin Chiral Metasurfaces by Twisted Stacking," Adv. Mater. **35**, 2206141 (2023).

602. M. Cen, J. Wang, J. Liu, H. He, K. Li, W. Cai, T. Cao, and Y. J. Liu, "Ultrathin Suspended Chiral Metasurfaces for Enantiodiscrimination," Adv. Mater. **34**, 2203956 (2022).

603. J. Kim, A. S. Rana, Y. Kim, I. Kim, T. Badloe, M. Zubair, M. Q. Mehmood, and J. Rho, "Chiroptical Metasurfaces: Principles, Classification, and Applications," Sensors **21**, (2021).

604. M. L. Solomon, J. Hu, M. Lawrence, A. García-Etxarri, and J. A. Dionne, "Enantiospecific Optical Enhancement of Chiral Sensing and Separation with Dielectric Metasurfaces," ACS Photonics **6**, 43–49 (2019).

605. Y.-P. Jia, Y.-L. Zhang, X.-Z. Dong, M.-L. Zheng, J. Li, J. Liu, Z.-S. Zhao, and X.-M. Duan, "Complementary chiral metasurface with strong broadband optical activity and enhanced transmission," Appl. Phys. Lett. **104**, 11108 (2014).

606. Z. Wang, B. H. Teh, Y. Wang, G. Adamo, J. Teng, and H. Sun, "Enhancing circular dichroism by super chiral hot spots from a chiral metasurface with apexes," Appl. Phys. Lett. **110**, 221108 (2017).

607. D. Beutel, P. Scott, M. Wegener, C. Rockstuhl, and I. Fernandez-Corbaton, "Enhancing the optical rotation of chiral molecules using helicity preserving all-dielectric metasurfaces," Appl. Phys. Lett. **118**, 221108 (2021).

608. S. Droulias and L. Bougas, "Absolute Chiral Sensing in Dielectric Metasurfaces Using Signal Reversals," Nano Lett. **20**, 5960–5966 (2020).

609. Y. Lim, I. C. Seo, S.-C. An, Y. Kim, C. Park, B. H. Woo, S. Kim, H.-R. Park, and Y. C. Jun, "Maximally Chiral Emission via Chiral Quasibound States in the Continuum," Laser Photon. Rev. **17**, 2200611 (2023).

610. J. Ni, S. Liu, G. Hu, Y. Hu, Z. Lao, J. Li, Q. Zhang, D. Wu, S. Dong, J. Chu, and C.-W. Qiu, "Giant Helical Dichroism of Single Chiral Nanostructures with Photonic Orbital Angular Momentum," ACS Nano **15**, 2893–2900 (2021).

611. F. de Juan, A. G. Grushin, T. Morimoto, and J. E. Moore, "Quantized circular photogalvanic effect in Weyl semimetals," Nat. Commun. **8**, 15995 (2017).

612. E. J. König, H.-Y. Xie, D. A. Pesin, and A. Levchenko, "Photogalvanic effect in Weyl semimetals," Phys. Rev. B **96**, 75123 (2017).

613. C.-K. Chan, N. H. Lindner, G. Refael, and P. A. Lee, "Photocurrents in Weyl semimetals," Phys. Rev. B **95**, 41104 (2017).

614. N. Nagaosa, T. Morimoto, and Y. Tokura, "Transport, magnetic and optical properties of Weyl materials," Nat. Rev. Mater. **5**, 621–636 (2020).

615. C. Le and Y. Sun, "Topology and symmetry of circular photogalvanic effect in the chiral multifold semimetals: a review," J. Phys. Condens. Matter **33**, 503003 (2021).

616. Z. Ni, K. Wang, Y. Zhang, O. Pozo, B. Xu, X. Han, K. Manna, J. Paglione, C. Felser, A. G. Grushin, F. de Juan, E. J. Mele, and L. Wu, "Giant topological longitudinal circular photo-galvanic effect in the chiral multifold semimetal CoSi," Nat. Commun. **12**, 154 (2021).

617. S. Körber, L. Privitera, J. C. Budich, and B. Trauzettel, "Interacting topological frequency converter," Phys. Rev. Res. **2**, 22023 (2020).

618. F. Nathan, I. Martin, and G. Refael, "Topological frequency conversion in a driven dissipative quantum cavity," Phys. Rev. B **99**, 94311 (2019).

619. S. V. B. Jensen, N. Tancogne-Dejean, A. Rubio, and L. B. Madsen, "Beyond Electric-Dipole Treatment of Light-Matter Interactions in Materials: Nondipole Harmonic Generation in Bulk Si," arXiv Prepr. arXiv2410.18547 (2024).

620. A. D. Bandrauk and H. Z. Lu, "Molecules in intense laser fields: Beyond the dipole approximation," Phys. Rev. A **73**, 13412 (2006).

621. A. D. Bandrauk, F. Fillion-Gourdeau, and E. Lorin, "Atoms and molecules in intense laser fields: gauge invariance of theory and models," J. Phys. B At. Mol. Opt. Phys. **46**, 153001 (2013).

622. M. F. Ciappina, J. Biegert, R. Quidant, and M. Lewenstein, "High-order-harmonic generation from inhomogeneous fields," Phys. Rev. A **85**, 33828 (2012).

623. M. Kozlov, O. Kfir, A. Fleischer, A. Kaplan, T. Carmon, H. G. L. Schwefel, G. Bartal, and O. Cohen, "Narrow-bandwidth high-order harmonics driven by long-duration hot spots," New J. Phys. **14**, 63036 (2012).

624. D. B. Miloševič and W. Becker, "Relativistic high-order harmonic generation," J. Mod. Opt. **50**, 375–386 (2003).

625. T. Higuchi, C. Heide, K. Ullmann, H. B. Weber, and P. Hommelhoff, "Light-field-driven currents in graphene," Nature **550**, 224–228 (2017).

626. A. Schiffrin, T. Paasch-Colberg, N. Karpowicz, V. Apalkov, D. Gerster, S. Mühlbrandt, M. Korbman, J. Reichert, M. Schultze, S. Holzner, J. V Barth, R. Kienberger, R. Ernstorfer, V. S. Yakovlev, M. I. Stockman, and F. Krausz, "Optical-field-induced current in dielectrics," Nature **493**, 70–74 (2013).

627. A. Ali, C. Wang, J. Cai, and K. J. Karki, "Probing Silicon Carbide with Phase-Modulated Femtosecond Laser Pulses: Insights into Multiphoton Photocurrent," ACS Photonics **11**, 1502–1507 (2024).

628. M. Hofherr, S. Häuser, J. K. Dewhurst, P. Tengdin, S. Sakshath, H. T. Nembach, S. T. Weber, J. M. Shaw, T. J. Silva, H. C. Kapteyn, M. Cinchetti, B. Rethfeld, M. M. Murnane, D. Steil, B. Stadtmüller, S. Sharma, M. Aeschlimann, and S. Mathias, "Ultrafast optically induced spin transfer in ferromagnetic alloys," Sci. Adv. **6**, eaay8717 (2024).

629. N. M. Allafi, M. H. Kolodrubetz, M. Bukov, V. Oganesyan, and M. Yarmohammadi, "Spin high harmonic generation through terahertz laser-driven phonons," Phys. Rev. B **110**, 64420 (2024).

630. G. Wang, C. Li, and P. Cappellaro, "Observation of Symmetry-Protected Selection Rules in Periodically Driven Quantum Systems," Phys. Rev. Lett. **127**, 140604 (2021).

631. L. Cruz-Rodriguez, D. Dey, A. Freibert, and P. Stammer, "Quantum phenomena in attosecond science," Nat. Rev. Phys. **6**, 691–704 (2024).

632. M. Even Tzur, M. Birk, A. Gorlach, M. Krüger, I. Kaminer, and O. Cohen, "Photon-statistics force in ultrafast electron dynamics," Nat. Photonics **17**, 501–509 (2023).

633. M. Lewenstein, M. F. Ciappina, E. Pisanty, J. Rivera-Dean, P. Stammer, T. Lamprou, and P. Tzallas, "Generation of optical Schrödinger cat states in intense laser–matter interactions," Nat. Phys. **17**, 1104–1108 (2021).

634. P. Stammer, J. Rivera-Dean, A. Maxwell, T. Lamprou, A. Ordóñez, M. F. Ciappina, P. Tzallas, and M. Lewenstein, "Quantum Electrodynamics of Intense Laser-Matter Interactions: A Tool for Quantum State Engineering," PRX Quantum **4**, 10201 (2023).

635. A. Rasputnyi, Z. Chen, M. Birk, O. Cohen, I. Kaminer, M. Krüger, D. Seletskiy, M. Chekhova, and F. Tani, "High-harmonic generation by a bright squeezed vacuum," Nat. Phys. **20**, 1960–1965 (2024).

636. M. E. Tzur, M. Birk, A. Gorlach, I. Kaminer, M. Krüger, and O. Cohen, "Generation of squeezed high-order harmonics," Phys. Rev. Res. **6**, 33079 (2024).

637. T. Zahavy, A. Dikopoltsev, D. Moss, G. I. Haham, O. Cohen, S. Mannor, and M. Segev, "Deep learning reconstruction of ultrashort pulses," Optica **5**, 666–673 (2018).

638. T. Pfeifer, M. Wollenhaupt, and M. Lein, "Ultrafast artificial intelligence: machine learning with atomic-scale quantum systems," New J. Phys. **26**, 93018 (2024).

639. N. D. Klimkin, Á. Jiménez-Galán, R. E. F. Silva, and M. Ivanov, "Symmetry-aware deep neural networks for high harmonic spectroscopy in solids," Opt. Express **31**, 20559–20571 (2023).

640. S. Sun, L. He, C. Xu, Y. Deng, P. Lan, and P. Lu, "Double-blind decoupling of molecular rotation and high-order harmonic generation with a neural network," Phys. Rev. A **109**, 33105 (2024).

641. J.-Z. Yan, S.-S. Zhao, W.-D. Lan, S.-Y. Li, S.-S. Zhou, J.-G. Chen, J.-Y. Zhang, and Y.-J. Yang, "Calculation of high-order harmonic generation of atoms and molecules by combining time series prediction and neural networks," Opt. Express **30**, 35444–35456 (2022).

642. "Dynamical Symmetries and Selection Rules in Nonlinear Optics - Webpage Community Resource," https://oren.net.technion.ac.il/database-of-dynamical-symmetries-and-selection-rules-in-nonlinear-optics/.